\documentclass[traditabstract]{aa}
\usepackage{graphicx}
\usepackage{txfonts}
 \begin{document}

%
%
 \newcommand{\nc}{\newcommand}

 \nc{\bea}{\begin{eqnarray}}
 \nc{\eea}{\end{eqnarray}}
 \nc{\beq}{\begin{equation}}
 \nc{\eeq}{\end{equation}}
 \nc{\nn}{\nonumber}
 \nc{\ve}[1]{{\bf #1}}
 \nc{\bP}{\mathbf{P}}
 \nc{\bM}{\mathbf{M}}
 \nc{\bMi}{{\mathbf{M}^{-1}}}
 \nc{\bF}{\mathbf{F}}
 \nc{\bZ}{\mathbf{Z}}
 \nc{\bI}{\mathbf{I}}
 \nc{\bA}{\mathbf{A}}
 \nc{\bR}{\mathbf{R}}
 \nc{\bC}{\mathbf{C}}
 \nc{\Cw}{{\mathbf{C}_w}}
 \nc{\Cwi}{{\mathbf{C}_w^{-1}}}
 \nc{\bD}{\mathbf{D}}
 \nc{\bDi}{{\mathbf{D}^{-1}}}
 \nc{\bB}{\mathbf{B}}
 \nc{\by}{\mathbf{y}}
 \nc{\bs}{\mathbf{s}}
 \nc{\bn}{\mathbf{n}}
 \nc{\bnc}{{\mathbf{n}_c}}
 \nc{\bd}{\mathbf{d}}
 \nc{\bm}{\mathbf{m}}
 \nc{\bmin}{{\mathbf{m}_\mathrm{in}}}
 \nc{\mout}{{\mathbf{m}_\mathrm{out}}}
 \nc{\bw}{\mathbf{w}}
 \nc{\ba}{\mathbf{a}}
 \nc{\ain}{{\mathbf{a}_\mathrm{in}}}
 \nc{\aout}{{\mathbf{a}_\mathrm{out}}}
 \nc{\aref}{{\mathbf{a}_r}}
 \nc{\nbase}{n_\mathrm{base}}
 \nc{\tbase}{t_\mathrm{base}}
 \nc{\tgen}{t_\mathrm{gen}}
 \nc{\fk}{f_k}
 \nc{\fc}{f_c}
 \nc{\fmin}{f_\mathrm{min}}
 \nc{\fgen}{f_\mathrm{gen}}
 \nc{\fsample}{f_\mathrm{sample}}
 \nc{\tsample}{t_\mathrm{sample}}
 \nc{\thetas}{\theta_s}
 \nc{\fspin}{f_\mathrm{spin}}
 \nc{\etal}{{et al.}}
 \nc{\half}{{\textstyle \frac12}}
 \nc{\Nside}{N_\mathrm{side}}
 \nc{\nhit}{n_\mathrm{hit}}
 \nc{\nhitp}{n_{\mathrm{hit},p}}
 \nc{\diag}{\mbox{diag}}


\title{Destriping CMB temperature and polarization maps}

\author{H. Kurki-Suonio\inst{1,2}, E. Keih\"{a}nen\inst{1}, R. Keskitalo\inst{1,2},
T. Poutanen\inst{2,1,3}, A.-S. Sirvi\"{o}\inst{1}, D. Maino\inst{4},
C. Burigana\inst{5,6}}

\offprints{H.\ Kurki-Suonio, \email{hannu.kurki-suonio@helsinki.fi}}

\institute{%
 University of Helsinki, Department of Physics,
 P.O. Box 64, FIN-00014 Helsinki, Finland
 \and
 Helsinki Institute of Physics,
 P.O. Box 64, FIN-00014 Helsinki, Finland
 \and
 Mets\"{a}hovi Radio Observatory, Helsinki University of Technology,
 Mets\"{a}hovintie 114, FIN-02540 Kylm\"{a}l\"{a}, Finland
 \and
 Dipartimento di Fisica, Universit\'a di Milano,
 Via Celoria 16, I-20131, Milano, Italy
 \and
 INAF/IASF-BO, Istituto di Astrofisica Spaziale e Fisica
Cosmica di Bologna,
    Via Gobetti 101, I-40129, Bologna, Italy
    \and Dipartimento di Fisica, Universit\`a degli Studi di Ferrara,
Via Saragat 1, I-44100 Ferrara, Italy
 }

\date{April 21, 2009}

\abstract{We study destriping as a map-making method for
temperature-and-polarization data for cosmic microwave background
observations. We present a particular implementation of destriping
and study the residual error in output maps, using simulated data
corresponding to the 70 GHz channel of the {\sc Planck} satellite,
but assuming idealized detector and beam properties. The relevant
residual map is the difference between the output map and a binned
map obtained from the signal $+$ white noise part of the data
stream. For destriping it can be divided into six components:
unmodeled correlated noise, white noise reference baselines,
reference baselines of the pixelization noise from the signal, and
baseline errors from correlated noise, white noise, and signal.
These six components contribute differently to the different angular
scales in the maps.  We derive analytical results for the first
three components. This study is related to {\sc Planck} LFI
activities.

 }

\authorrunning{Kurki-Suonio et al.}
\maketitle


\section{Introduction}
\label{sec:intro}

Construction of sky maps from the time-ordered data (TOD) is an
important part of the data analysis of cosmic microwave background
(CMB) surveys.  For large surveys like {\sc Planck}\footnote{\tt
http://www.rssd.esa.int/index.php?project=PLANCK} (Planck
Collaboration \cite{Bluebook}), this is a computationally demanding
task.  Methods which aim at finding the optimal minimum-variance map
(Wright \cite{Wri96}, Borrill \cite{Bor99}, Dor\'{e} et
al.~\cite{Dor01}, Natoli et al.~\cite{Nat01}, Yvon \& Mayet
\cite{Yvo05}, de Gasperis et al.~\cite{deG05}) are computationally
heavy and require large computers.  Also, a faster method is needed
for Monte Carlo studies to assess systematic effects, noise biases,
and error estimates.

Destriping (Burigana et al.~\cite{Burigana99}, Delabrouille
\cite{Delabrouille98}, Maino et al.~\cite{Maino99, Maino02}, Revenu
et al.~\cite{Revenu00}, Sbarra et al.~\cite{Sba03}, Poutanen et
al.~\cite{Pou04}, Keih\"{a}nen et al.~\cite{Keihanen04}) is a fast
map-making method that removes correlated low-frequency noise from
the TOD utilizing crossing points, i.e., the same locations on the
sky observed at different times. Correlated noise is modeled as a
sequence of (``uniform'') baselines, i.e., constant offsets in the
TOD. High-frequency noise (frequency of the same order or higher
than the inverse of the baseline length) cannot be modeled this way.
Thus the method assumes that the high-frequency part of the noise is
uncorrelated (white noise).

In some implementations, a set of base functions (e.g. low order
Legendre polynomials) is used instead of just the uniform baseline
(Delabrouille \cite{Delabrouille98}, Maino et al.~\cite{Maino02},
Keih\"{a}nen et al.~\cite{Keihanen04,Madam05}), or a spline is
fitted to the TOD (Ganga \cite{Ganga94}).

In this paper we describe one destriping implementation for making
temperature and polarization maps of the sky and study the residual
errors in the maps. This implementation was originally known as the
``Polar'' code, and used in the map-making comparison studies of the
{\sc Planck} CTP Working Group (Poutanen et al. \cite{Cambridge},
Ashdown et al. \cite{Helsinki,Paris,Trieste}). Polar has now been
merged into the ``Madam'' destriping code. The novel feature in
Madam was the introduction of an optional noise prior (noise filter)
that utilizes prior information on the noise power spectrum
(Keih\"{a}nen et al.~\cite{Madam05}). Polar corresponds to Madam
with the noise prior turned off.  The results presented in this
paper were obtained with the Madam code, with the noise prior turned
off. We briefly comment on the effect of the noise prior in
Sect.~\ref{sec:noise_prior}. The use and effect of the noise prior
will be described in detail in Keih\"{a}nen et al.~(\cite{Madam09}).

Destriping errors have been previously analyzed by Stompor \& White
(\cite{Sto04}) and Efstathiou (\cite{E05,E07}).

The TOD can be considered as a sum of signal $+$ white noise $+$
correlated (``$1/f$'') noise. If there were no correlated noise, the
optimal way to produce a map would be a simple binning of the TOD
samples onto map pixels. (We do not address here the question of
correcting for the effect of the instrument beam. ``Deconvolution''
map-making methods that correct for the effect of the beam shape
have been developed (Burigana \& Sa\'{e}z \cite{Bur03}, Armitage \&
Wandelt \cite{Arm04}, Harrison et al.~\cite{Har08}), but tend to be
computationally very resource intensive. They also alter the noise
properties of the maps in a way that is difficult to follow in CMB
angular power spectrum estimation.) Thus the task of a map-making
method is to remove the correlated noise as well as possible, with
as little effect on the signal and white noise as possible.  The
difference of the output map from the binned signal $+$ white noise
map is thus the residual map to consider to judge the quality of the
output map. We divide this residual into six components: unmodeled
$1/f$ noise,
 $1/f$ baseline error, white noise reference baselines,
white noise baseline error, pixelization noise reference baselines,
and pixelization noise baseline error. We study the nature of each
component, and its dependence on the baseline length.

We have used simulated data corresponding to 1 year of observations
with 4 {\sc Planck} LFI 70 GHz detectors (two horns, each with two
orthogonally polarized detectors).

For simplicity, we did not include foregrounds in the signal (see
Ashdown et al.~\cite{Paris,Trieste} for effect of foreground signal)
or such systematic effects as beam asymmetries, sample integration,
cooler noise, or pointing errors (see Ashdown et
al.~\cite{Trieste}).  Even in Ashdown et al.~(\cite{Trieste}) the
simulated data used was still fairly idealized.  We are currently
working on more realistic simulations.

In Sect.~\ref{sec:choices} we discuss some early-stage design
choices made in the development of our map-making method.
Sect.~\ref{sec:method} contains the derivation and description of
the method. Sect.~\ref{sec:simulation} describes the simulated data
used to test the method. In Sect.~\ref{sec:time_domain} we analyze
residual errors in the time domain, and in
Sect.~\ref{sec:map_domain} in the map domain. In
Sect.~\ref{sec:knee_frequency} we discuss the effect of the noise
knee frequency, and in Sect.~\ref{sec:noise_prior} we give a preview
of results obtained when a noise prior is added to the method.  We
mainly consider maps made from a full year of data, but in
Sect.~\ref{sec:partial_surveys} we discuss maps made from shorter
time segments. In Sect.~\ref{sec:conclusion} we summarize our
conclusions.

\section{Design choices}
\label{sec:choices}

\subsection{Ring set or not}

For a {\sc Planck}-like scanning strategy, where the detectors scan
the same circle on the sky many times before the spin axis of the
satellite is repointed, an intermediate data structure can be
introduced between the TOD and the frequency map.  The circles from
one repointing period can be coadded to a ring, i.e., averaged to
appear just as a single sweep of the circle.  In this context it is
natural to choose one baseline per ring.  Destriping is then
performed on this ring set, instead of the original uncoadded TOD.
This reduces the memory and computing time requirements by a large
factor.

If the scanning is \emph{ideal}, i.e., the observations (samples)
from the different circles of the same ring fall on exactly the same
locations on the sky, destriping coadded rings is equivalent to
destriping the uncoadded TOD with baseline length equal to the
repointing period (i.e., one baseline per ring). In this case it is
also almost equal (for map-making purposes) to destriping the
uncoadded TOD with baseline length equal to the spin period (i.e.
one baseline per circle), see Sect.~\ref{sec:map_unmodeled}.

In reality, the spin axis will nutate with some small amplitude, so
that the different circles will not scan exactly the same path on
the sky.  The spin rate is also not exactly constant, and the
detector sampling frequency is not synchronized with the spin.
Binning the samples first into a ring (``phase binning'') (van
Leeuwen et al.~\cite{vanL02}) and then repixelizing the ring pixels
into map pixels after destriping may then introduce some extra
smoothing of the data.

We have chosen to sidestep this intermediate structure and to assign
the samples directly to map pixels. The baseline length is then not
necessarily tied to scan circles and rings, and also data taken
during the repointing maneuvers can be used.  For short baselines no
data compression is possible and map-making is done from the full
TOD, requiring large computer memory. For long baselines (many scan
circles) the data can be compressed by binning observations directly
to map pixels.  For baselines one repointing period (one ring) long
a similar data compression is achieved as by using the intermediate
ring set structure.  Whereas the phase-binned ring set is still
closely connected to the time domain, our ``pixel binning'' destroys
the time-ordered structure of the ring, and therefore works only
with uniform baselines.

This way we have achieved a versatile destriping method, where the
baseline length is an adjustable parameter.  Shorter baselines can
be used in large computers for higher accuracy, whereas longer
baselines require less memory and computing time and can be used in
medium-sized computers and for Monte Carlo studies.  The baseline
length is not tied to the scanning strategy, and our destriping
method can be applied to any scanning strategy that has crossing
points, not just to a {\sc Planck}-like scanning strategy.

However, for a {\sc Planck}-like scanning strategy there is a
certain advantage in choosing the baseline length so that an integer
number of baselines fits to one repointing period. Baseline segments
that extend to two different repointing periods are avoided this
way. This is mainly an issue for long baselines (not very much
shorter than the repointing period).  For baselines shorter than the
spin period there seems to be some advantage in choosing the
baseline length so that an integer number of baselines fits to one
spin period.  See Keih\"{a}nen et al.~(\cite{Madam09}).  In this
paper we only consider such choices for baseline length.

\subsection{Crossing points and signal error}

The baselines are estimated from \emph{crossing points}, i.e.,
observations falling on the same map pixels at different times.
Samples are assigned to pixels based on the pointing of the detector
beam center.  The beam center may still point at a different
location within the same map pixel for different samples.  We do not
attempt to correct for this effect and this results in a ``signal
error'' in our output maps. The signal error due to in-pixel
differences in beam pointing could be largely eliminated in another
kind of destriping implementation, where the scanning circles are
treated as exact geometrical curves (instead of just a sequence of
map pixels), and the observations are interpolated to the exact
crossing points of these lines (Revenu et al.~\cite{Revenu00}). In
this case only actual crossings of the scan circles contribute to
baseline determination, whereas in our implementation it is enough
that two paths pass through the same pixel without actually crossing
there.  The latter situation is very common, since successive
circles are almost parallel.

However, in a realistic situation there are other contributions to
signal error that could not be eliminated this way.  One such
contribution is the different beam orientations of the different
observations of the crossing point, as real beams are not circularly
symmetric.  In Ashdown et al.~(\cite{Trieste}) elliptic beams were
considered for the {\sc Planck} 30 GHz channel and it was found that
this had a contribution to the signal error, which was of comparable
size or larger.


\section{Destriping technique}
\label{sec:method}

\subsection{Derivation}

The destriping method can be derived from a maximum-likelihood
analysis of an idealized model of observations.  The signal observed
by a detector sensitive to one linear polarization direction is
proportional to
 \beq
    s = I + Q\cos2\psi + U\sin2\psi \,,
 \label{eq:IQU}
 \eeq
where $\psi$ is the polarization angle of the detector, and $I$,
$Q$, and $U$ are the Stokes parameters of the radiation coming from
the observation direction.  In the idealized model the time-ordered
data (TOD), a vector $\by$ consisting of $n_t$ samples, is
 \beq
    \by = \bP\bmin+\bn
 \label{eq:pointing}
 \eeq
where $\bmin$ (the ``input map'') represents the sky idealized as a
map of $n_p$ sky pixels, $\bP$ is the pointing matrix, and $\bn$ is
a vector of length $n_t$ representing the detector noise.  For
observations with multiple detectors, the TODs from the individual
detectors are appended end-to-end to form the full TOD vector $\by$.

Since we are dealing with polarization data, the map $\bm$ is an
object with $3n_p$ elements; for each sky pixel the elements are the
$I$, $Q$, and $U$ Stokes parameters. The pointing matrix $\bP$ is of
size $(n_t,3n_p)$. Each row has $3$ nonvanishing elements $(1,
\cos2\psi, \sin2\psi)$ at the location corresponding to the sky
pixel in which the detector beam center falls for the sample in
question (the sample ``hits'' the pixel).  We do not make any
attempt at deconvolving the detector beam.  Thus the map $\bm$
represents the sky smoothed with the detector beam
and the pixel window function. The pointing matrix spreads the map
into a signal TOD $\bP\bm$.

We divide the TOD into $n_b$ segments of equal length $\nbase$; $n_t
= n_b\nbase$.   For each segment we define an offset, called
\emph{baseline}. The baselines model the low-frequency correlated
noise component, ``$1/f$ noise'', which we want to remove from the
data, and we approximate the rest of the noise as white.  Thus our
idealized noise model is
 \beq
    \bn = \bF\ain + \bw
 \label{idealized_noise}
 \eeq
where the vector $\ain$ (of length $n_b$) contains the baseline
amplitudes and the matrix $\bF$, of size $(n_t,n_b)$, spreads them
into the baseline TOD, which contains a (different) constant value
$a_b$ ($b = 1,\ldots,n_b$) for each baseline segment.  Each column
of $\bF$ contains $\nbase$ elements of $1$ corresponding to the
baseline in question, and the rest of the matrix elements are $0$.
The vector $\bw$ represents white noise, and is assumed to be the
result of a Gaussian random process, where the different samples in
$\bw$ are uncorrelated,
  \beq
     \langle w_t w_{t'} \rangle = \sigma_t^2 \delta_{tt'} \,.
  \eeq
Here $\langle \ \rangle$ denotes expectation value.  The white noise
(time-domain) covariance matrix
  \beq
     \Cw = \langle \bw \bw^T \rangle
  \label{eq:cw}
  \eeq
is thus diagonal (with elements $\sigma_t^2$), but not necessarily
uniform (the white noise variance $\sigma_t^2$ may vary from sample
to sample).

Given the TOD $\by$, and assuming we know the white noise variance
$\bC_w$, we want to find the maximum likelihood map $\mout$.  We
assume no prior knowledge of the baseline amplitudes $\ba$, i.e.,
they are assigned uniform prior probability.  (The variant of the
method, where such prior knowledge is used, is described in
Keih\"{a}nen et al.~\cite{Madam09}.)

Given the input map $\bm$ and the baseline amplitudes $\ba$, the
probability of the data $\by$ is
 \beq
   P(\by|\bm,\ba) = \left(\det 2\pi\Cw\right)^{-1/2}
   \exp\left(-\half \bw^T\Cwi\bw\right)
 \label{eq:like}
 \eeq
where $\bw = \by - \bF\ba - \bP\bm$.  This is interpreted as the
likelihood of $\ba$ and $\bm$, given the data $\by$ (Press et
al.~\cite{NumRec}). Maximizing the likelihood is equivalent to
minimizing the logarithm of its inverse. We obtain the chi-squared
function
 \beq
    \chi^2 = -2\ln P = \left(\by - \bF\ba - \bP\bm\right)^T \Cwi
                      \left(\by - \bF\ba - \bP\bm\right)
 \label{eq:chisq}
 \eeq
to be minimized.  (We dropped the constant prefactor of
Eq.~(\ref{eq:like}).) We want to minimize this with respect to both
$\ba$ and $\bm$.

Minimization of Eq.~(\ref{eq:chisq}) with respect to $\bm$ gives the
maximum-likelihood map
 \beq
    \bm = \left(\bP^T\Cwi\bP\right)^{-1}
    \bP^T\Cwi\left(\by-\bF\ba\right)
 \label{eq:mmax}
 \eeq
for a given set of baseline amplitudes $\ba$.

The symmetric non-negative definite matrix
 \beq
    \bM \equiv \bP^T\Cwi\bP \,,
 \label{eq:Mdef}
 \eeq
which operates in the map space, is $3\times3$ block diagonal, one
block $\bM_p$ for each pixel $p$:
 \beq
    \bM_p = \left( \begin{array}{ccc}
    \sum_t\frac{1}{\sigma_t^2}
    & \sum_t\frac{\cos 2\psi_t}{\sigma_t^2}
    & \sum_t\frac{\sin2\psi_t}{\sigma_t^2} \\
    \sum_t\frac{\cos2\psi_t}{\sigma_t^2}
    & \sum_t\frac{\cos^22\psi_t}{\sigma_t^2}
    & \sum_t\frac{\cos2\psi_t\sin2\psi_t}{\sigma_t^2} \\
    \sum_t\frac{\sin2\psi_t}{\sigma_t^2}
    & \sum_t\frac{\sin2\psi_t\cos 2\psi_t}{\sigma_t^2}
    & \sum_t\frac{\sin^22\psi_t}{\sigma_t^2}
    \end{array} \right) \,,
 \eeq
where the sums run over all samples $t$ that hit pixel $p$.
$\bM_p^{-1}$ is the white noise covariance matrix for the three
Stokes parameters $I$, $Q$, $U$ in pixel $p$. $\bM_p$ can only be
inverted if the pixel $p$ is sampled with at least 3 sufficiently
different polarization directions $\psi_t$, so that all Stokes
parameters can be determined.  This can be gauged by the condition
number of $\bM_p$. If the inverse condition number {\em rcond}
(ratio of smallest to largest eigenvalue) is below some
predetermined limit, the pixel $p$ is excluded from all maps (as are
pixels with no hits), and the samples that hit those pixels are
ignored in all TODs. Technically, this is done by setting
$\bM_p^{-1}= 0$ for such pixels and $\left(\Cwi\right)_{tt} = 0$ for
the corresponding samples. $\bM$ can then be easily inverted by
non-iterative means.

If all $\sigma_t$ are equal, $\sigma_t^2\,\bM_p(1,1)$, gives the
number of hits (observations) in pixel $p$.  Thus $\bM$ is sometimes
called the $N_\mathrm{obs}$ matrix.  The optimal distribution of
polarization directions $\psi_t$ measured from a pixel is one where
they are uniformly distributed over $180^\circ$ (Couchot et
al.~\cite{Couchot99}).  In this case $\bM_p =
(\nhitp/\sigma_t^2)\,\diag(1,1/2,1/2)$ and $\bM_p^{-1} =
(\sigma_t^2/\nhitp)\,\diag(1,2,2)$ giving the maximum possible value
\emph{rcond}~$= 0.5$.

Substituting Eq.~(\ref{eq:mmax}) back into Eq.~(\ref{eq:chisq}) we
get this into the form
 \beq
    \chi^2 = \left(\by-\bF\ba\right)^T \bZ^T\Cwi\bZ
             \left(\by-\bF\ba\right) \,,
 \label{eq:chisq2}
 \eeq
where we have defined
 \beq
    \bZ \equiv \bI - \bP\bM^{-1}\bP^T\Cwi \,.
 \label{eq:Zdef}
 \eeq
Here $\bI$ is the unit matrix. The matrix $\bZ$ operates in TOD
space and is a projection matrix, $\bZ^2$ = $\bZ$. If all $\sigma_t$
are equal, $\bZ$ is symmetric.  In general, $\bC_w^{-1}\bZ$ is
symmetric, so that
 \beq
    \bC_w^{-1}\bZ = \bZ^T\Cwi = \bZ^T\Cwi\bZ \,.
 \label{eq:Zsym}
 \eeq

We minimize Eq.~(\ref{eq:chisq2}) with respect to $\ba$ to obtain
the maximum-likelihood estimate of the baseline amplitudes $\aout$.
It is the solution of the equation
 \beq
    \left(\bF^T\Cwi\bZ\bF\right)\ba =
    \bF^T\Cwi\bZ\by \,,
 \label{eq:base_eq}
 \eeq
where we have used Eq.~(\ref{eq:Zsym}).

The matrix
 \beq
    \bD \equiv \bF^T\Cwi\bZ\bF
 \label{eq:Ddef}
 \eeq
on the left-hand side of Eq.~(\ref{eq:base_eq}) operates in the
baseline space. It is symmetric but singular. Eq.~(\ref{eq:base_eq})
has a solution only if its right-hand side is orthogonal to the null
space of $\bD$.  The solution becomes unique when we require it to
be orthogonal to the null space too.

The null space of $\bD$ contains the vector that gives all baselines
the same amplitude. This represents the inability to detect a
constant offset of the entire noise stream $\bn$, because it has the
same effect on $\by$ as a constant shift in the $I$ of the entire
$\bmin$ (the monopole). This is of no concern (but should be kept in
mind) since the goal is to measure the CMB {\em anisotropy} and
polarization, not its mean temperature. If the baselines are
sufficiently connected by crossing points (two different baseline
segments of the TOD hitting the same pixel), there are no other kind
of vectors in the null space, so that the dimension of the null
space is one.  The right-hand side of Eq.~(\ref{eq:base_eq}) is
orthogonal to this one-dimensional null space, and thus
Eq.~(\ref{eq:base_eq}) can now be solved. In practice it is solved
by the conjugate gradient method. If the initial guess is orthogonal
to the null space, the method converges to a solution that is also
orthogonal to the null space. Normally we start with the zero vector
as the initial guess to guarantee this.  This means that the average
of the solved baseline amplitudes is zero.  Strictly speaking this
holds exactly only when no pixels are excluded from the baseline
determination due to their poor \emph{rcond}.

We write the solution of Eq.~(\ref{eq:base_eq}) as
 \beq
     \aout = \bDi\bF^T\Cwi\bZ\by \,.
 \label{eq:aout}
 \eeq
$\bDi$ is interpreted as the inverse in this orthogonal subspace.
$\bD$ and $\bDi$ will act in this subspace only.

Using the maximum likelihood baselines from Eq.~(\ref{eq:aout}) in
Eq.~(\ref{eq:mmax}) we get the output map of the destriping method:
 \beq
    \mout = \bMi
    \bP^T\Cwi\left(\by-\bF\aout\right) \,.
 \label{eq:mout}
 \eeq
Eqs.~(\ref{eq:aout}) and (\ref{eq:mout}) summarize the destriping
method.

Implementation details are discussed in Keih\"{a}nen et
al.~(\cite{Madam09}).

\subsection{Description}

Let us review the different operations involved:

$\bP^T\Cwi$ acts on a TOD $\by$ to produce from it a {\em sum map}
$\bP^T\Cwi\by$ where each pixel has Stokes parameters representing a
sum over observations that hit the pixel,
 \bea
    I_p & = & \sum_{t\in p}\frac{1}{\sigma_t^2}y_t \\
    Q_p & = & \sum_{t\in p}\frac{1}{\sigma_t^2}\cos2\psi_t y_t \\
    U_p & = & \sum_{t\in p}\frac{1}{\sigma_t^2}\sin2\psi_t y_t
 \eea

Instead of a sum, we should take the average of observations.  This
is accomplished by $\bMi\bP^T\Cwi$, which corresponds to solving the
Stokes parameters from the observations hitting this pixel; i.e.,
without regard for pixel-to-pixel noise correlations. The
observations are just weighted by the inverse white noise variance.
The resulting map is called the naive map, or the {\em binned map}.
We shall use the shorthand notation
 \beq
     \bB \equiv \bMi\bP^T\Cwi \,.
 \label{eq:Bdef}
 \eeq
$\bB$ acts on a TOD to produce from it a binned map. Note that
$\bB\bP = \bI$.

(It may be better to use the same $\sigma_t$ for the two
polarization directions of the same horn, to avoid polarization
artifacts from systematic effects (Leahy et
al.~\cite{prelaunch_pol}).  In case their true noise levels are
different, destriping allows also the option of using equal
$\sigma_t$ in solving baselines, Eq.~(\ref{eq:aout}), but the actual
$\sigma_t$ in the final binning to the map, Eq.~(\ref{eq:mout}).  We
do not study this issue in this paper, as we used simulated data
with a constant $\sigma_t$.)

We can now see that the effect of
 \beq
    \bZ = \bI - \bP\bB
 \label{eq:Zshort}
 \eeq
on a TOD is to bin it to a map, read a TOD out of this map, and
subtract it from the original TOD.  Thus $\bZ\by$ represents an
estimate of the noise part of $\by$.  Acting on a TOD constructed
from a map as $\bP\bm$, $\bZ$ returns zero, $\bZ\bP\bm = 0$.

Likewise, $\bF^T\Cwi$ acts on a TOD to sum up the samples of each
baseline segment, weighting each sample by $\sigma_t^{-2}$.

The effect of the matrix
 \beq
    \bD = \bF^T\Cwi\bZ\bF = \bF^T\Cwi\bF - \bF^T\Cwi\bP\bB\bF
 \label{eq:Dshort}
 \eeq
on a baseline amplitude vector $\ba$ is to produce a TOD $\bF\ba$
containing just these baselines, make a noise estimate $\bZ\bF\ba$
from this baseline TOD, and calculate then the weighted sum
$\bF^T\Cwi\bZ\bF$ of this noise estimate for each baseline.  Thus
the content of Eq.~(\ref{eq:base_eq}) is to find such a set of
baseline amplitudes that these noise estimate sums are the same for
the baseline TOD $\bF\ba$ as for the actual input TOD $\by$.  Thus
the solved baselines $\aout$ represent the estimated average noise
for each baseline segment of the TOD.

These baselines are then subtracted from the TOD to produce the {\em
cleaned} TOD $\by - \bF\aout$, which is then binned to produce the
output map
 \beq
     \mout = \bB\left(\by-\bF\aout\right)
 \eeq
in Eq.~(\ref{eq:mout}).

In a good scanning of the sky the number of hits in each pixel is
large. From Eq.~(\ref{eq:Zshort}) we see that $\bZ$ contains two
parts. The first part $\bI$ gives each row $t$ a large diagonal
element $1$ corresponding to the TOD sample this row is acting on.
The second part gives this row a large number of small nonzero
elements corresponding to all samples $t'$ that hit this same pixel.
The sum of these elements is $-1$ so that the row sum is zero. Thus
the first and second parts make an equally large contribution, but
the second part comes in many small pieces.

The matrix $\bD$ has a similar structure. The first part (see
Eq.~\ref{eq:Dshort}) is diagonal containing the sum $\sum_{t\in b}
1/\sigma_t^2 $ over all samples in the baseline segment $b$.  The
second part gives to each row $b$ a nonzero element for each
baseline $b'$ that has a crossing point with $b$.  For a good
scanning each baseline has a large number of crossing points, so
that this second part contributes a large number of small elements
to each row.

We define a shorthand notation for the matrix
 \beq
    \bA \equiv \bDi\bF^T\Cwi\bZ \,,
 \label{eq:Adef}
 \eeq
which appears in Eq.~(\ref{eq:aout}) as $\aout = \bA\by$.  This
matrix acts on a TOD and produces from it the set of baseline
amplitudes according to the destriping method, Eq.~(\ref{eq:aout}).
Note that $\bA\bF = \bI$, in the sense that $\bA\bF\ba = \ba$ for
any set of baseline amplitudes, whose average is zero (so that $\ba$
is orthogonal to the null space).  Since $\bZ$ is a projection
matrix, $\bA\bZ = \bA$.

Note that all the matrices introduced (see
Table~\ref{table:matrices}) are constructed from $\bI$, $\bP$,
$\Cw$, and $\bF$. The adjustable parameter in the destriping method
is the baseline length $\nbase$, which affects the matrix $\bF$.
$\bP$ is determined by the scanning strategy and map pixelization,
and $\Cw$ by detector noise properties.

\begin{table}[!tbp]
 \begin{center}
 \begin{tabular}{llll}
 matrix & size & Eq. & comment\\
 \hline
 $\bP$  &  $n_t\times 3n_p$ & (\ref{eq:pointing}) & pointing matrix\\
 $\bF$   & $n_t \times n_b$ & (\ref{idealized_noise}) & baselines to TOD \\
 $\Cw \equiv \langle \bw \bw^T \rangle$  & $n_t \times n_t$ & (\ref{eq:cw}) & white noise cov. \\
 $\bM \equiv \bP^T\Cwi\bP$  & $3n_p \times 3n_p$ & (\ref{eq:Mdef}) & $N_\mathrm{obs}$ matrix \\
 $\bB \equiv \bMi\bP^T\Cwi$ & $3n_p \times n_t$ & (\ref{eq:Bdef}) & bin TOD to a map \\
 $\bZ \equiv \bI - \bP\bB$ & $n_t\times n_t$ &
 (\ref{eq:Zdef}) & \\
 $\bD \equiv \bF^T\Cwi\bZ\bF$ & $n_b\times n_b$ & (\ref{eq:Dshort}) & \\
 $\bA \equiv \bDi\bF^T\Cwi\bZ$ & $n_b\times n_t$ & (\ref{eq:Adef}) & solve baselines \\
 $\bR \equiv \left(\bF^T\Cwi\bF\right)^{-1}\bF^T\Cwi$ & $n_b\times
 n_t$ & (\ref{eq:Rdef}) & reference baselines \\
 \hline
 \end{tabular}
 \end{center}
 \caption{
Table of matrices. The square matrices $\Cw$, $\bM$, $\bD$, and
$\Cwi\bZ$ are symmetric, $\Cw$ is diagonal, and $\bM$ $3\times3$
block diagonal. $\bZ$ is a projection matrix. $\bD$ is singular, and
$\bDi$ is its inverse in the subspace orthogonal to its null space.
The third column refers to the equation in which the matrix was
introduced.
 }
 \label{table:matrices}
\end{table}

\subsection{Destriping error}

One can easily show that
 \beq
    \bD_{bb'} = \frac12{\frac{\partial^2(\chi^2)}{\partial a_b \partial
    a_{b'}}}_{|\ba = \aout, \bm = \mout} \,,
 \eeq
i.e, the matrix $\bD$ of Eq.~(\ref{eq:Ddef}) is the Fisher matrix of
the baselines. Its inverse gives the covariance of the baseline
error $\Delta\ba \equiv \aout-\ain$. In particular,
 \beq
    \left(\bD^{-1}\right)_{bb} = \langle\Delta a_b^2 \rangle \,.
 \label{avar}
 \eeq

Ignoring the nondiagonal terms we get an approximation
 \beq
    \langle\Delta a_b^2 \rangle \approx
    \left(\sum_{t\in b}\frac{1}{\sigma_t^2}\right)^{-1} \,.
 \label{apxavar1}
 \eeq
Assuming the white noise variance stays constant, $\sigma_t =
\sigma$, over a given baseline, this becomes
 \beq
    \langle\Delta a_b^2 \rangle \approx \frac{\sigma^2}{\nbase} \,.
 \label{apxavar}
 \eeq

We can understand the approximate result (\ref{apxavar}) as follows.
Destriping solves the baseline amplitude $a_b$ from the differences
between samples from baseline segment $b$ and from other baseline
segments that hit the same pixel. Both the baselines $\ain$ and the
white noise contribute to these differences.  Without the presence
of white noise the baselines could be solved exactly, $\aout = \ain$
(in the idealized model considered in this section), so the error
comes from the white noise in the samples.  Typically there are many
more differences involved than there are samples in the baseline
(number of hits per pixel $\gg 2$), but samples from the baseline
segment $b$ contribute only $\nbase$ uncorrelated random variables,
with variance $\sigma^2$, and their contribution corresponds to the
result (\ref{apxavar}). The total of samples from baseline segment
$b$ contribute equally to its amplitude determination than do the
total of samples from other baseline segments that hit the same
pixels. But the latter contribution comes from a much larger number
of uncorrelated random variables, with a similar variance, so it
gets averaged out by the larger number and makes a much smaller
contribution to the baseline error variance (\ref{avar}).  This
contribution from the white noise from the crossing samples is
ignored in (\ref{apxavar1}) and (\ref{apxavar}). However, it should
add to the baseline error variance, so that Eqs.~(\ref{apxavar1})
and (\ref{apxavar}) are underestimates.  See also Efstathiou
(\cite{E05,E07}).

The approximation (\ref{apxavar}) corresponds to the white noise
reference baseline contribution discussed in
Sect.~\ref{sec:time_domain}. We see in Sect.~\ref{sec:map_domain}
that, while this approximation is good in the time domain, it is not
that good in the map domain.

\section{Simulation}
\label{sec:simulation}

\begin{figure}[!tbp]
 \begin{center}
    \includegraphics*[width=0.5\textwidth]
    {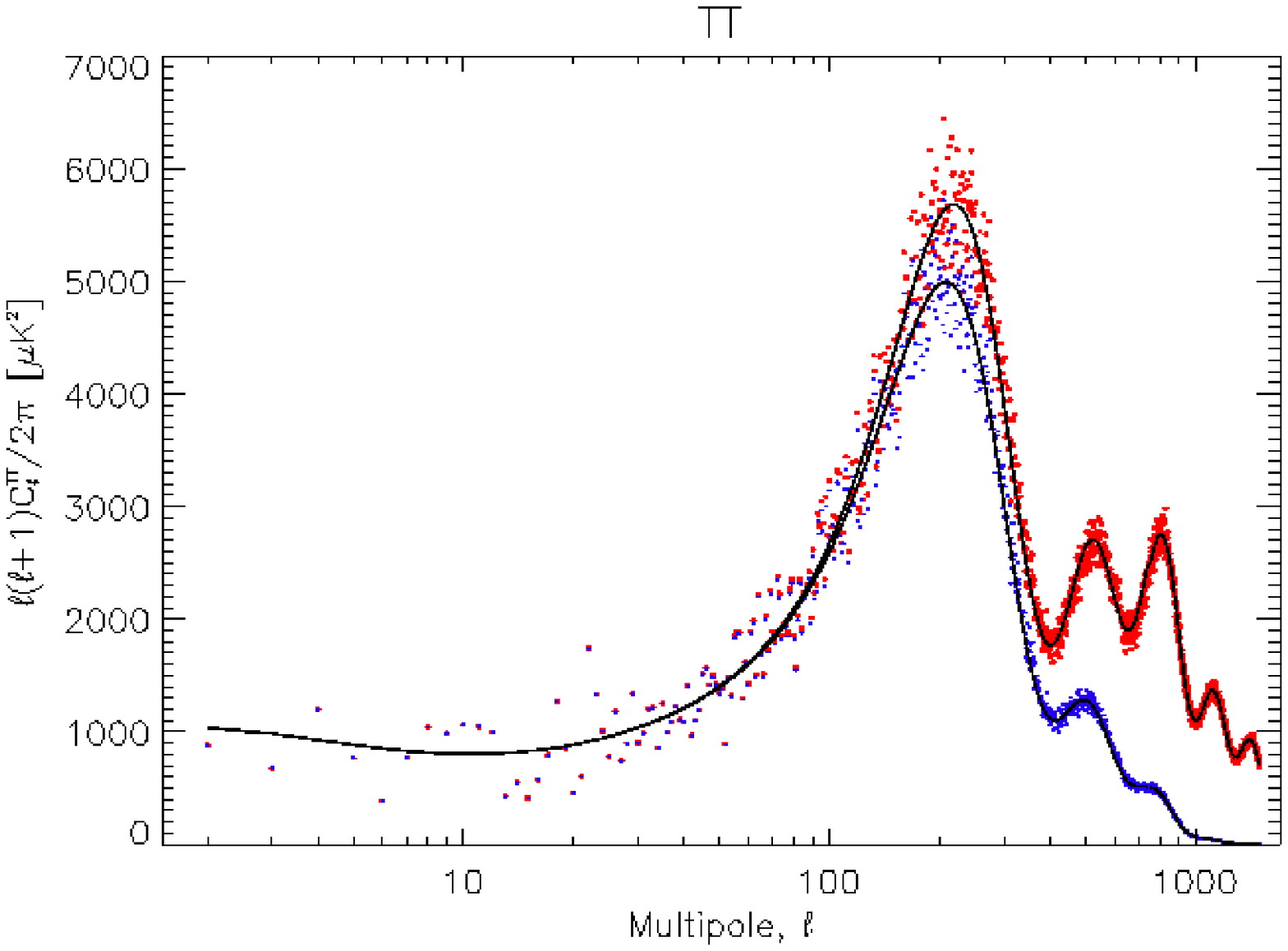} \\
    \vspace{-0.3cm}
    \includegraphics*[trim=0 0 0 40, clip, width=0.5\textwidth]
    {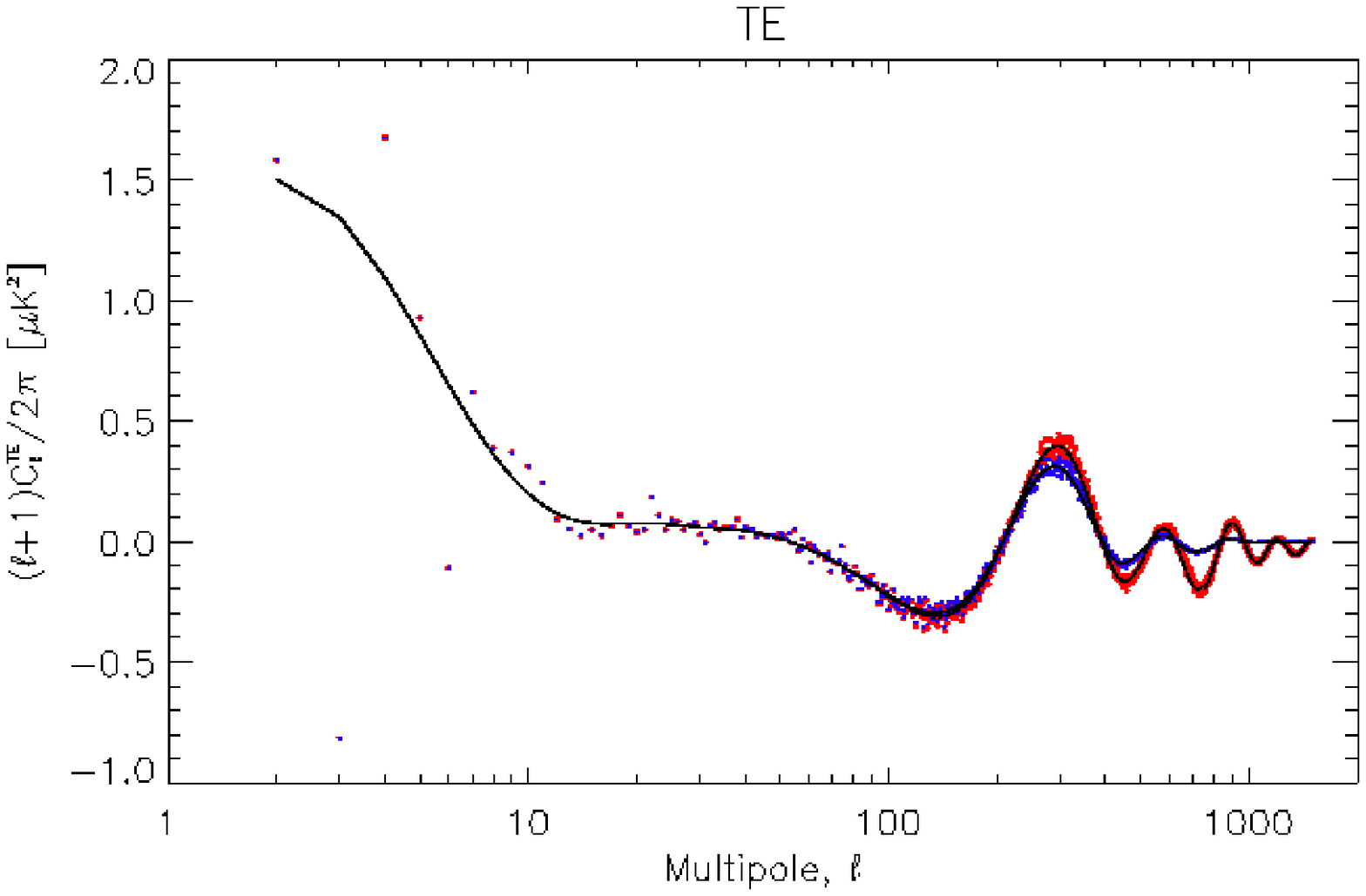} \\
    \vspace{-0.25cm}
    \includegraphics*[trim=0 0 0 145, clip, width=0.5\textwidth]
    {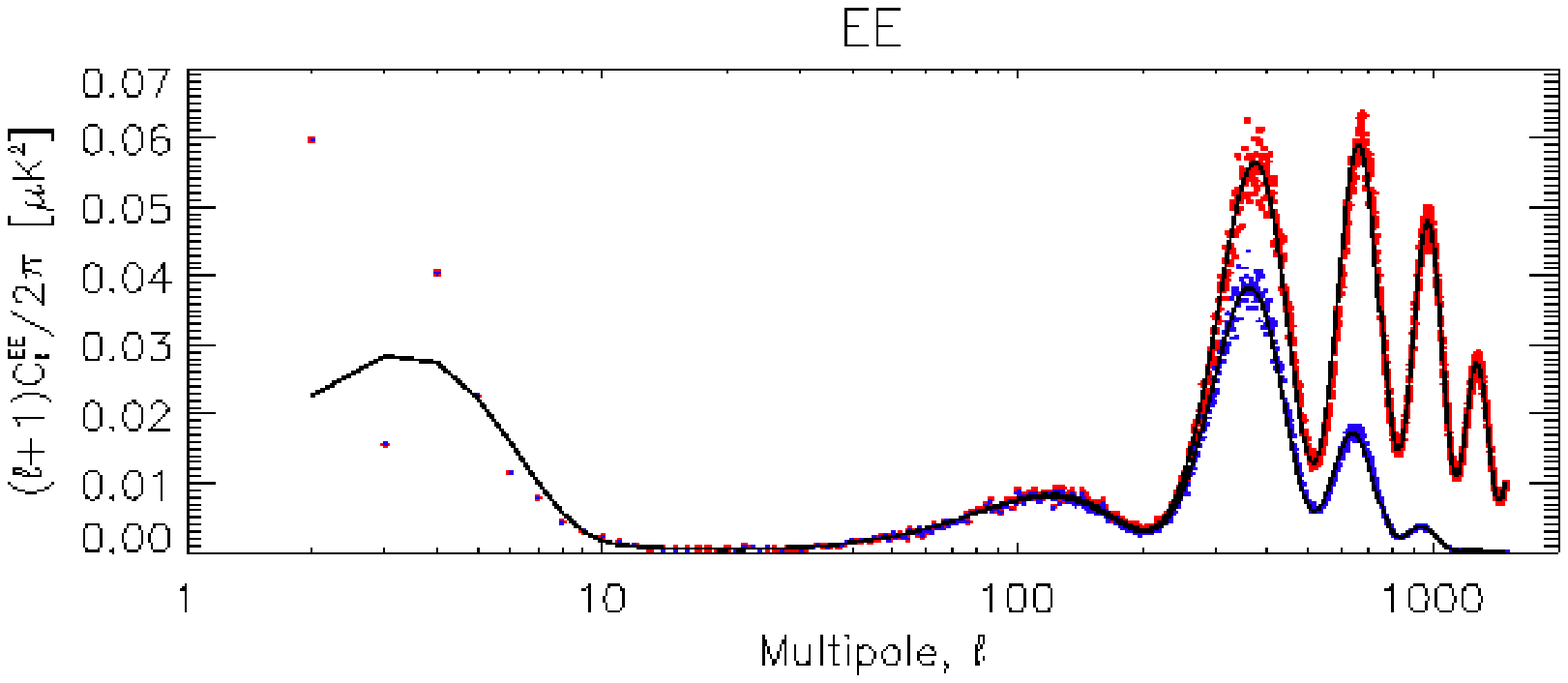} \\
 \caption{The input angular power spectrum.
\emph{Black lines} show the theoretical $C_\ell$ produced by CAMB,
and the beam-and-pixel-smoothed version of it.  \emph{Red dots} show
the $C_\ell$ of the $a_{\ell m}$ realization. \emph{Blue dots} show
the $C_\ell$ calculated from the binned noiseless map. (A version of
this paper with higher-quality figures is available at
http://www.helsinki.fi/$\sim$tfo\_cosm/tfo\_planck.html )
 }
 \label{fig:input_cl}
 \end{center}
\end{figure}

\begin{figure}[!tbp]
 \begin{center}
   \includegraphics*[trim=120 0 70 0, clip, width=0.28\textwidth,angle=90]
    {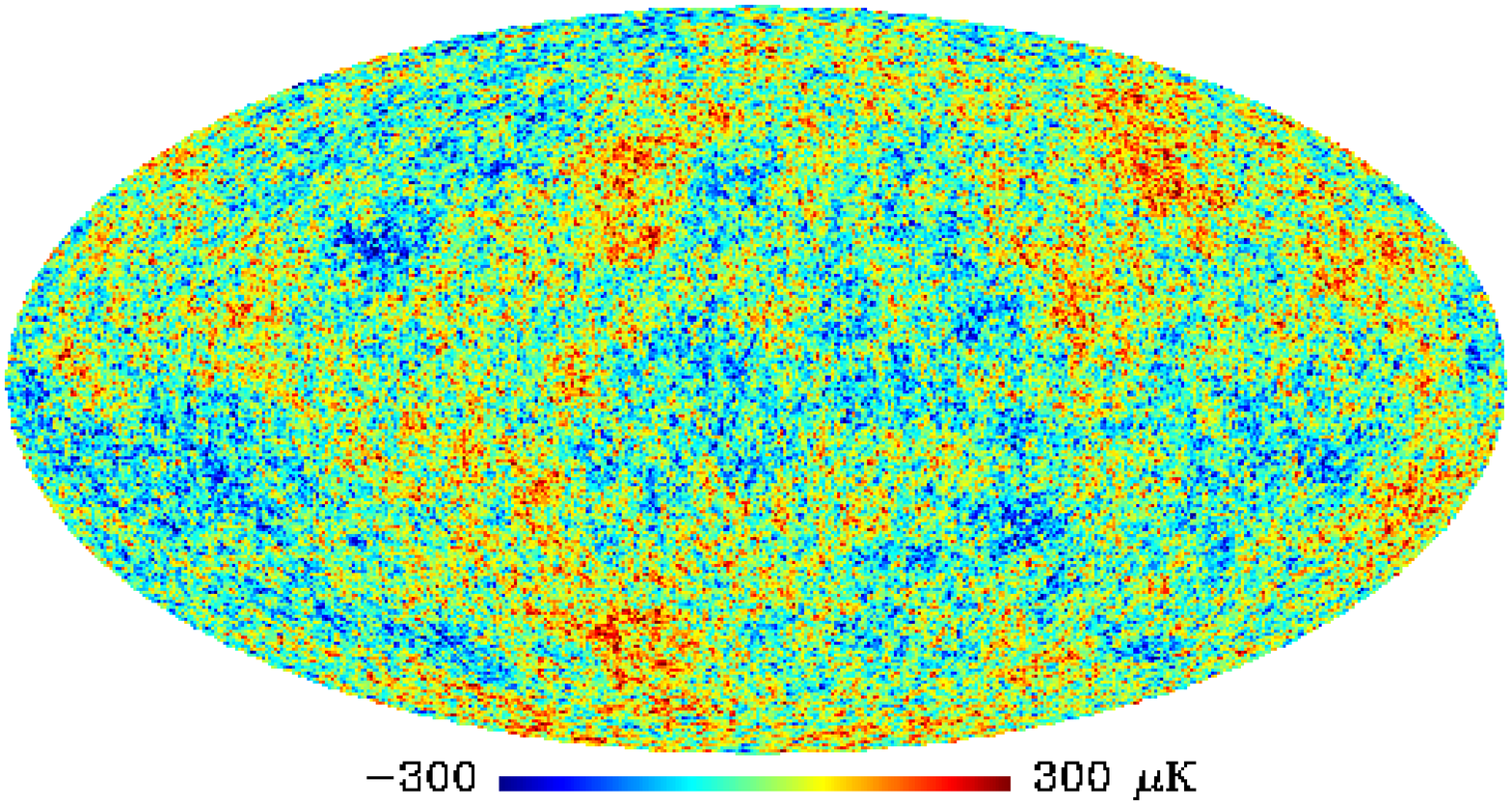} \\
    \vspace{-0.2cm}
    \includegraphics*[trim=120 0 70 0, clip, width=0.28\textwidth,angle=90]
    {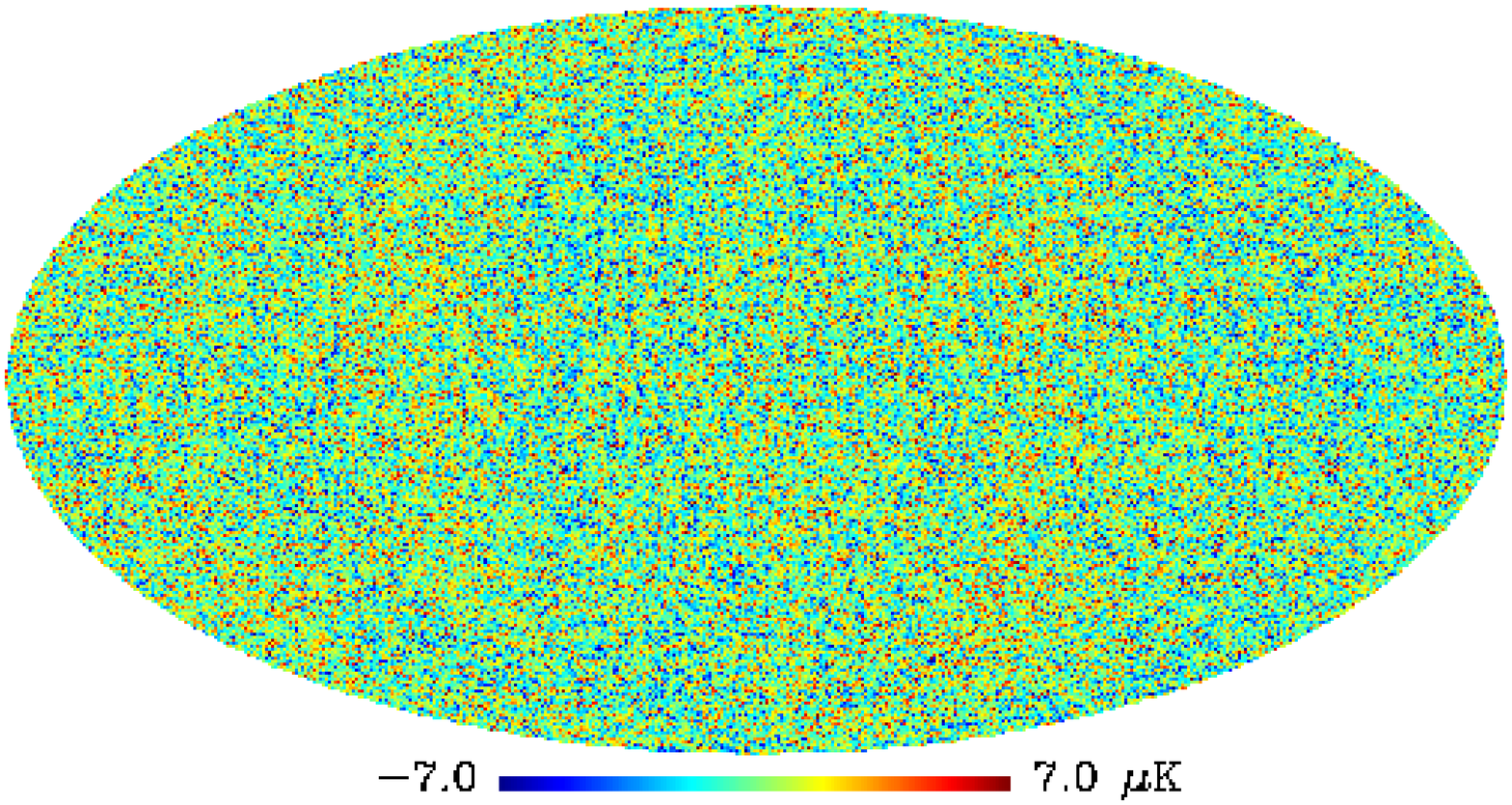} \\
    \vspace{-0.2cm}
 \caption{Binned CMB signal map, $I$
and $Q$, full sky. }
 \label{fig:signal_maps}
 \end{center}
\end{figure}

\begin{figure}[!tbp]
  \begin{center}
    \vspace{-0.5cm}
    \includegraphics*[trim=100 70 70 250, clip, width=0.24\textwidth]
    {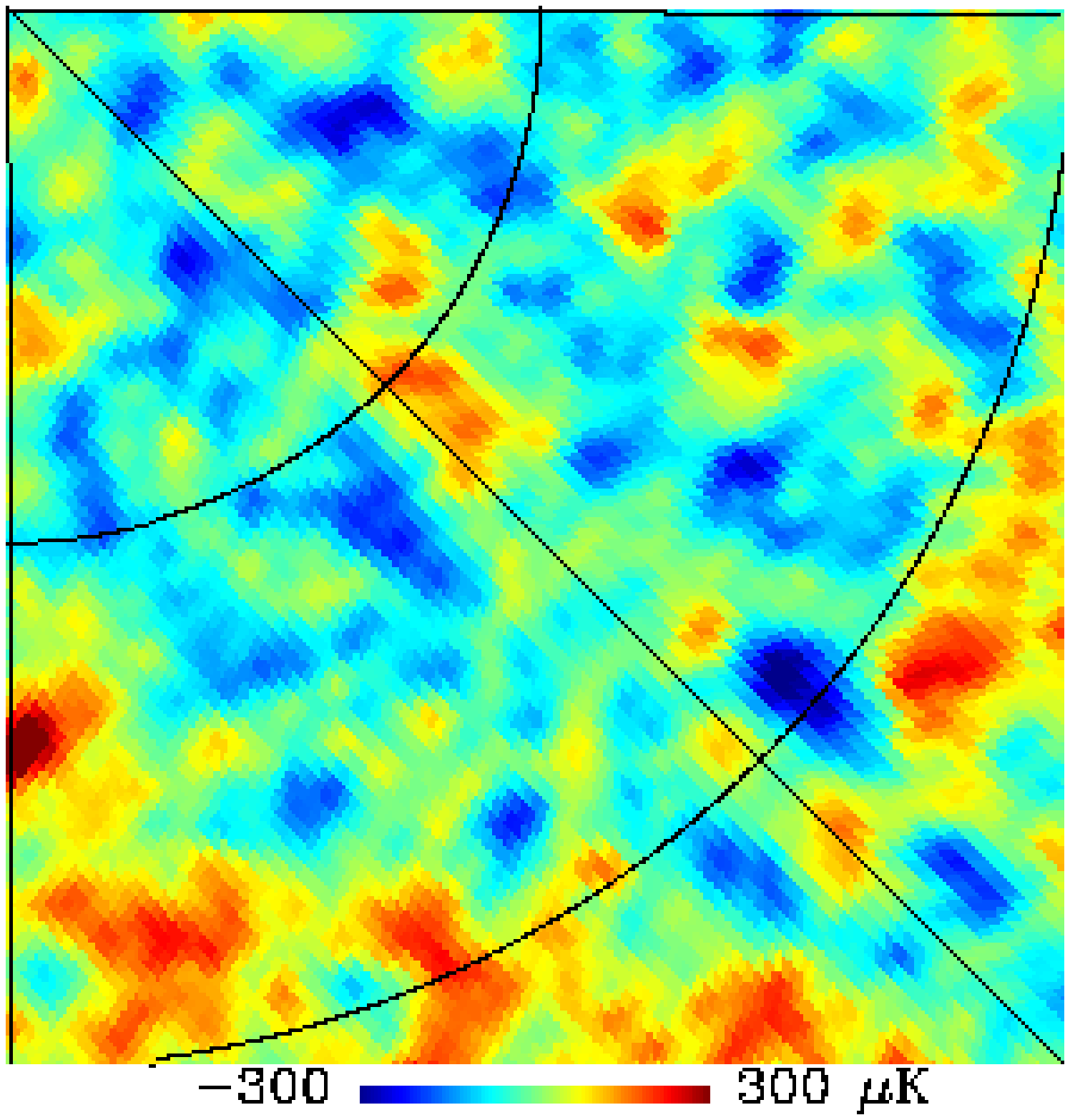}
    \includegraphics*[trim=100 70 70 250, clip, width=0.24\textwidth]
    {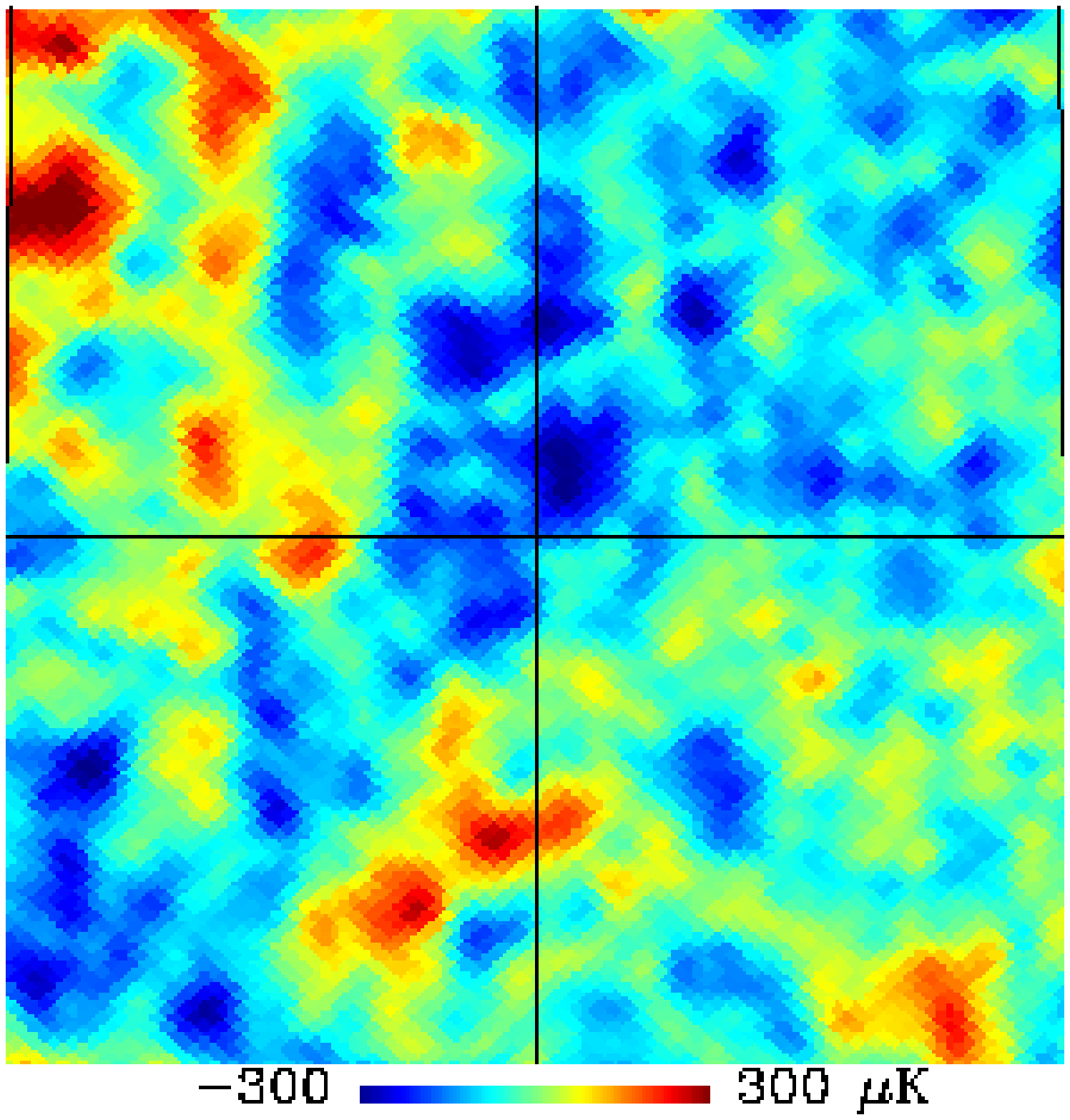} \\
    \vspace{-0.2cm}
    \caption{
Binned CMB temperature ($I$) signal maps from two
$10^\circ\times10^\circ$ regions.
 \emph{Left column:} Near the ecliptic North Pole (centered at
$\theta = 7^\circ$, $\phi = -90^\circ$).
 \emph{Right column:} Near the ecliptic (centered at $\theta =
 85^\circ$, $\phi = 5^\circ$).
    }
    \label{fig:10by10_signal_maps}
  \end{center}
\end{figure}

\begin{figure}[!tbp]
  \begin{center}
    \includegraphics*[trim=100 70 70 250, clip, width=0.24\textwidth]
    {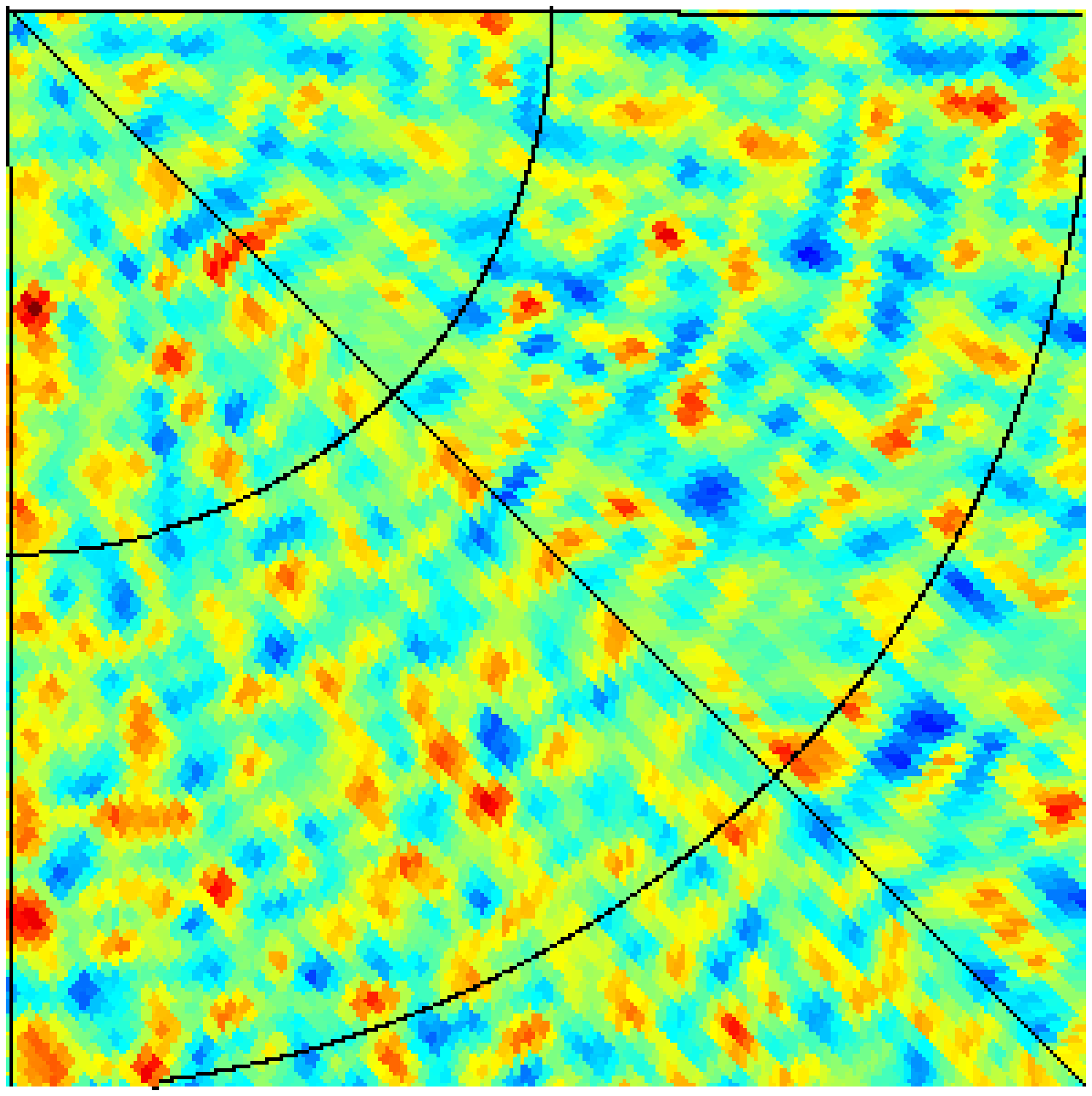}
    \includegraphics*[trim=100 70 70 250, clip, width=0.24\textwidth]
    {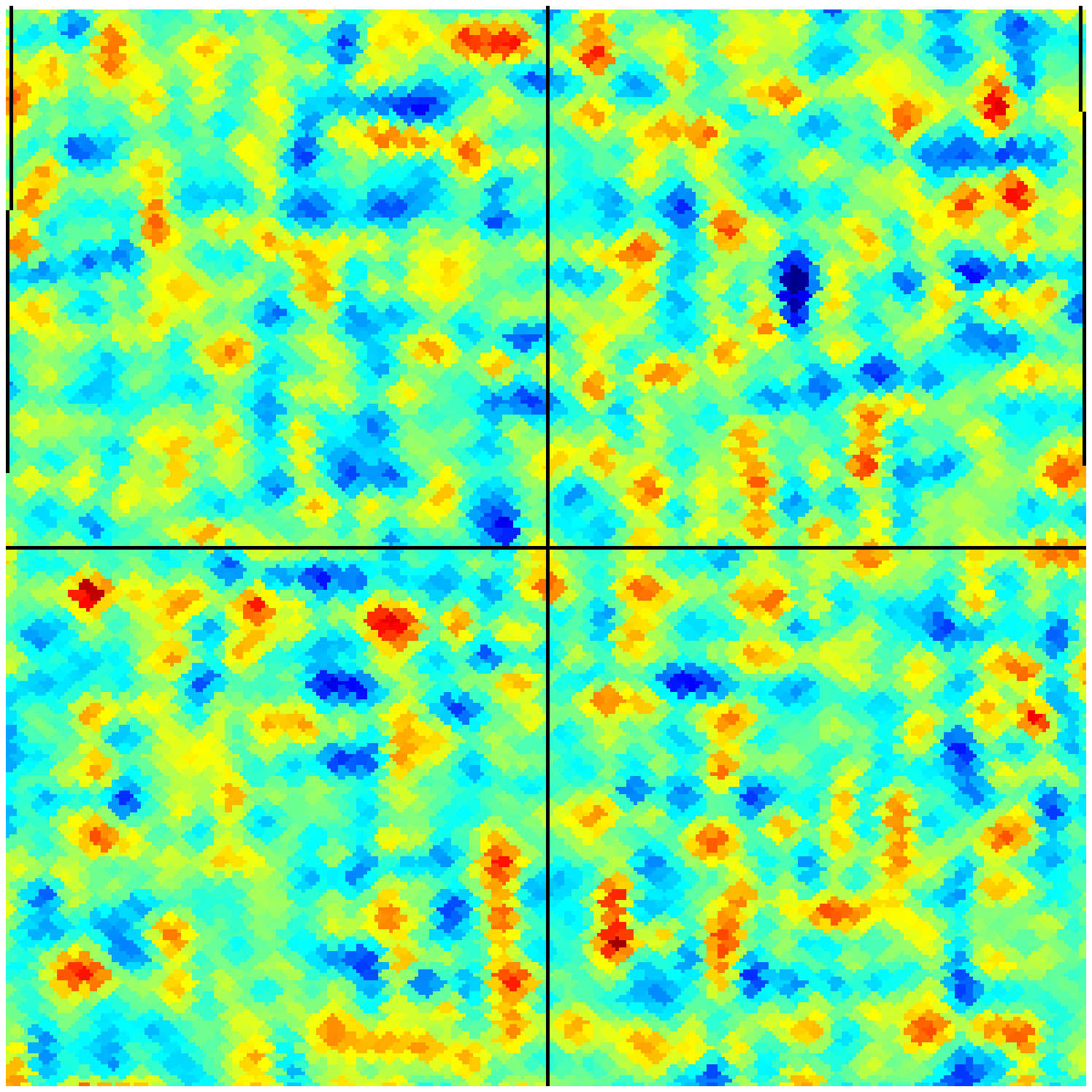} \\
    \vspace{-0.3cm}
    \includegraphics*[trim=100 70 70 250, clip, width=0.24\textwidth]
    {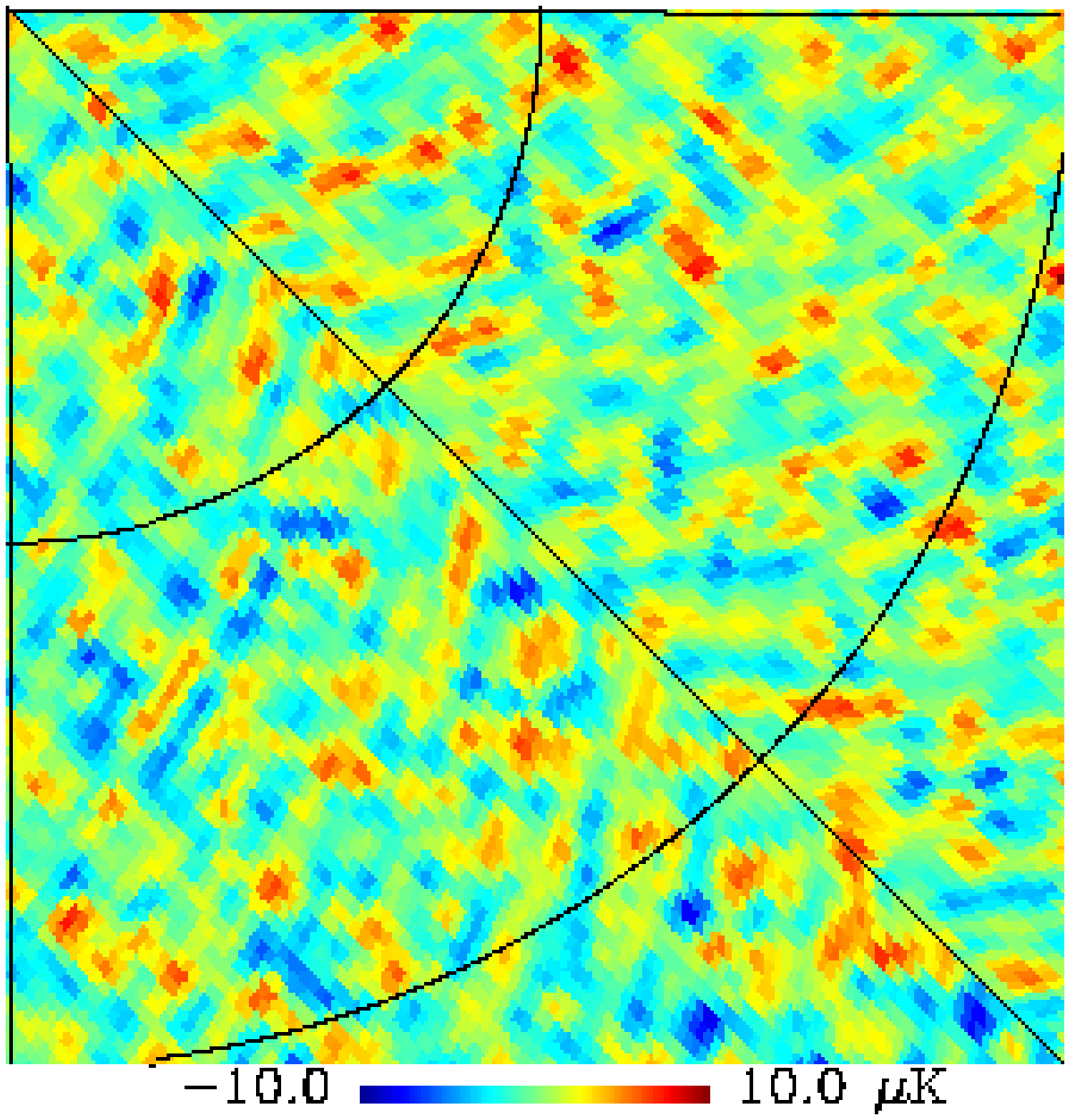}
    \includegraphics*[trim=100 70 70 250, clip, width=0.24\textwidth]
    {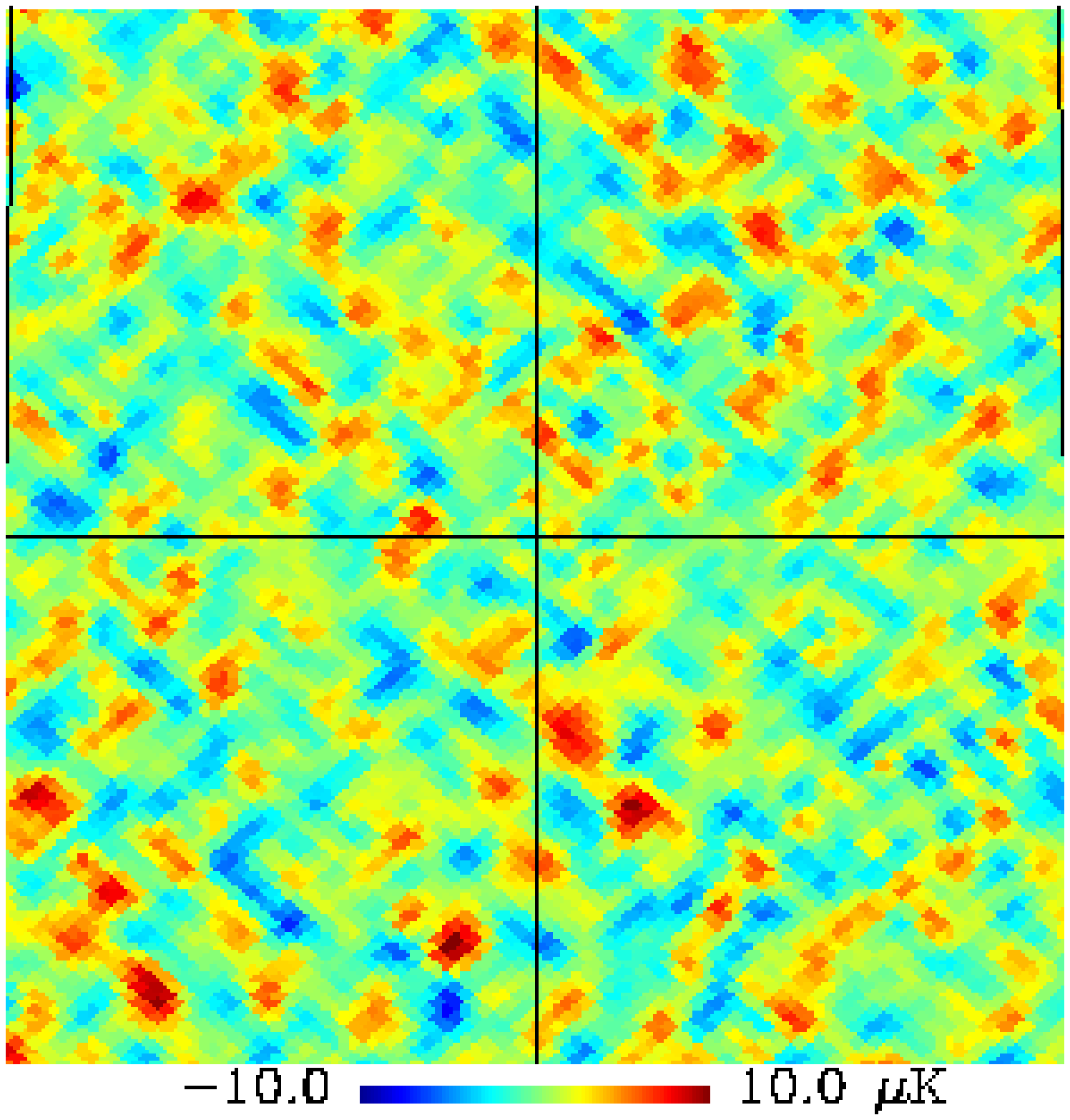}\\
    \vspace{-0.2cm}
    \caption{ Same as Fig.~\ref{fig:10by10_signal_maps}, but for the
Stokes parameters $Q$ (\emph{top}) and $U$ (\emph{bottom}). Since
the signal contains only $E$ mode polarization, $Q$ shows structures
elongated along lines of latitude and longitude, whereas $U$ shows
structures elongated $45^\circ$ away from them.
    }
    \label{fig:10by10_signal_QU_maps}
  \end{center}
\end{figure}

In this study we tested the destriping method using simulated data
that is more realistic than the model used to derive the method in
Sect.~\ref{sec:method}.  The TOD was produced using {\sc Planck}
Level-S simulation software (Reinecke et al. \cite{LevelS}) as a sum
of signal, white noise, and correlated noise (called also $1/f$
noise),
 \beq
    \by = \bs + \bn = \bs + \bw + \bnc \,.
 \eeq
We produced the three time streams, $\bs$, $\bw$, and $\bnc$
separately, so we could analyze the effect of each component on the
destriping error. The TOD represented 1 full year (366 d) of data
from 4 polarized {\sc Planck} 70 GHz detectors (LFI-19a, LFI-19b,
LFI-22a, LFI-22b).

\subsection{Signal}

We considered the CMB signal only; no foregrounds were included in
the simulation. Detector pointings $\theta_t, \phi_t, \psi_t$ for
each sample $t$ were produced to imitate a realistic scanning
strategy.  We used a set of input spherical harmonic coefficients
$a_{\ell m}^T$, $a_{\ell m}^E$ to represent the sky.  The signal
sample $s_t$ was then produced by the convolution of a circularly
symmetric Gaussian beam (fwhm $= 12.68'$) centered at this pointing
with the input $a_{\ell m}^{T,E}$ (Wandelt \& G\'{o}rski
\cite{totconvWG}, Challinor \cite{totconvC}). Thus the $I$, $Q$, and
$U$ (Eq.~\ref{eq:IQU}) of the signal parts of different samples
hitting the same pixel are different, as $\theta_t, \phi_t$ can vary
within the pixel.

 To produce the input $a_{\ell m}^{T,E}$ we used the
CAMB\footnote{http://camb.info} code to produce the theoretical
angular power spectra $C^{TT}_\ell$, $C^{TE}_\ell$, $C^{EE}_\ell$
for the Friedmann-Robertson-Walker universe with cosmological
parameter values $\Omega_0 = 1$, $\Omega_\Lambda = 0.7$, $\omega_m =
0.147$, $\omega_b = 0.022$, $\tau = 0.1$, and with scale-invariant
($n = 1$) adiabatic primordial scalar perturbations with amplitude
$5\times10^{-5}$ for the curvature perturbations. A realization
$\{a^T_{\ell m},a^E_{\ell m}\}$ was then produced from these
spectra.  Effects of gravitational lensing were ignored, and
therefore there is no $B$ mode polarization in the input spectrum.

Fig.~\ref{fig:input_cl} shows the input $C_\ell$ as well as the
$C_\ell$ of the binned (noiseless) signal map $\bB\bs$ made from the
simulated TOD.

Fig.~\ref{fig:signal_maps} shows the binned signal maps $\bB\bs$,
and Figs.~\ref{fig:10by10_signal_maps}  and
\ref{fig:10by10_signal_QU_maps} zoom into two
$10^\circ\times10^\circ$ regions to reveal small-scale detail. In
this paper we keep using these same two $10^\circ\times10^\circ$
regions, one near the ecliptic north pole, one near the ecliptic, to
show map detail.

\subsection{Scanning}

 \begin{figure}[!tbp]
  \begin{center}
    \includegraphics*[trim=100 70 70 250, clip, width=0.24\textwidth]
    {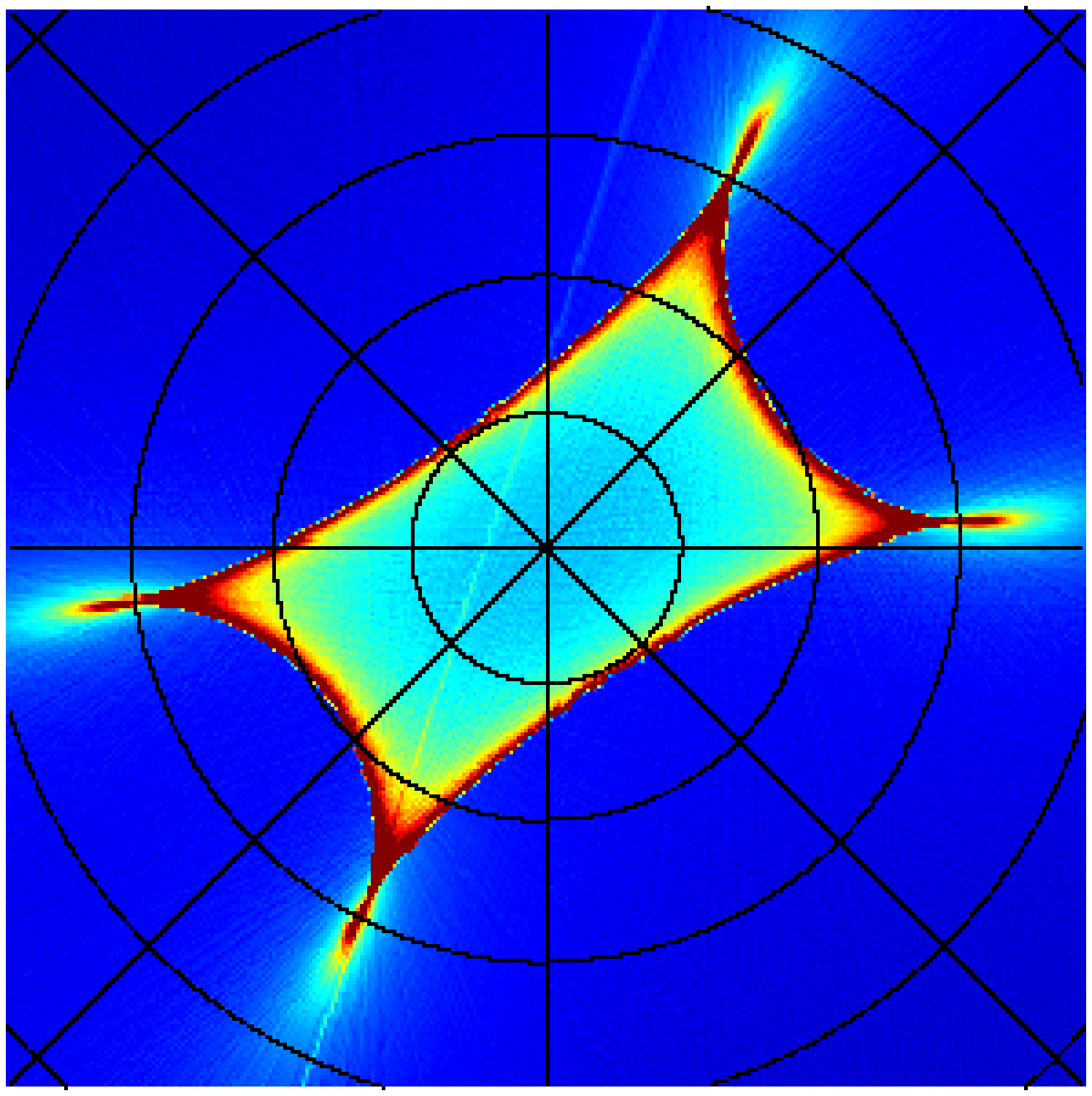}
    \includegraphics*[trim=100 70 70 250, clip, width=0.24\textwidth]
    {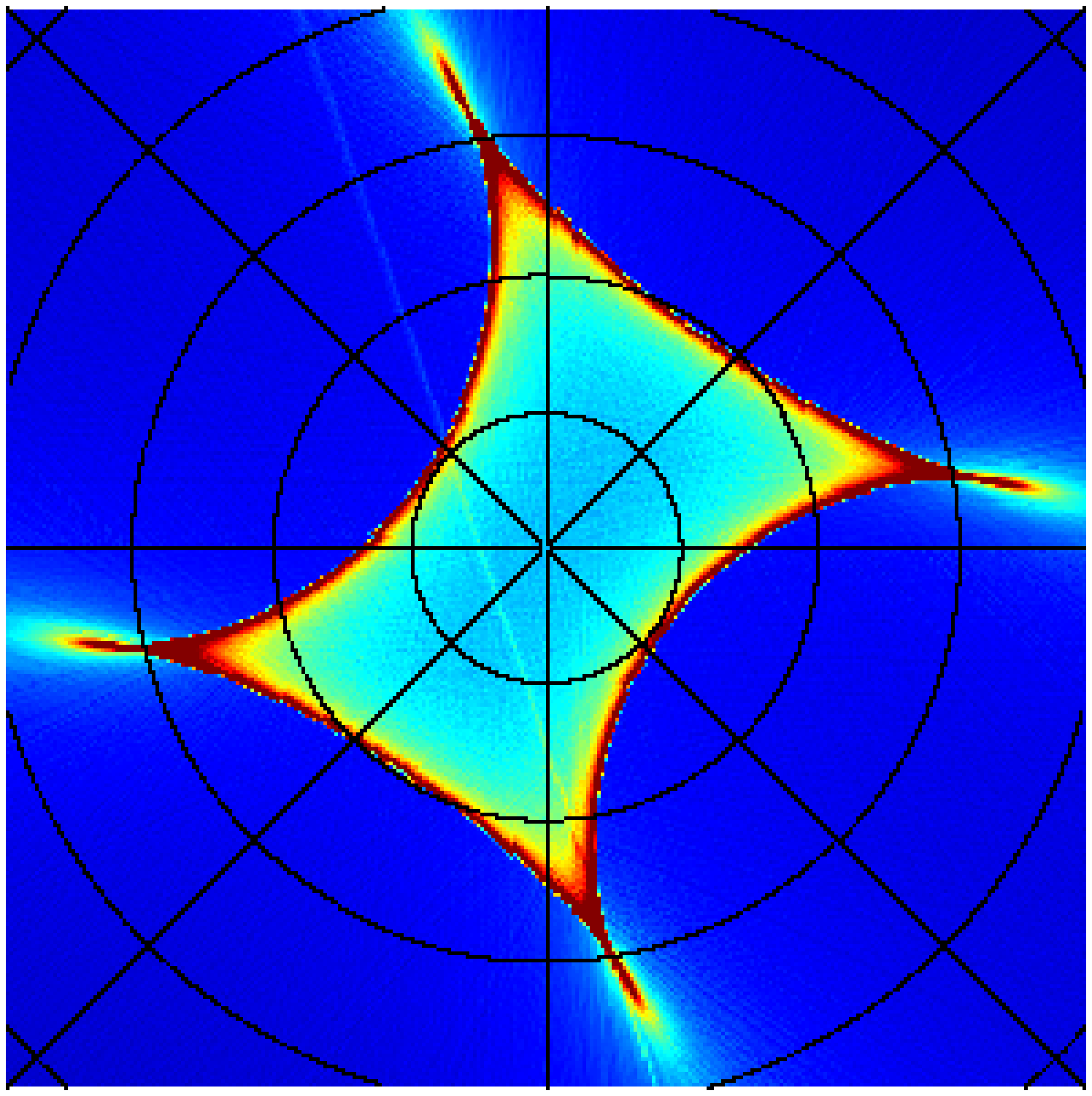}
    \vspace{-0.5cm}
    \caption{
Hit map for the 1-year simulation.  We show regions around the
ecliptic North Pole (\emph{left}) and South Pole (\emph{right}). The
color scale is linear and goes from zero (blue) to 50 000 (red).
Lines of latitude are drawn at $5^\circ$ intervals.
    }
    \label{fig:hit_map_at_poles}
  \end{center}
\end{figure}

 \begin{figure}[!tbp]
 \begin{center}
  \includegraphics*[width=0.50\textwidth]{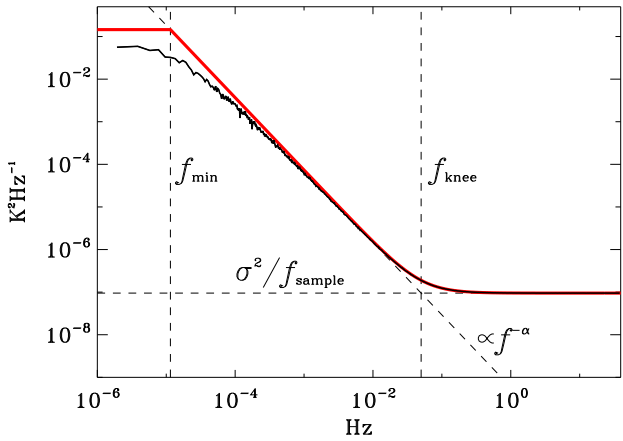}
 \caption{Power spectrum of the noise stream $\bn$ (\emph{black}).
 The spectrum was calculated from the detector 19a noise stream by
dividing it into 6-day pieces, obtaining their spectra separately,
taking their average, and binning the spectrum into 800 logarithmic
bins, to get a smooth curve. The \emph{red} curve is the noise
model.
 }
 \label{fig:noise_spectrum}
 \end{center}
 \end{figure}

The sky scanning strategy was cycloidal (Dupac \& Tauber
\cite{scan_stgy}): the satellite spin axis was repointed at 1 h
intervals, causing it to make a clockwise circle (radius
$7.5^\circ$) around the anti-Sun direction in 6 months.

Random errors ($\mbox{rms} = 1.3'$) were added to the repointing.
Between repointings the spin axis nutated at an amplitude related to
the repointing error. The mean nutation amplitude was $1.6'$. The
nutation was dominated by a combination of two periods, $\sim 45$ s
and $\sim 90$ s. In reality, the repointing errors and nutation
amplitudes are expected to be smaller. Thus effects of nutation and
small-scale variations in the map pixel hit count appear somewhat
exaggerated in this study.

The satellite rotated clockwise (i.e., spin vector pointing away
from the Sun) at about 1 rpm ($\fspin \approx 1/60$ Hz); spin rate
variations (rms $0.1^\circ/\mbox{s}$) around this nominal rate were
chosen randomly at each repointing.  (In reality, the spin rate
variations are expected to be smaller.) Coupled with the 60 s spin
period, the 45 s and 90 s nutation periods produce a 3-min
periodicity in the detector scanning pattern.

\begin{figure}[!tbp]
  \begin{center}
    \includegraphics*[trim=100 70 70 250, clip, width=0.24\textwidth]
    {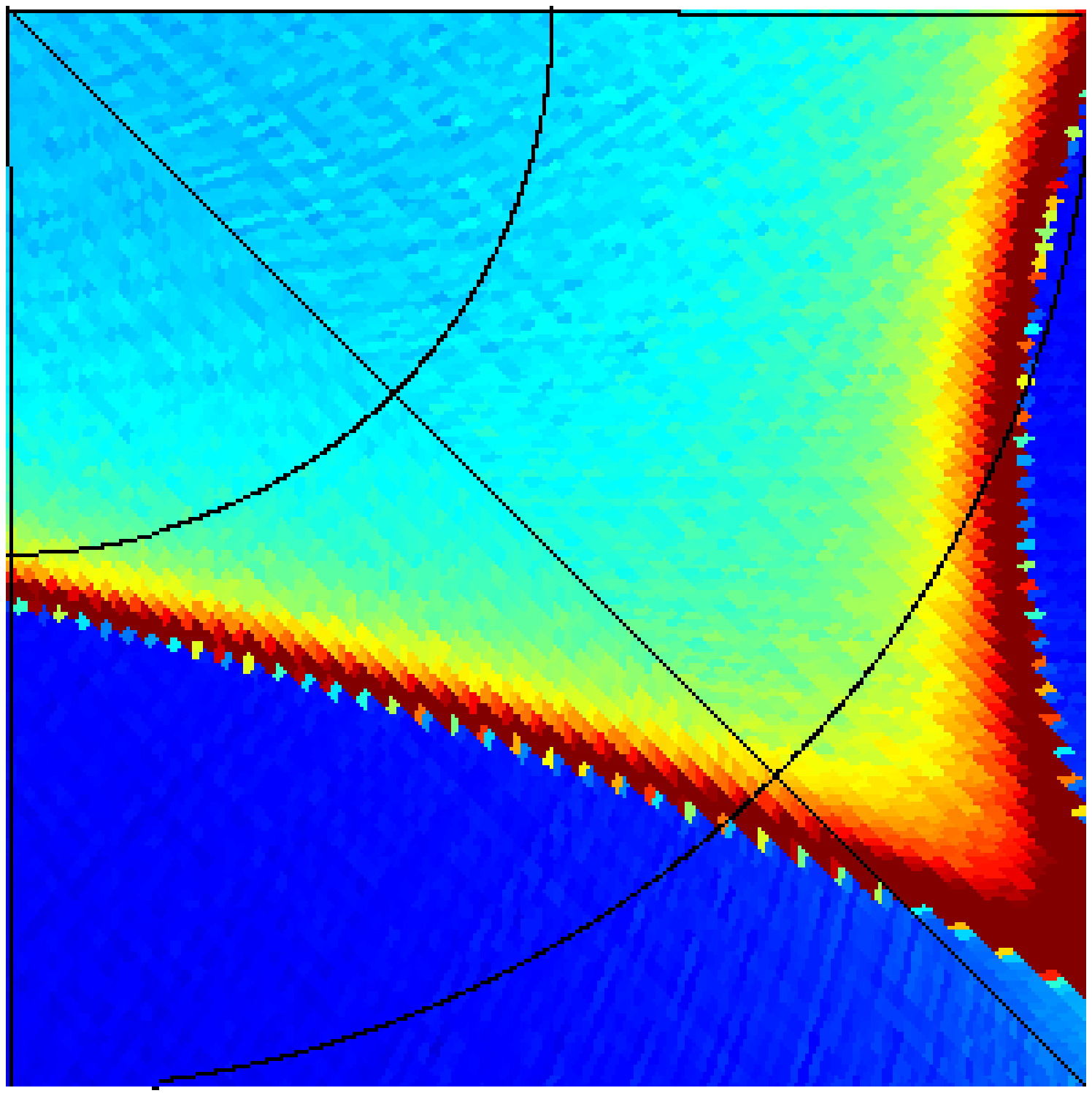}
    \includegraphics*[trim=100 70 70 250, clip, width=0.24\textwidth]
    {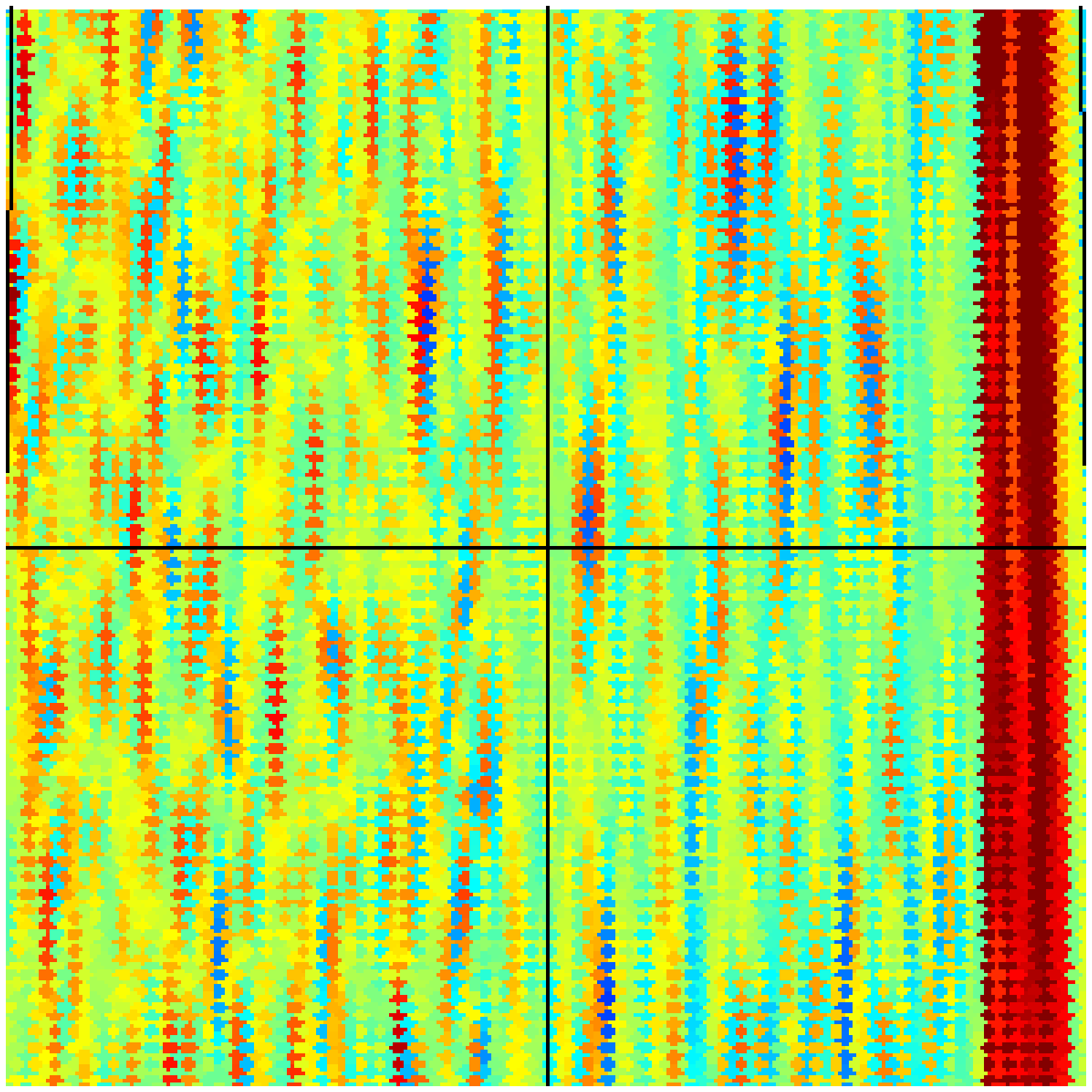}
    \vspace{-0.5cm}
    \caption{Hit maps of the two $10^\circ\times10^\circ$ regions.
The color scale is linear and goes from 0 to 50 000 in the
\emph{left} plot, and from 1000 to 3000 in the \emph{right} plot.
The 1-year survey begins and ends near the right edge of the right
plot, where the first and last rings overlap to produce a higher hit
count.
    }
    \label{fig:10by10_hit_maps}
  \end{center}
\end{figure}

We simulated 4 detectors corresponding to 2 horns (19 and 22). The
detectors were pointed $\theta_\mathrm{det} = 87.77^\circ$ away from
the spin axis, causing them to draw almost great circles on the sky,
the ``22'' trailing the ``19'' detectors by $3.1^\circ$ , but
following the same path. The a and b detectors of each horn shared
the same pointing but had different polarization directions by
exactly $90^\circ$.  The polarization directions of the 19 and 22
detector pairs differed from each other by $44.8^\circ$.

The cycloidal scanning causes the detector scanning rings to form
caustics around the ecliptic poles, where nearby scanning rings
cross (see Fig.~\ref{fig:hit_map_at_poles}).  A large number of ring
crossings cluster at the four corners of these caustics. For
destriping, such a clustering, where very many crossing points fall
on the same pixel, is disadvantageous, since there is less
independent information available for solving the baselines of these
rings. The clustering occurs when the curvature of the path of
satellite pointing on the sky equals the curvature of the scanning
circle. The curvature changes sign when the spin axis is close to
the ecliptic, but slightly north of it, and the clusterings near the
north and south ecliptic poles occur a little bit before and after
that. Conversely, the crossing points are spread more widely along
the caustics when the spin axis is near its north or south extrema.

The sampling frequency was $\fsample = 76.8$ Hz, and each sample was
simulated as an instantaneous measurement (no integration along scan
direction). The baseline length $\nbase$, is given in time units as
$\tbase = \nbase/f_\mathrm{sample}$ in the following. The beam
center moves on the sky at an angular velocity $\omega_\mathrm{scan}
= 360^\circ f_\mathrm{spin} \sin \theta_\mathrm{det} \approx
6^\circ/\mbox{s}$, so that one baseline corresponds to a path of
length $\theta_\mathrm{base} = \omega_\mathrm{scan}\tbase$ on the
sky.  The samples are separated by $\thetas =
\omega_\mathrm{scan}/f_\mathrm{sample} \approx 4.68'$ on the sky.
The length of the simulated TOD was $n_t =
4\times366\times24\times3600\,\mbox{s}\,\times\fsample =
9\,714\,401\,280$.

\subsection{Noise}

The noise part was produced as a sum of white and correlated noise,
 \beq
    \bn = \bw + \bnc
 \eeq
where the correlated part ($1/f$ noise) was produced by a
stochastic-differential-equation (SDE) method, that produces noise
whose power spectrum is approximately of the form $P_c(f) \propto
f^{-\alpha}$. It is not of the form $\bF\ba$, but contains a part
that cannot be modeled with baselines.

The white noise rms was set to $\sigma = 2700\,\mu$K (thermodynamic
scale for CMB anisotropies), corresponding to the {\sc Planck} 70
GHz goal sensitivity (Planck Collaboration \cite{Bluebook}). The
$1/f$ noise was simulated with slope $-\alpha = -1.7$ and $\fmin =
1.15\times10^{-5}$ Hz (period of one day), so that the power
spectrum was flat for $f < \fmin$. See
Fig.~\ref{fig:noise_spectrum}. Since the white and $1/f$ streams
were produced separately, the knee frequency $\fk$ (where the white
and $1/f$ noise powers are equal) could be adjusted by multiplying
the $1/f$ stream with different factors. We used $\fk = 50$ mHz as
the reference case, representing a conservative upper limit for the
{\sc Planck} 70 GHz detectors (Burigana et al.~\cite{Burigana97},
Seiffert et al.~\cite{Seiffert02}, Tuovinen \cite{Tuovinen}), but
consider also $\fk = 25$ mHz in Sects.~\ref{sec:knee_frequency} and
\ref{sec:noise_prior} .

The power spectrum of the $1/f$ noise is thus approximately
 \beq
    P_c(f) = \left\{ \begin{array}{ll}
    \frac{\sigma^2}{f_c}
    \left(\frac{\fk}{f}\right)^\alpha & f \geq \fmin \\
    \frac{\sigma^2}{f_c}
    \left(\frac{\fk}{\fmin}\right)^\alpha & f \leq
    \fmin \,.
    \end{array} \right.
 \label{oof_noise_model}
 \eeq
where $f_c = \fsample/2 = 38.4$ Hz is the Nyquist (critical)
frequency.

The $1/f$ noise was generated in $\tgen = 15.25$ d pieces. The
actual statistics calculated from the simulated $1/f$ stream were:
mean $= -9.86\,\mu$K, stdev $= \sigma_c = 1868.5\,\mu$K, so that
$\sigma_c^2 = 0.48\sigma^2$. As can be seen from
Fig.~\ref{fig:noise_spectrum}, the simulated $1/f$ stream power
falls below the noise model for the lowest frequencies. A sizable
part of the variance in the $1/f$ stream comes from the very lowest
frequencies.

The nonzero mean of the $1/f$ stream has to be taken into account
when comparing solved baselines to the input $1/f$ stream, since the
destriping method sets the average of the solved baselines to zero.

\subsection{Maps}

The maps were produced in the
HEALPix\footnote{http://healpix.jpl.nasa.gov} pixelization
(G\'{o}rski et al.~\cite{Gorski05}) in ecliptic coordinates. We used
the $\Nside = 512$ resolution for all maps, corresponding to $n_p =
3\,145\,728$ for the full sky, with square root of pixel solid angle
$\theta_p \equiv \Omega_p^{1/2} = 6.87'$. For the full-year TOD the
polarization of each pixel was well sampled; the lowest {\em rcond}
was 0.422. The mean \emph{rcond} was 0.492 and the maximum 0.49999.
The hit count $\nhit$ (number of hits per pixel) varied from 818 to
273480 for the full 1-year simulation. The mean number of hits was
$\langle \nhit \rangle \equiv n_t/n_p = 3088.125$.  The mean inverse
hit count was $\langle \nhit^{-1} \rangle = 0.000418 = (0.0204)^2
\approx 1/2391$. Fig.~\ref{fig:hit_map_at_poles} shows the hit count
in the regions around the ecliptic poles, where it varies a lot.
Fig.~\ref{fig:10by10_hit_maps} shows the hit count in the two
$10^\circ\times10^\circ$ regions.

\begin{figure}[!tbp]
  \begin{center}
    \includegraphics*[trim=100 70 70 250, clip, width=0.24\textwidth]
    {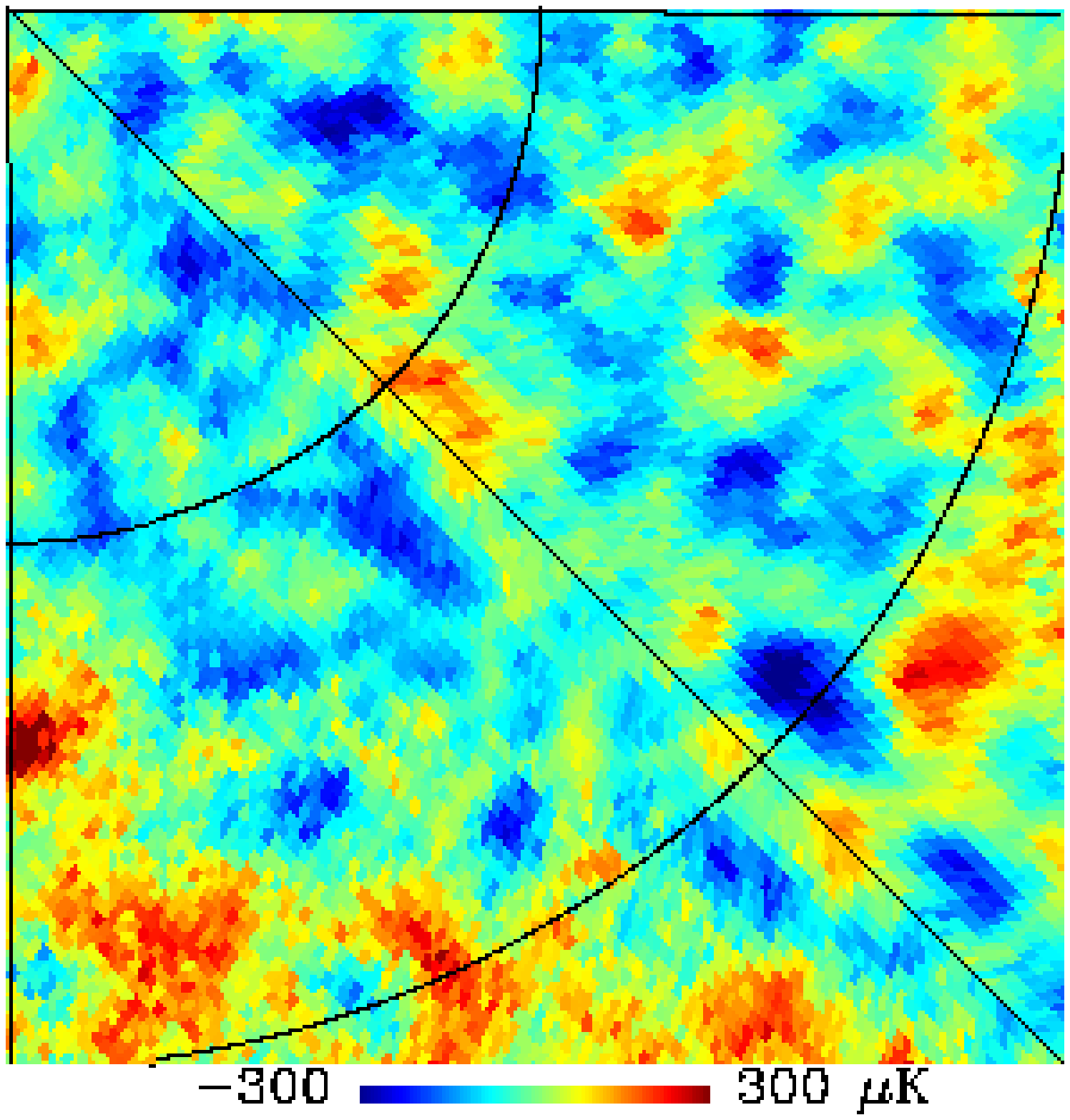}
    \includegraphics*[trim=100 70 70 250, clip, width=0.24\textwidth]
    {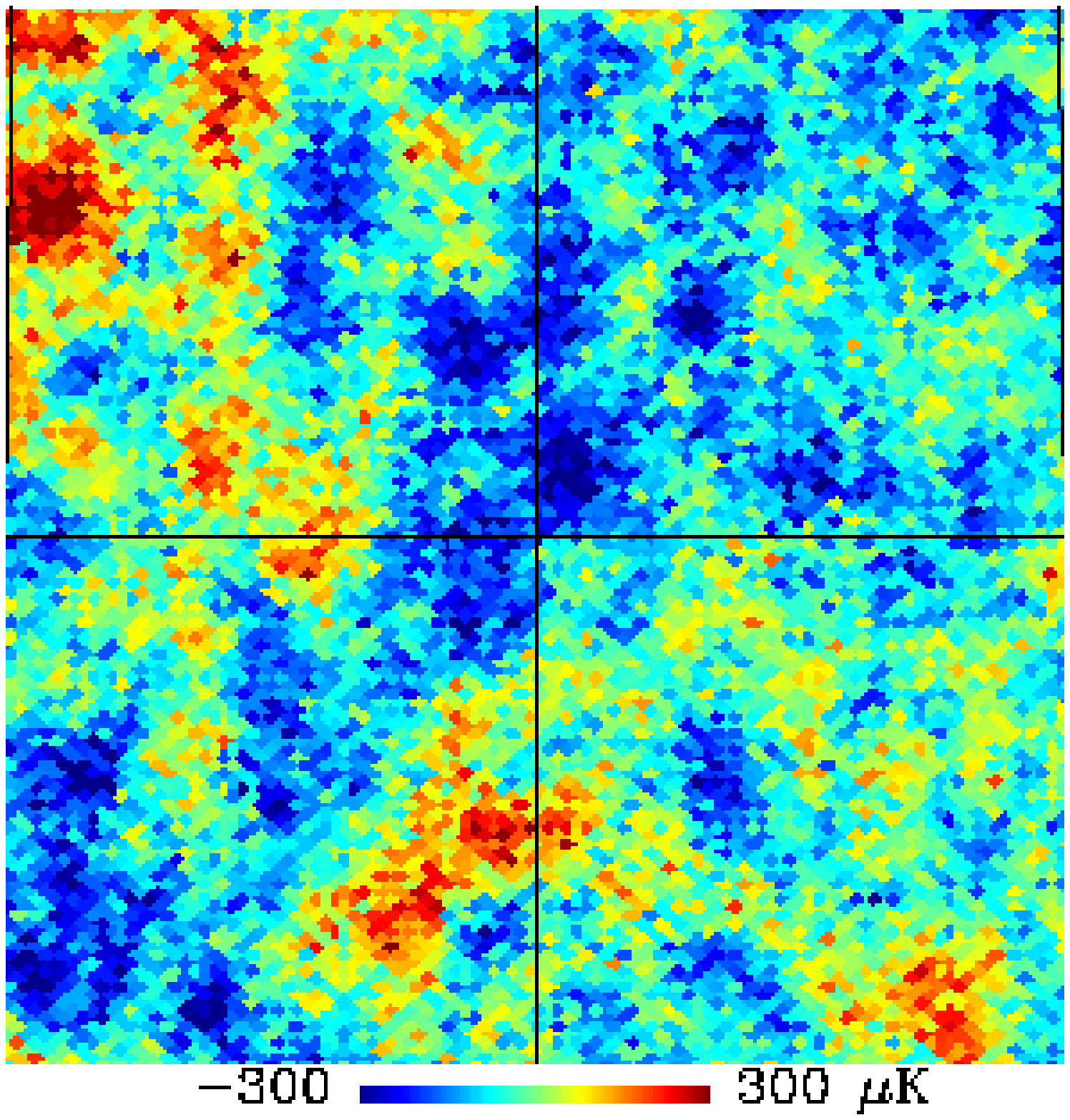}
    \vspace{-0.2cm}
    \caption{
Destriped (output) temperature map (one year survey, 15 s baselines)
for the two $10^\circ\times10^\circ$ regions.
    }
    \label{fig:output_I_10by10_map}
  \end{center}
\end{figure}

\begin{figure} [!tbp]
  \begin{center}
    \includegraphics*[trim=100 70 70 250, clip, width=0.24\textwidth]
    {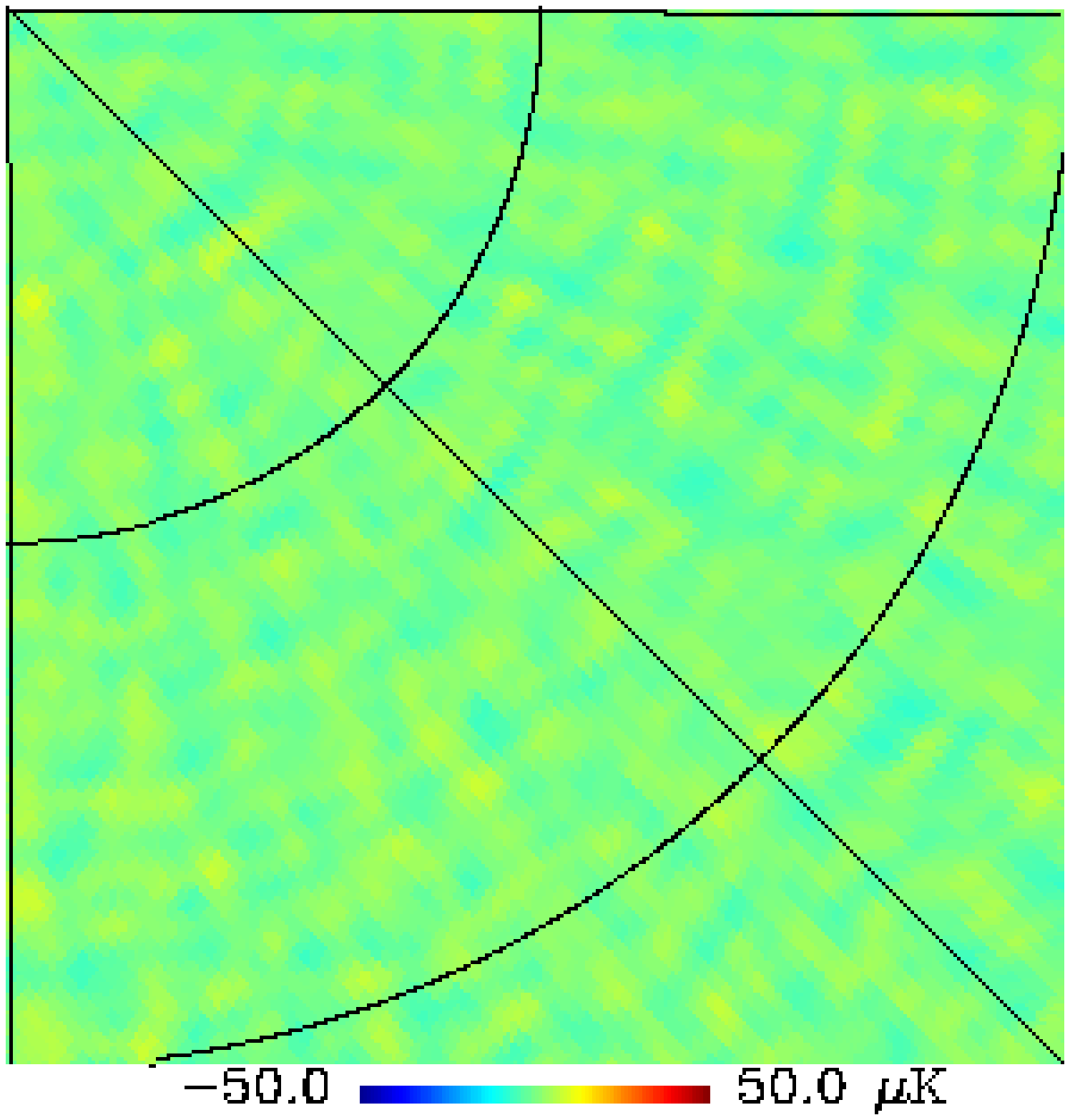}
    \includegraphics*[trim=100 70 70 250, clip, width=0.24\textwidth]
    {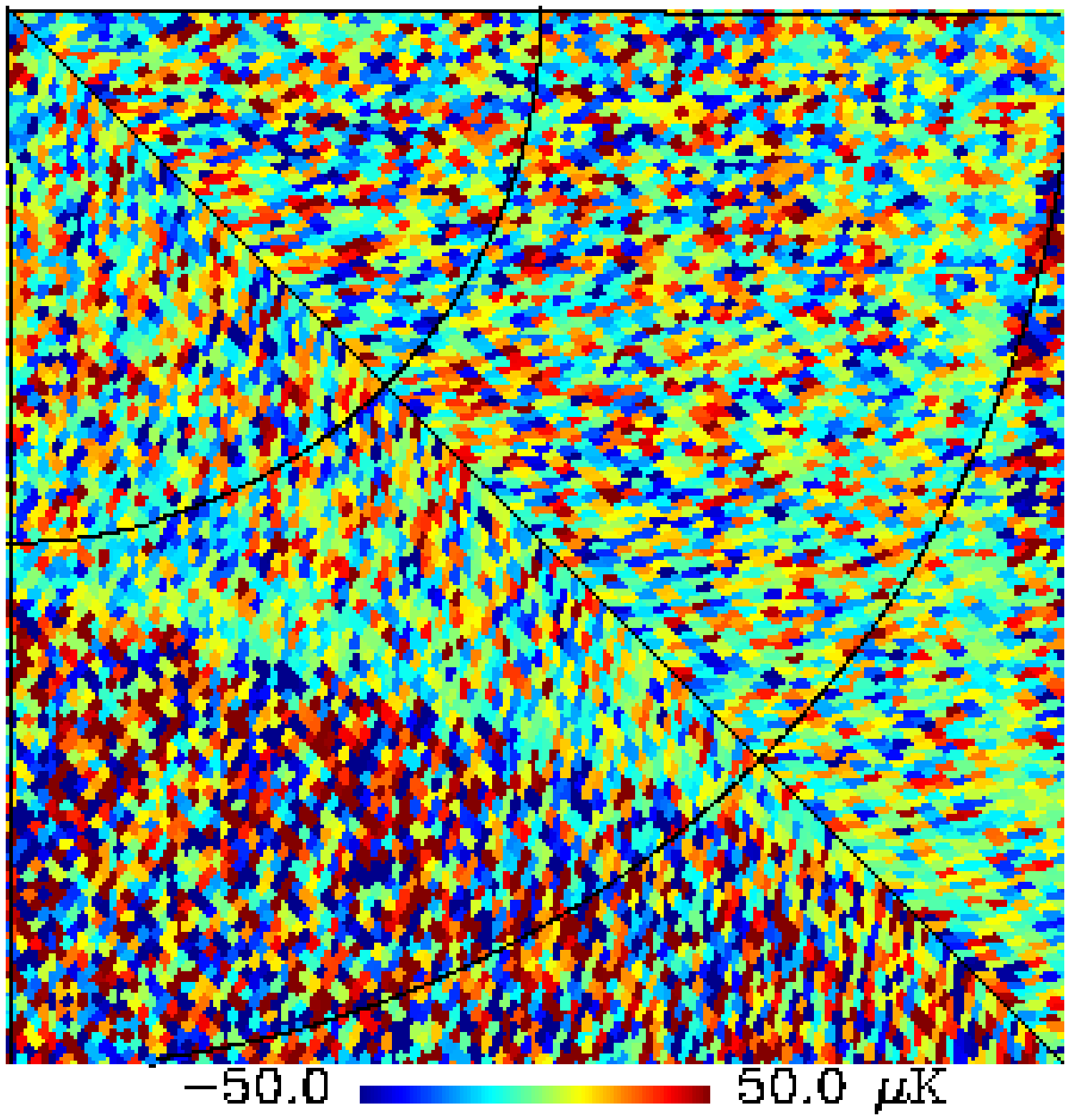}
    \vspace{-0.2cm}
    \caption{
Even near the ecliptic poles, where the noise in the output map is
the lowest, the pixel-scale noise from four 70 GHz detectors is
higher than the CMB polarization signal in the map. \emph{Left:}
Binned signal Q map. \emph{Right:} Output Q map (15 s baselines).
(Both plots are from the same pole region.)
    }
    \label{fig:output_Q_10by10_map}
  \end{center}
\end{figure}

Fig.~\ref{fig:output_I_10by10_map} shows the output temperature map
$\mout$ for the case $\tbase = 15$ s.  The visual appearance is the
same for other baseline lengths. To see differences one has to look
at residual maps, see Sect.~\ref{sec:map_domain}. Polarization maps
(Fig.~\ref{fig:output_Q_10by10_map}) are dominated by small-scale
noise.  The most obvious map-making related feature in the output
maps is the reduction of noise where the hit count is larger.  Other
effects are more subtle and are analyzed in the following sections.

 \section{Time domain}
 \label{sec:time_domain}

We now analyze the application of the destriping method to this kind
of data. We consider the case of the full 1-year survey with $\fk =
50$ mHz noise, except for Sect.~\ref{sec:knee_frequency}, where we
discuss the effect of changing the knee frequency, and for
Sect.~\ref{sec:partial_surveys}, where we consider maps made from
shorter pieces of the TOD. Consider this first in the time domain,
i.e., look at the cleaned TOD
 \beq
    \bd \equiv \by - \bF\aout \,.
 \eeq

We assume that the two noise streams, $\bw$ and $\bnc$ are
independent Gaussian random processes.

Since the destriping method is linear, we can divide the baseline
amplitudes obtained by Eq.~(\ref{eq:aout}) into the parts coming
from the different TOD components,
 \beq
    \aout = \bA\by = \bA\bs + \bA\bw + \bA\bnc \,.
 \eeq
Likewise, the cleaned TOD can be divided into five terms,
 \beq
    \bd = \bs + \bw  - \bF\bA\bw + \left(\bnc -
    \bF\bA\bnc\right)- \bF\bA\bs \,.
 \label{cleaned_tod}
 \eeq
The first term is the signal and the second term is the white noise.
We call the third term {\em white noise baselines}, the fourth term
(in parenthesis) {\em residual $1/f$ noise}, and the fifth term {\em
signal baselines}.

The TOD vector $\bZ\bs$ appearing in the signal baseline term
 \beq
    \bF\bA\bs \equiv \bF\bDi\bF^T\Cwi\bZ\bs
 \eeq
is the {\it pixelization noise} (Dor\'{e} et al.~\cite{Dor01}).  It
is the noise estimate we get from the signal TOD (which contains no
noise). Signal gradients (of the beam-smoothed input sky) within a
map pixel are the origin of the pixelization noise.

\subsection{Reference baselines}
\label{sec:ref_baselines}

We define the \emph{reference baselines} $\bR\bn$ of a noise stream
$\bn$ as the weighted averages of each baseline segment, i.e.,
matrix $\bR$ is defined as
 \beq
    \bR \equiv \left(\bF^T\Cwi\bF\right)^{-1}\bF^T\Cwi \,.
 \label{eq:Rdef}
 \eeq
Note that $\bR\bF = \bI$.

The reference baselines of the $1/f$ noise,
 \beq
    \aref \equiv \bR\bnc \,,
 \label{aref}
 \eeq
can be
viewed as the ``goal'' of baseline estimation.  Subtracting them
from the full TOD gives us a TOD stream
 \beq
    \by - \bF\aref = \bs + \bw + \left(\bnc-\bF\aref\right) \,,
 \eeq
which contains, beside the signal and the white noise, only the part
$\left(\bnc-\bF\aref\right)$ of correlated noise that cannot be
represented in terms of baselines.  We call this \emph{unmodeled}
$1/f$ noise.

The actual cleaned TOD that results from destriping, can now be
written as
 \beq
    \bd = \by - \bF\aout = \bs + \bw - \bF\bA\bw +
    \left(\bnc - \bF\aref\right) - \bF\left(\bA\bnc-\aref\right)
    - \bF\bA\bs \,,
 \eeq
where the residual $1/f$ noise is split into the unmodeled $1/f$
noise and the $1/f$ \emph{baseline error}, $\bA\bnc-\aref =
\left(\bA-\bR\right)\bnc$. Thus, altogether, there are three
contributions to missing the goal of baseline estimation,
 \beq
    \aout - \aref = \bA\bw + \left(\bA-\bR\right)\bnc + \bA\bs \,,
 \label{miss_goal}
 \eeq
white noise baselines, $1/f$ baseline error, and signal baselines.

Unlike the solved baseline contributions $\bA\bn$, the reference
baselines $\bR\bn$ do not involve the pointing matrix (except for
the case of pixelization noise), so the differences between them are
related to how the scanning strategy connects the baselines with
crossing points. Thus, for analyzing errors, it is useful to
separate also the white noise baselines into white noise reference
baselines and white noise baseline error, $\bA\bw = \bR\bw +
(\bA-\bR)\bw$; and likewise the signal baselines into reference
baselines of pixelization noise and signal baseline error $\bA\bs =
\bR\bZ\bs + (\bA-\bR)\bZ\bs$.

The white noise baseline error stream $(\bA-\bR)\bw$ is uncorrelated
with the white noise reference baseline stream $\bR\bw$.  To show
this, we note that
 \bea
    \bR\Cw\bA^T & = &
    \bR\Cw\left(\bI-\bB^T\bP^T\right)\Cwi\bF\bD^{-1}
    \nn\\
    & = & \bD^{-1} - \bR\bP\bB\bF\bD^{-1} \nn\\
    & = & \bD^{-1} -
    \left(\bF^T\Cwi\bF\right)^{-1}\left(\bF^T\Cwi\bF-\bD\right)\bD^{-1}
    \nn\\
    & = & \left(\bF^T\Cwi\bF\right)^{-1} = \bR\Cw\bR^T \,,
 \eea
so that
 \beq
    \left\langle(\bR\bw)\left((\bA-\bR)\bw\right)^T\right\rangle =
    \bR\Cw\bA^T - \bR\Cw\bR^T = 0
    \,.
 \eeq

\subsection{Approximation to solved baselines}

The solved baselines $\aout$ can now be written as
 \bea
    \aout & = & \aref + \bA\bs + \bA\bw + \bA(\bnc - \bF\aref) \nonumber\\
          & = & \aref +
          \left(\bF^T\Cwi\bZ\bF\right)^{-1}\bF^T\Cwi\bZ\bs +
          \nonumber\\ & & \mbox{} +
          \left(\bF^T\Cwi\bZ\bF\right)^{-1}\bF^T\Cwi\bZ\bw + \nonumber\\
          & & \mbox{} +
    \left(\bF^T\Cwi\bZ\bF\right)^{-1}\bF^T\Cwi\bZ(\bnc-\bF\aref)
 \eea
(up to an overall constant), so that
 \bea
    \bF^T\Cwi\bZ\bF(\aout-\aref) & = &
 \bF^T\Cwi\bZ\bs + \bF^T\Cwi\bZ\bw + \nonumber\\
 & & \mbox{}+ \bF^T\Cwi\bZ(\bnc-\bF\aref) \,.
 \eea
For the TOD streams $\bF(\aout-\aref)$, $\bw$, and $\bnc-\bF\aref$
there should be no significant correlations between the samples from
different circles that hit the same pixel.  (If the baseline length
is longer than one scanning circle, this statement is limited to the
samples that come from different baseline segments for
$\bF(\aout-\aref)$ and $\bnc-\bF\aref$.) Therefore, for a large
number of hits, the effect of the $\bP\bB$ part of $\bZ$ tends to
average out, leaving the $\bI$ part dominant, and we can approximate
$\bZ \approx \bI$, leading to $\bA \approx \bR$, for these terms. We
get the approximation
 \beq
    \aout \approx \aref + \bR\bw + \bR\bZ\bs \,.
 \label{approx}
 \eeq
Comparing to Eq.~(\ref{miss_goal}), we note that in
Eq.~(\ref{approx}) white noise baselines are approximated by white
noise reference baselines, $1/f$ baseline error is ignored, and
signal baselines are approximated by the reference baselines of
pixelization noise. Thus, for a good scanning, the solved baselines
$\aout$ should track the reference baselines of instrument noise +
pixelization noise.

\subsection{Division}
\label{sec:division}

The effect of destriping on the white noise is just harmful for the
maps (we elaborate on this in Sect.~\ref{sec:map_domain}), so the
relevant time domain residual is
 \bea
    \bd-\bs-\bw & = & \bnc-\bF\aout \nn\\
    & = & -\bF\bA\bw +(\bnc-\bF\bA\bnc) -\bF\bA\bs
    \,.
 \label{time_split}
 \eea
It consists of three components:
 \begin{enumerate}
 \item white noise baselines,
 \item residual $1/f$ noise,
 \item and signal baselines,
 \end{enumerate}
which are uncorrelated with each other.

Each component can be further divided into two parts:
 \bea
    \bd-\bs-\bw & = & -\bF\bR\bw - \bF(\bA-\bR)\bw \nn\\
    & & + (\bnc-\bF\aref) - \bF(\bA\bnc-\aref) \nn\\
    & & - \bF\bR\bZ\bs - \bF(\bA-\bR\bZ)\bs \,.
 \label{time_six}
 \eea
We call these six components:
 \begin{description}
 \item[$1\alpha$)] white noise reference baselines
 \item[$1\beta$)] white noise baseline error
 \item[$2\alpha$)] unmodeled $1/f$ noise
 \item[$2\beta$)] $1/f$ baseline error
 \item[$3\alpha$)] reference baselines of pixelization noise
 \item[$3\beta$)] signal baseline error
 \end{description}

Of these components, $1\alpha$, $1\beta$, 2 and 3 are uncorrelated
with each other.  This division forms the basis of the discussion in
the rest of this paper. Approximation (\ref{approx}) corresponds to
ignoring the $\beta$-components.  Components $1\alpha$ and $2\alpha$
are independent of scanning strategy, which makes their properties
easier to understand. The $\beta$-components are related to how the
baselines are solved using crossing points and thus couple to the
scanning strategy in a complicated manner.

We turn now to our results with simulated data to see how this comes
out in practice, first for the noise part, and then for the signal
part.

\begin{figure}[!tbp]
  \begin{center}
    \includegraphics*[width=0.5\textwidth]{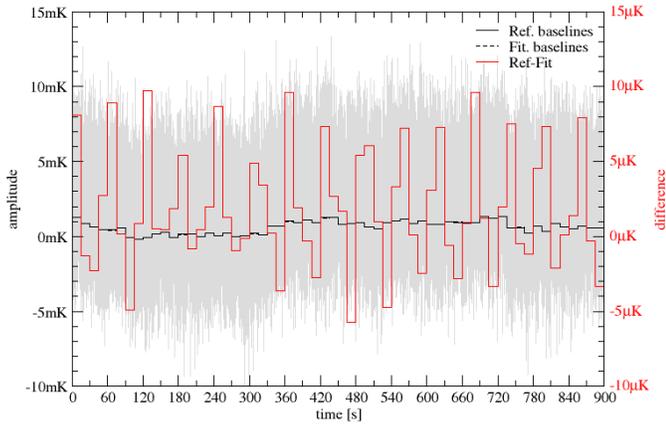}
  \end{center}
  \caption{First 15 minutes of the ($1/f$ + white) noise stream
   ({\em grey}) and its reference ({\em black solid}) and solved
({\em black dashed}) $\tbase = 15$ s baselines. Note that these two
baseline curves are nearly on top of each other. The difference of
the curves is plotted as the {\em red} curve and corresponds to the
scale on the right. This represents the error in the approximation
(\ref{approx}), except for the signal contribution to it.
 }
  \label{fig:baselines_noise_15s}
\end{figure}

In Fig.~\ref{fig:baselines_noise_15s} we show a part of the
simulated ($1/f$ + white) noise stream $\bn$, both the reference and
solved baselines, $\bF\bR\bn$ and $\bF\bA\bn$, and their difference
$\bF(\bR-\bA)\bn = \bF(\bA-\bR)\bnc + \bF(\bA-\bR)\bw$, for $\tbase
= 15$ s.

We consider now separately  the white noise and $1/f$ noise parts.

\begin{figure}[!tbp]
  \begin{center}
   \includegraphics*[width=0.5\textwidth]{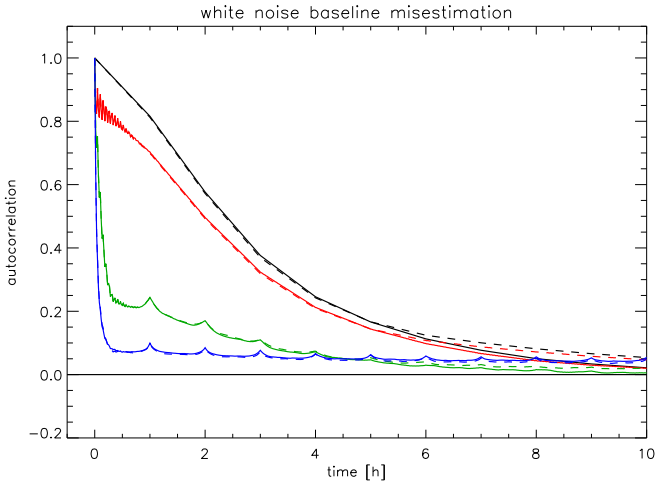}
  \end{center}
  \caption{
Autocorrelation function for the white noise baseline error
$(\bA-\bR)\bw$ for $\tbase = 1$ hour (\emph{black}), 1 min
(\emph{red}), 15 s (\emph{green}), 2.5 s (\emph{blue}).  The
\emph{solid} lines are for the ``temperature'' baselines $(a+b)/2$
and the \emph{dashed} lines are for the ``polarization'' baselines
$(a-b)/2$.  For the $\tbase = 15$ s and 2.5 s cases, only lags that
are multiples of 1 min ($\approx$ the spin period) are included.
 }
  \label{fig:white_aucorr}
\end{figure}

\begin{figure}[!tbp]
  \begin{center}
   \includegraphics*[width=0.5\textwidth]{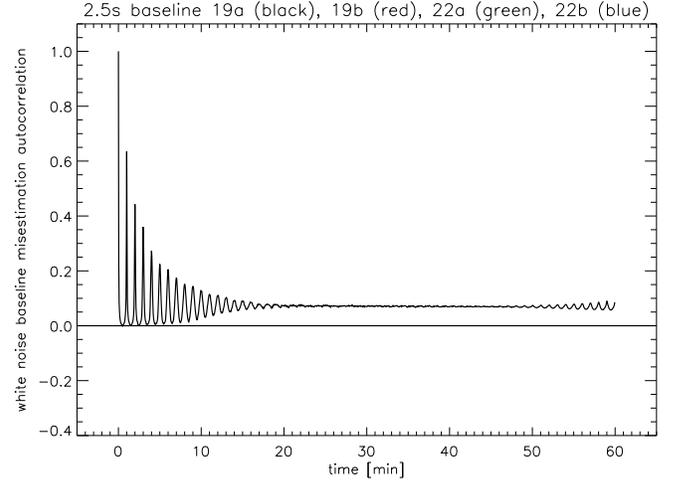}
  \end{center}
  \caption{
Autocorrelation function for the white noise baseline error
$(\bA-\bR)\bw$ for $\tbase = 2.5$ s and detector 19a. For short
lags, only baselines whose lag is a multiple of 1 min ($\approx$ the
spin period) show significant correlation with each other.  For
longer lags this distinction disappears due to random spin rate
variations. Since these variations have a rms which is about $1/60$
of the spin rate, for lags around 30 min any one of the baseline
segments within a spin period is about equally likely to land on a
given location of the scan circle.  The correlation between two
baselines that land on the same location of the circle is much
larger (presumably similar to the $\tbase = 1$ min and 1 hour
cases), but the way we calculate the autocorrelation function (in
time domain, not in the spin phase domain) is not able to pick this
out.
 }
  \label{fig:white_aucorr_2}
\end{figure}

\subsection{White noise baselines}
\label{sec:white_noise_baselines}

\begin{table}[!tbp]
 \begin{center}
 \begin{tabular}{rrrr} 
 $\tbase$ & Eq.~(\ref{apxavar}) & $\sigma_{wr}$ (reference) & $\sigma_{wb}$ (solved) \\
 \hline
 2.5 s   & 194.856 & 194.845 & 195.115 \\ 
 15 s    & 79.550 & 79.551 & 79.631 \\ 
 1 min   & 39.775 & 39.766 & 39.808 \\ 
 1 h     & 5.135 & 5.130 & 5.399 \\ 
 \hline
 \end{tabular}
 \end{center}
 \caption{
Statistics of the white noise baselines. We give the standard
deviation (square root of the variance) of the baselines (in
$\mu$K).
 }
 \label{table:wbl}
\end{table}

\begin{table}[!tbp]
 \begin{center}
 \begin{tabular}{llll}
 $\tbase$ & $a$ and $b$ & $(a+b)/2$ & $(a-b)/2$ \\
 \hline
 2.5 s   & 10.244 & 5.916 & 8.354 \\
 15 s    & 3.595  & 2.063 & 2.944 \\
 1 min   & 1.854  & 1.066 & 1.517 \\
 1 h     & 1.719  & 0.989 & 1.406 \\
 \hline
 \end{tabular}
 \end{center}
 \caption{
Statistics of the white noise baseline error. We give the standard
deviation $\sigma_{we}$ in $\mu$K.
 }
 \label{table:wblmis}
\end{table}

\begin{table}[!tbp]
 \begin{center}
 \begin{tabular}{llll}
 $\tbase$ & $(a+b)/2$ & $(a-b)/2$ & overlap\\
 \hline
 2.5 s   & 0.826 & -0.003 & 0.793 \\
 15 s    & 0.901 & 0.005  & 0.966 \\
 1 min   & 0.925 & -0.011 & 0.991 \\
 1 h     & 0.998 & -0.013 & 0.99986 \\
 \hline
 \end{tabular}
 \end{center}
 \caption{
Correlations between detectors 19 and 22 for white noise baseline
error.
 }
 \label{table:wblmiscor}
\end{table}

From Tables~\ref{table:wbl} and \ref{table:wblmis} we see that the
white noise baselines track their reference baselines well. The
white noise reference baselines are just white noise themselves,
their variance $\sigma_{wr}^2$ down from the white noise variance
$\sigma^2$ by $\nbase$.  The white noise baseline variance
 \beq
    \sigma_{wb}^2 \approx \sigma_{wr}^2 =
    \frac{\sigma^2}\nbase
    = \left(\frac{f_x}{f_c}\right)\sigma^2
    = \frac{\sigma^2}{f_s\tbase}  \,,
 \label{sigmawb}
 \eeq
is slightly larger.

In Table~\ref{table:wblmis} we show the statistics of white noise
baseline error.  We show it also for the average and the difference
between the two polarization directions $a$ and $b$, which represent
the contribution of this effect to the temperature and polarization
measurements.

The white noise reference baselines are completely uncorrelated with
each other. This is not true for the solved white noise baselines.
The difference, the baseline errors, show significant
autocorrelation, see Figs.~\ref{fig:white_aucorr} and
\ref{fig:white_aucorr_2}, and correlation between detectors 19 and
22, see Table~\ref{table:wblmiscor}. Although the white noise
baseline variance is not much larger than the white noise reference
baseline variance, these correlations make the difference between
them important.

These correlation properties are easy to understand.  While the
amplitude of the reference baseline arises from the noise of the
baseline segment itself, the error $(\bA-\bR)\bw$ is caused by the
noise in the crossing baselines.  Since the baseline segments that
are separated from each other by an integer number of spin periods,
and not too many pointing periods, cross almost the same set of
other baseline segments, often in the same pixels, their baseline
errors $(\bA-\bR)\bw$ are strongly correlated with each other. Also,
since the corresponding baseline segments from the horns 19 and 22
are only $\omega_\mathrm{scan}3.1^\circ$ = 0.517 s shifted from each
other, they crossed almost the same set of other baseline segments,
and are thus strongly correlated. The overlap fraction
$1-\mbox{0.517 s}/\tbase$ is given in Table~\ref{table:wblmiscor}.
Note that this number does not take into account that the hits from
the two horns may still be distributed differently to the pixels in
the overlap region.

More exactly, the ``temperature'' combinations $(a+b)/2$ of
$(\bA-\bR)\bw$ are strongly correlated between 19 and 22, whereas
the ``polarization'' combinations $(a-b)/2$ are not. The
polarization directions of the $ab$ pairs 19 and 22 differ from each
other by $44.8^\circ$ making their polarization measurements
$(a-b)/2$ almost orthogonal. Thus they also pick almost orthogonal
error combinations from the crossing baseline segments, and remain
uncorrelated.  This also explains why the stdev of the
``polarization baseline error'' is larger than the ``temperature''
one in Table~\ref{table:wblmis}. In effect, only half of the
crossing baseline pairs $ab$ contribute to determining an $(a-b)/2$
baseline combination, so the number of degrees of freedom is down by
$1/2$ and the variance thus larger by 2.

\subsection{Residual $1/f$ noise.}
\label{sec:residual_oof}

\begin{figure}[!tbp]
  \begin{center}
    \includegraphics*[width=0.5\textwidth]{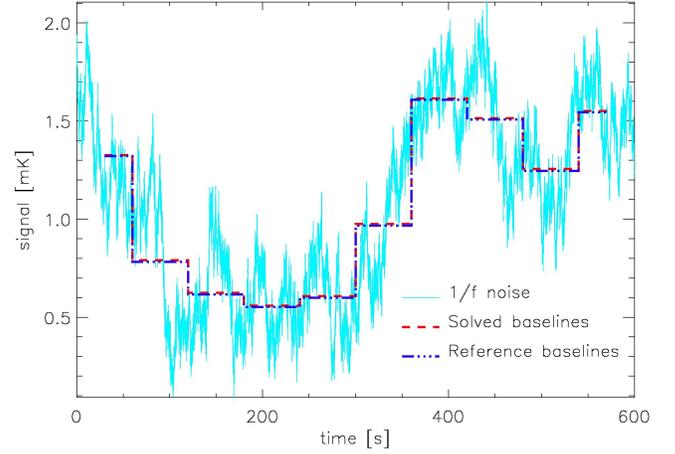}
  \end{center}
  \caption{
First 10 minutes of the $1/f$ noise stream $\bnc$ together with its
solved and reference baselines for $\tbase = 1$ min.  The difference
between the $1/f$ noise and its reference baselines is the unmodeled
$1/f$ noise.
 }
  \label{fig:baselines_oof}
\end{figure}

\begin{figure}[!tbp]
  \begin{center}
    \includegraphics*[width=0.5\textwidth]{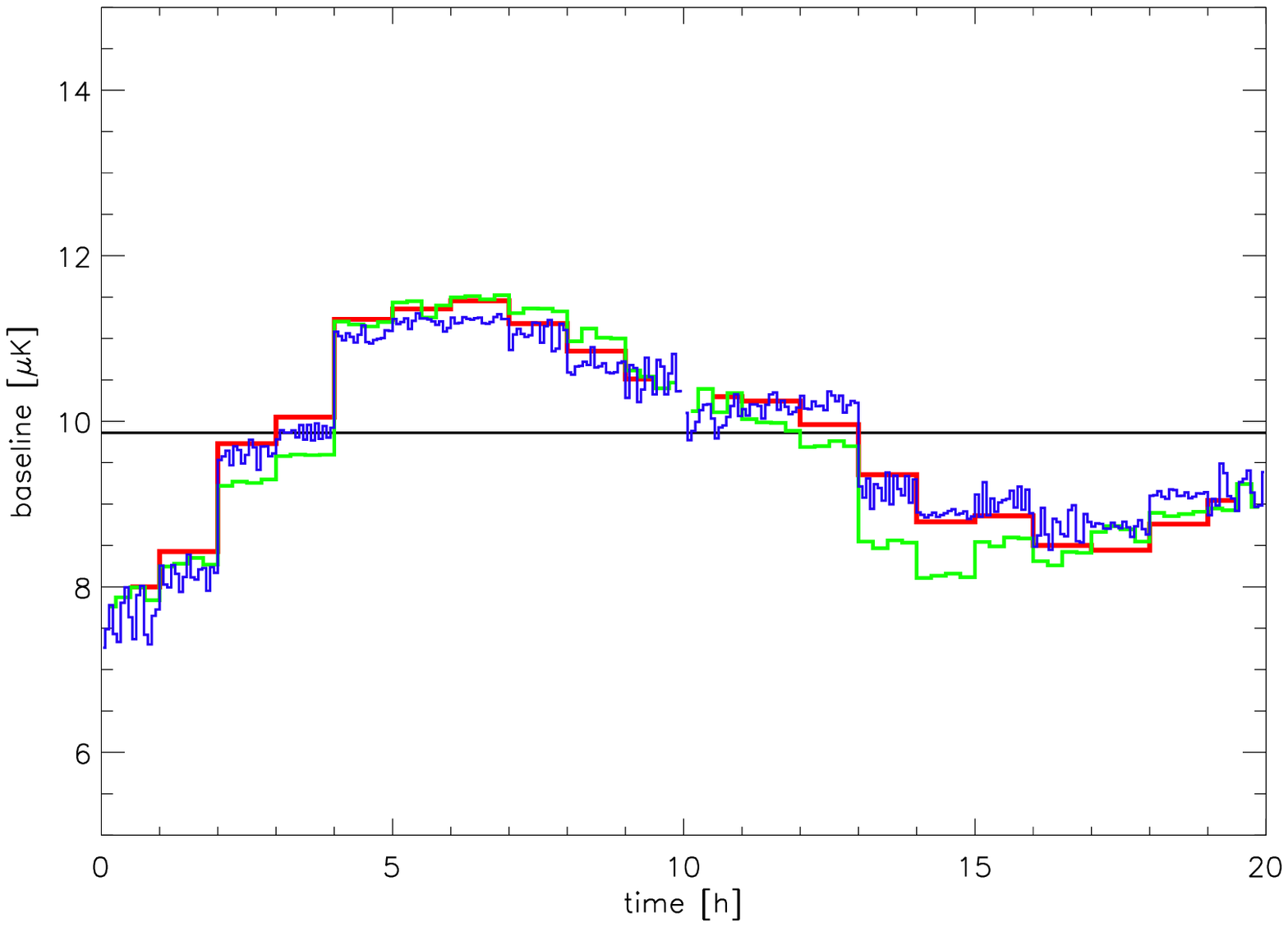}
    \includegraphics*[width=0.5\textwidth]{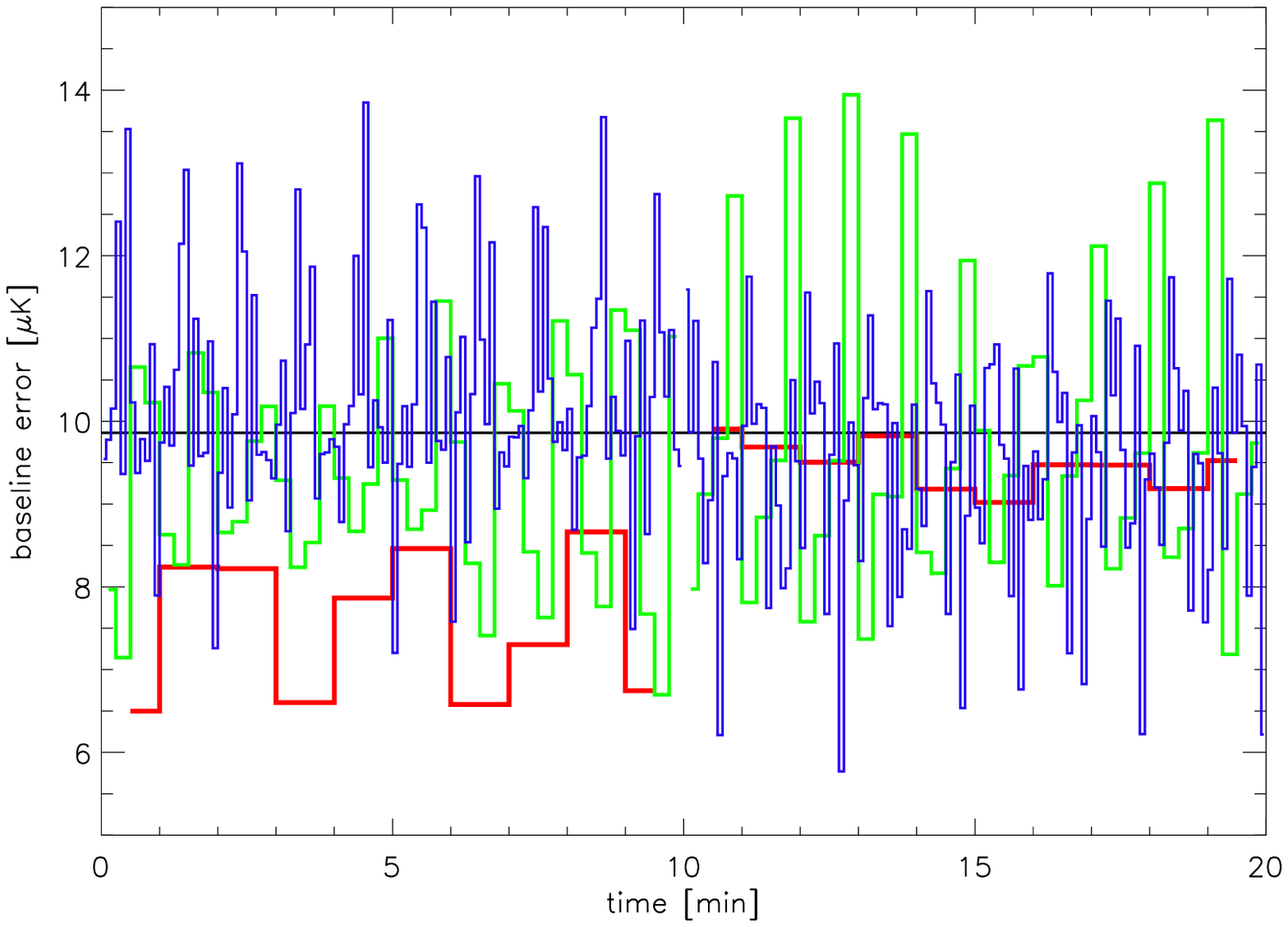}
  \end{center}
  \caption{
The $1/f$ baseline error, i.e., the difference between the solved
and reference $1/f$ baselines $\bF\bA\bnc-\bF\aref$.  The \emph{top}
panel shows the case for $\tbase= 1$h (\emph{red}), 15 min
(\emph{green}), and 4 min (\emph{blue}) from the first 10 hours and
from another set of 10 hours from the second month of the
simulation. The \emph{bottom} panel shows the case for $\tbase=
1$min (\emph{red}), 15 s (\emph{green}), and 5 s (\emph{blue}) from
the first 10 minutes and from another set of 10 minutes from the
second month of the simulation. The horizontal line at $9.86 \mu$K
shows the mean difference that is due to the nonzero mean of the
$1/f$ noise. Only the deviations from this mean are significant.
 }
  \label{fig:oof_baseline_error}
\end{figure}

Fig.~\ref{fig:baselines_oof} shows the $1/f$ part of the noise
$\bnc$, its reference baselines $\aref$ and the solved $1/f$
baselines, $\bA\bnc$. The difference between these two sets of
baselines, $1/f$ baseline error, is shown in
Fig.~\ref{fig:oof_baseline_error}.

The $1/f$ noise can be separated into reference baselines and
unmodeled $1/f$ noise, $\bnc = \bF\aref + (\bnc-\bF\aref)$. When we
separate the $1/f$ baseline error, $(\bA-\bR)\bnc$ into the
corresponding components,
 \beq
    (\bA-\bR)\bnc = (\bA-\bR)\bF\aref + \bA(\bnc-\bF\aref) \,,
 \eeq
we note that the first term on the right hand side gives the same
contribution to each baseline ($ = -$ the mean of the $1/f$ noise),
and is thus irrelevant.  Thus the $1/f$ baseline error arises from
the unmodeled $1/f$ noise, in the same manner as white noise
baseline error $(\bA-\bR)\bw$ arises from white noise. (The
reference baselines of unmodeled $1/f$ noise are zero).

\subsubsection{Unmodeled $1/f$ noise}

\begin{figure}[!tbp]
  \begin{center}
    \includegraphics*[width=0.5\textwidth]{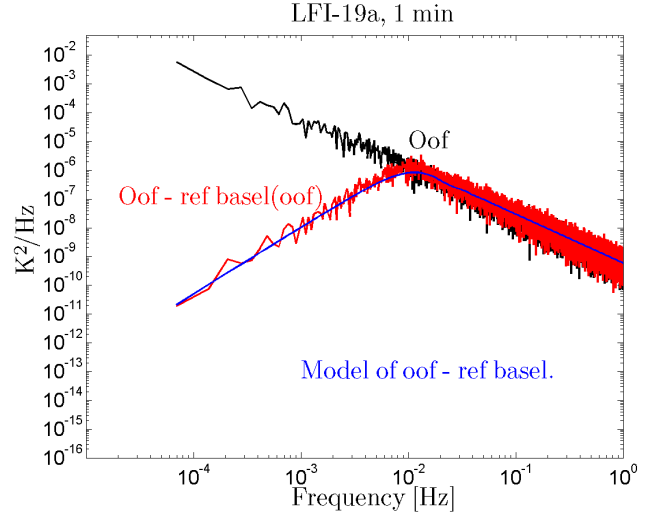}
  \end{center}
  \caption{
Power spectrum $P_c(f)$ of the $1/f$ noise (\emph{black}) and its
unmodeled part $P_u(f)$ (\emph{red}), for $\tbase = 1$ min.  The
\emph{blue} line is an analytical estimate corresponding to
Eqs.~(\ref{oof_noise_model}), (\ref{Pu_power}), and
(\ref{transferHf}).
  }
  \label{fig:unmod_pow_spec}
\end{figure}

\begin{figure}[!tbp]
  \begin{center}
   \includegraphics*[width=0.5\textwidth]{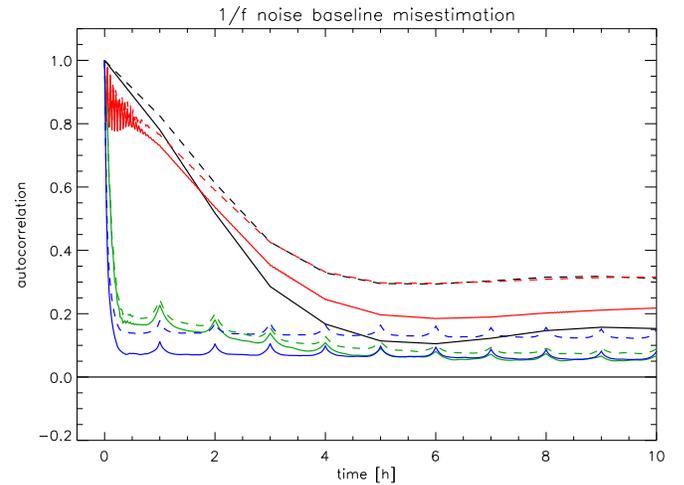}
  \end{center}
  \caption{
Same as Fig.~\ref{fig:white_aucorr}, but now for the $1/f$ noise
baseline error.
 }
  \label{fig:oof_aucorr}
\end{figure}

Since baselines can only model frequencies $\leq f_x \equiv
1/(2\tbase)$ and do it better for lower frequencies, the power
spectrum of the unmodeled $1/f$ noise $\bnc - \bF\aref$ is equal to
that of $\bnc$ for $f \gg f_x$ and falls rapidly towards smaller $f$
for $f < f_x$. See Fig.~\ref{fig:unmod_pow_spec}.

Since the reference baselines $\aref$ are obtained from the stream
$\bnc$, the power spectrum $P_u(f)$ of the unmodeled $1/f$ noise
$\bnc - \bF\aref$ can be obtained from the power spectrum $P_c(f)$
of $\bnc$ through a transfer function,
 \beq
    P_u(f) = H(f)P_c(f) \,.
 \label{Pu_power}
 \eeq
This transfer function is
 \beq
    H(f) = \left[1- \frac {\sin^2 (\pi f \tbase)} {(\pi f \tbase)^2} \right]^2
    \,.
 \label{transferHf}
 \eeq
For $f \ll f_x$, $H(f) \propto f^4$, so the slope of $P_u(f)$ is
$4-\alpha$ for $f \ll f_x$.

We get a rough estimate of the total power in $\bnc - \bF\aref$ by
integrating the original $P_c(f)$ from $f_x$ to $f_c$ and
$(f/f_x)^4P_c(f)$ from $\fmin$ to $f_x$,
 \bea
    \sigma_u^2 & \approx
    & \int_{\fmin}^{f_x} \left(\frac{f}{f_x}\right)^4 P_c(f) df
    + \int_{f_x}^{f_c} P_c(f) df \nonumber \\
    & \approx & \left\{\frac{1}{5-\alpha}+\frac{1}{\alpha-1}\left[1 -
    \left(\frac{f_x}{f_c}\right)^{\alpha-1}\right]\right\}\sigma^2
    \frac{\fk}{f_c}\left(\frac{\fk}{f_x}\right)^{\alpha-1} \nonumber
    \\
    & = & \sigma^2\left(\frac{\fk}{f_c}\right)^\alpha\left\{\frac{1}{5-\alpha}
    \left(\frac{f_c}{f_x}\right)^{\alpha-1}
    +\frac{1}{\alpha-1}\left[\left(\frac{f_c}{f_x}\right)^{\alpha-1}
    -1\right]\right\}  \,.
  \label{sigu_apx}
 \eea
For $\alpha = 1$ this becomes
 \beq
    \sigma_u^2 \approx \sigma^2\left(\frac{\fk}{f_c}\right)\left[\frac14
    +\ln\frac{f_c}{\fk}\right] \,.
 \label{sigu_apx_1}
 \eeq
For our case, $\alpha = 1.7$, $\sigma = 2700 \mu$K, and we can
approximate
 \beq
    1 - \left(\frac{f_x}{f_c}\right)^{\alpha-1} \approx 1 \,,
 \eeq
so that Eq.~(\ref{sigu_apx}) gives
 \beq
    \sigma_u \approx 57.3 \mu\,\mbox{K}\, \tbase(\mbox{s})^{0.35} \propto
\tbase^{(\alpha-1)/2} \,.
 \label{sigu_apx_num}
 \eeq
In Table~\ref{table:Pu_power} we compare this approximation to a
numerical integration of Eq.~(\ref{Pu_power}) and the actual
variance of $\bnc - \bF\aref$ from our simulated data.  We see that
Eq.~(\ref{sigu_apx_num}) is an underestimate.  An exception to this
is the 1-hour-baseline case; this is explained by the simulated
noise having less power at low frequencies than the noise model.

\begin{table}[!tbp]
 \begin{center}
 \begin{tabular}{llll}
 $\tbase$ & Eq.~(\ref{sigu_apx_num}) & Eq.~(\ref{Pu_power}) &
 actual \\
 \hline
 7.5 s   & 115.9 & 103.1 & 121.7 \\
 15 s    & 147.8 & 131.7 & 154.1 \\
 1 min   & 240.0 & 214.4 & 250.3 \\
 1 h     & 1006.0 & 897.3 & 938.0 \\
 \hline
 \end{tabular}
 \end{center}
 \caption{
The standard deviation $\sigma_u$ (square root of the variance) of
the unmodeled $1/f$ stream. Units are $\mu$K.  The third column is
from a numerical integration of Eq.~(\ref{Pu_power}). The last
column is the actual $\sigma_u$ calculated from the first 16 hours
of the simulated LFI-19a $1/f$ stream.
 }
 \label{table:Pu_power}
\end{table}

\subsubsection{$1/f$ baseline error}
\label{sec:oof_error}

Since the unmodeled $1/f$ noise is correlated, the properties of
$1/f$ baseline errors differ from white noise baseline errors. Over
timescales $\ll \tbase$ the unmodeled $1/f$ is positively
correlated, but since it averages to zero over each baseline
segment, there is a net anticorrelation.

From Fig.~\ref{fig:oof_baseline_error} we see for short baselines a
spin-synchronous pattern. For baselines of $\tbase = 1$ min, the
remaining pattern shows a 3-minute period that comes from spin-axis
nutation. For even longer baselines the long-time-scale correlation
shows up clearly. $1/f$ baseline error decreases for longer
baselines, but much less steeply than white noise baseline error.
See Table~\ref{table:oofbl_error}. Fig.~\ref{fig:oof_aucorr} shows
the $1/f$ baseline error autocorrelation. For white noise baseline
error we noted that we get a correlation for nearby baselines since
they often cross the same other baseline in the same pixel.  Since
unmodeled $1/f$ noise is positively correlated over short
timescales, it is not necessary for this crossing to occur in the
same pixel to get the positive correlation.  Therefore we now get
positive correlations for even longer timescales. This makes $1/f$
baseline error more important than its small variance suggests,
since it is not averaged away when binned onto the map.

\begin{figure}[!tbp]
  \begin{center}
    \includegraphics*[width=0.5\textwidth]{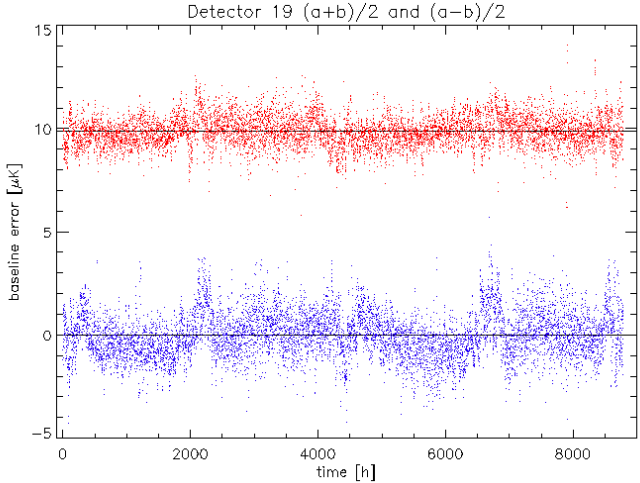}
  \end{center}
  \caption{
The $1/f$ baseline error, $\bF\bA\bnc-\bF\aref$, for 1-hour
baselines for the full 1-year simulation. We show the sum
(\emph{red}) and the difference (\emph{blue}) between the two
polarization directions. The times when ring crossings occur near
the corners of the caustics around the ecliptic poles are centered
roughly around hours 240, 1950, 4610, and 6340. Correlations in
baseline amplitudes are enhanced near these times.
 }
  \label{fig:oof_be_1_year}
\end{figure}

In Fig.~\ref{fig:oof_be_1_year} we show the $1/f$ baseline error for
$\tbase = 1$ h over the full year of the simulation for the detector
pair 19. The effect of the 6-month period of the cycloidal scanning
is clearly visible.

\begin{table}[!tbp]
 \begin{center}
 \begin{tabular}{llll}
 $\tbase$ & $a$ and $b$ & $(a+b)/2$ & $(a-b)/2$ \\
 \hline
 5 s     & 1.790 & 0.995 & 1.486 \\
 15 s    & 1.894 & 1.075 & 1.557 \\
 1 min   & 1.404 & 0.765 & 1.177 \\
 1 h     & 1.360 & 0.726 & 1.142 \\
 \hline
 \end{tabular}
 \end{center}
 \caption{
Statistics of the $1/f$ baseline error. We give the standard
deviation $\sigma_{ub}$ of the baseline difference $\bA\bnc -
\bR\bnc$ in $\mu$K.
 }
 \label{table:oofbl_error}
\end{table}

\subsection{Noise power spectra}
\label{sec:noise_spectra}

\begin{figure}[!tbp]
  \begin{center}
    \includegraphics*[width=0.5\textwidth]
      {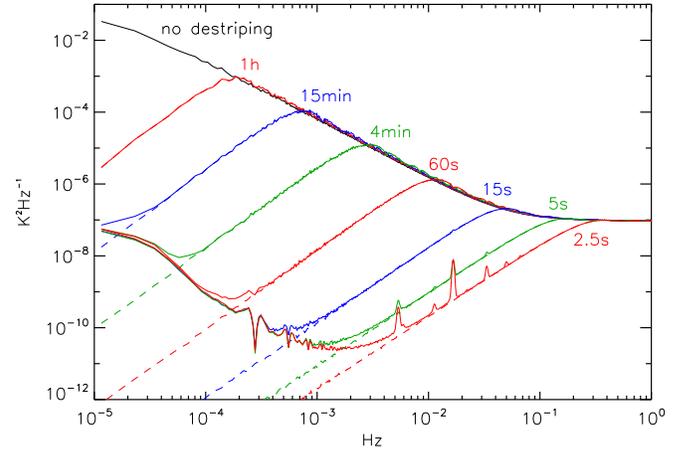}
    \caption[TOD spectra]{Effect of baseline subtraction on the
noise power spectrum.  The {\em solid black} line is the spectrum of
the original noise stream $\bn$.  The \emph{colored} lines show the
spectra of cleaned noise streams for different baseline lengths.
\emph{Solid} lines are for the case of subtracting the solved
baselines, $\bn-\bF\bA\bn$, and the \emph{dashed} lines for
subtracting reference baselines, $\bn-\bF\bR\bn$.
    }
    \label{fig:TOD_spectra}
  \end{center}
\end{figure}

We show the power spectra of the cleaned (destriped) noise streams
$\bn-\bF\bA\bn$ for different baseline lengths in
Fig.~\ref{fig:TOD_spectra}.  They are compared to the spectrum of
the original noise stream $\bn$ and the spectra where noise
reference baselines are subtracted instead, $\bn-\bF\bR\bn$.

Subtraction of baselines suppresses noise at $f \lesssim 1/\tbase$.
For lower frequencies the noise is suppressed more, as baselines can
model the lower frequencies better.  When reference baselines are
subtracted, the noise power keeps going down toward lower
frequencies; however the solved baselines seem to be able to
suppress noise power about 6 orders of magnitude only.

For shorter baselines, spectral features appear at special
frequencies. They do not appear when reference baselines are
subtracted, so they are clearly related to the scanning strategy.
1) There are peaks at $f =$ 1/min and $f =$ 1/(3 min) corresponding
to the spin and nutation frequencies, and their harmonics.  2) There
is a notch in power at $f =$ 1/h, corresponding to the repointing
period, and its harmonics. These are easy to understand:

The solved baselines come from a noise estimate based on subtracting
from each sample the average of all samples that hit the same pixel.
Consider an ideal scanning where the same pixel sequence is hit
during each spin period within a repointing period:

1) If the spin period is equal to or a multiple of the period of a
noise frequency component, all samples hitting the same pixel during
a given repointing period get the same value from this noise
component.  Thus this noise component can be detected as noise (and
not signal) only by comparing hits from different repointing
periods, resulting in a much poorer noise estimate.

2) On the other hand, if the repointing period is equal to or a
multiple of the noise period, but the spin period is not, then the
different samples hitting the same pixel average to zero.  This
noise component is then recognized as noise in its entirety, and the
solved baseline becomes equal to the reference baseline.

Since in our simulation the scanning deviated from ideal, these
features are weakened, but still clearly visible.

\begin{figure}[t]
  \begin{center}
    \includegraphics*[width=0.5\textwidth]
      {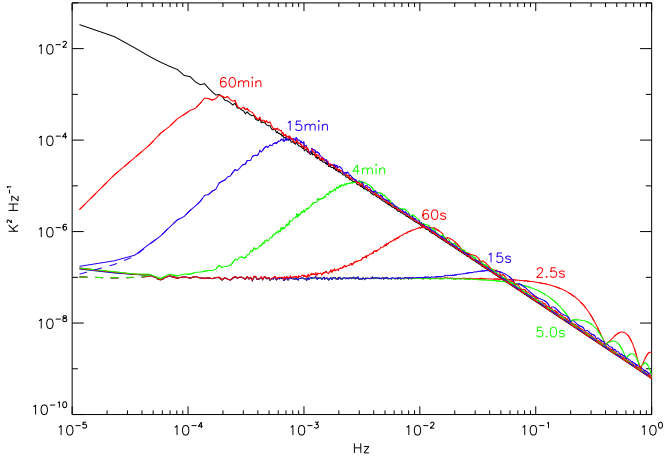}
    \caption[oof minus full baselines]{The effect of subtracting both
the white noise and the $1/f$ baselines on the power spectrum of the
$1/f$ noise.  The \emph{solid black} line is the original $1/f$
spectrum. The \emph{solid colored} lines show the power spectra of
$\bnc-\bF\bA\bn$ and the \emph{dashes} lines the power spectra of
$\bnc-\bF\bR\bn$.
    }
    \label{fig:oof_minus_full_baselines}
  \end{center}
\end{figure}

As mentioned in Sect.~\ref{sec:intro} and Sect.~\ref{sec:division}
and elaborated in Sect.~\ref{sec:map_domain}, the relevant residual
noise is the stream $\bnc-\bF\bA\bn$ where both white noise and
$1/f$ baselines are subtracted from the $1/f$ noise stream. This is
shown in Fig.~\ref{fig:oof_minus_full_baselines}.  The subtraction
of the white noise baselines has added power to the cleaned $1/f$
stream. At low frequencies the spectrum appears now white, since the
white noise reference baseline stream is white at timescales longer
than $\tbase$. The power of the baseline error rises towards the
lowest frequencies and shows up below $5\times10^{-5}$ Hz. The
subtraction of uniform baselines from the noise stream has a
chopping effect that transforms a part of the low frequency, $f <
1/\tbase$, power into high frequency, $f \gtrsim 1/\tbase$.

\subsection{Optimal baseline length}
\label{sec:optimal}

What is the optimal baseline length to use?  A partial answer can be
found by minimizing the variance of the residual correlated noise,
i.e., the residual noise minus the original white noise,
 \beq
    \bnc-\bF\bA\bn = (\bnc-\bF\bR\bnc) - \bF\bR\bw - \bF(\bA-\bR)\bw
    - \bF(\bA-\bR)\bnc
 \eeq
The variances of the four components are $\sigma_u^2$,
$\sigma_{wr}^2$, $\sigma_{we}^2$, and $\sigma_{ub}^2$.  The first
three components are uncorrelated, so that their total variance is
$\sigma_{wr}^2 + \sigma_{we}^2 + \sigma_u^2$.  Since $\sigma_{wm}^2$
and $\sigma_{ub}^2$ are much smaller than $\sigma_{wb}^2 +
\sigma_u^2$, we make the approximation where we drop the last two
components.

Using Eqs.~(\ref{sigmawb}) and (\ref{sigu_apx}) we find that
$\sigma_{wr}^2 + \sigma_u^2$ is minimized for
 \beq
    f_x = \frac{1}{2\tbase} =
    \left(1+\frac{\alpha-1}{5-\alpha}\right)^{1/\alpha} \fk
 \label{opt_tbase}
 \eeq
for which
 \beq
   \frac{\sigma_{wr}^2 + \sigma_u^2}{\sigma^2}  \approx
   \left(\frac{4}{5-\alpha}\right)^{1/\alpha}\frac{\alpha}{\alpha-1}
   \left(\frac{\fk}{f_c}\right) -
   \frac{1}{\alpha-1}\left(\frac{\fk}{f_c}\right)^\alpha \,.
 \eeq

For $\alpha = 1$ this becomes
 \beq
    f_x = \frac{1}{2\tbase} = \fk
 \label{opt_tbase_1}
 \eeq
and
 \beq
   \frac{\sigma_{wr}^2 + \sigma_u^2}{\sigma^2}  \approx
   \left(\frac54 + \ln \frac{f_c}{\fk}\right)\frac{\fk}{f_c} \,.
 \eeq

For $\fk = 50$ mHz and $\alpha = 1.7$ this gives $\tbase = 8.93$ s,
for which $\sqrt{\sigma_{wr}^2 + \sigma_u^2} = 0.05936\sigma = 160.3
\,\mu$K. This result does not take into account baseline errors and
signal baselines (see Sect.~\ref{sec:signal_baselines}).

However, because of the different correlation properties of the
different residuals, they have a different impact in map-making, and
it is not enough to consider the time-domain variance. So we need to
return to this issue later, after we have studied the residuals in
the map domain and their angular power spectra.

\subsection{Pixelization noise and signal baselines}
\label{sec:signal_baselines}

\begin{figure}[!tbp]
  \begin{center}
    \includegraphics*[width=0.5\textwidth]{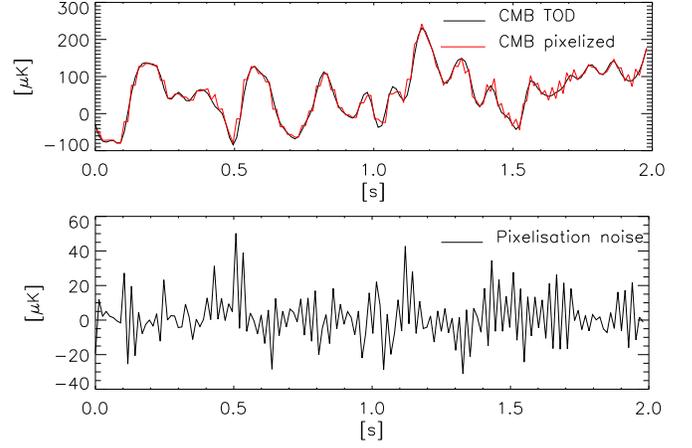}
  \end{center}
  \caption{
Pixelization noise. In the top panel we show a piece of the original
signal stream $\bs$ (\emph{black}) and the pixelized signal stream
$\bP\bB\bs$ (\emph{red}).  Their difference $\bZ\bs = \bs -
\bP\bB\bs$ is the pixelization noise (\emph{bottom panel}).
  }
  \label{fig:pixelization_noise}
\end{figure}

\begin{figure}[!tbp]
  \begin{center}
    \includegraphics*[width=0.5\textwidth]{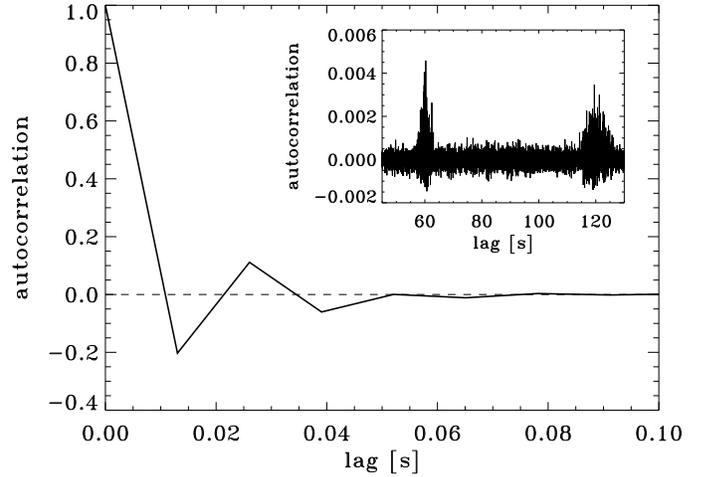}
  \end{center}
  \caption{
Autocorrelation function of the pixelization noise.  Note the
alternating correlation and anticorrelation for small lags, and the
correlations at 1 minute lag, when the scanning returns to the same
location on the sky. Due to spin rate variations the correlations
are spread around the 1 minute value, and more around 2 minute.
  }
  \label{fig:pixelization_autoc}
\end{figure}

Pixelization noise, $\bZ\bs \equiv \bs - \bP\bB\bs$, arises from
signal gradients through a combination of pixelization, scanning
strategy, and sampling frequency.  These lead to correlations
between close-by samples, and also correlations between samples from
the same locations on the sky which are not close to each other in
time domain.

In Fig.~\ref{fig:pixelization_noise} we show a short piece of the
pixelization noise.  Its autocorrelation function is shown in
Fig.~\ref{fig:pixelization_autoc}.  We see that  neighboring samples
(lag $= 1 = (1/76.8)$ s) are anticorrelated, whereas we have a
positive correlation for lag $= 2$. The power spectrum is close to
that of white noise, except that there are some features near the
Nyquist frequency due to these correlations.

Comparing the pixel size $\theta_p = 6.87'$ to the sample separation
$\thetas = 4.68'$ and remembering that the scanning direction is
mostly close to the direction of the pixel diagonal (both are often
close to the ecliptic meridians), so that the pixel geometry tends
to repeat at $ \sim \sqrt{2}\theta_p = 9.71'$ intervals, close to
$2\thetas = 9.38'$, we see that a pair of samples with lag 2 tends
to land in about the same relative location within their respective
pixels. With this small separation there is a positive correlation
in the CMB signal gradient (smoothed with the beam). These combine
to give a positive correlation between lag 2 samples. On the other
hand, neighboring samples often land at opposite sides of the same
pixel, or of neighboring pixels, leading to a negative correlation
in their pixelization noise. These correlations would be different
for different ratios of sample separation to pixel size.

Since the pixelization noise arises from the signal, we analyze it
in the combinations $\bs_+ \equiv (\bs_a+\bs_b)/2$ (``temperature'')
and $\bs_- \equiv (\bs_a-\bs_b)/2$ (``polarization''), so that
 \beq
    s_+ = I  \qquad\mbox{and}\qquad s_- = Q\cos2\psi_a +
    U\sin2\psi_a  \,.
 \eeq
Each of the $\bs_+$ and $\bs_-$ time streams contains $n_t/2$
samples.  For the ideal detectors considered here, $\bs_+$ arises
from gradients in the intensity of the signal, and $\bs_-$ from
gradients in the polarization of the signal.  We assume the signal
is statistically isotropic, and that $C_\ell^{EB} = 0$.

We find that the expectation value for the variance of the
pixelization noise is
 \bea
    \sigma_p^2(\mbox{temp})
    & \approx & (\Gamma-\Lambda)\sum_\ell
    \frac{(2\ell+1)\ell(\ell+1)}{8\pi}B_\ell^2C_\ell^{TT} \nn\\
    \sigma_p^2(\mbox{pol})
    & \approx & (\Gamma-\Lambda)\sum_\ell
    \frac{(2\ell+1)\left[(\ell-2)(\ell+3)+2\right]}{16\pi} \times
    \nn\\
    & & \times B_\ell^2\left(C_\ell^{EE}+C_\ell^{BB}\right)
 \label{sigp_anal}
 \eea
where
 \beq
    \Gamma  =  \frac{1}{n_t}\sum_p\sum_{t\in p}|\mathbf{r}_t|^2
    \,, \qquad
    \Lambda  =  \frac{1}{n_t}\sum_p
    \frac{1}{n_{\mathrm{hit},p}}\sum_{tt'\in p}\mathbf{r}_t\cdot
    \mathbf{r}_{t'}\,,
 \eeq
and $\mathbf{r}_t \equiv \hat{\mathbf{n}}_t-\hat{\mathbf{u}}_p $.
Here $\hat{\mathbf{n}}_t$ is the unit vector giving the pointing
$(\theta_t,\phi_t)$ of sample $t$, and $\hat{\mathbf{u}}_p$ is the
unit vector pointing to the center of the pixel $p$ hit by sample
$t$.  The approximation in Eq.~(\ref{sigp_anal}) is the small-pixel
approximation $|\mathbf{r}_t| \ll 1$.  For $\sigma_p(\mbox{pol})$ we
have also assumed an optimal distribution of polarization directions
$\psi_t$ within each pixel.

This result, Eq.~(\ref{sigp_anal}), can be compared to the variance
of the signal itself
 \bea
    \sigma_s^2(\mbox{temp})
    & \equiv & \langle I(\hat{\mathbf{n}})I(\hat{\mathbf{n}})\rangle
    = \sum_\ell \frac{2\ell+1}{4\pi} C_\ell^{TT} \nn\\
   \sigma_s^2(\mbox{pol})
    & \equiv & \langle Q(\hat{\mathbf{n}})Q(\hat{\mathbf{n}})\rangle
    = \langle U(\hat{\mathbf{n}})U(\hat{\mathbf{n}})\rangle \nn\\
    & = & \sum_\ell \frac{2\ell+1}{8\pi}
    \left(C_\ell^{EE}+C_\ell^{BB}\right) \,.
   \label{sigs_anal}
 \eea

Assuming the pixels are perfect squares, in the limit of a large
number of hits $n_{\mathrm{hit},p}$ uniformly distributed over the
pixel we get
 \beq
    \Gamma \approx \frac{\Omega_p}{6} \qquad\mbox{and}\qquad
    \Lambda \ll \Gamma
 \eeq
Since HEALPix pixels are not square, but somewhat elongated, we
expect the actual $\Gamma-\Lambda$ to be somewhat larger.  In
principle it can be calculated from the pointing data for a chosen
pixelization.

For our input $C_\ell$, setting $\Gamma-\Lambda = \Omega_p/6$,
Eqs.~(\ref{sigp_anal}) and (\ref{sigs_anal}) give
 $\sigma_s(\mbox{temp}) = 100.792\,\mu$K,
 $\sigma_p(\mbox{temp}) = 13.332\,\mu$K,
 $\sigma_s(\mbox{pol}) = 2.530\,\mu$K, and
 $\sigma_p(\mbox{pol}) = 0.781\,\mu$K.

The actual pixelization noise level in the simulation was
 $\sigma_p(\mbox{temp}) = 13.516\,\mu$K and
 $\sigma_p(\mbox{pol}) = 0.776\,\mu$K.

\begin{figure}[!tbp]
  \begin{center}
    \includegraphics*[trim=100 70 70 250, clip, width=0.24\textwidth]
    {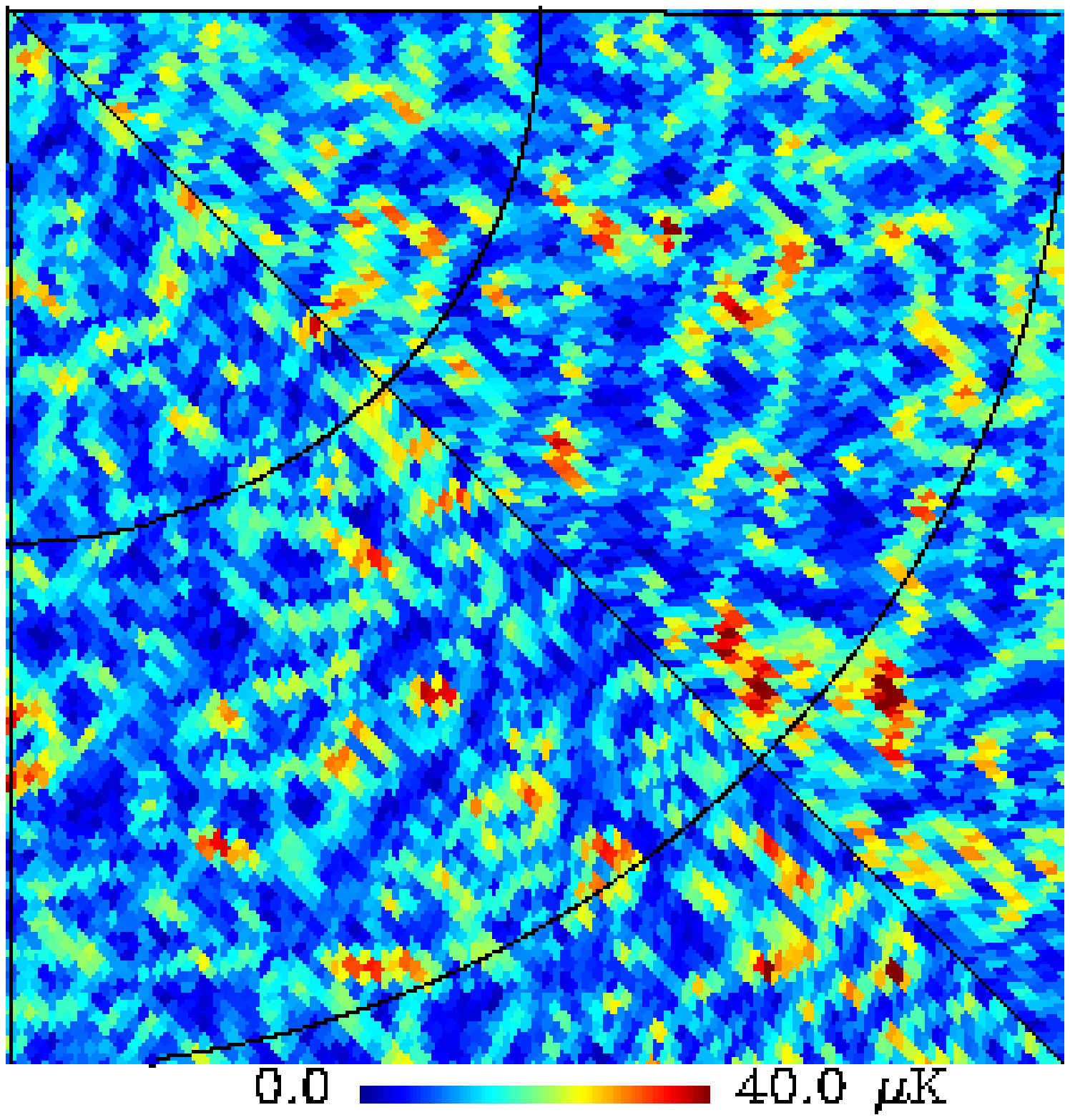}
    \includegraphics*[trim=100 70 70 250, clip, width=0.24\textwidth]
    {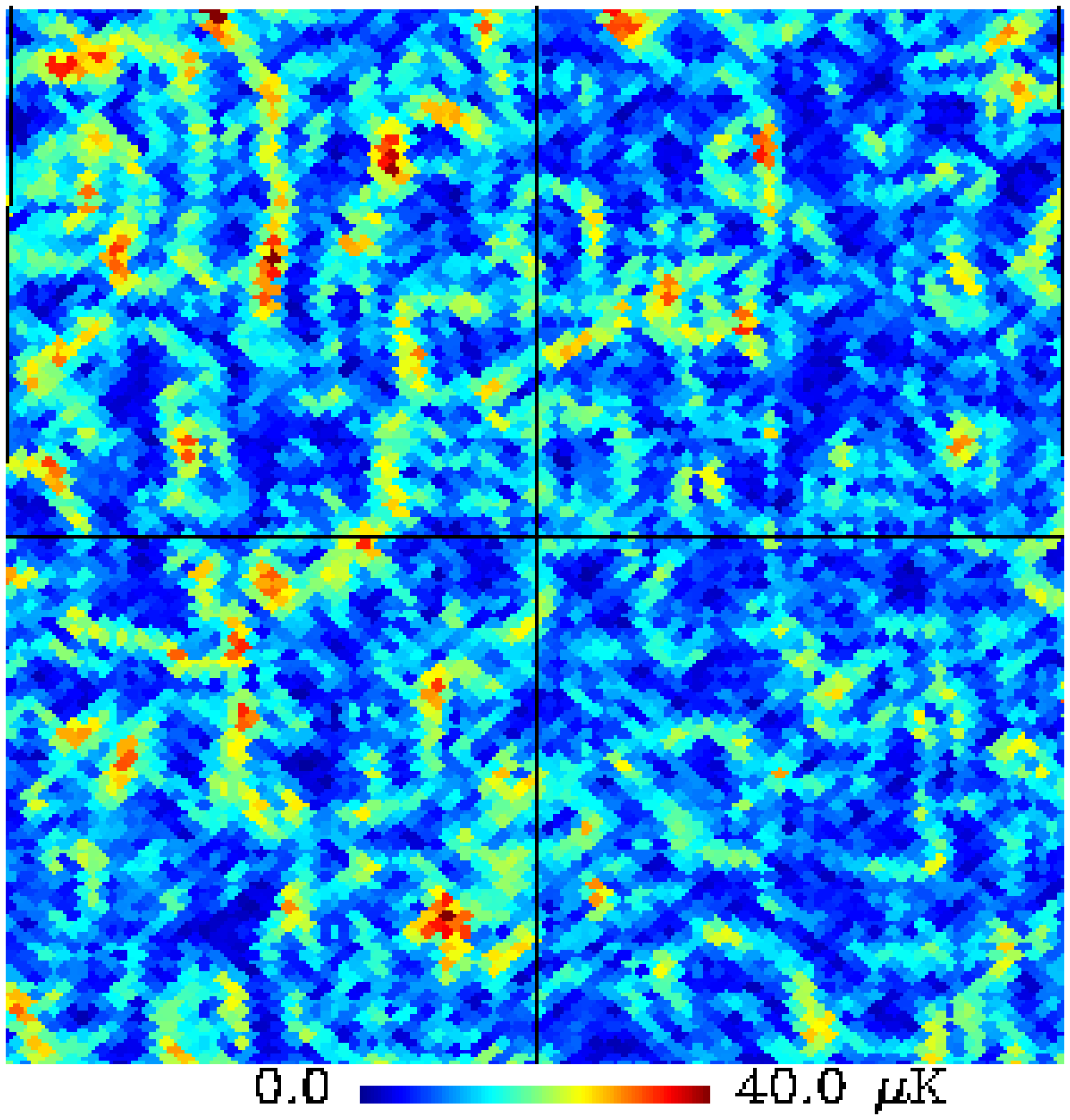}
    \vspace{-0.2cm}
    \caption{
Pixelization noise rms in map pixels for the two
$10^\circ\times10^\circ$ regions.
    }
    \label{fig:pix_std}
  \end{center}
\end{figure}

By definition, a binned map of pixelization noise vanishes,
$\bB\bZ\bs = \bB\bs - \bB\bP\bB\bs = 0$.  Instead we can make a map
of the rms of the pixelization noise at each pixel, by squaring each
element of $\bZ\bs$, binning the resulting time stream into a map,
and taking the square root of each $I$ pixel.  See
Fig.~\ref{fig:pix_std}.  Comparing to
Fig.~\ref{fig:10by10_signal_maps} we see that pixels of large
pixelization noise tend to ``outline'' hot and cold spots of the
signal.

Signal baselines arise from the pixelization noise in the same
manner as white noise baselines arise from white noise, and we can
divide them into the reference baselines of pixelization noise, and
signal baseline errors, $\bA\bs = \bR\bZ\bs + (\bA-\bR\bZ)\bs$.
Approximating the pixelization noise as white, we get an estimate
for the standard deviation (stdev) of $\bR\bZ\bs$
 \bea
    \sigma_{pr}(\mathrm{temp}) & \approx & \frac{\sigma_p(\mbox{temp})}{\sqrt{\nbase}}
    \approx 1.521\, \mu\mbox{K}\, \tbase(\mbox{s})^{-1/2} \,, \nn\\
    \sigma_{pr}(\mathrm{pol}) & \approx & \frac{\sigma_p(\mbox{pol})}{\sqrt{\nbase}}
    \approx 0.0891\, \mu\mbox{K}\, \tbase(\mbox{s})^{-1/2} \,.
 \label{sigma_sb_approx}
 \eea

\begin{figure}[!tbp]
  \begin{center}
   \includegraphics*[width=0.5\textwidth]{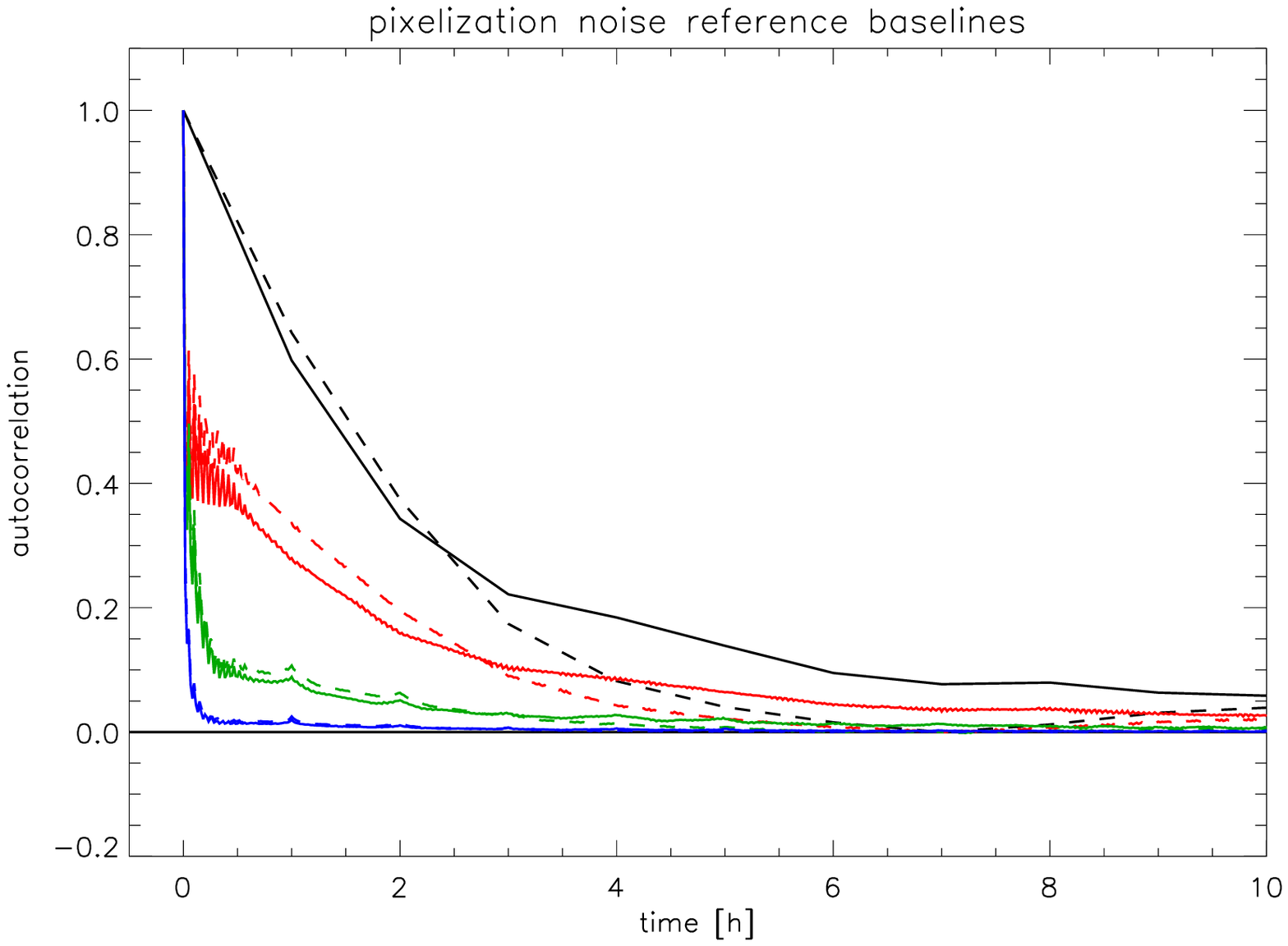}
   \includegraphics*[width=0.5\textwidth]{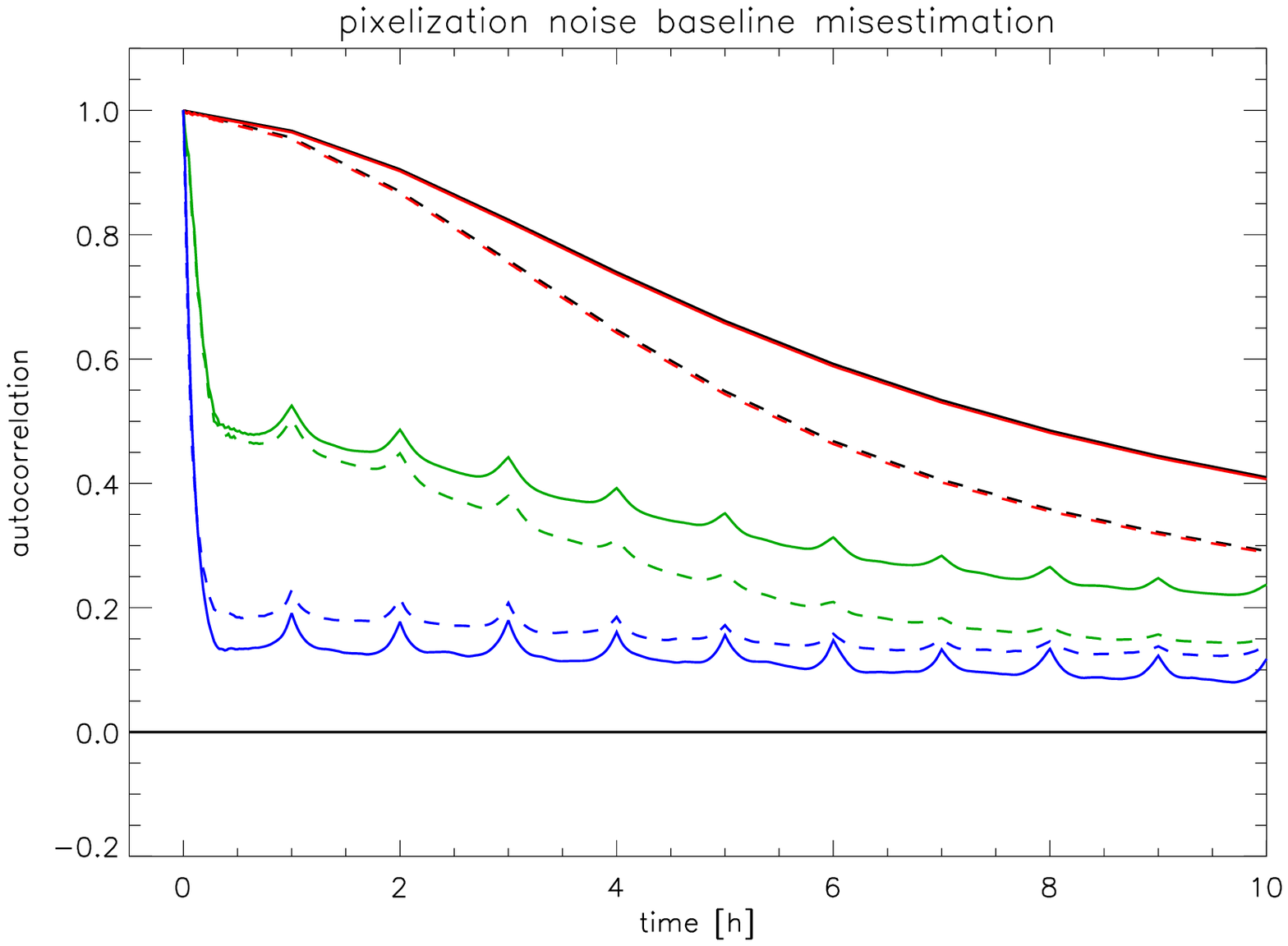}
  \end{center}
  \caption{
Correlations in the signal baselines. The \emph{top} panel shows the
autocorrelation function of the reference baselines $\bR\bZ\bs$ of
pixelization noise, and the \emph{bottom} panel that of the signal
baseline error $(\bA-\bR\bZ)\bs$. The colors and line styles are as
in Fig.~\ref{fig:white_aucorr}.
 }
  \label{fig:sig_aucorr}
\end{figure}

\begin{table}[!tbp]
 \begin{center}
 \begin{tabular}{llllll}
 $\tbase$ & Eq.~(\ref{sigma_sb_approx}) & $\bR\bZ\bs$ &
 $(\bA-\bR\bZ)\bs$ & $\bA\bs$ &  $\bA\bs$ (G) \\
 \hline
 2.5 s  & 0.962 & 0.972 & 0.672 & 1.297 \\
 5 s    & 0.680 & 0.686 & 0.300 & 0.831 \\
 15 s   & 0.393 & 0.396 & 0.121 & 0.461 \\
 1 min  & 0.196 & 0.200 & 0.075 & 0.248 & 0.243 \\
 1 h    &       & 0.136 & 0.075 & 0.199 & 0.196 \\
 \hline
 2.5 s  & 0.0564 & 0.0560 & 0.0277 & 0.0701 \\
 5 s    & 0.0399 & 0.0396 & 0.0145 & 0.0476 \\
 15 s   & 0.0230 & 0.0229 & 0.0068 & 0.0269 \\
 1 min  & 0.0115 & 0.0115 & 0.0043 & 0.0144 & 0.0147 \\
 1 h    &        & 0.0083 & 0.0043 & 0.0120 & 0.0124\\
 \hline
\end{tabular}
 \end{center}
 \caption{Statistics of the different contributions to the signal baselines.
We give the standard deviation (square root of the variance) in
$\mu$K of the reference baselines of pixelization noise, the signal
baseline error, and the solved signal baselines.  The top half of
the table is for the temperature stream $(a+b)/2$ and the bottom
half for the polarization stream $(a-b)/2$. The column marked
``(G)'' was obtained using galactic coordinates for the HEALPix
maps.}
 \label{table:sbl}
\end{table}

In Table~\ref{table:sbl} we show the stdev of the signal baselines
$\bA\bs$  and their two contributions $\bR\bZ\bs$ and
$(\bA-\bR\bZ)\bs$, and compare them to the estimate
(\ref{sigma_sb_approx}) for $\bR\bZ\bs$. We see the estimate is
quite good for $\bR\bZ\bs$.  The estimate is slightly off mainly
because the actual pixelization noise was slightly larger for
temperature and slightly smaller for polarization than the
analytical estimate (\ref{sigp_anal}).

In Fig.~\ref{fig:sig_aucorr} we show the autocorrelation functions
of the reference baselines of pixelization noise and the signal
baseline error. Unlike for white noise, now also the reference
baselines are correlated. This correlation extends over several
repointing periods. These correlations enhance the relevance of the
signal baselines in the map domain.

\section{Map domain}
\label{sec:map_domain}

Destriping is a linear process. The output map can therefore be
viewed as a sum of component maps, each component map being a result
of destriping one individual TOD component: signal, $1/f$ noise and
white noise:
 \beq
    \mout = \bB(\bs-\bF\bA\bs) + \bB(\bnc-\bF\bA\bnc)
    + \bB(\bw-\bF\bA\bw) \,.
 \eeq

As the three TOD components are statistically independent, so are
the corresponding component maps. The expectation value of the map
rms is thus obtained as the root sum square (rss) of the expectation
values of the rms of the component maps.

In this paper we have considered a single 1-year realization of each
TOD component only. There are likely to be random correlations
between the component maps. However, the output map and residual map
rms we get when we produce maps from the full TOD, agree to better
than 0.5\% with the rss of the rms of the corresponding maps
produced from component TODs.

\subsection{Linearity}

\begin{figure}[t]
  \begin{center}
    \includegraphics*[width=0.5\textwidth]
      {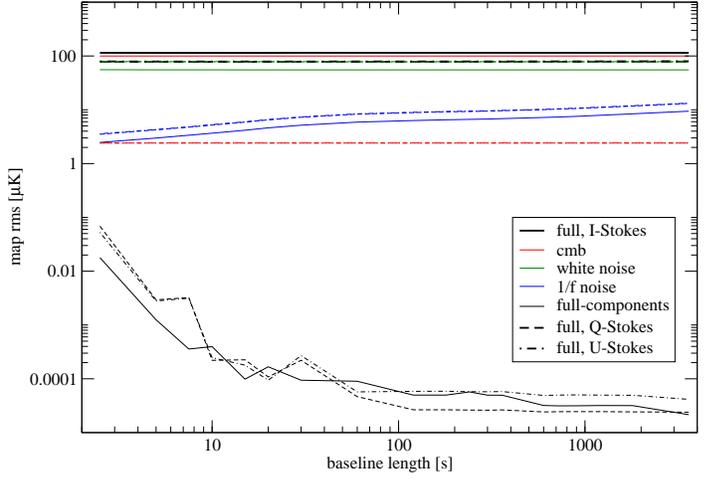}
    \caption[Linearity test]{
      Linearity of destriping.  The three \emph{thin black} lines
at the bottom shows the rms difference between the ``full'' map,
obtained from the sum of the component TODs, and the sum of the
component maps.  This is compared to the signal and noise levels in
the maps; different \emph{colors} showing the different components
and the \emph{thick black} line the full map; different \emph{line
styles} correspond to the three Stokes parameters.
    }
    \label{fig:linearity_test}
  \end{center}
\end{figure}

In practice the linearity of destriping is affected by the numerical
accuracy of the baseline calculation, which depends on the
convergence criterion for the conjugate gradient iteration.  We
checked the linearity by comparing the sum of the component maps to
a map obtained directly from the sum of the component TODs.  The
difference is well below the nK level, except for baselines shorter
than 10 s. See Fig.~\ref{fig:linearity_test}.

\subsection{Splitting the residual map into components}

\begin{figure}[!tbp]
  \begin{center}
    \includegraphics*[width=0.5\textwidth]
             {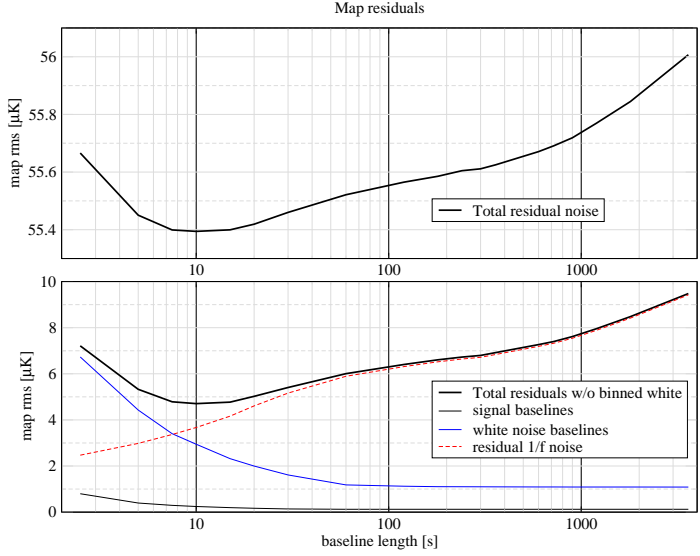}
    \caption[Signal components of the output map]{
The residual temperature map rms as a function of baseline length.
The {\em top} panel is for $\mout - \bB\bs$, which is dominated by
the binned white noise map, whose rms is $55.19\:\mu$K independent
of $\tbase$. The {\em bottom} panel shows the other three
components, and their root sum square, which is the rms of the
residual map $\mout - \bB\bs - \bB\bw$.
 }
    \label{fig:component_residuals}
  \end{center}
\end{figure}

\begin{figure}[!tbp]
  \begin{center}
    \includegraphics*[width=0.5\textwidth]
             {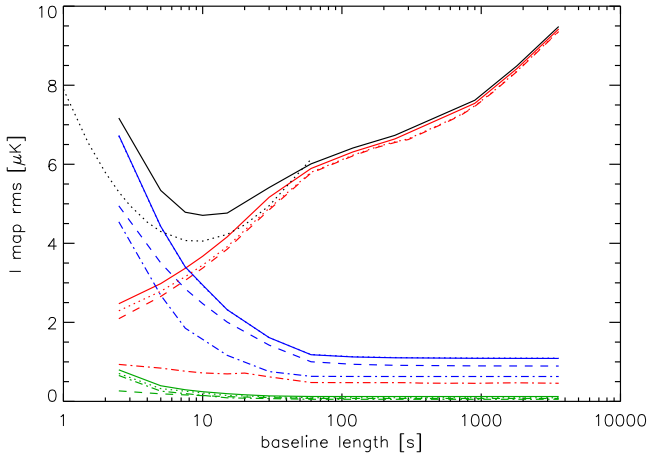}
    \caption{Same as the bottom panel of
Fig.~\ref{fig:component_residuals}, but now we have further
separated the residual $1/f$ noise into unmodeled $1/f$ noise
(\emph{dashed}) and $1/f$ baseline error (\emph{dash-dotted}); the
white noise baselines into reference baselines (\emph{dashed}) and
baseline error (\emph{dash-dotted}); and signal baselines (shown
here in \emph{green}) into reference baselines of pixelization noise
(\emph{dashed}) and baseline error (\emph{dash-dotted}). The
\emph{dotted} lines show the rss of the \emph{dashed} and
\emph{dash-dotted} lines.  For white noise it falls on the solid
line, showing that the two white noise residual map components are
uncorrelated.  For $1/f$ noise and pixelization noise the two
components are positively correlated. The \emph{black dotted} line
is the analytical approximation Eq.~(\ref{map_rms_apx}).
 }
    \label{fig:comp_res_in_6}
  \end{center}
\end{figure}

\begin{figure}[!tbp]
  \begin{center}
    \includegraphics*[width=0.5\textwidth]
      {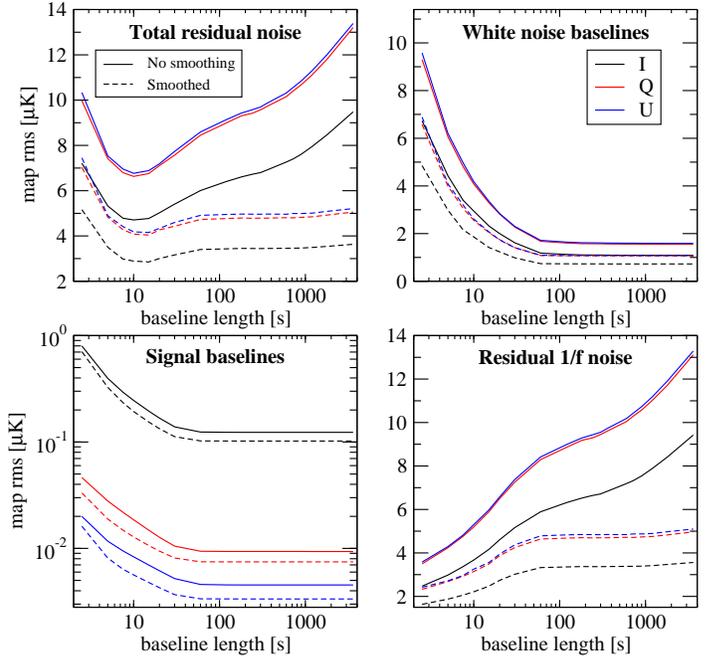}
    \caption[Smoothed residual rms]{The rms of the residual $I$, $Q$, and $U$ maps,
and their different components. The overall residual power in a
destriped polarization map is higher than in a temperature map. The
CMB signal residual in turn is much weaker. Note how Q is larger
than U in the signal baselines. Smoothing the residual maps with a
Gaussian beam (\emph{dashed lines}) removes excess power at sub-beam
scales.
    }
    \label{fig:smoothed_comp_res}
  \end{center}
\end{figure}

\begin{figure}[!tbp]
 \begin{center}
    \includegraphics*[trim=120 0 70 0, clip, width=0.28\textwidth,angle=90]
    {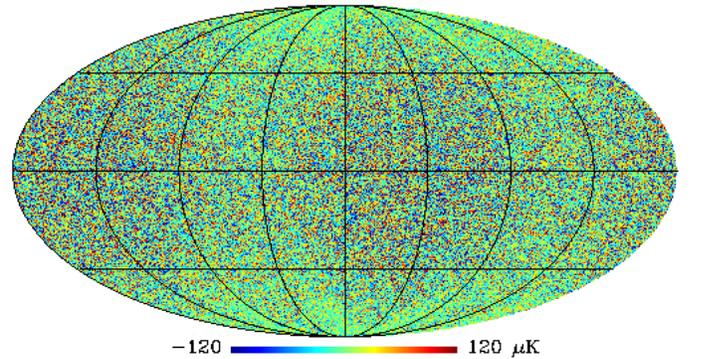} \\
    \vspace{-0.2cm}
    \caption{
The residual $I$ map including binned white noise, $\mout - \bB\bs$,
for $\tbase = 1$ min. It looks the same for other baseline lengths.
Also the binned white noise map $\bB\bw$ looks the same, as it is
the dominant component in the residual map. The $Q$ and $U$ maps
look the same, but have a larger amplitude.
    }
    \label{fig:residual_I_map}
  \end{center}
\end{figure}

The three component maps can each be further divided into a binned
map and a baseline map.

The correlation properties of these components are different in the
map domain from the TOD domain.

While the white noise baselines $\bF\bA\bw$ are correlated with the
white noise $\bw$ in the TOD, the correlation vanishes in the map:
Writing
 \bea
    \bB\bw & = & \bMi\bP^T\Cwi\bw \\
    \bB\bF\bA\bw & = &
    \bMi\bP^T\Cwi\bF\bDi\bF^T\Cwi\left(\bI-\bP\bMi\bP^T\Cwi\right)\bw
 \eea
we get that
 \bea
    \langle(\bB\bw)(\bB\bF\bA\bw)^T\rangle & = &
    \bMi\bP^T\Cwi\langle\bw\bw^T\rangle
    \left(\bI-\Cwi\bP\bMi\bP^T\right) \nonumber\\
    & & \mbox{} \Cwi\bF\bDi\bF^T\Cwi\bP\bMi \,.
 \eea
By substituting  $\langle\bw\bw^T\rangle = \Cw$ one readily sees
that the correlation vanishes,
 \beq
    \langle (\bB\bw)(\bB\bF\bA\bw)^T\rangle =0 \,.
 \eeq

Because the binned white noise map and the white noise baseline map
are independent, the (expectation value of) the rms of the residual
white noise map $\bB\bw - \bB\bF\bA\bw$  is obtained as the rss of
the rms of the two maps.

This means, that while subtracting the white noise baselines removes
power from the TOD, it adds power to the map. If the noise is pure
white, naive binning produces a better map than destriping.
 On the other hand,
baselines of the $1/f$ noise are correlated with the $1/f$ noise
itself both in the TOD and in the map.

We can use the rms value taken over all the residual map pixels as a
figure-of-merit for the map-making method.  We calculate it
separately for the three Stokes parameters, $I$ (temperature), $Q$,
and $U$.  \emph{Note that we subtract the $I$ monopole from the
residual map before calculating the rms}, since it is irrelevant for
CMB anisotropy and polarization studies, and destriping leaves a
spurious $I$ monopole in the map. See
Figs.~\ref{fig:component_residuals}, \ref{fig:comp_res_in_6}, and
\ref{fig:smoothed_comp_res}.

\begin{figure}[!tbp]
  \begin{center}
    \includegraphics*[trim=120 0 70 0, clip, width=0.28\textwidth,angle=90]
    {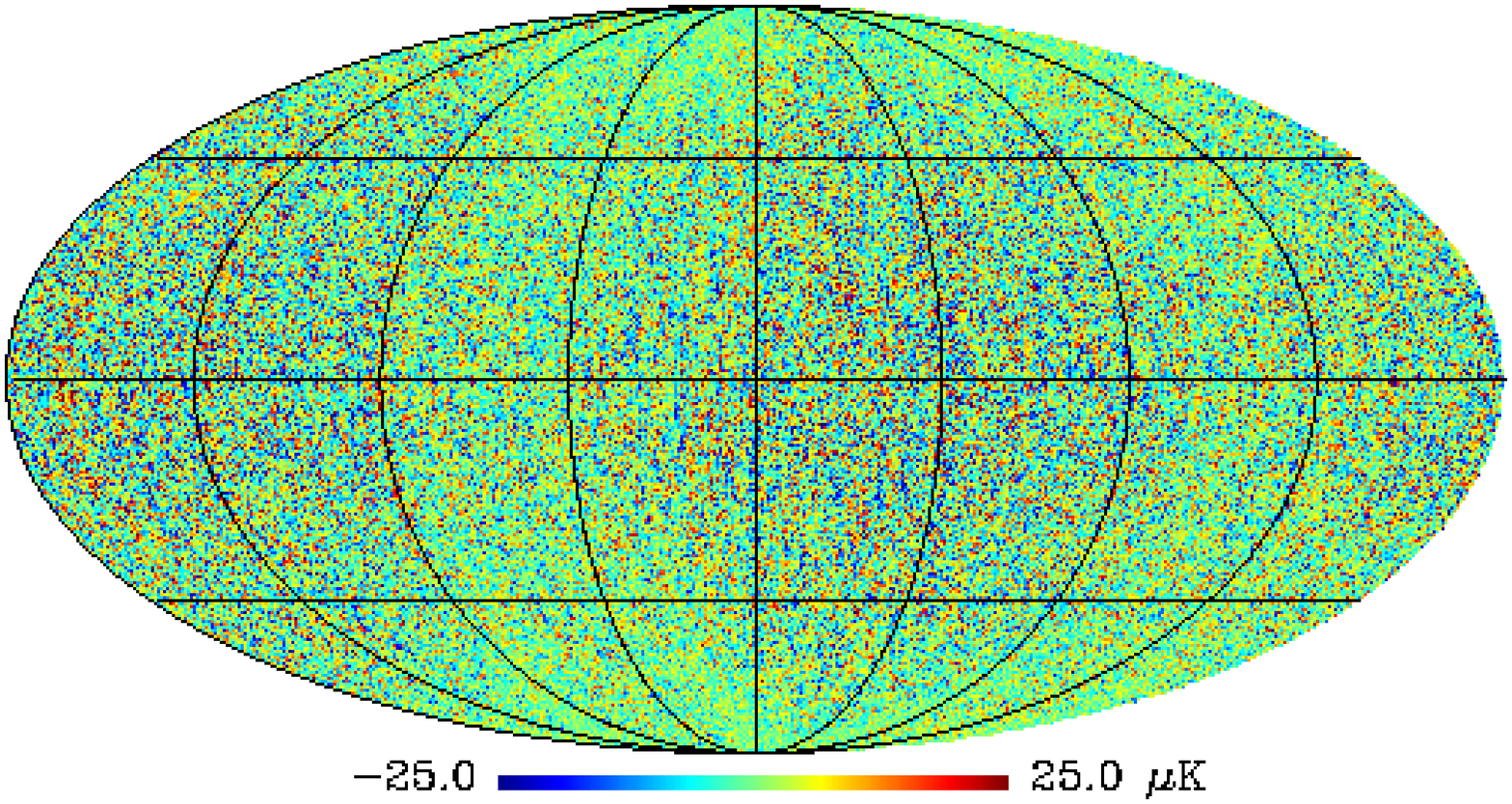} \\
    \vspace{-0.2cm}
    \includegraphics*[trim=120 0 70 0, clip, width=0.28\textwidth,angle=90]
    {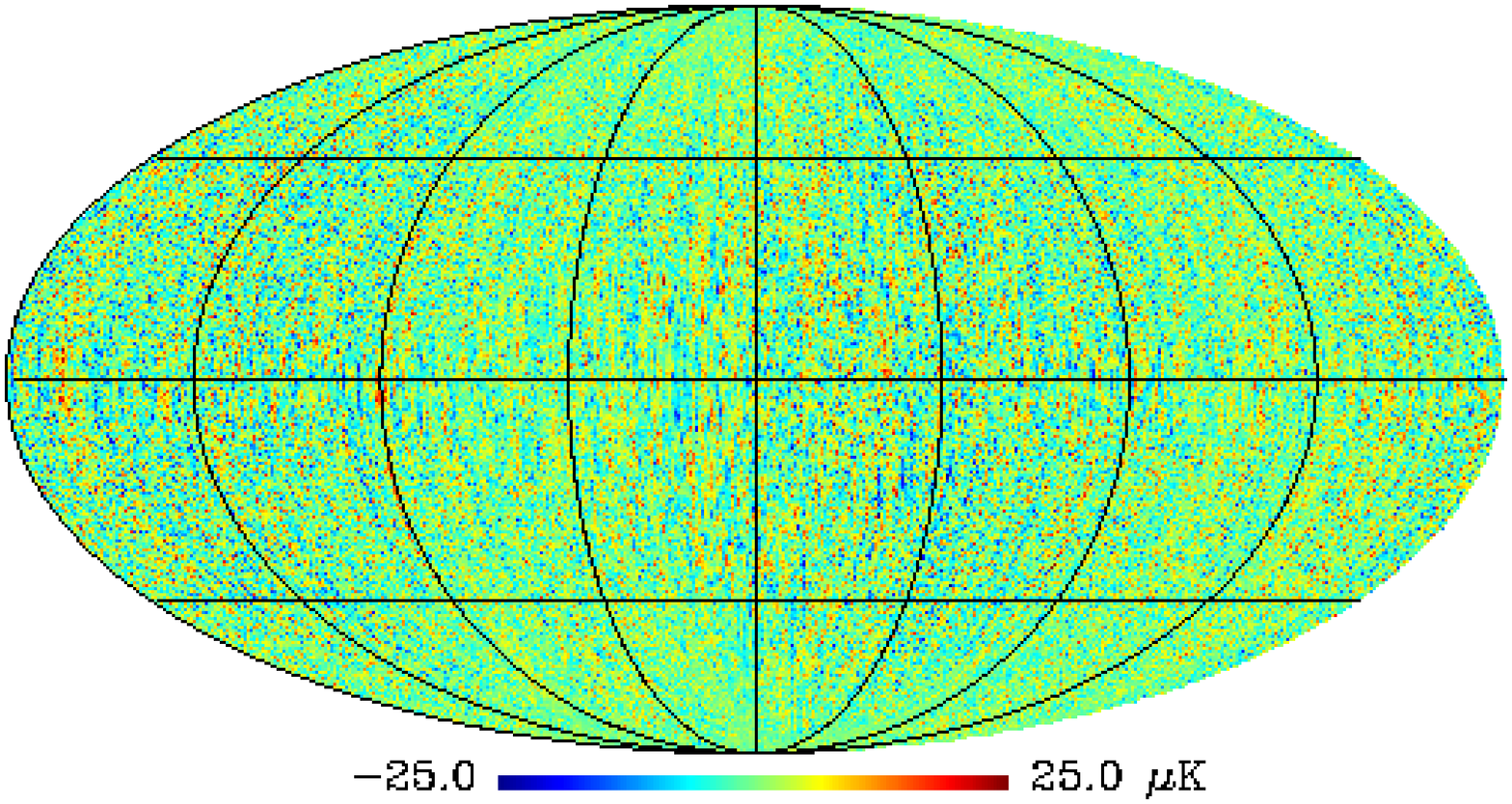} \\
    \vspace{-0.2cm}
    \includegraphics*[trim=120 0 70 0, clip, width=0.28\textwidth,angle=90]
    {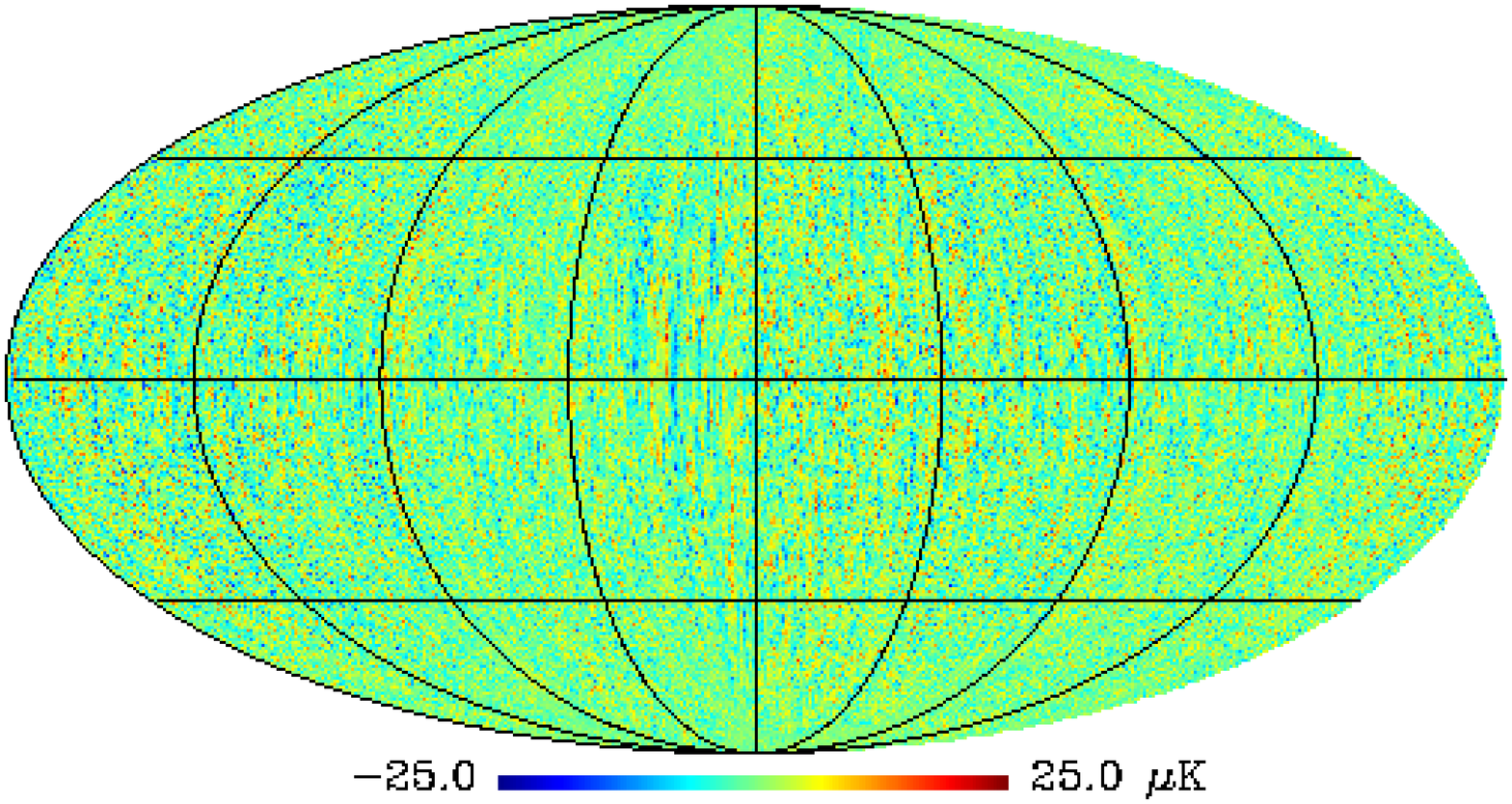} \\
    \vspace{-0.2cm}
    \includegraphics*[trim=120 0 70 0, clip, width=0.28\textwidth,angle=90]
    {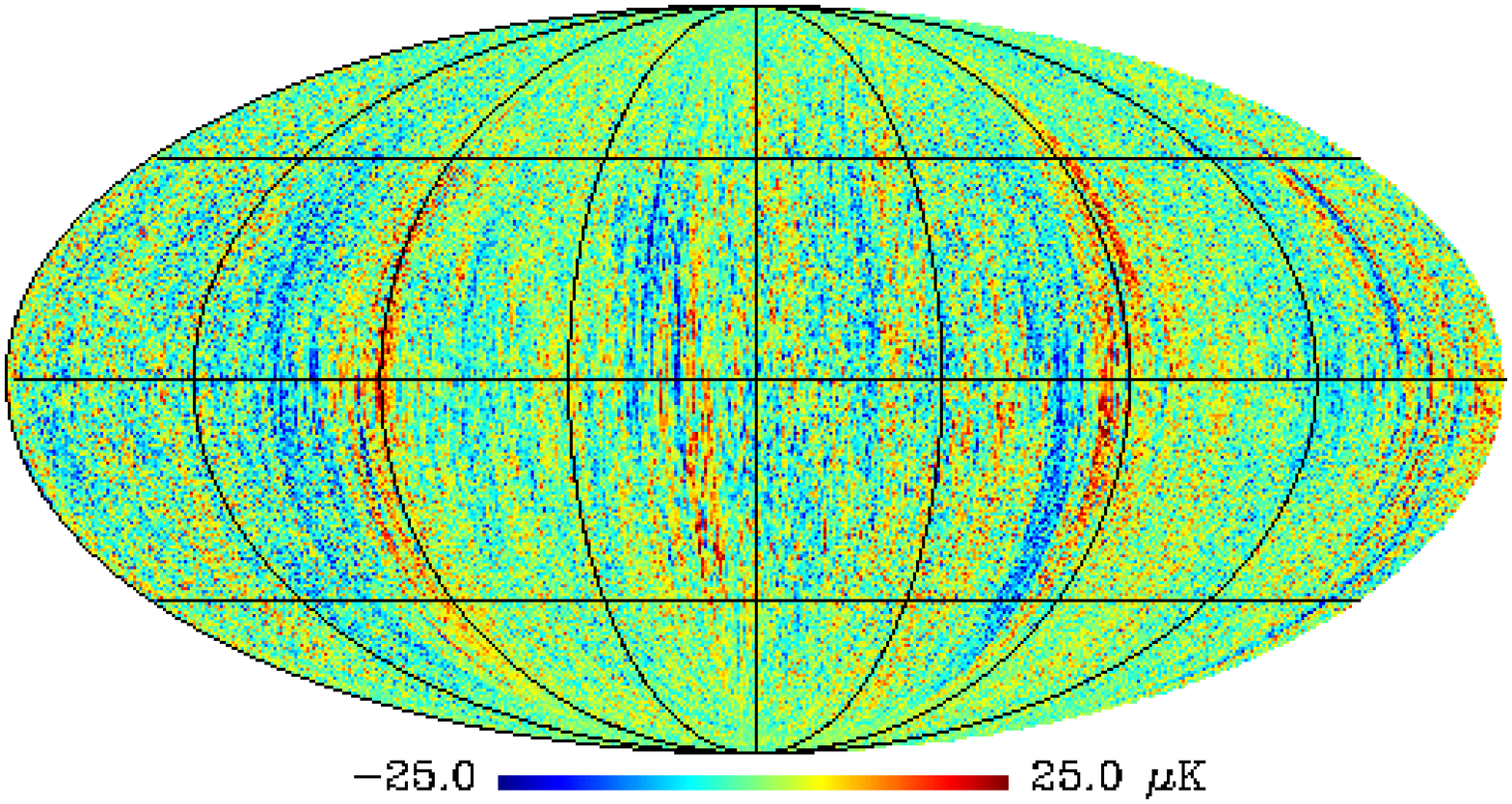} \\
    \vspace{-0.2cm}
    \caption{
The residual $I$ map (with binned white noise subtracted), $\mout -
\bB\bs - \bB\bw$, for different baseline lengths (\emph{from top
down:}) 1 hour, 1 minute, 15 seconds, and 2.5 seconds. The $Q$ and
$U$ maps look the same with $\sim\!\sqrt{2}$ larger amplitude.
    }
    \label{fig:res_I_maps_for_diff_tbase}
  \end{center}
\end{figure}

\begin{figure}[!tbp]
  \begin{center}
    \includegraphics*[trim=100 70 70 250, clip, width=0.24\textwidth]
    {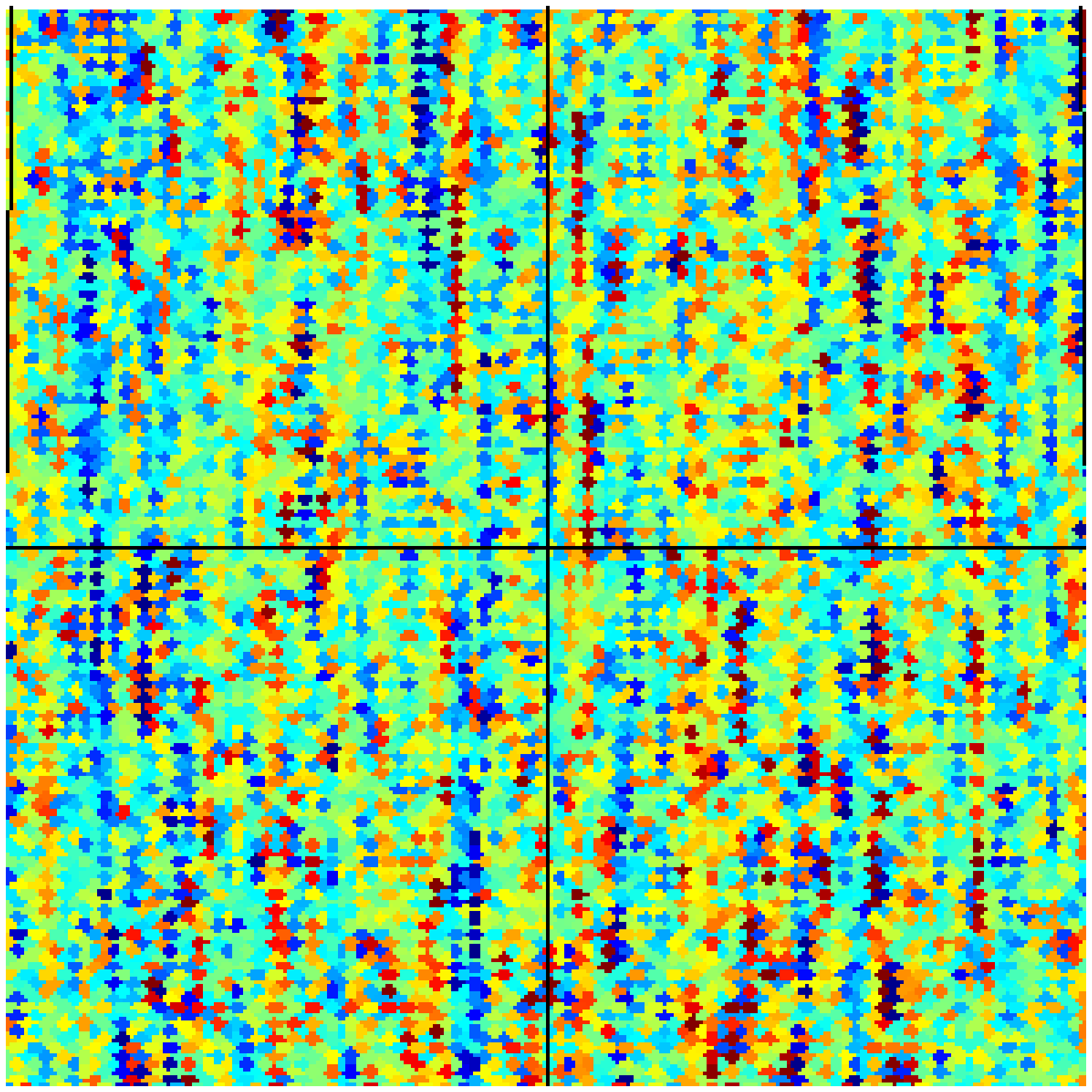}
    \includegraphics*[trim=100 70 70 250, clip, width=0.24\textwidth]
    {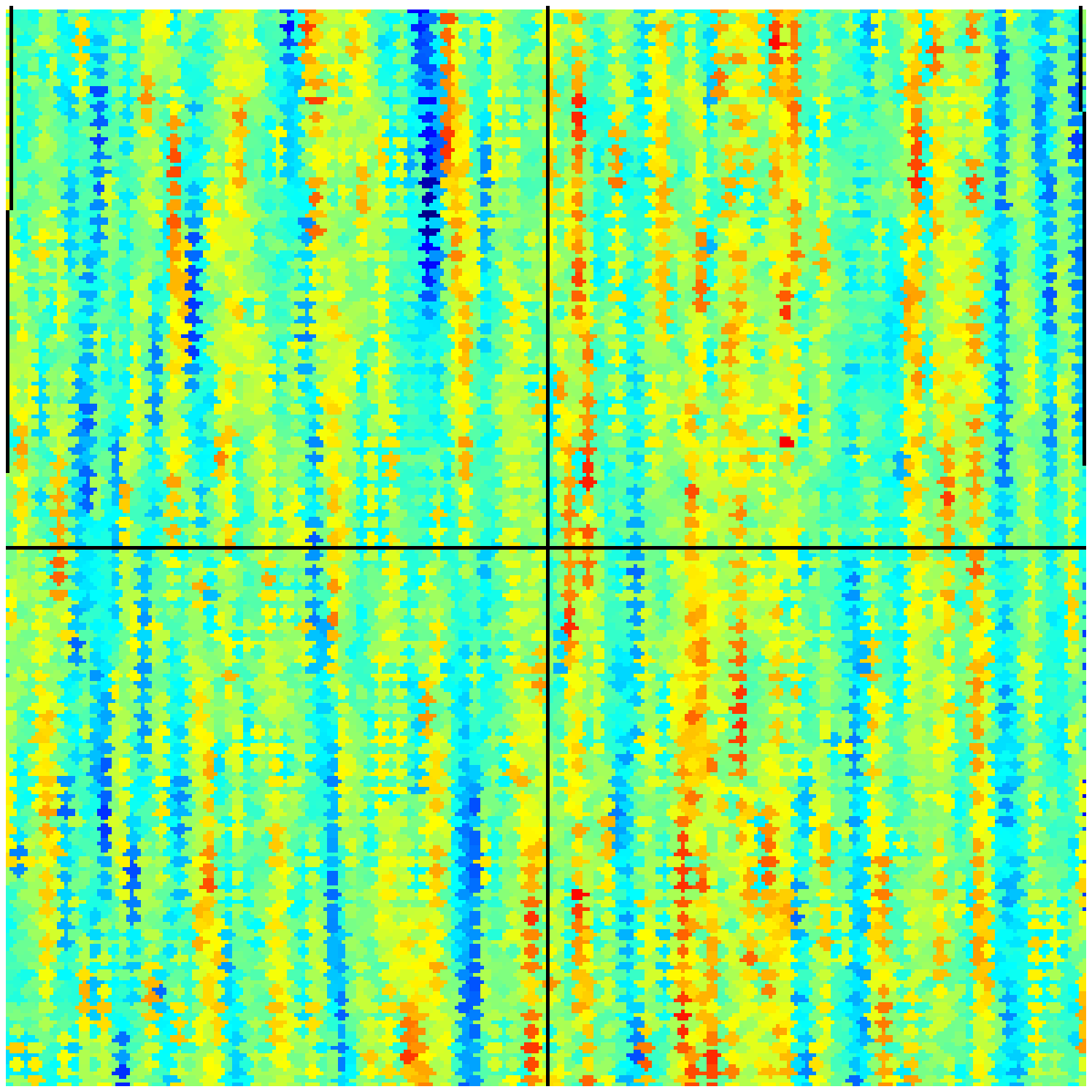} \\
    \vspace{-0.3cm}
    \includegraphics*[trim=100 70 70 250, clip, width=0.24\textwidth]
    {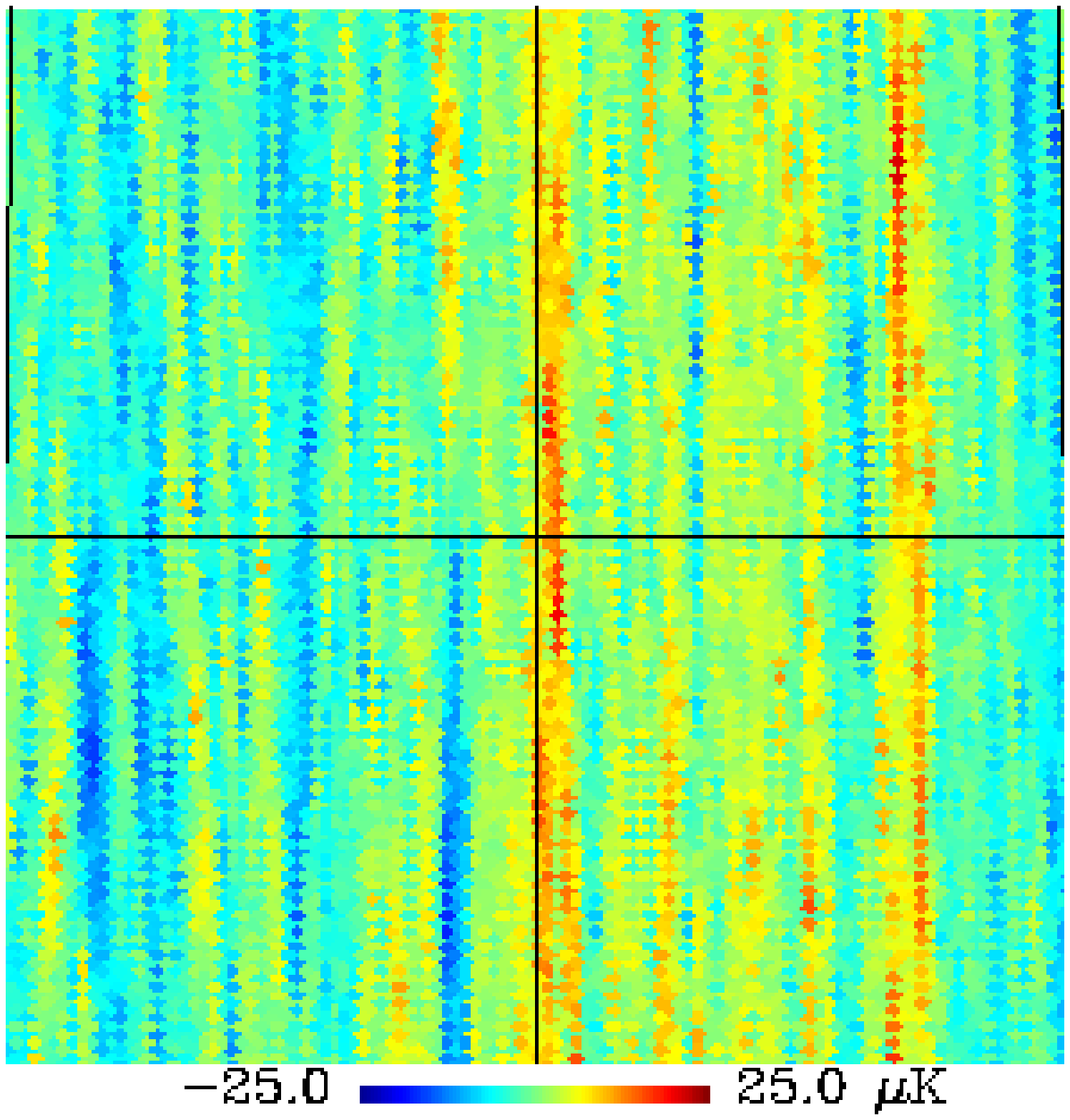}
    \includegraphics*[trim=100 70 70 250, clip, width=0.24\textwidth]
    {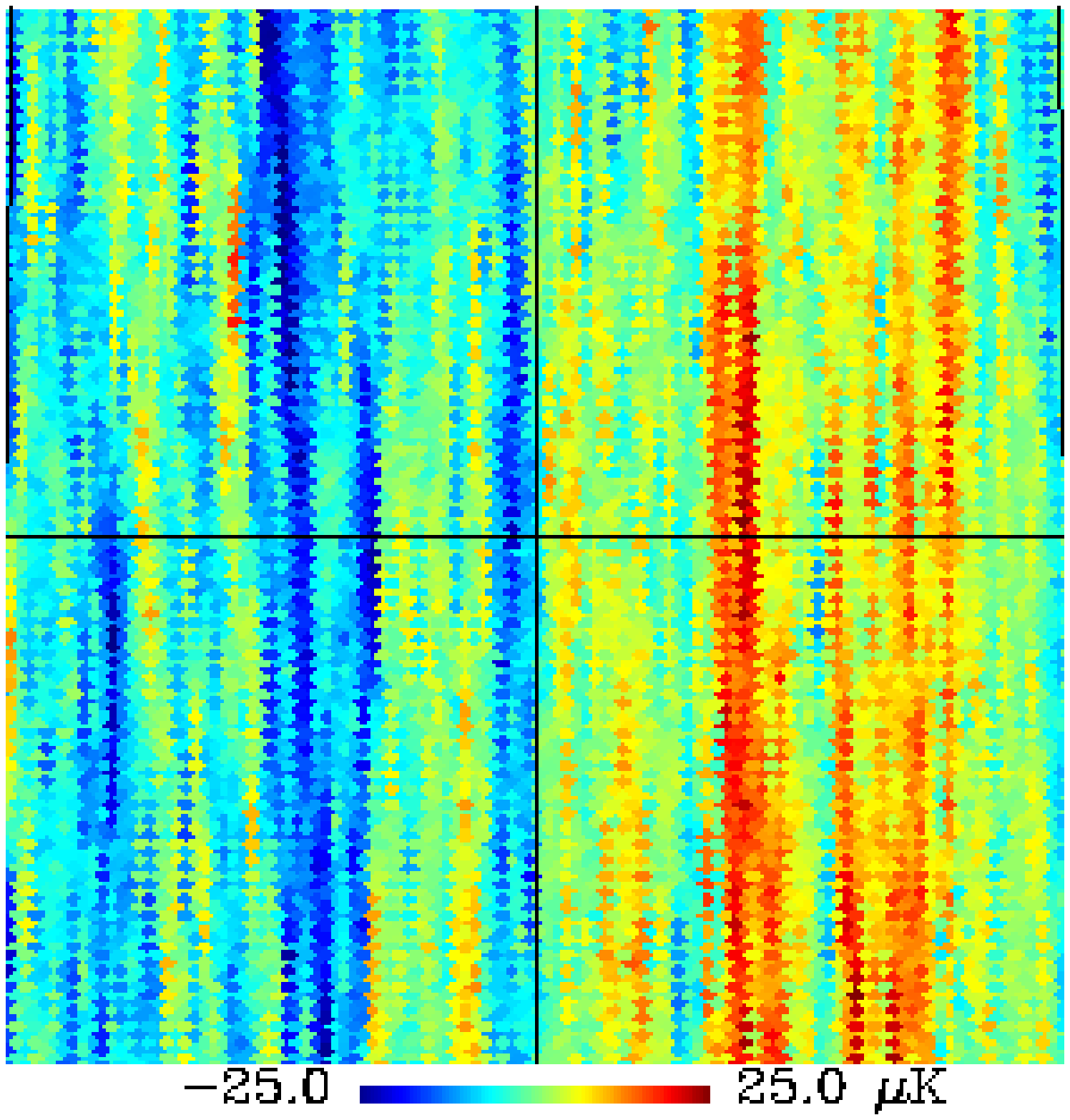} \\
    \vspace{-0.2cm}
    \caption{Same as Fig.~\ref{fig:res_I_maps_for_diff_tbase}
but for the $10^\circ\times10^\circ$ region near the ecliptic.
    }
    \label{fig:res_I_10by10_maps_for_db}
  \end{center}
\end{figure}

\begin{figure}[!tbp]
  \begin{center}
    \includegraphics*[trim=100 70 70 250, clip, width=0.24\textwidth]
    {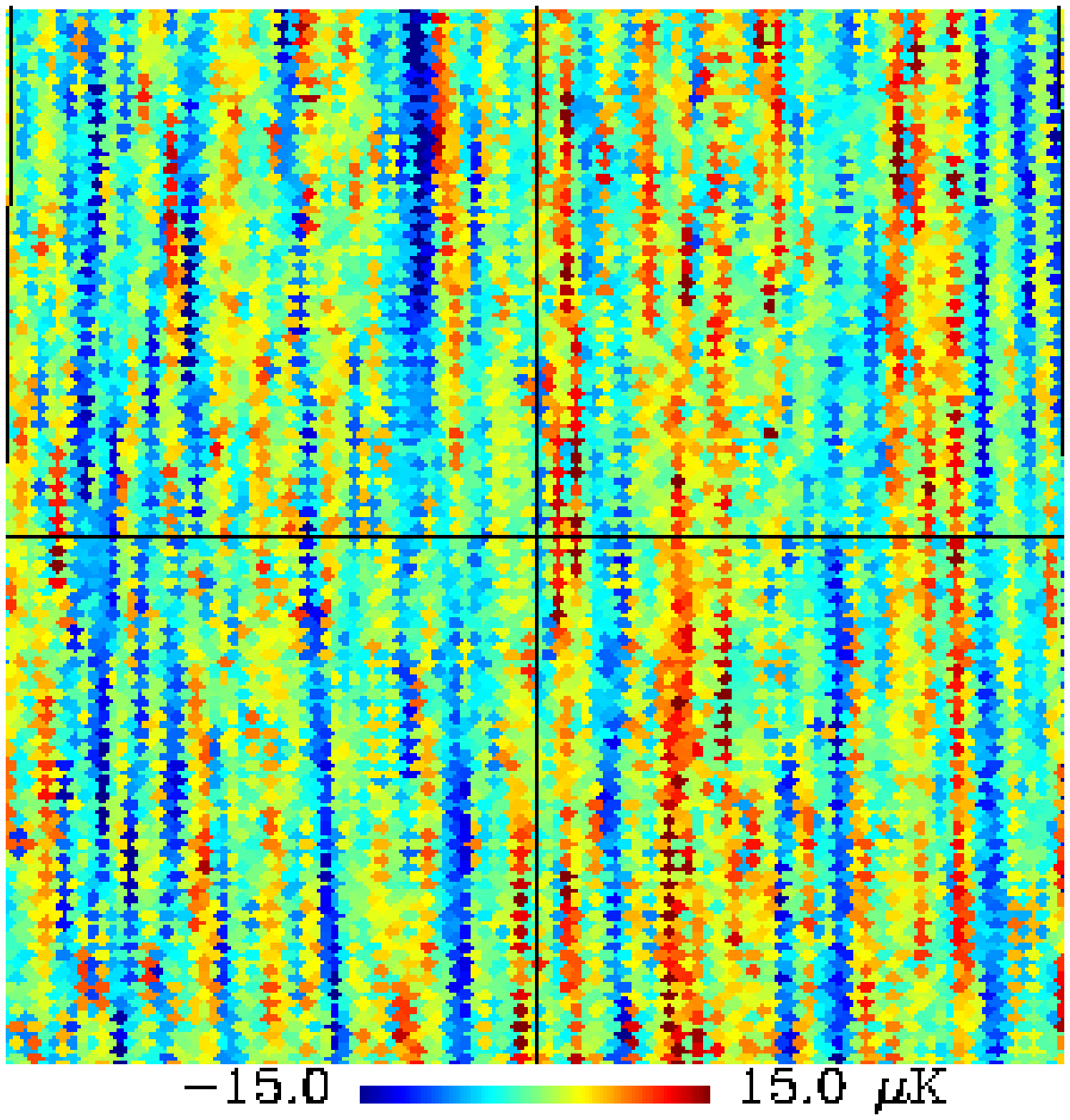}
    \includegraphics*[trim=100 70 70 250, clip, width=0.24\textwidth]
    {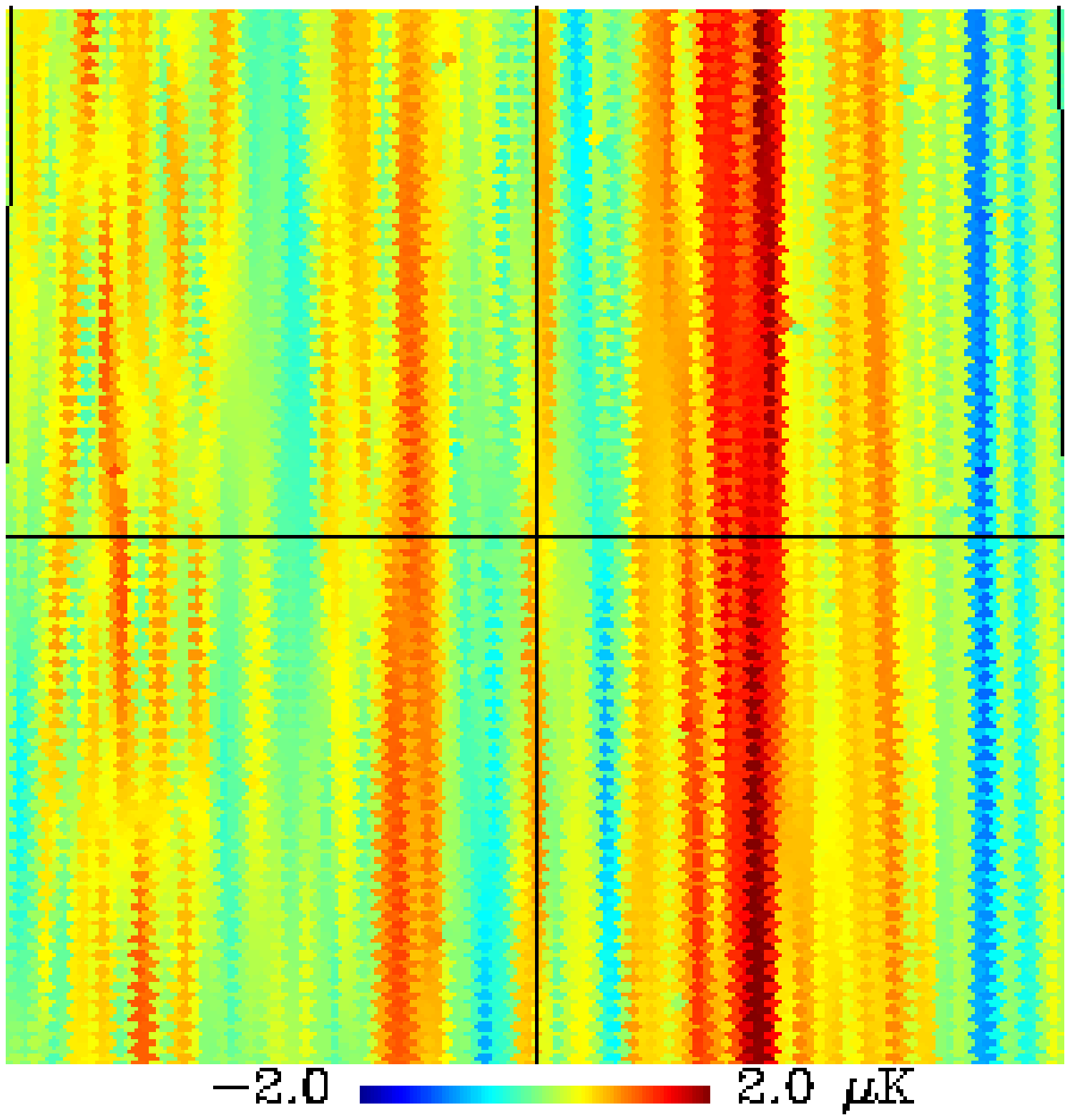} \\
    \includegraphics*[trim=100 70 70 250, clip, width=0.24\textwidth]
    {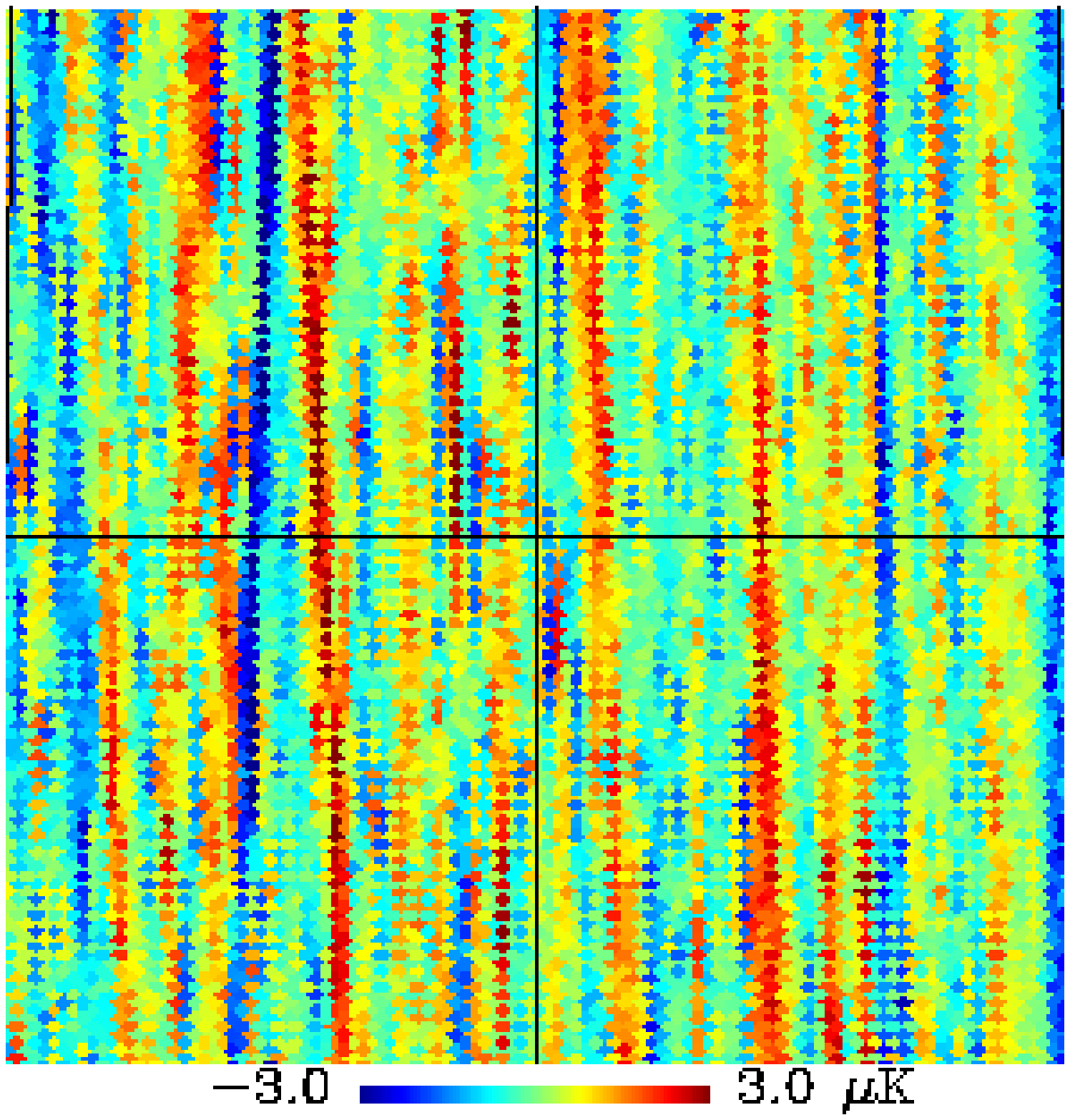}
    \includegraphics*[trim=100 70 70 250, clip, width=0.24\textwidth]
    {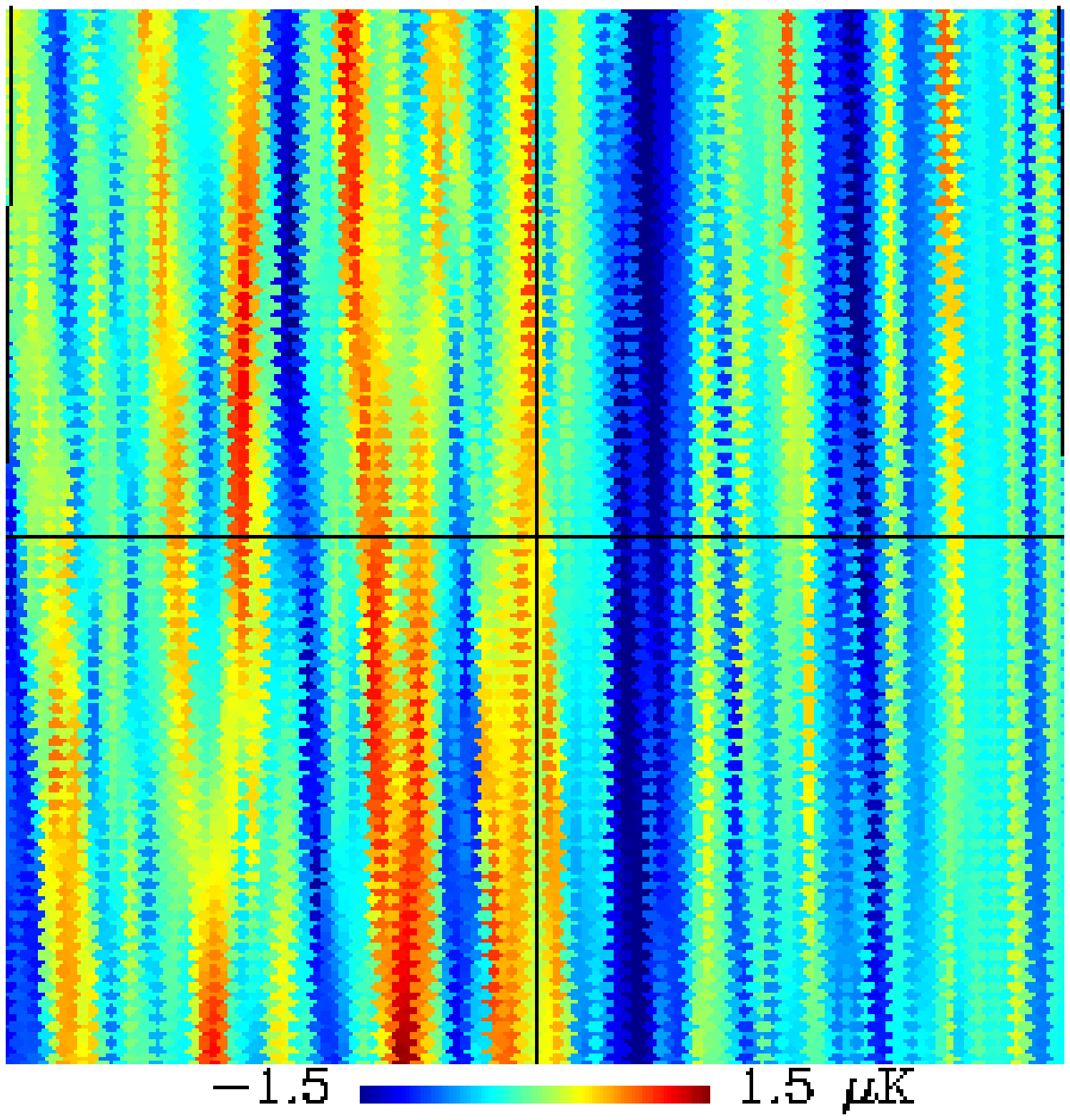} \\
    \includegraphics*[trim=100 70 70 250, clip, width=0.24\textwidth]
    {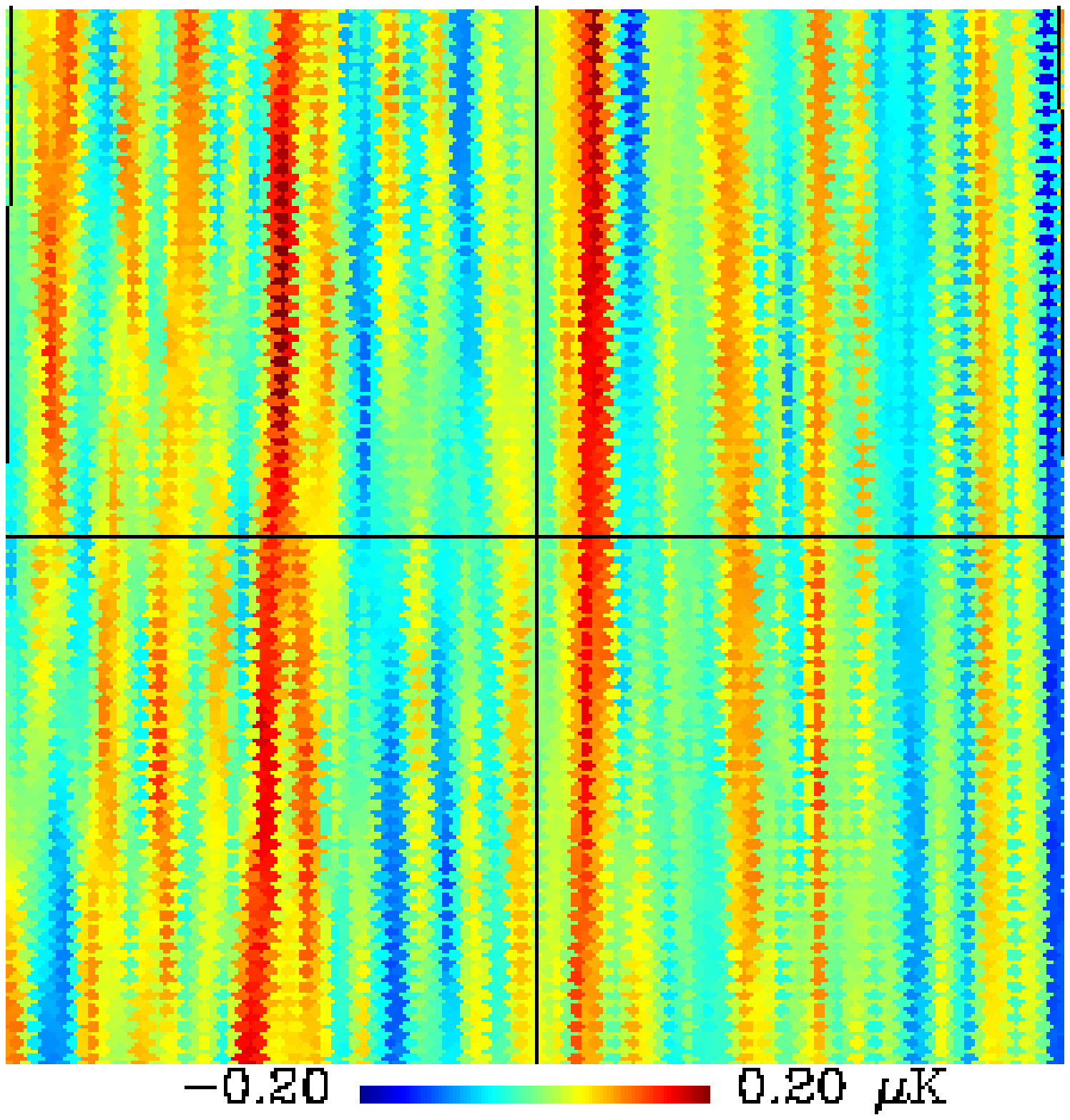}
    \includegraphics*[trim=100 70 70 250, clip, width=0.24\textwidth]
    {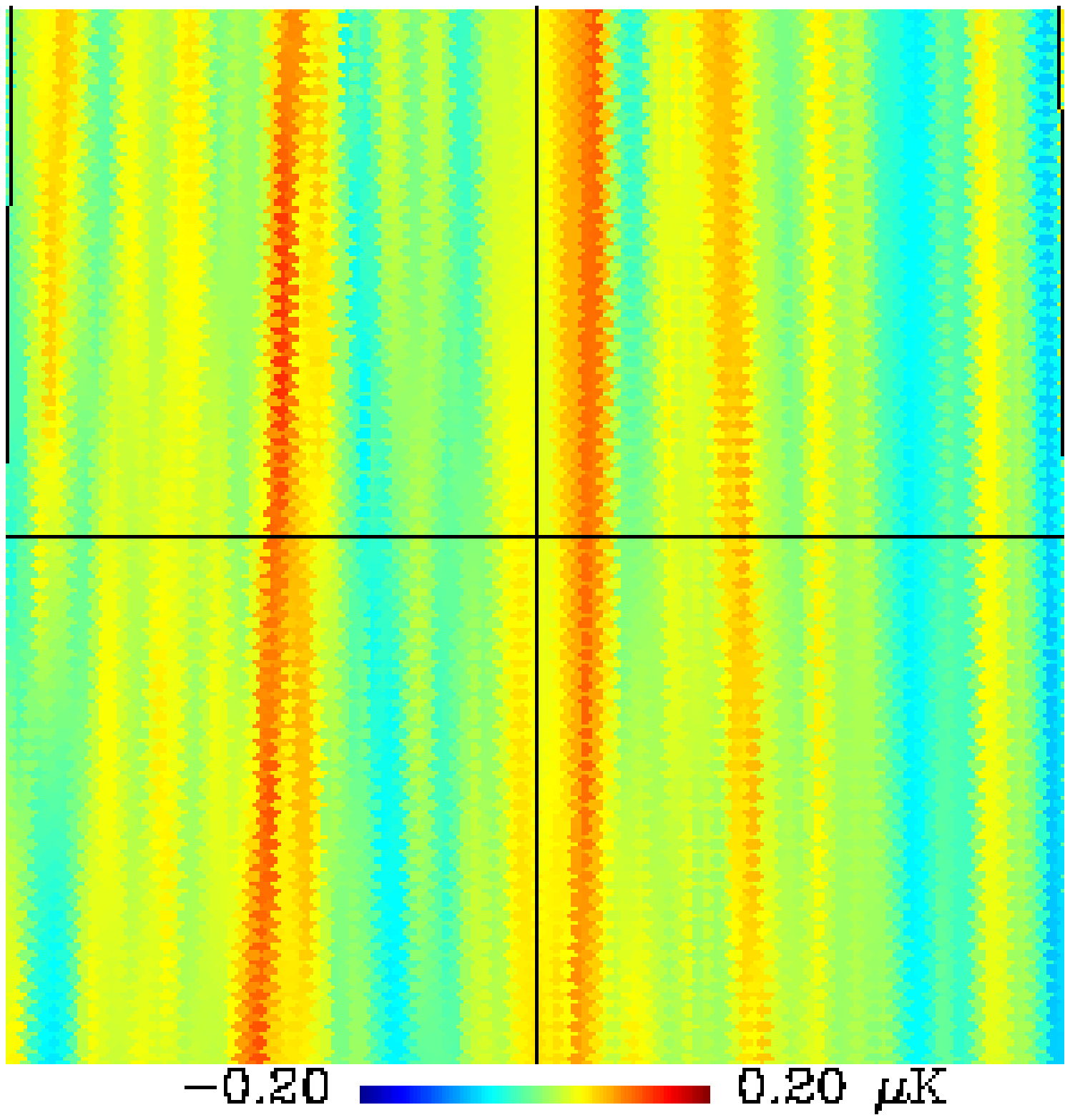} \\
    \vspace{-0.2cm}
    \caption{The six components of the residual $I$ map
shown for the $10^\circ\times10^\circ$ region near the ecliptic
($\tbase = 1$ min).
 \emph{Top left:} Unmodeled $1/f$ noise $\bB(\bnc-\bF\aref)$.
 \emph{Top right:} $1/f$ baseline error $\bB\bF(\bA\bnc-\aref)$.
 \emph{Middle left:} White noise reference baselines $-\bB\bF\bR\bw$.
 \emph{Middle right:} White noise baseline error
 $-\bB\bF(\bA-\bR)\bw$.
 \emph{Bottom left:} Pixelization noise reference baselines $-\bB\bF\bR\bZ\bs$.
 \emph{Bottom right:} Signal baseline error
 $-\bB\bF(\bA-\bR\bZ)\bs$.
    }
    \label{fig:six_comp_res_I_10by10_maps}
  \end{center}
\end{figure}

The difference between the output map and the binned signal map
 \beq
    \mout - \bB\bs = \bB\bw - \bB\bF\bA\bw + \bB\left(\bnc -
    \bF\bA\bnc\right) -\bB\bF\bA\bs
 \eeq
is the residual map including binned white noise. We see that it can
be divided into four uncorrelated contributions: the binned white
noise map, the white noise baseline map, the residual $1/f$ noise
map, and the signal baseline map.

The $\mout - \bB\bs$ $I$ map is shown in
Fig.~\ref{fig:residual_I_map}. This map is dominated by the binned
white noise map, which is independent of the baseline length.
Therefore the visual appearance of the residual map is the same for
all baseline lengths, as it looks the same as the white noise map.
Also the residual $Q$ and $U$ maps look the same, just with
$\sim\!\sqrt{2}$ larger amplitude.

The binned white noise map is independent of the baseline length,
and it is an unavoidable component of the residual map, uncorrelated
with the other components. Therefore we focus on the rest of the
residual map, without the binned white noise component, i.e.,
 \beq
    \mout - \bB\bs -\bB\bw = - \bB\bF\bA\bw + \bB\left(\bnc -
    \bF\bA\bnc\right) -\bB\bF\bA\bs \,.
 \label{residualmbw}
 \eeq
For the rest of this paper, the \emph{residual map} without further
qualification, refers to this map. See
Figs.~\ref{fig:res_I_maps_for_diff_tbase} and
\ref{fig:res_I_10by10_maps_for_db}.

We divide this further into
 \bea
    \mout - \bB\bs - \bB\bw & = & - \bB\bF\bR\bw - \bB\bF(\bA-\bR)\bw \nn\\
    & & + \bB(\bnc-\bF\aref) - \bB\bF(\bA\bnc-\aref) \nn\\
    & & - \bB\bF\bR\bZ\bs - \bB\bF(\bA-\bR\bZ)\bs \,.
 \label{map_six}
 \eea
in analogy with Eq.~(\ref{time_six}). Of these six components, the
unmodeled $1/f$ map $\bB(\bnc-\bF\aref)$ is correlated with the
$1/f$ baseline error map $\bB\bF(\bA\bnc-\aref)$, and the
pixelization noise reference baseline map $\bB\bF\bR\bZ\bs$ is
correlated with the signal baseline error map
$\bB\bF(\bA-\bR\bZ)\bs$. Otherwise the components are uncorrelated
with each other. The division is useful, since the different
components have a different structure in the map domain. The
unmodeled $1/f$ map consists of different noise frequencies with
half-wavelength mostly of the same order or shorter than the
baseline (Fig.~\ref{fig:unmod_pow_spec}).  The other 5 components
consist of baselines laid on the map.  For the white noise reference
baseline map, these baselines are uncorrelated with each other, the
other 4 components have different levels of correlation.
 In Fig.~\ref{fig:six_comp_res_I_10by10_maps} we show each of
 the six components for a $10^\circ\times10^\circ$ region of the $I$
 map.

In Fig.~\ref{fig:comp_res_in_6}, we show separately the rms of these
components.

\subsection{Analytical estimates}

\subsubsection{White noise baselines}

Since the white noise reference baselines are uncorrelated with each
other, the white noise reference baseline map rms can be estimated.
If all hits to a pixel came from a different baseline, we could
treat them as white noise, with the white noise reference baseline
variance $\sigma_{wr}^2 = \sigma^2/\nbase$.  Thus the variances of
$I_p$, $Q_p$, and $U_p$ would be just
 $(M_p)^{-1}(1,1)/\nbase \approx \sigma^2/\nbase/\nhitp$,
 $(M_p)^{-1}(2,2)/\nbase \approx 2\sigma^2/\nbase/\nhitp$, and
 $(M_p)^{-1}(3,3)/\nbase \approx 2\sigma^2/\nbase/\nhitp$ .

For baseline lengths less than or equal to the spin period, $\tbase
\leq 1 $min, this holds if the sample separation is larger than the
pixels, $\thetas \gg \theta_p$. However, in our case $\thetas <
\theta_p$, and two or three successive samples may hit the same
pixel. These successive hits are then almost always from the same
baseline, and, for the baseline components, fully correlated, i.e.,
equal.

Denote by $nf_n$ the fraction of hits to a pixel that belong to a
sequence of exactly $n$ successive hits to the same pixel, i.e,
there are $f_n\nhitp$ such sequences, and $\sum_n nf_n = 1$.  The
variance of $I_p$ is then
 \beq
   \langle I_p^2 \rangle =
   \left(\sum_n(n^2f_n)\right)\frac{\sigma_{wb}^2}{\nhitp} \,,
 \eeq
where $\sum_n(n^2f_n) \geq 1$.

With Eq.~(\ref{sigmawb}) we have for the I map rms
 \beq
    I_\mathrm{rms}
    \approx \sqrt{\left\langle \frac{\sum n^2f_n}{\nhit}\right\rangle}
    \frac{\sigma}{\sqrt{\nbase}} \,.
 \label{Iapx_wr}
 \eeq

To get an estimate for $\langle\sum_n(n^2f_n)\rangle$ we assume a
square pixel and a uniform hit probability distribution within the
pixel area.

For scanning in the direction of the pixel diagonal, for $\thetas >
\sqrt{2}\theta_p$ there are no successive hits to the same pixel and
$f_1 = \sum_n(n^2f_n) = 1$. For $\theta_p/\sqrt{2} < \thetas <
\sqrt{2}\theta_p$ there can be a maximum of two hits, with $\langle
f_2\rangle = \left(1-r/\sqrt{2}\right)^2$ and
 \beq
    \sum_n n^2f_n = 3 - 2\sqrt{2}r + r^2  \,;
 \eeq
and with $\sqrt{2}\theta_p/3 < \thetas < \theta_p/\sqrt{2}$ a
maximum of three hits, with $\langle f_1\rangle = r^2$, $\langle
f_2\rangle = 3\sqrt{2}r - 1 - 7r^2/2$ and
 \beq
    \sum_n(n^2f_n) = 5 - 6\sqrt{2}r + 5r^2 \,;
 \eeq
where $r \equiv \thetas/\theta_p$.

For scanning in the direction of the pixel side, two hits, but no
more, are possible for $\theta_p/2 < \thetas < \theta_p$ with $f_2 =
1 - r$.

For our case, $ \sqrt{2}/3 < 1/2 < r = 0.682 < 1/\sqrt{2}$, which
gives
 \beq
    \left\langle \sum_n(n^2f_n)\right\rangle = 1.54 =
    (1.24)^2
 \label{successive}
 \eeq
for scanning in the diagonal direction, and $\langle
\sum_n(n^2f_n)\rangle = 1.64 = (1.28)^2$ for the pixel side
direction.

However, HEALPix pixels are not square, but can be significantly
elongated. In principle, the mean value $\langle \frac{\sum
n^2f_n}{\nhit} \rangle$ could be calculated from the pointing data
and the chosen pixelization.  Here we just take it by comparing the
actual $I_\mathrm{rms}$ from the maps (see Table~\ref{table:wbrmap})
to Eq.~(\ref{Iapx_wr}).  This gives (from $\tbase = 15$ s and 1 min)
$\left\langle\sum n^2f_n/\nhit\right\rangle \approx 0.000636\approx
1.52\langle\nhit^{-1}\rangle \approx (0.025)^2$. We denote
 \beq
    \biggl\langle\biggl\langle\sum n^2f_n\biggr\rangle\biggr\rangle
    \equiv \left\langle \frac{\sum n^2f_n}{\nhit}\right\rangle \bigg/
    \langle\nhit^{-1}\rangle
    \approx 1.52.
 \eeq
Eq.~(\ref{successive}) is quite close to this.

For the other Stokes parameters, we expect
 \beq
    Q_\mathrm{rms} \approx U_\mathrm{rms} \approx \sqrt{2}I_\mathrm{rms}
 \label{QUapx_wb}
 \eeq
where the approximation corresponds to assuming an ideal
distribution of polarization directions.  We expect this
approximation to be good for the full year data, since most pixels
have \emph{rcond} values close to 0.5.

For baselines longer than the spin period, contributions to a pixel
from successive scan circles tend to come from the same baseline, so
the reduction in the white noise baseline variance is canceled by
the reduction in the number of contributing baselines. Thus the map
rms from white noise baselines is almost flat between $\tbase = 1$
min and $\tbase = 1$ h.

Although in the time domain the white noise baseline error is much
smaller than white noise reference baselines, their correlations
make them important in the map domain.  Assuming the correlation
between $(a+b)/2$ baselines contributing to the same pixel were $c$,
the expected variance of the white noise baseline error map would be
 \beq
    \langle I_{s-r}^2 \rangle = \langle I_\mathrm{ref}^2 \rangle
    \frac{2\sigma_{we+}^2}{\sigma_{wr}^2}\left(1+\frac{c}{\langle\nhit^{-1}\rangle}\right)
    \,,
 \label{bslcorr}
 \eeq
where $\sigma_{we+}$ is the stdev of the $(a+b)/2$ white noise
baseline errors, given in Table~\ref{table:wblmis}. Assuming
Eq.~(\ref{bslcorr}) to hold for our maps, we have solved for $c$ for
different $\tbase$ in Table~\ref{table:wbsrmap}. These numbers can
be compared to Fig.~\ref{fig:white_aucorr} and
Table~\ref{table:wblmiscor}.

 Since the
white noise reference baselines and the white noise baseline errors
are uncorrelated, the full white noise baseline map rms is close to
the rss of these two components.  See Table~\ref{table:wblmap}.

\begin{table}[!tbp]
 \begin{center}
 \begin{tabular}{lllll}
 $\tbase$ & $I_\mathrm{rms}$ & $ \langle\langle\sum_n n^2f_n\rangle\rangle$
 & $Q_\mathrm{rms}$ & $U_\mathrm{rms}$ \\
 \hline
 2.5 s   & 4.947 & 1.542 & 6.974 & 7.072 \\
 15 s    & 2.006 & 1.521 & 2.846 & 2.914 \\
 1 min   & 1.003 & 1.522 & 1.435 & 1.466 \\
 1 h     & 0.895 &       & 1.289 & 1.320 \\
 \hline
 \end{tabular}
 \end{center}
 \caption{
Statistics of the white noise reference baseline maps. The map rms
values given are in  $\mu$K. The third column is an estimate of
$\langle \frac{\sum n^2f_n}{\nhit} \rangle /
\langle{\nhit^{-1}}\rangle$ obtained by comparing $I_\mathrm{rms}$
to Eq.~(\ref{Iapx_wr}).
 }
 \label{table:wbrmap}
\end{table}

\begin{table}[!tbp]
 \begin{center}
 \begin{tabular}{lllll}
 $\tbase$ & $I_\mathrm{rms}$ & $c$
 & $Q_\mathrm{rms}$ & $U_\mathrm{rms}$ \\
 \hline
 2.5 s   & 4.543 & 0.168 & 6.223 & 6.433 \\
 15 s    & 1.165 & 0.092 & 1.664 & 1.666 \\
 1 min   & 0.634 & 0.102 & 0.881 & 0.904 \\
 1 h     & 0.630 &       & 0.875 & 0.898 \\
 \hline
 \end{tabular}
 \end{center}
 \caption{
Statistics of the white noise baseline error maps. The map rms
values given are in  $\mu$K. The third column is an estimate of the
correlation between baselines contributing to the same pixel
obtained by comparing $I_\mathrm{rms}$ to Eq.~(\ref{bslcorr}).
 }
 \label{table:wbsrmap}
\end{table}

\begin{table}[!tbp]
 \begin{center}
 \begin{tabular}{llll}
 $\tbase$ & $I_\mathrm{rms}$ & $Q_\mathrm{rms}$ & $U_\mathrm{rms}$ \\
 \hline
 2.5 s   & 6.729 & 9.303 & 9.580 \\
 15 s    & 2.319 & 3.291 & 3.356 \\
 1 min   & 1.181 & 1.678 & 1.715 \\
 1 h     & 1.088 & 1.552 & 1.590 \\
 \hline
 \end{tabular}
 \end{center}
 \caption{
Statistics of the white noise baseline maps. The map rms values
given are in  $\mu$K.
 }
 \label{table:wblmap}
\end{table}

\subsubsection{Unmodeled $1/f$ noise}
\label{sec:map_unmodeled}

Most of the power in unmodeled $1/f$ noise is in frequencies near
$1/(2\tbase)$.  Therefore, for $\theta_p \ll \theta_\mathrm{base}$
successive hits to the same pixel should be almost fully correlated.
When $\tbase$ is much below the spin period, hits from different
spin periods should be almost uncorrelated. Thus we can estimate the
unmodeled $1/f$ noise map rms in the same manner as the white noise
baseline map, as
 \beq
    I_\mathrm{rms}
    \approx \sqrt{\left\langle \frac{\sum n^2f_n}{\nhit}\right\rangle}
    \,\sigma_{u}
    \approx 0.025\sigma_{u} \,.
 \label{Iapx_u}
 \eeq
When $\tbase$ is comparable to the spin period, there will be
correlations (positive or negative) between hits from nearby spin
periods. We compare Eq.~(\ref{Iapx_u}) to the actual binned maps of
unmodeled $1/f$ noise in Table~\ref{table:umap}.  We see that for
$\tbase = 1$ min, Eq.~(\ref{Iapx_u}) is an overestimate, indicating
that there are negative correlations between hits from nearby spin
periods.

\begin{table}[!tbp]
 \begin{center}
 \begin{tabular}{llllll}
 $\tbase$ & Eqs.~(\ref{sigu_apx_num},\ref{Iapx_u}) &
 Eq.~(\ref{Iapx_u}) & $I_\mathrm{rms}$ & $Q_\mathrm{rms}$ & $U_\mathrm{rms}$ \\
 \hline
 7.5 s   & 2.994 & 3.081 & 3.062 & 4.314 & 4.358 \\
 15 s    & 3.727 & 3.901 & 3.852 & 5.450 & 5.526 \\
 1 min   & 6.053 & 6.313 & 5.778 & 8.128 & 8.225 \\
 1 h     &       &       & 9.357 & 13.02 & 13.178 \\
 \hline
 \end{tabular}
 \end{center}
 \caption{
Statistics of the unmodeled $1/f$ noise maps. The map rms values
given are in $\mu$K. The second and third columns are estimates for
the $I$ rms based on Eq.~(\ref{Iapx_u}), with $\left\langle
\frac{\sum n^2f_n}{\nhit}\right\rangle$ taken from
Table~\ref{table:wbrmap}. The third column uses the actual
$\sigma_u$, whereas the second column uses the analytical estimate
(\ref{sigu_apx_num}).
 }
 \label{table:umap}
\end{table}

\subsubsection{Ideal scanning}
\label{sec:ideal_scanning}

We define \emph{ideal scanning} so that the pointings from the
different scan circles of the same repointing period fall on top of
each other, i.e., there is no nutation and the sampling is
synchronized with the spin period.  In this case, the part of the
unmodeled $1/f$ noise for long baselines ($\tbase$ a multiple of the
spin period) that is modeled by 1 min baselines gets totally
averaged out, so that the contribution from unmodeled $1/f$ noise to
residual maps would stay constant from $\tbase = 1$ min to $\tbase =
1$ h. In our nonideal case, some of this noise leaks out, so that
the unmodeled contribution rises slowly in this range also. See
dashed red line in Fig.~\ref{fig:comp_res_in_6}.

Likewise, for an ideal scanning, the white noise reference baselines
make an equal contribution to the map for any baseline length that
is an integer multiple of the spin period, and fits into the
repointing period an integer number of times.

\subsubsection{Total noise}
\label{sec:total_noise}

From Fig.~\ref{fig:comp_res_in_6} we see that the two dominant
contributions to the residual maps are the white noise reference
baselines and the unmodeled $1/f$ noise.  For both of them we have
analytical estimates, and both of them map from time domain to map
domain in roughly the same way.  Thus we get an analytical estimate
for the residual $I$ map rms by multiplying the $\sqrt{\sigma_{wr}^2
+ \sigma_u^2}$ estimate from Eqs.~(\ref{sigmawb}) and
(\ref{sigu_apx}) with $\sqrt{\left\langle \frac{\sum
n^2f_n}{\nhit}\right\rangle} \approx 0.025$ for $\tbase \leq 1$ min.
When $\alpha > 1$ and $\tbase \gg \tsample$, so that
$(f_x/\fc)^\alpha \ll 1$, we have
 \bea
     I_\mathrm{rms}^2 & \approx & \sigma^2\left\langle\left\langle\sum n^2
     f_n\right\rangle\right\rangle\langle\nhit^{-1}\rangle \biggl[\frac{1}{\fsample\tbase} + \nn\\
     & &
     + \left(\frac{2\fk}{\fsample}\right)\left(2\fk\tbase\right)^{\alpha-1}
     \left(\frac{4}{(5-\alpha)(\alpha-1)}\right)\biggr] \,.
 \label{map_rms_apxgen}
 \eea
This gives for our case ($\sigma = 2700 \mu$K, $\langle\langle\sum
n^2 f_n\rangle\rangle = 1.52$, $\nhit^{-1} = 0.000418$, $\fsample =
76.8$ Hz, $\alpha = 1.7$),
 \beq
    I_\mathrm{rms}^2 \approx 60.3\tbase^{-1} + 2.08\tbase^{0.7}
 \label{map_rms_apx}
 \eeq
where $\tbase$ is given in seconds. (For $\tbase \geq 1$ min, our
analytical treatment can just estimate that the map rms should stay
constant from $\tbase = 1$ min to $\tbase = 1$ h.)

However, this estimate is not as good as in the time domain, since
the importance of the neglected components, the baseline errors, has
grown dramatically when going from the time domain to the map
domain. See Fig.~\ref{fig:comp_res_in_6}. Since these components
rise towards shorter baselines, the residual map rms is minimized at
a somewhat larger $\tbase$ than Eq.~(\ref{opt_tbase}) gives.

\subsection{Pixelization noise and signal baselines}

\begin{figure}[!tbp]
  \begin{center}
    \includegraphics*[trim=100 70 70 250, clip, width=0.24\textwidth]
    {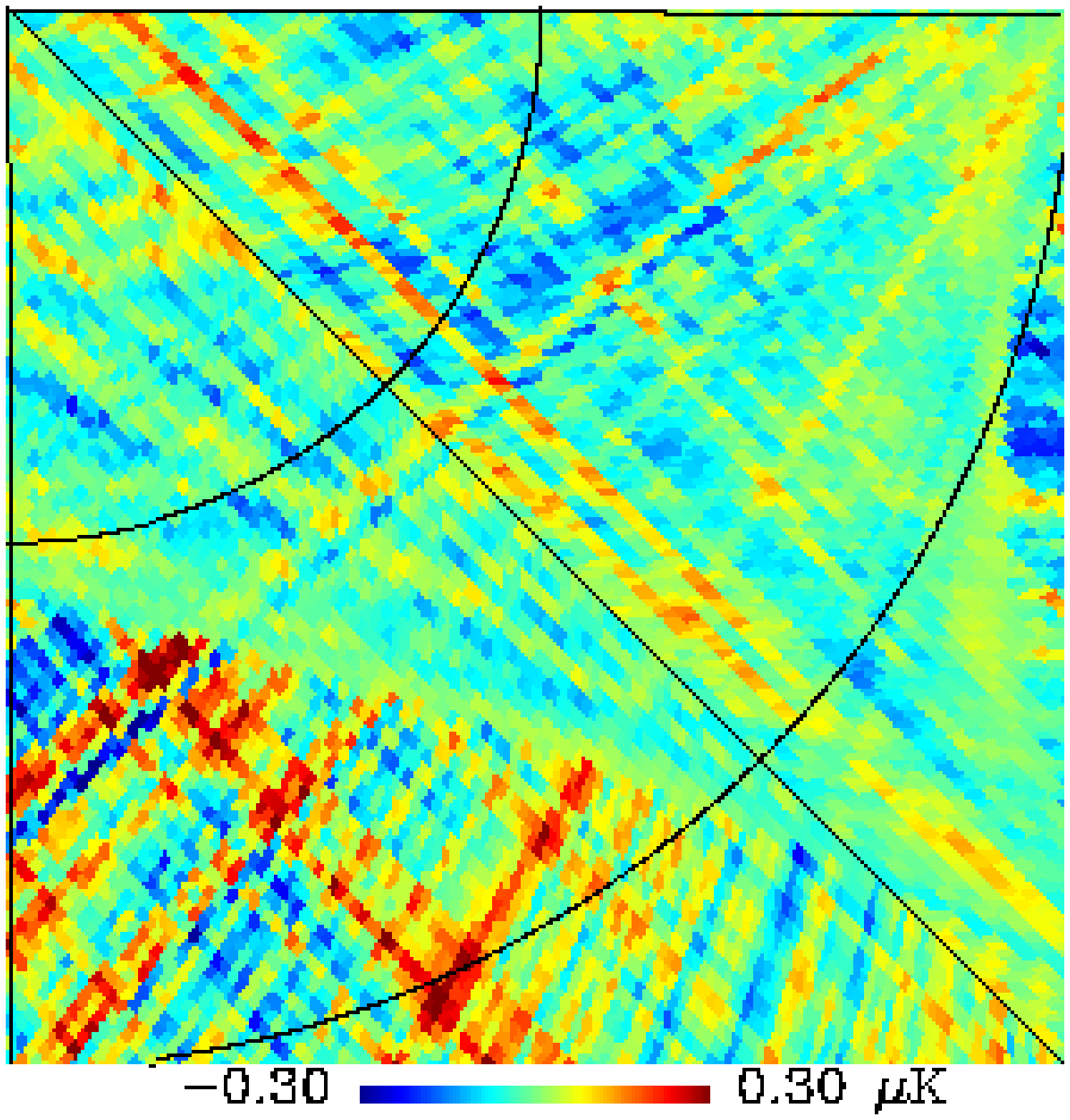}
    \includegraphics*[trim=100 70 70 250, clip, width=0.24\textwidth]
    {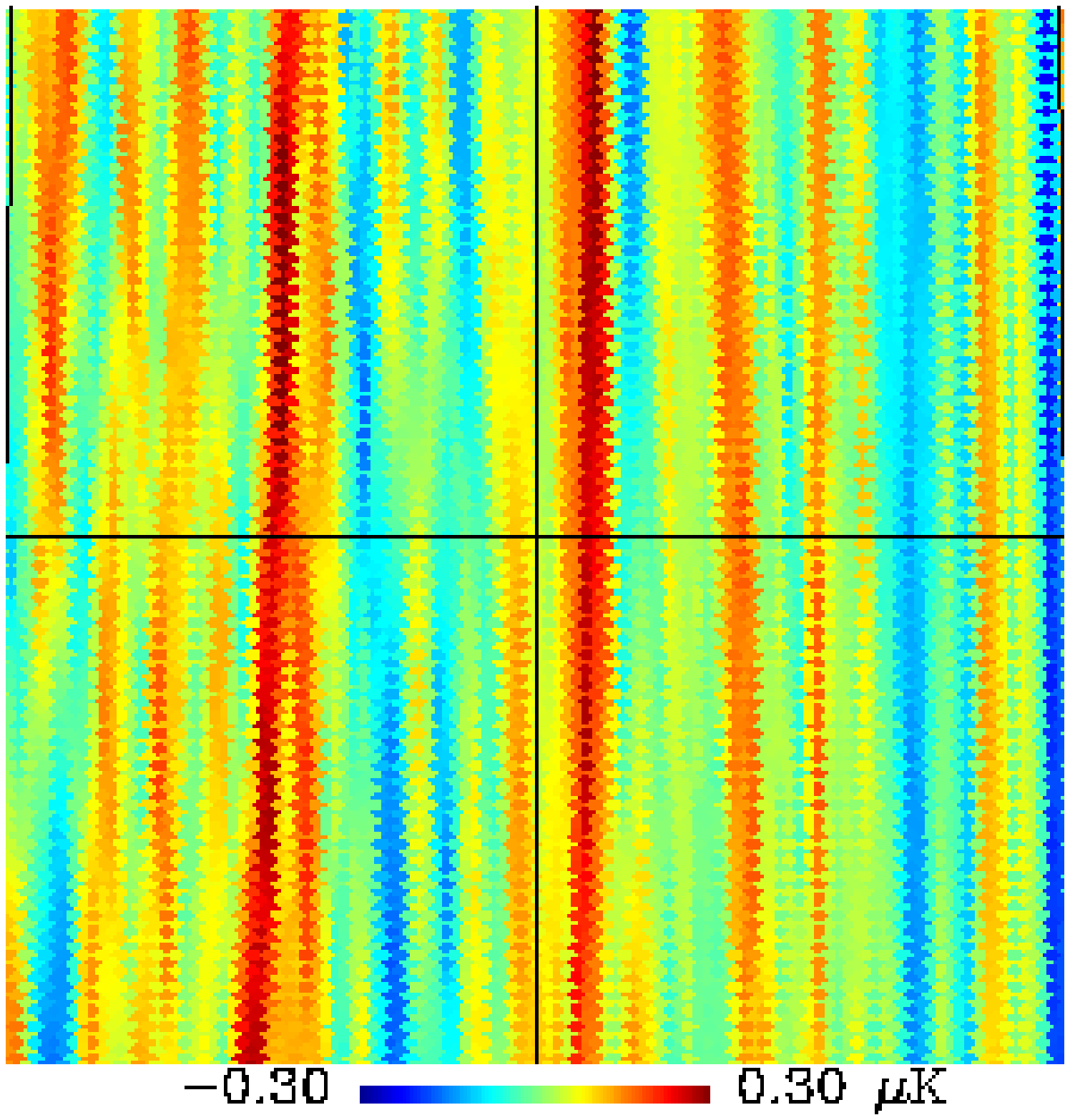} \\
    \includegraphics*[trim=100 70 70 250, clip, width=0.24\textwidth]
    {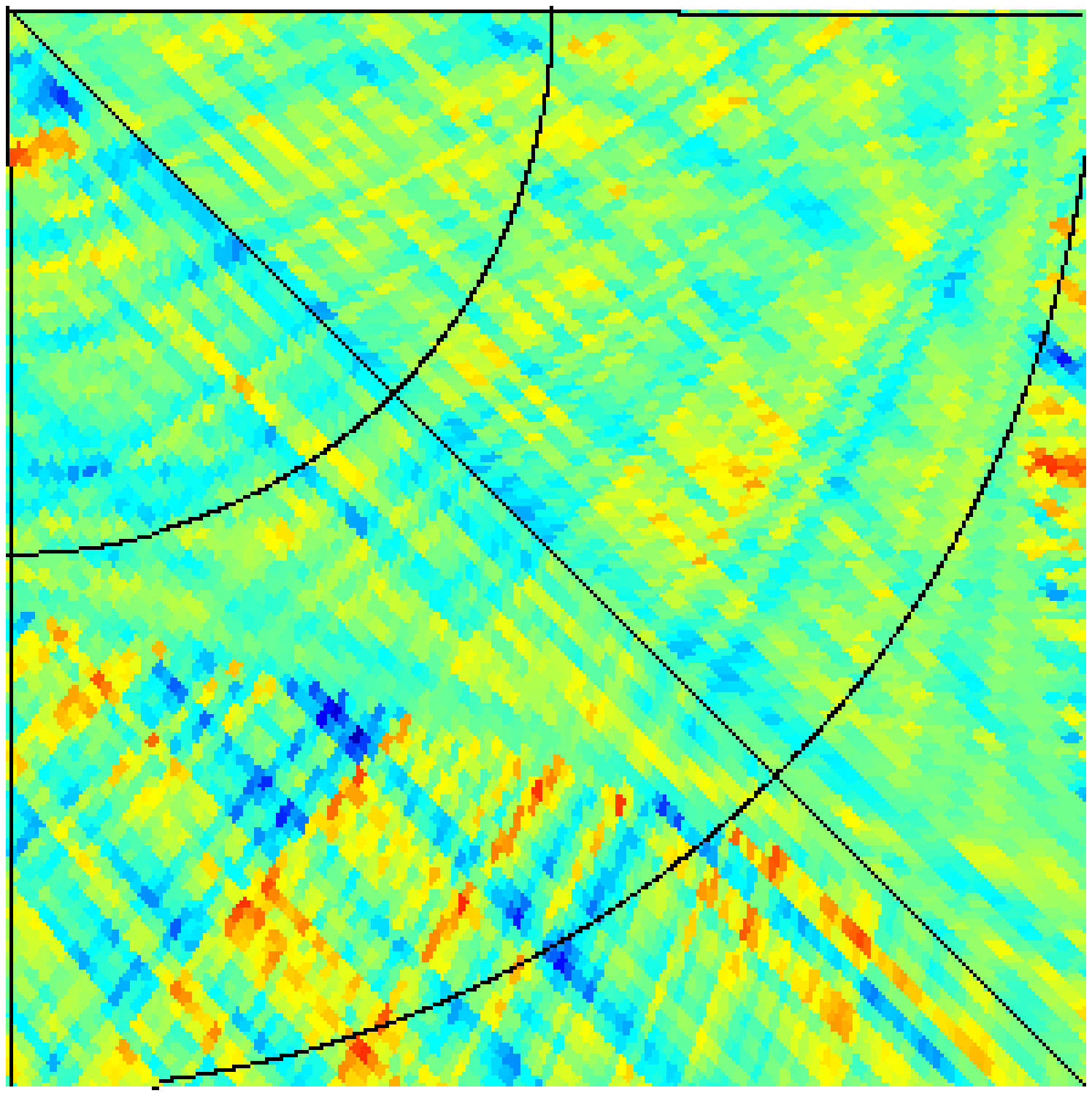}
    \includegraphics*[trim=100 70 70 250, clip, width=0.24\textwidth]
    {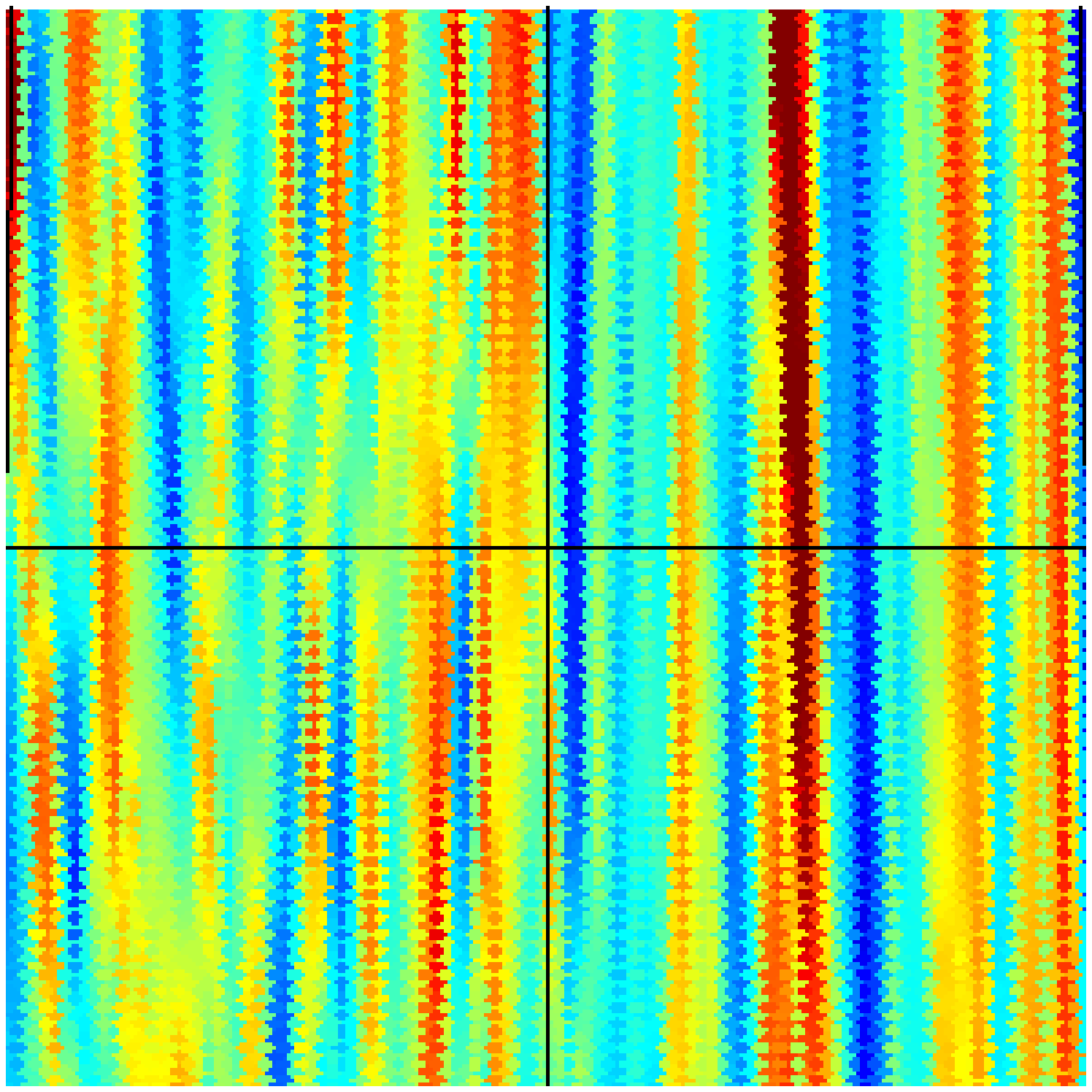} \\
    \vspace{-0.3cm}
    \includegraphics*[trim=100 70 70 250, clip, width=0.24\textwidth]
    {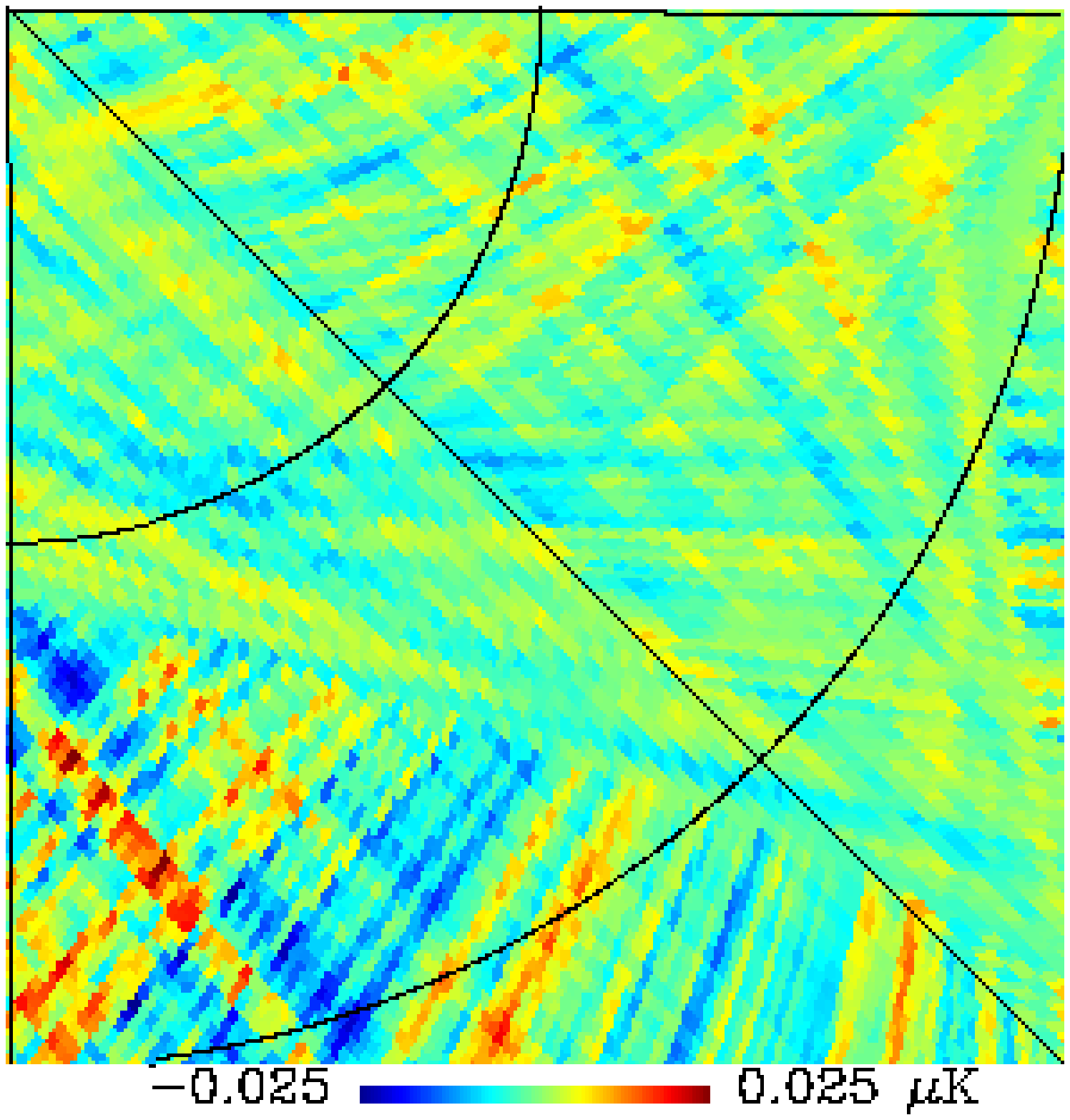}
    \includegraphics*[trim=100 70 70 250, clip, width=0.24\textwidth]
    {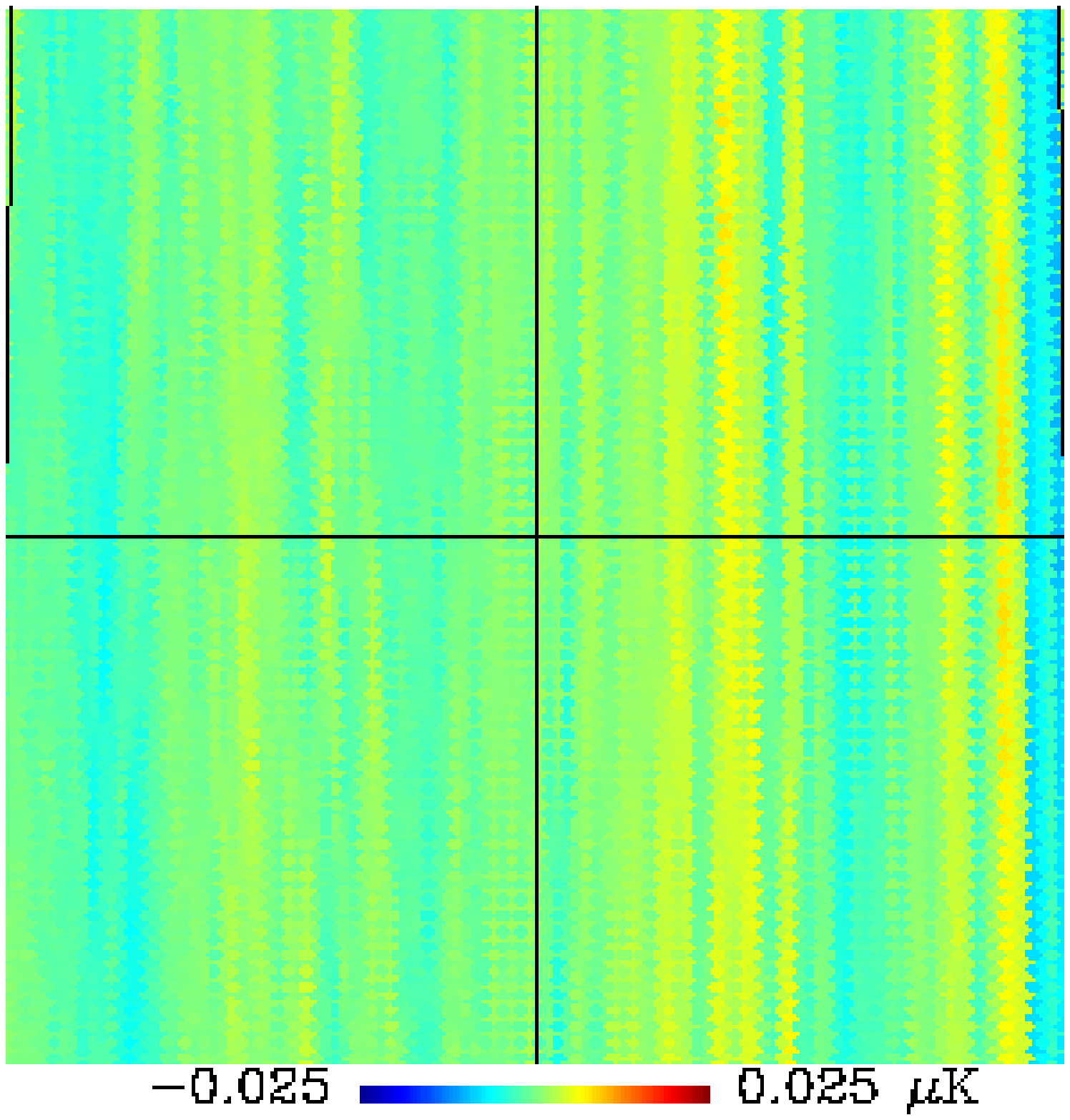} \\
    \vspace{-0.2cm}
    \caption{
The signal baseline $I$ (\emph{top}), $Q$ (\emph{middle}), and $U$
(\emph{bottom}) maps $-\bB\bF\bA\bs$ ($\tbase = 1$ min) for the two
$10^\circ\times10^\circ$ regions.
    }
    \label{fig:sbl_IQU_10by10_map}
  \end{center}
\end{figure}

For pixelization noise, already the reference baselines are strongly
correlated (see Fig.~\ref{fig:sig_aucorr}) and therefore their map
rms cannot be estimated like for white noise reference baselines and
unmodeled $1/f$ noise. Due to these correlations their impact in the
map level is significantly larger than their small variance in the
time level (see Table~\ref{table:sbl}) would indicate.  Instead, for
both the pixelization noise reference baselines and the signal
baseline error, the situation is similar to white noise and $1/f$
noise baseline error.

For the residual $1/f$ and white noise baselines, the $Q$ and $U$
maps look the same as the $I$ maps, just with a factor
$\sim\!\sqrt{2}$ larger amplitude, since they originate from the
same time-domain noise, which is independent for each detector.

For the signal baselines (see Fig.~\ref{fig:sbl_IQU_10by10_map}) the
situation is, however, different, since they originate from the
signal, where $Q$ and $U$ are much smaller than $I$.  We also note
that $Q$ is much larger than $U$, although they are of same
magnitude in the signal. This is related to the coordinate
dependence of the definition of the Stokes parameters $Q$ and $U$
together with a combination of factors in our study. First, the
signal contains only $E$ mode polarization, which means that $Q$ has
structures along the coordinate lines, whereas $U$ has structures
oriented $45^\circ$ from them.  Second, we are using ecliptic
coordinates, and we have employed a scanning strategy, where the
scanning goes almost parallel to the lines of longitude for a large
part of the sky. The signal baselines originate from the signal
gradients within pixels. For a signal structure oriented along the
scanning direction, the signal gradient structure remains similar
for a sequence of pixels along the scanning.  Thus the measurement
differences between different scans through these pixels are similar
for a sequence of pixels, favoring their misinterpretation as noise
baselines.

\begin{table}[!tbp]
 \begin{center}
 \begin{tabular}{llllll}
 $\tbase$ & coord. &
 $I$ rms & $Q$ rms & $U$ rms & $P$ rms \\
 \hline
 1 min   & E & 123.3 & 9.4 & 4.5 & 10.4 \\
 1 min   & G & 115.9 & 6.8 & 7.9 & 10.4 \\
 1 h     & E & 124.0 & 9.4 & 4.6 & 10.5 \\
 1 h     & G & 116.5 & 6.9 & 7.9 & 10.5 \\
 \hline
 \end{tabular}
 \end{center}
 \caption{
Effect of the coordinate system (E $=$ ecliptic, G $=$ galactic) on
the signal baseline maps.  The map rms is given in nK (not in
$\mu$K, like the other tables!), and $P$ stands for
$\sqrt{Q^2+U^2}$.
 }
 \label{table:sigblmap_crd}
\end{table}

To verify the effect of the coordinate system, we redid the $\tbase
= 1$ h and 1 min cases using galactic coordinates. See
Table~\ref{table:sigblmap_crd}. We see that the asymmetry between
$Q$ and $U$ largely disappears, but the total polarization signal
residual is not much affected. For the temperature residual we see a
small improvement. This is partly explained by the reduction of the
signal baseline variance, seen in Table~\ref{table:sbl}.

\subsection{Residuals at different angular scales}

 The residual map rms alone is a poor measure of the
quality of the output map. Since the nature of the residual
 (see Figs.~\ref{fig:res_I_maps_for_diff_tbase} and
 \ref{fig:res_I_10by10_maps_for_db})
is different for different baseline lengths, we need to look at the
structure of the different map residuals in more detail.

For long baselines, the residual mostly comes from the part of the
$1/f$ noise that cannot be modeled with baselines, and appears
mostly at very small angular scales on the map, near the pixel
scale; whereas for shorter baselines it comes from unwanted
baselines, which appear as larger scale structures.
 This can be seen
from Fig.~\ref{fig:smoothed_comp_res}, where we have smoothed the
residual map with the detector beam, before taking the rms. This
smoothing almost erases the difference between the $1/f$ residuals
for baseline lengths from $\tbase = 1$ min to 1 hour. This is
because the 1 min scanning circles fall almost on top of each other
during the 1 hour repointing period, and the width (due to nutation)
of the ring on the sky traced by the beam center during a repointing
period is less than the beam width. Beam-smoothing has much less
effect on the white noise baseline map and the signal baseline map.
Baseline lengths $\tbase = 10$ s to 15 s still give the smallest
total residuals, but the difference from longer baselines is much
reduced by beam-smoothing.  Since residuals at larger scales are for
most purposes more harmful than sub-beam residuals on the map, we
conclude that it is better to choose a somewhat longer baseline than
what would minimize the residual map rms.

\subsection{Angular power spectra of map residuals}

\begin{figure}[!tbp]
  \begin{center}
    \includegraphics*[width=0.5\textwidth]
       {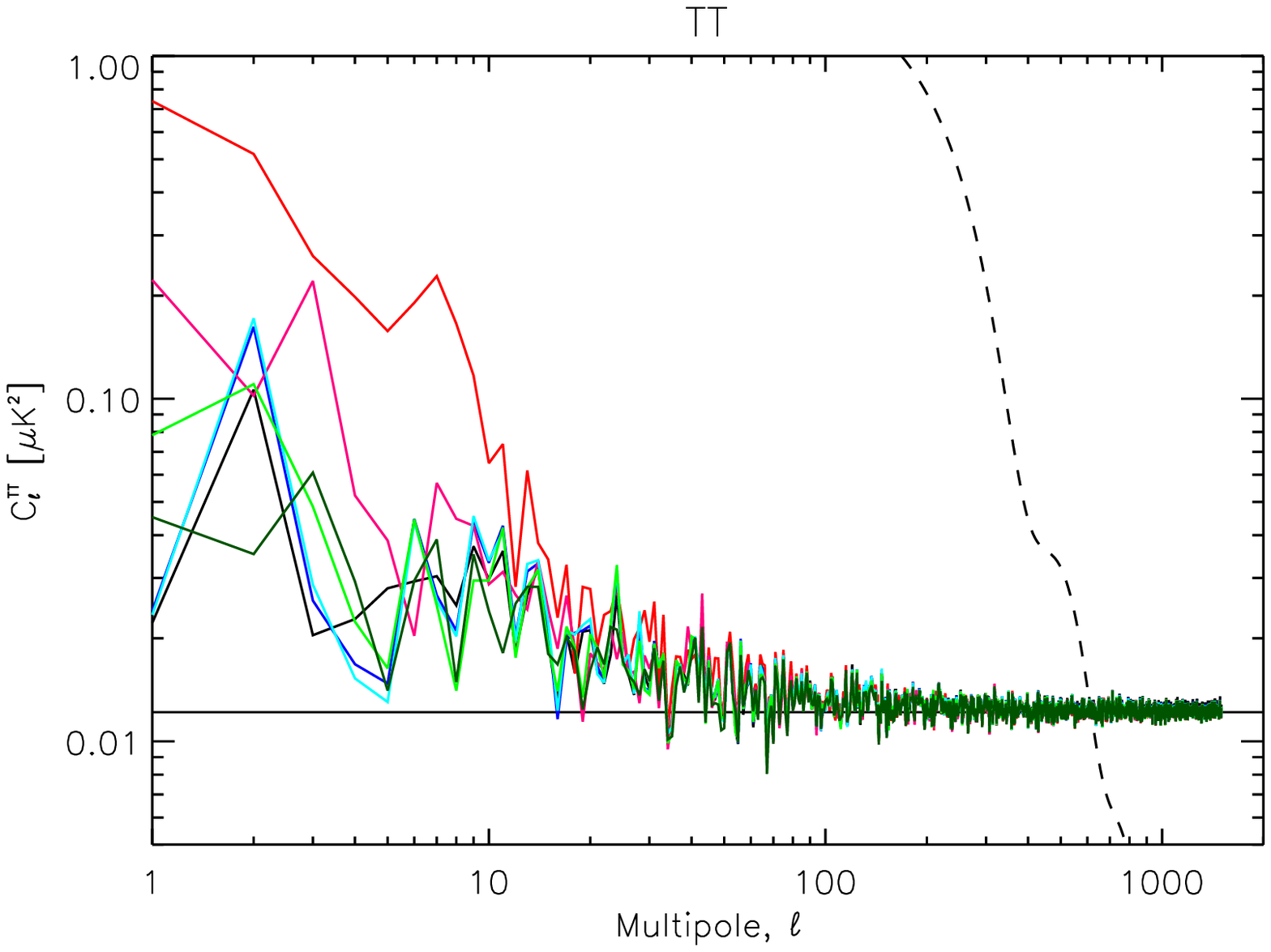}
    \includegraphics*[width=0.5\textwidth]
      {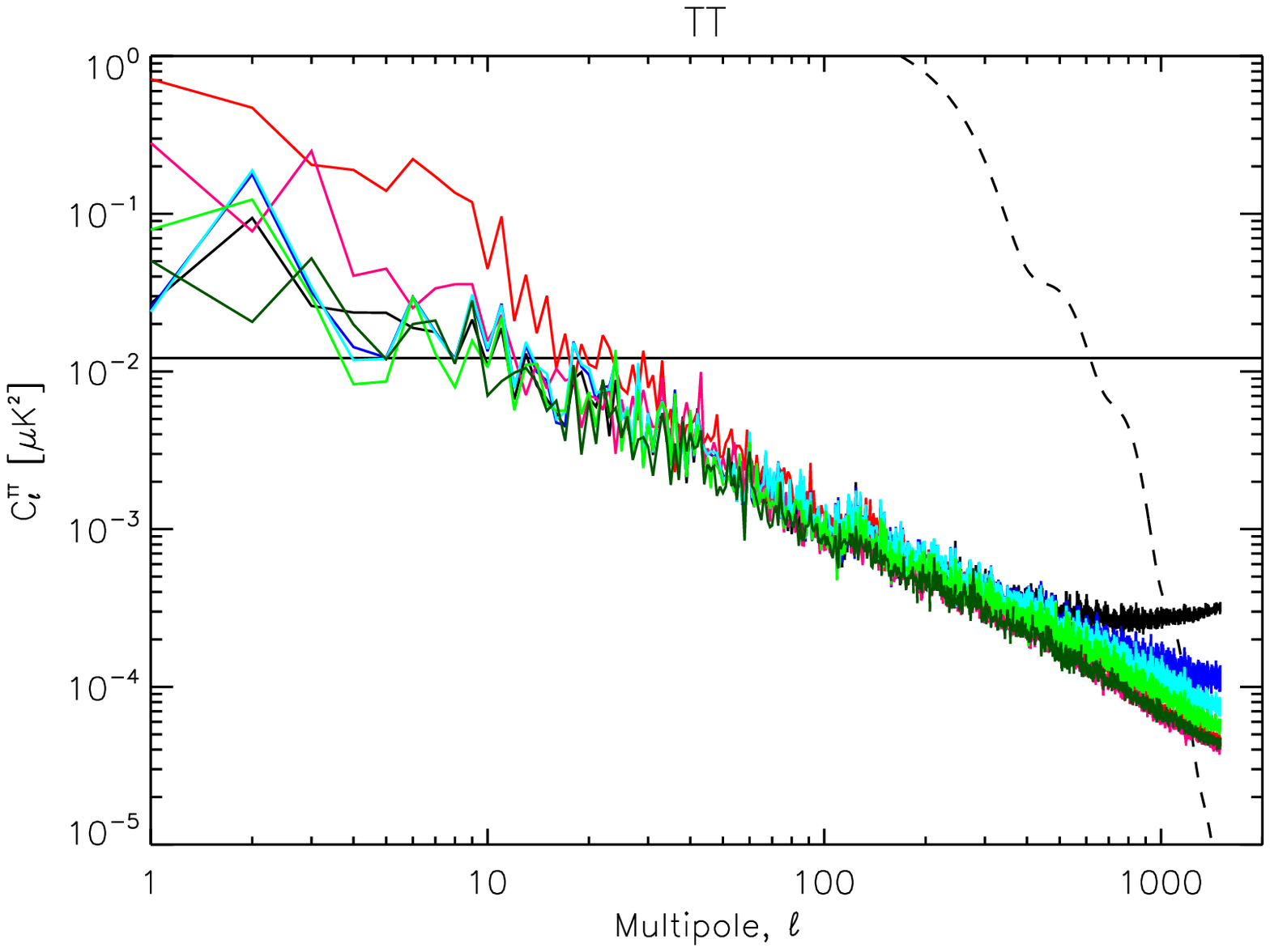}
    \caption{
Angular power spectra of the residual temperature maps for different
baseline lengths: 1 h (\emph{black}),
 4 min
(\emph{blue}), 1 min (\emph{light blue}), 30 s (\emph{light green}),
15 s (\emph{green}), 7.5 s (\emph{pink}), and 5 s (\emph{red}). The
\emph{black horizontal} line is the white noise level and the
\emph{black dashed} line is the theoretical CMB input spectrum
smoothed with the beam and pixel window functions.
 \emph{Top:} Residual map including white noise, $\mout-\bB\bs$.
 \emph{Bottom:} Residual map $\mout-\bB\bs-\bB\bw$ (binned white noise subtracted).
    }
    \label{fig:res_cl}
  \end{center}
\end{figure}

\begin{figure}[!tbp]
  \begin{center}
    \includegraphics*[width=0.5\textwidth]
      {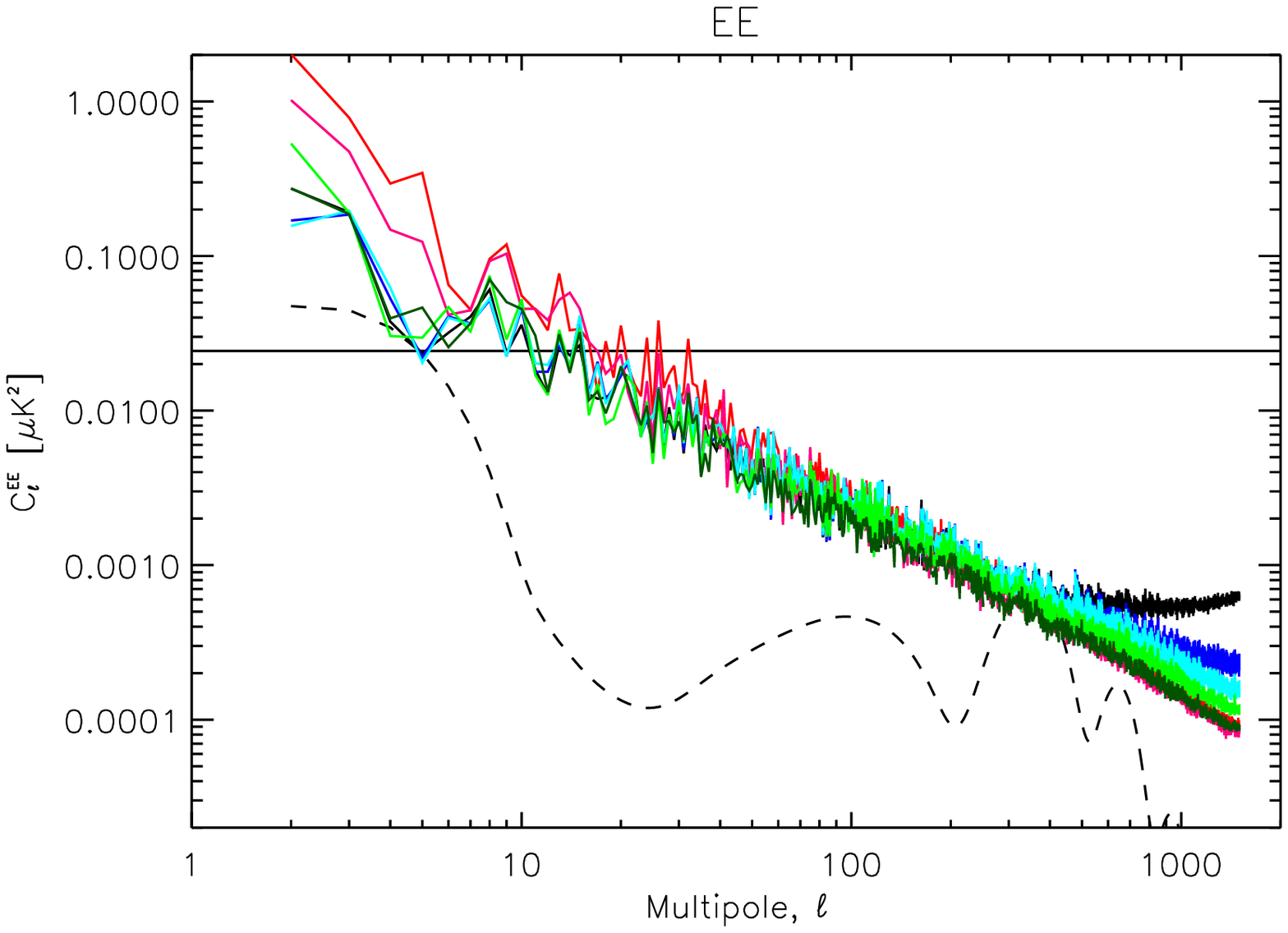}
    \includegraphics*[width=0.5\textwidth]
      {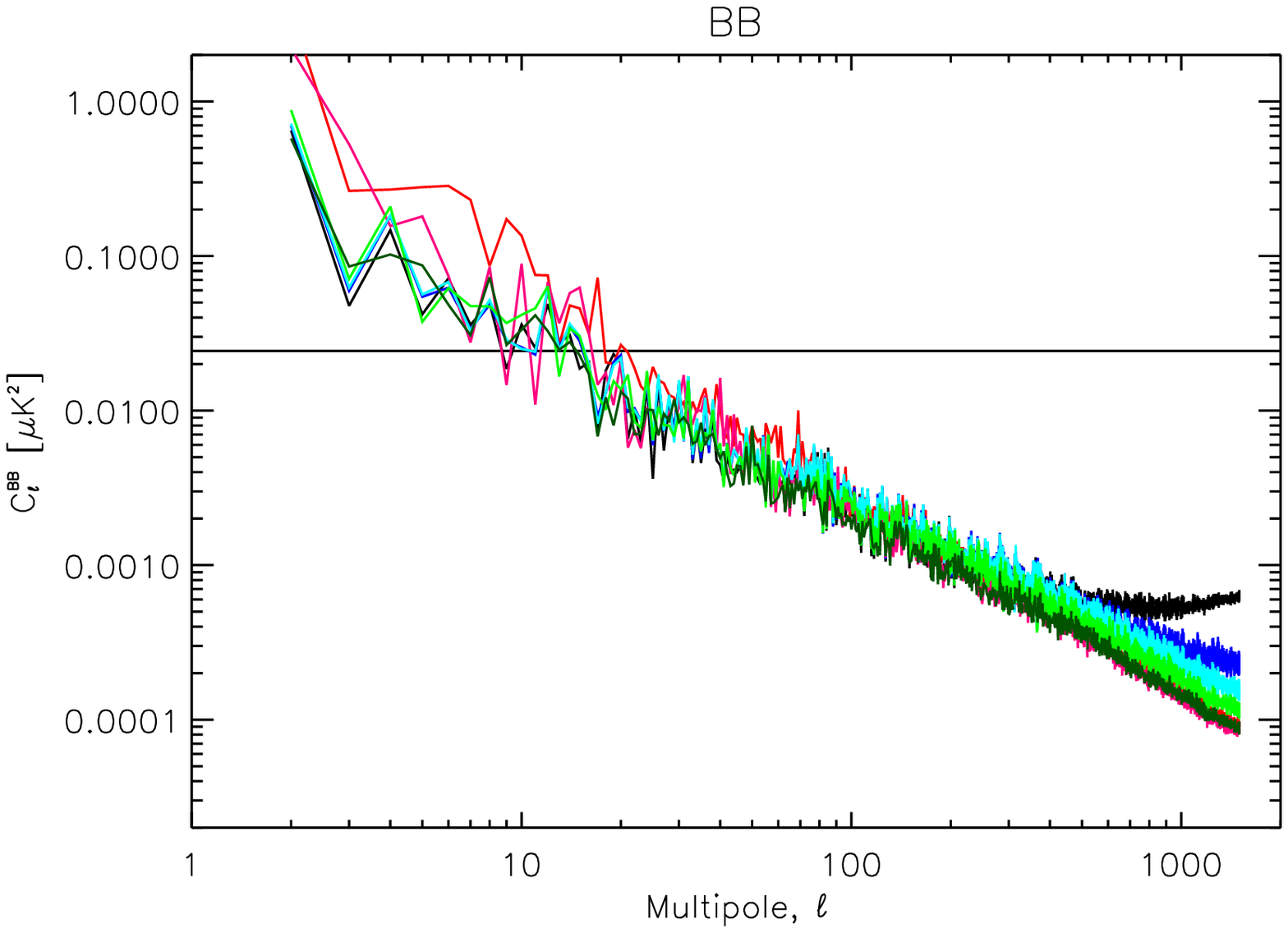}
    \caption{Same as Fig.~\ref{fig:res_cl}, but for the polarization
    $E$ and $B$ mode spectra.
    }
    \label{fig:res_cl_pol}
  \end{center}
\end{figure}

Since we are considering full-sky maps, their angular power spectra
$C_\ell^{XY}$ can be calculated directly from them (we used {\tt
anafast} of the HEALPix package).

We plot the angular power spectra of the residual maps in
Figs.~\ref{fig:res_cl} and \ref{fig:res_cl_pol} for different
baseline lengths.  It is clear that baselines shorter than $\tbase =
10$ s, lead to more large scale structure in the residuals.  Long
baselines lead to a high-$\ell$ tail in the residual that appears
much like white noise (flat $C_\ell$).

\begin{figure}[!tbp]
  \begin{center}
    \vspace{-0.2cm}
    \includegraphics*[width=0.42\textwidth]
      {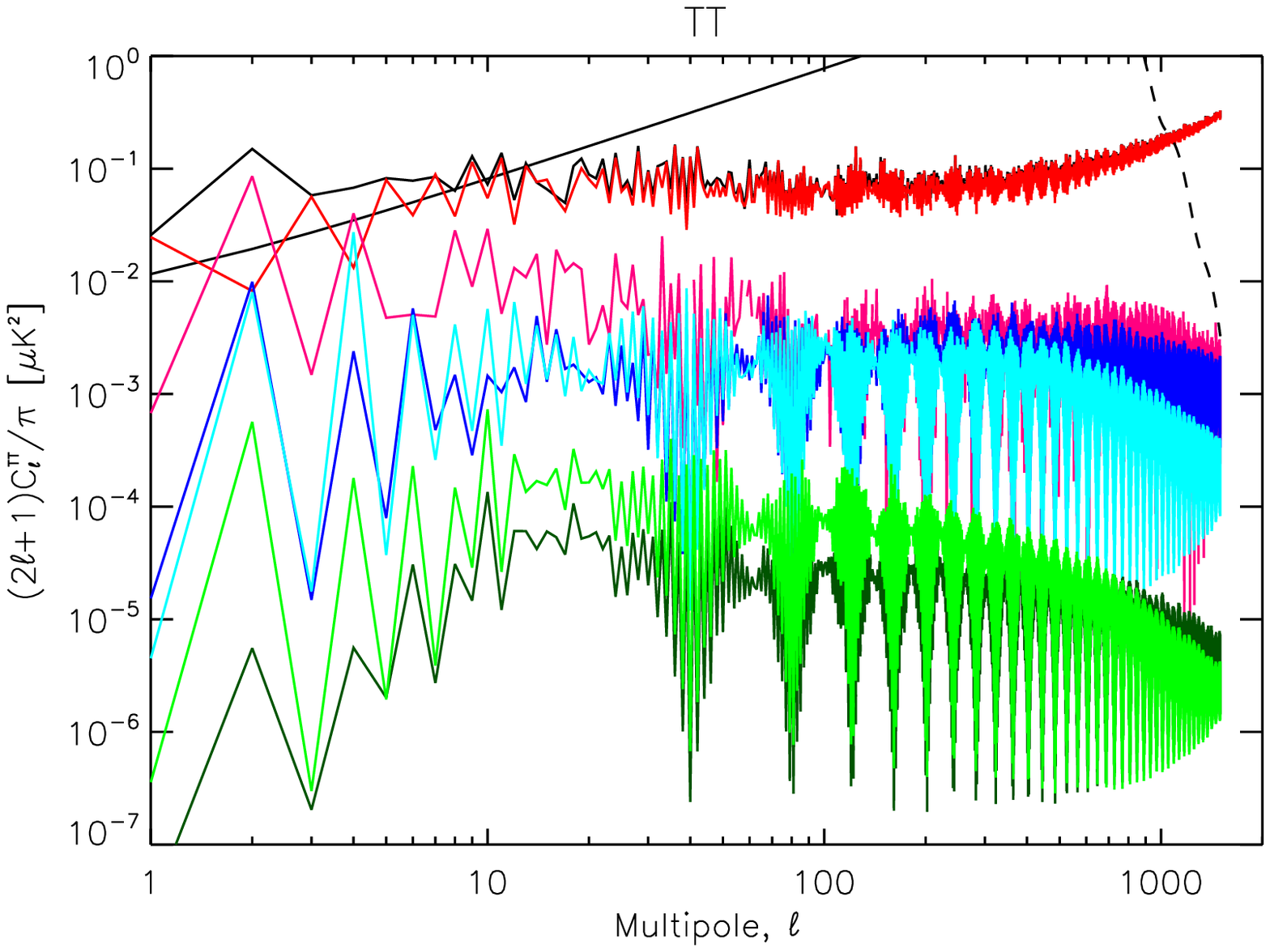}
    \vspace{-0.2cm}
    \includegraphics*[width=0.42\textwidth]
      {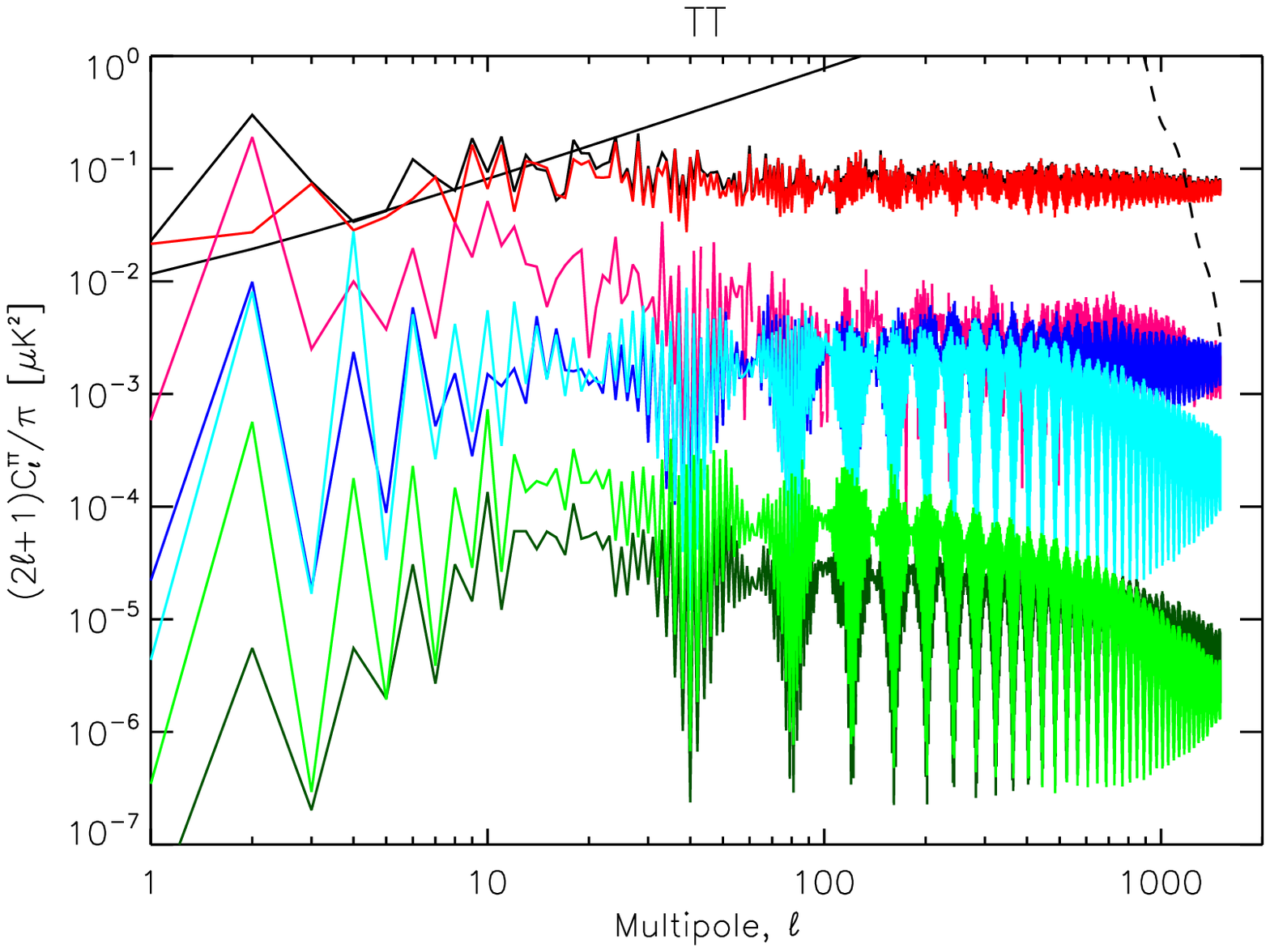}
    \vspace{-0.2cm}
    \includegraphics*[width=0.42\textwidth]
      {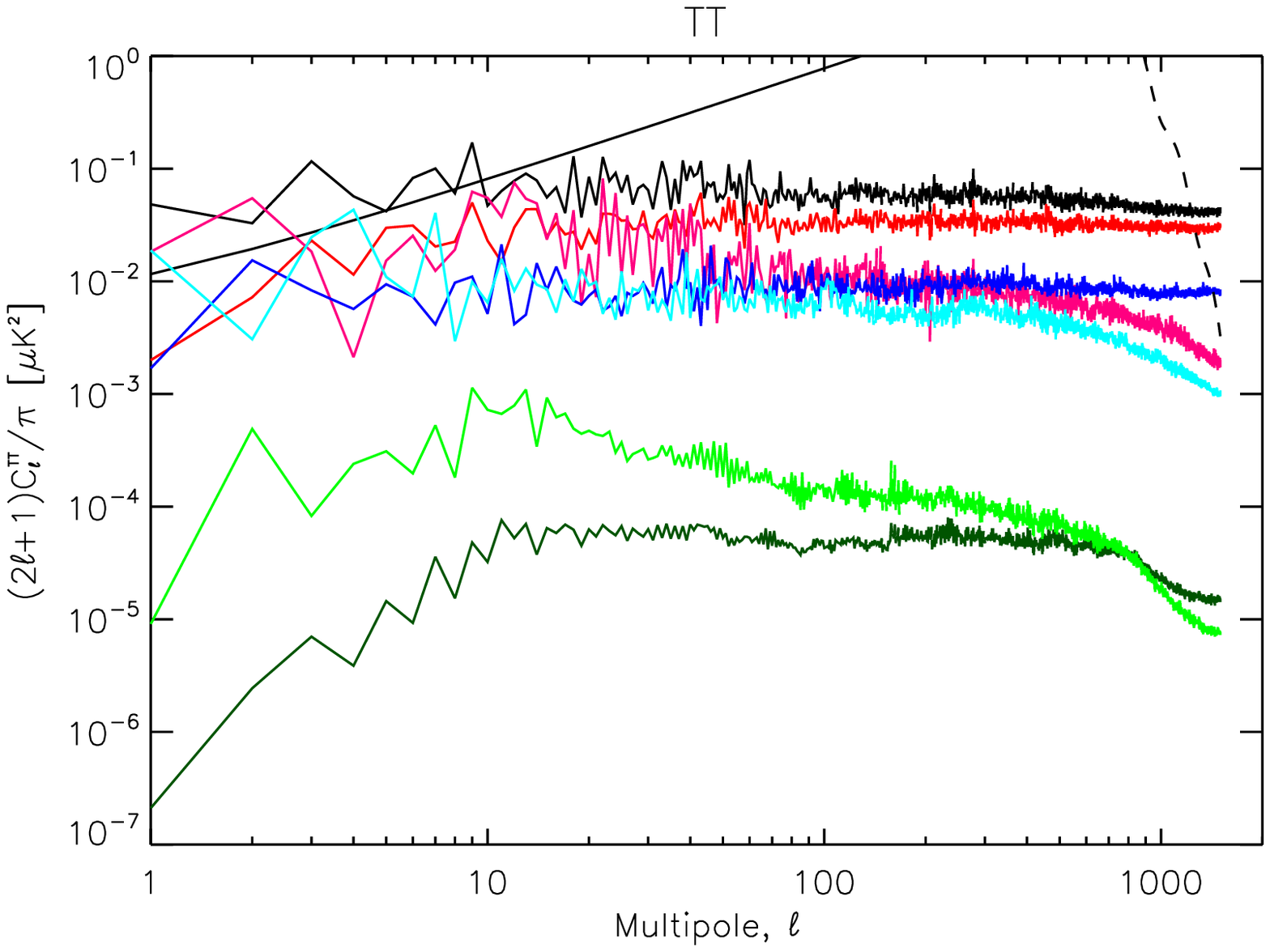}
    \vspace{-0.2cm}
    \includegraphics*[width=0.42\textwidth]
      {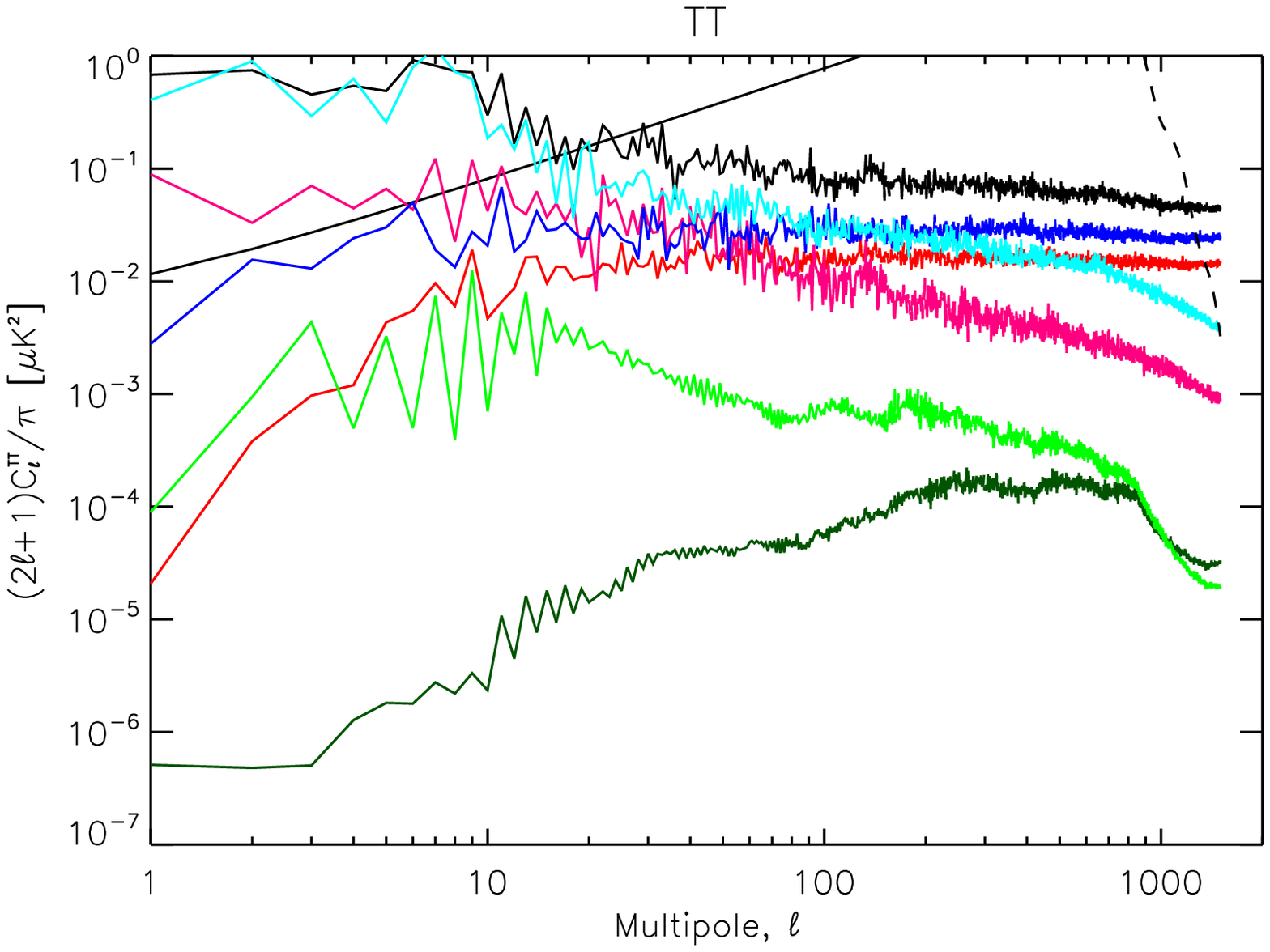}
    \vspace{-0.1cm}
    \caption{
Angular temperature power spectra $C_\ell^{TT}$ of different map
components for four different baseline lengths: 1 h (\emph{top
panel}), 1 min, 15 s, and 5 s (\emph{bottom panel}).
 \emph{Solid black:} Residual map (with binned white noise subtracted).
 This can be split into the following six components:
 \emph{Red:} Unmodeled $1/f$ noise.
 \emph{Pink:} Additional effect of $1/f$ baseline error.
 \emph{Blue:} White noise reference baselines
 \emph{Light blue:} White noise baseline error
 \emph{Green:} Reference baselines of pixelization noise.
 \emph{Light green:} Additional effect of signal baseline error.
The smooth black curve is the white noise level.
    }
    \label{fig:sixcomp_cl_TT}
  \end{center}
\end{figure}

\begin{figure}[!tbp]
  \begin{center}
    \vspace{-0.2cm}
    \includegraphics*[width=0.42\textwidth]
      {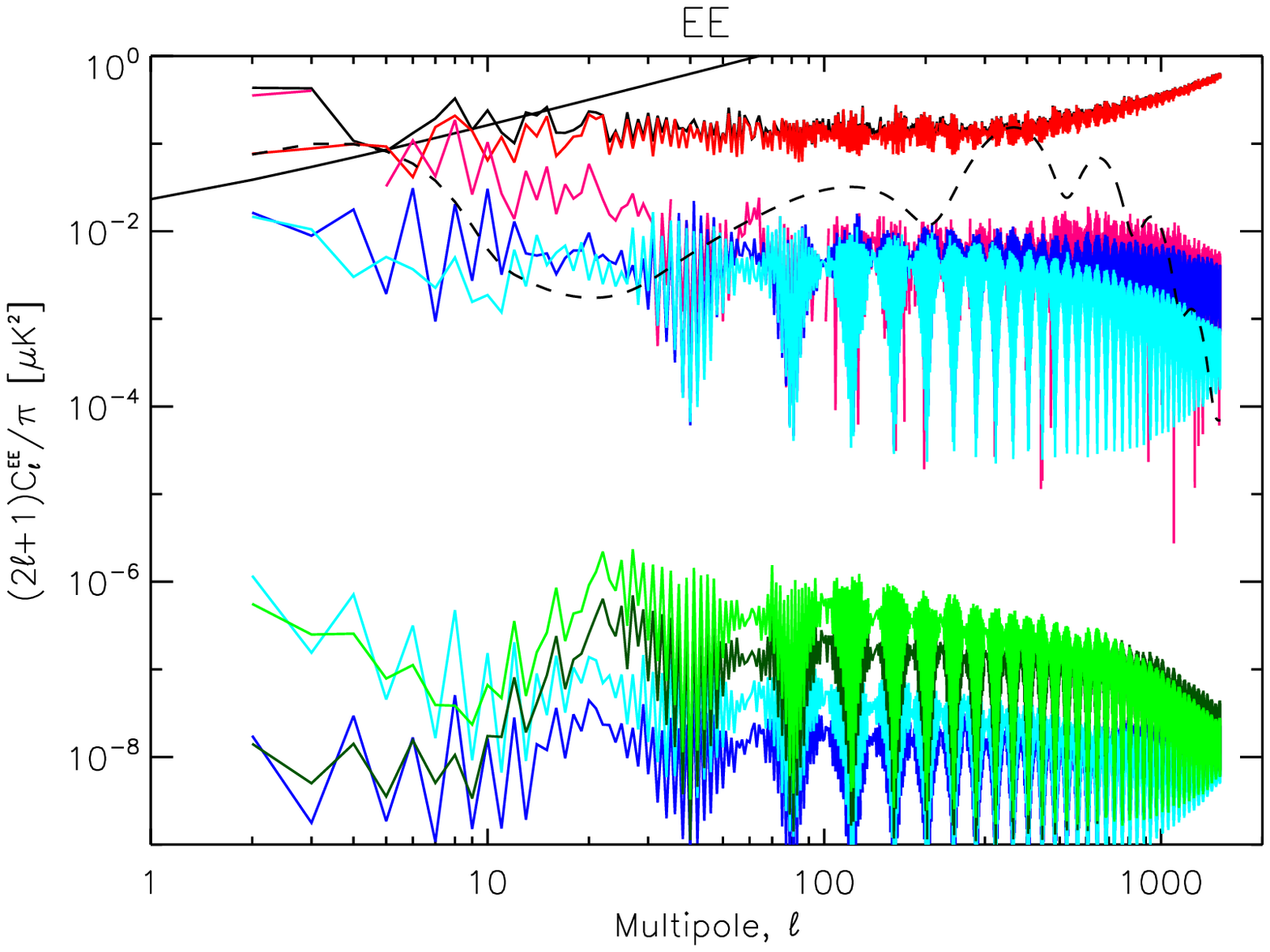}
    \vspace{-0.2cm}
    \includegraphics*[width=0.42\textwidth]
      {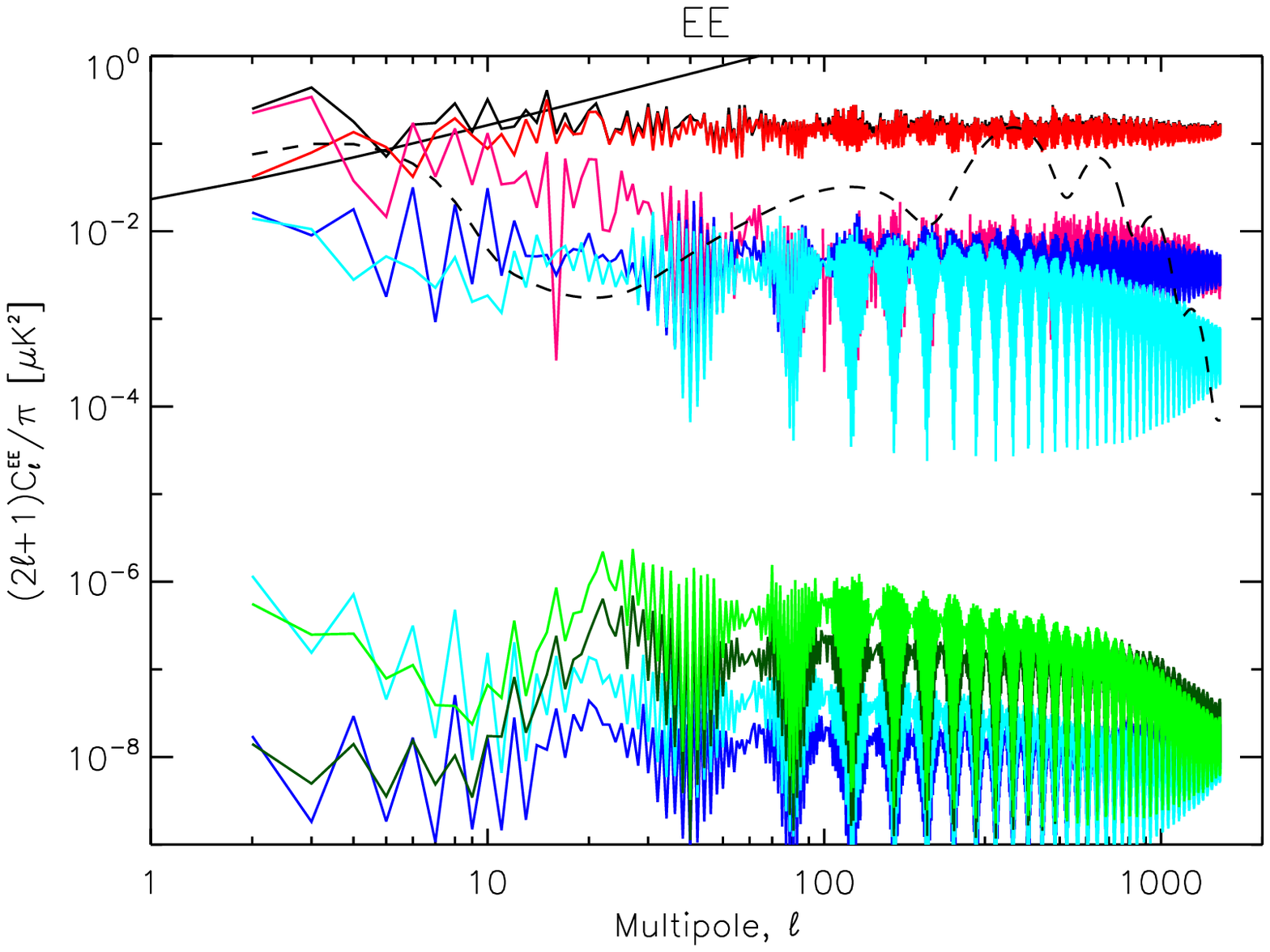}
    \vspace{-0.2cm}
    \includegraphics*[width=0.42\textwidth]
      {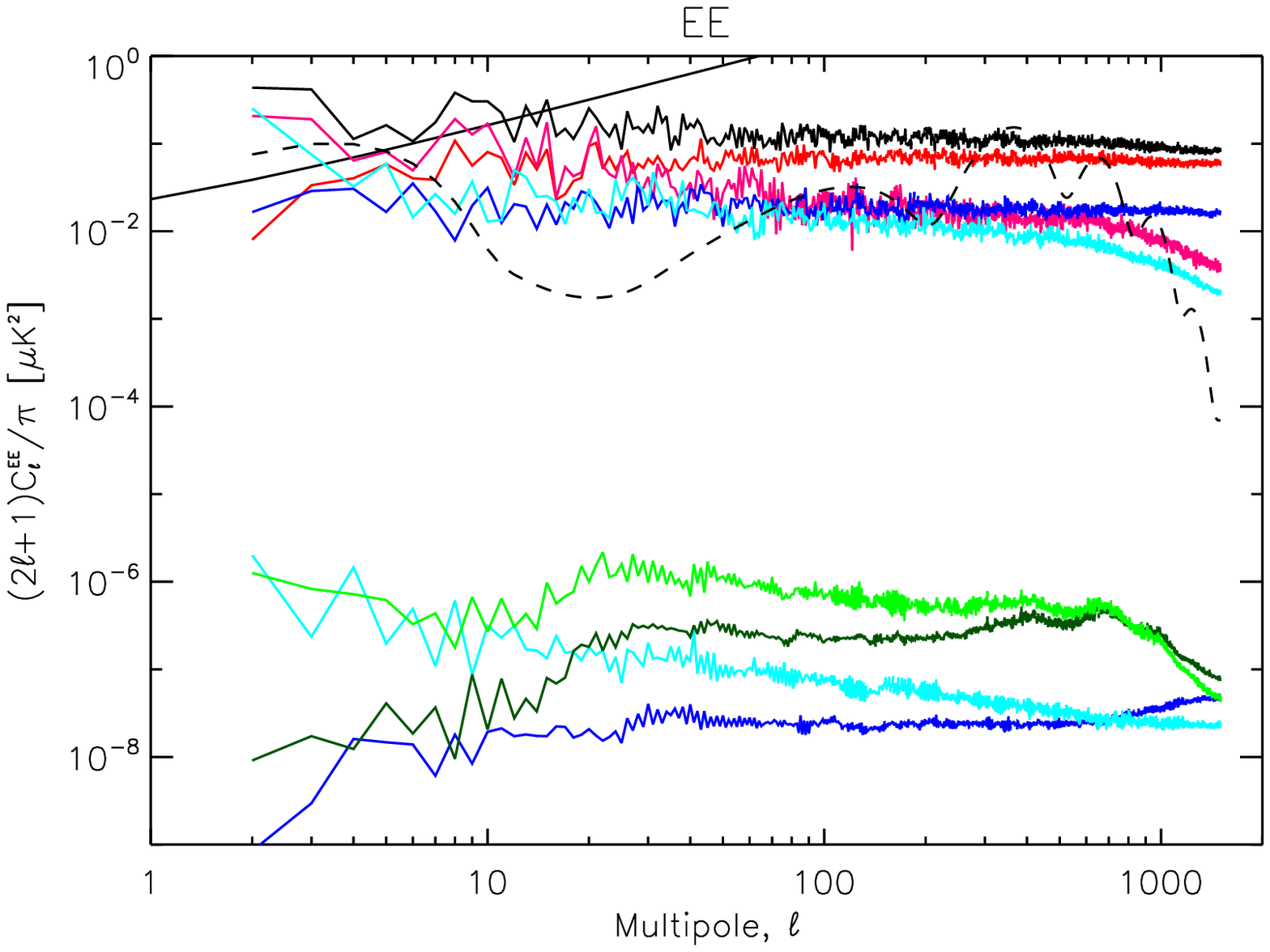}
    \vspace{-0.2cm}
    \includegraphics*[width=0.42\textwidth]
      {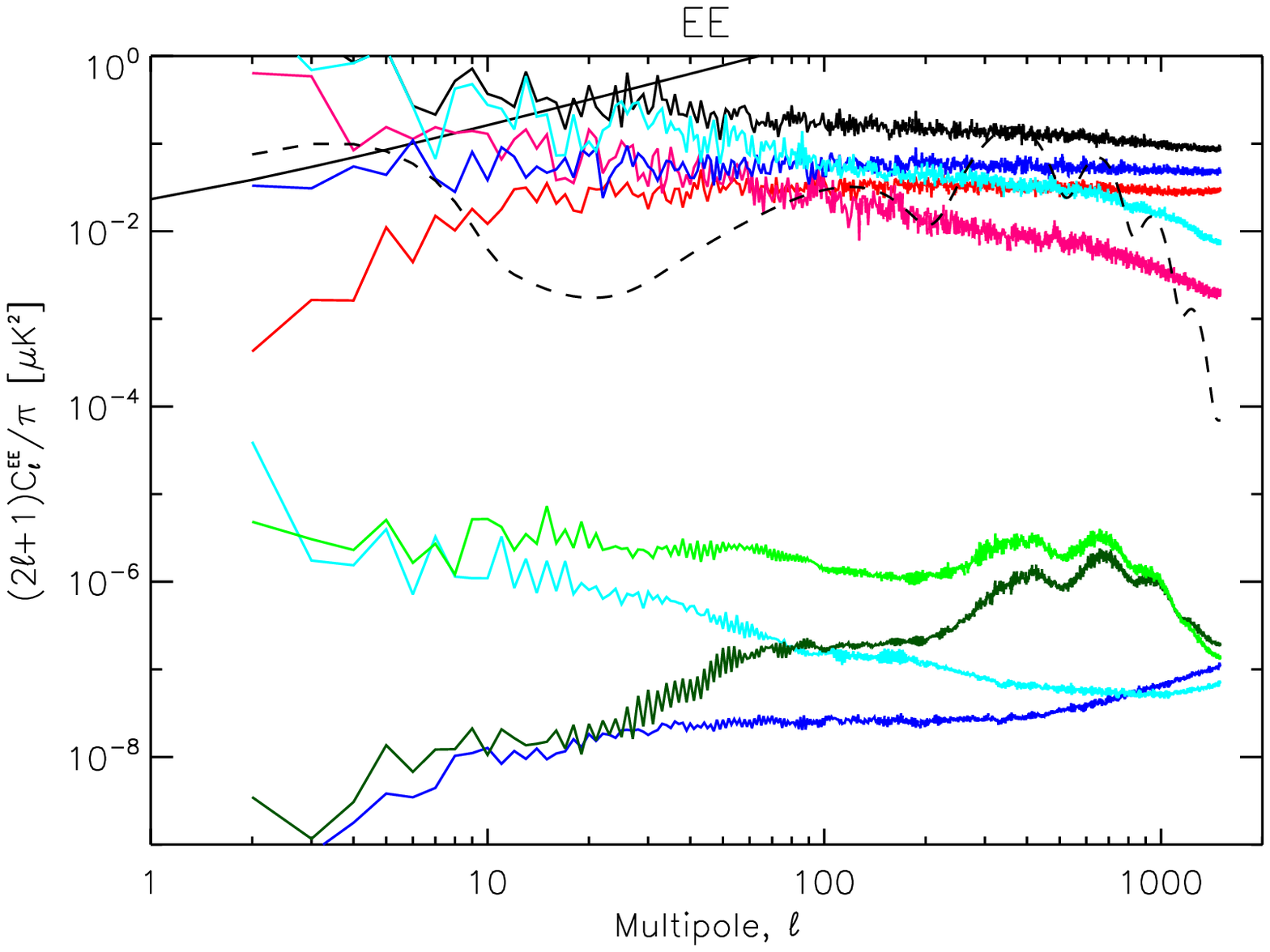}
    \vspace{-0.1cm}
    \caption{
Same as Fig.~\ref{fig:sixcomp_cl_TT} but for the $E$ mode
polarization spectrum $C_\ell^{EE}$.  We also show the two signal
baseline contributions of the $B$ mode polarization spectrum
$C_\ell^{BB}$.  They are the lower blue curves:
 \emph{Blue:} Reference baselines of pixelization noise.
 \emph{Light blue:} Additional effect of signal baseline error.
    }
    \label{fig:sixcomp_cl_EE}
  \end{center}
\end{figure}

In Figs.~\ref{fig:sixcomp_cl_TT} and \ref{fig:sixcomp_cl_EE} we show
angular power spectra $C_\ell^{TT}$ and $C_\ell^{EE}$ of different
map components for the cases $\tbase = 1$ h, 1min, 15s, and 5 s.

For baselines that are multiples of the spin period (1 min and 1 h)
we see the characteristic even--odd variation in the $C_\ell$ of the
baseline components.  If the scanning circles had the full
$90^\circ$ radius, they would contribute only to the even
multipoles.  In our case the circle radius is $87.77^\circ =
90^\circ - 2.23^\circ$, and therefore we see a beat pattern, where
the maximum even--odd multipole difference occurs at multipoles
$\ell$ that are near multiples of $90^\circ/2.23^\circ = 40.4$. For
the low $\ell$ of the unmodeled $1/f$ contribution we see the
opposite pattern, since the unmodeled $1/f$ contains mostly
frequencies which vary just over those timescales over which the
baseline contributions are constant.

The angular power spectrum of full-circle uncorrelated baselines
goes as
 \beq
    C_\ell \propto \frac{1}{2\ell+1} \,,
 \label{C_ell_fall}
 \eeq
(Eftstathiou \cite{E05}). Therefore we have plotted
$(2\ell+1)C_\ell/\pi$ in Figs.~\ref{fig:sixcomp_cl_TT} and
\ref{fig:sixcomp_cl_EE}. We see that Eq.~(\ref{C_ell_fall}) indeed
holds well for white noise reference baselines; even for $\tbase \ll
1/\fspin$, although these have less power at the lowest $\ell$.
Although the unmodeled $1/f$ contribution does not consist of
constant baselines, the correlations of the parts between different
baseline segments are weak (nonexistent for $\tbase \ll 1/\fspin$).
Therefore Eq.~(\ref{C_ell_fall}) holds fairly well for the unmodeled
$1/f$ also; except at low $\ell$ for short baselines, where there is
a lack of power since the unmodeled $1/f$ varies more rapidly along
the scan path; and for high $\ell$ for $\tbase \gg 1/\fspin$, which
have excess power at high $\ell$, related to imperfect superposition
of the different scan circles of the same ring, mainly due to
nutation.

The other contributions have different angular scale dependencies,
related to the correlations between baselines. We see that the
baseline error components have steeper spectra than the reference
baseline and unmodeled $1/f$ contributions. This makes them
important at large scales (low multipoles), where they are
comparable or even stronger than the white noise reference baseline
and unmodeled $1/f$ components, which dominate at high $\ell$ and
contribute most to the residual map rms.  If one considers just the
signal baselines, the baseline error completely dominates over the
reference baselines for short $\tbase$ and low $\ell$.

For long baselines ($\sim 1$ min or longer), the unmodeled $1/f$
noise dominates the residuals for $\ell > 10$, but the $1/f$
baseline error contribution can be comparable for $\ell < 10$.  For
shorter baselines, the white noise baselines become more important.

 The $C_\ell^{BB}$
spectra of different residual components look qualitatively like
$C_\ell^{EE}$, except for the signal baseline components, which have
less power, reflecting the lack of B-mode signal in the input.
Therefore we have not plotted the $C_\ell^{BB}$ spectra, except for
these signal baseline components, which we have included in
Fig.~\ref{fig:sixcomp_cl_EE} along with the $C_\ell^{EE}$ spectra.

\begin{figure}[!tbp]
  \begin{center}
    \vspace{-0.5cm}
    \includegraphics*[width=0.5\textwidth]
      {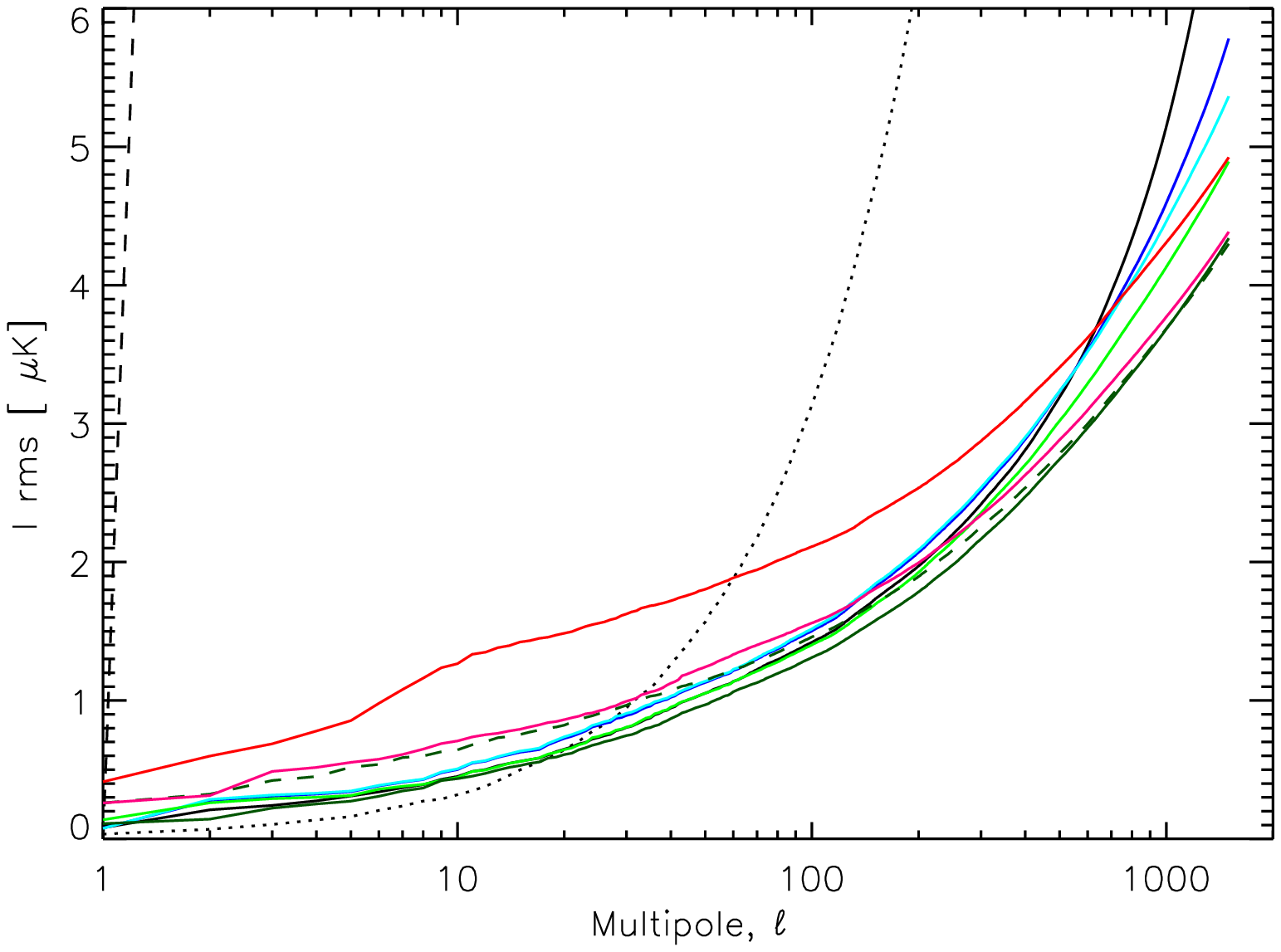}
    \vspace{-0.5cm}
    \includegraphics*[width=0.5\textwidth]
      {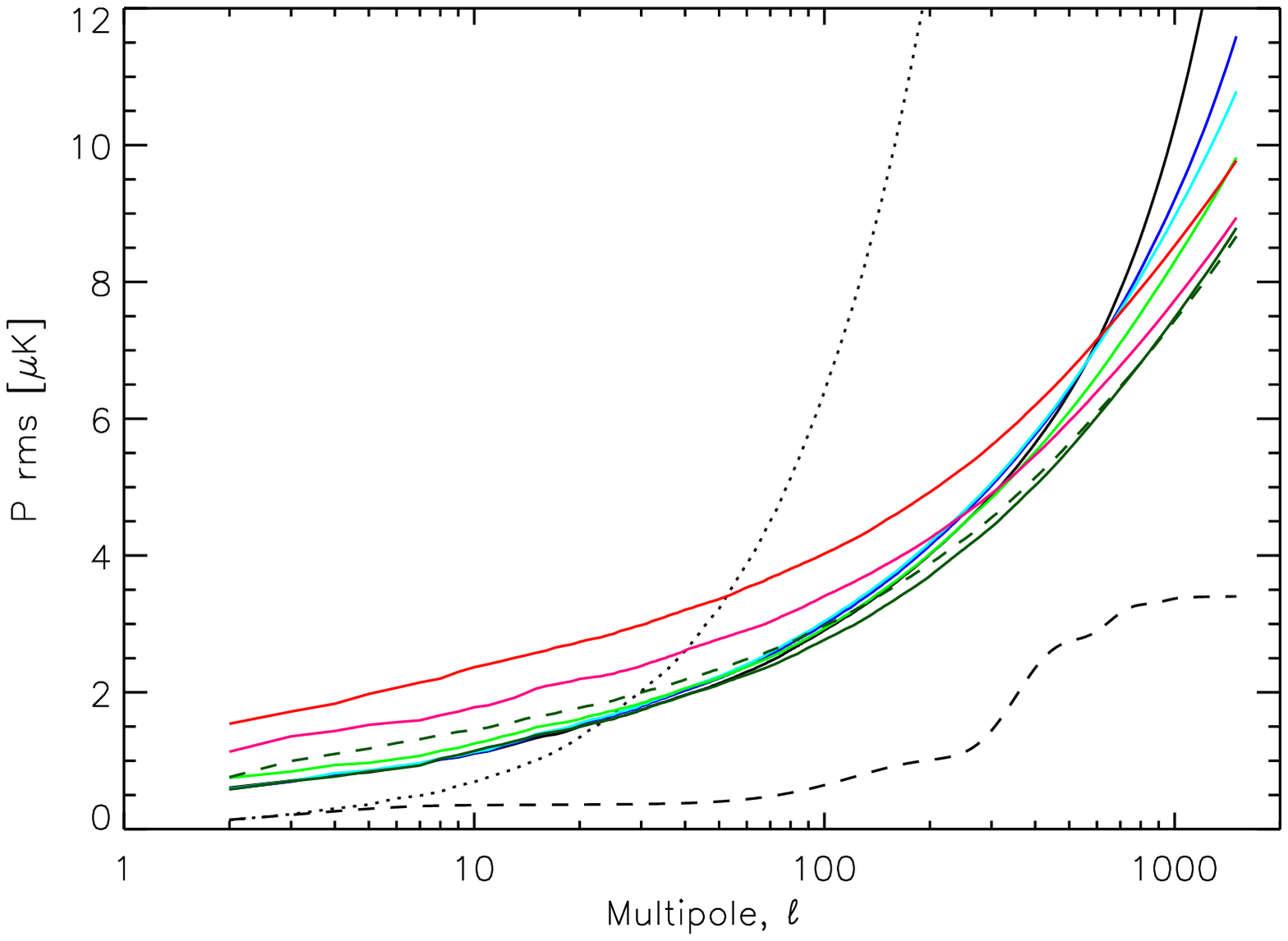}
    \vspace{-0.5cm}
    \includegraphics*[width=0.5\textwidth]
      {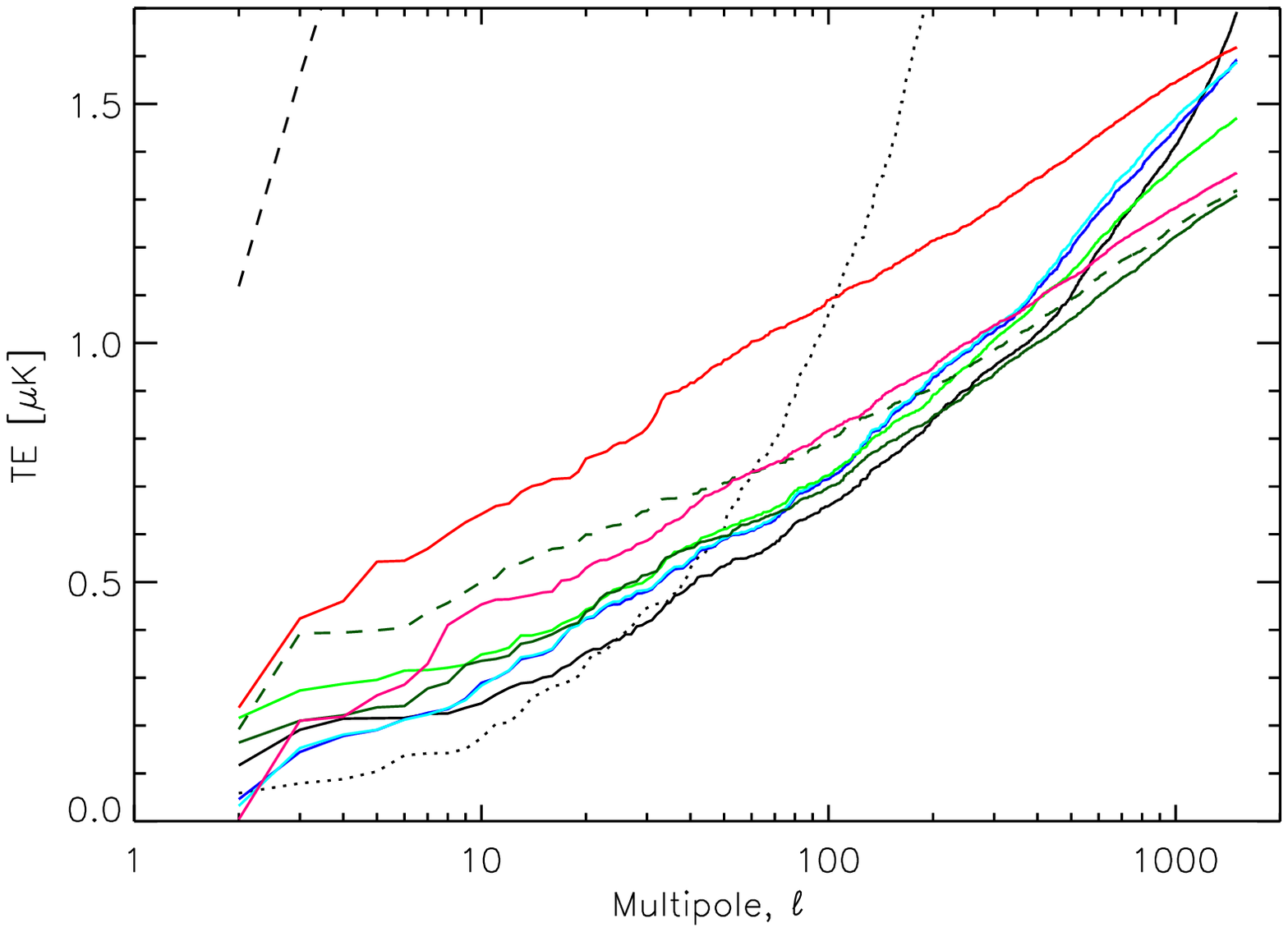}
    \vspace{-0.2cm}
    \caption{
Square roots of the cumulative angular power spectra (see
Eqs.~\ref{cum_Cell} and \ref{cum_CTE}) for the residual temperature
(\emph{top} panel) and polarization (\emph{middle}) maps and $TE$
correlation (\emph{bottom}), up to a given multipole $\ell$, plotted
as a function of $\ell$ for different baseline lengths: 1 h
(\emph{black}), 4 min (\emph{blue}), 1 min (\emph{light blue}), 30 s
(\emph{light green}), 15 s (\emph{green}), 10 s (\emph{green
dashed}), 7.5 s (\emph{pink}), and 5 s (\emph{red}).
    }
    \label{fig:cum_cl}
  \end{center}
\end{figure}

Because of large $\ell$-to-$\ell$ variations in the residual
$C_\ell$, these plots are difficult to read.  Therefore we also plot
(Fig.~\ref{fig:cum_cl}) square roots of the cumulative angular power
spectra,
 \beq
    \sqrt{\sum_{\ell' = 1}^\ell \frac{2\ell' + 1}{4\pi} C_{\ell'}^{TT}} \,,
    \qquad
    \sqrt{\sum_{\ell' = 2}^\ell \frac{2\ell' + 1}{4\pi} \left(C_{\ell'}^{EE}
    + C_{\ell'}^{BB}\right)} \,,
 \label{cum_Cell}
 \eeq
which give the total contribution to the residual $I$ and $P$ map
rms from multipoles up to $\ell$, and
 \beq
    \sqrt{\sum_{\ell' = 2}^\ell \frac{2\ell' + 1}{4\pi}
    \big|C_{\ell'}^{TE}\big|}
    \,.
 \label{cum_CTE}
 \eeq
The beam fwhm $\theta = 12.68'$ corresponds to multipole $\ell =
180^\circ/\theta \sim 850$, so we are more concerned about the
behavior up to this $\ell$ than above it.

Fig.~\ref{fig:cum_cl} provides probably the most concise meaningful
comparison of the quality of maps vs baseline length.
 The $\tbase = 15$ s case appears
the best in terms of cumulative residual power in the map for the
relevant multipoles. Although $\tbase = 10$ s produces a smaller
residual map rms, it is only because of its small sub-beam-scale
residuals. Interestingly, the $\tbase = 1$ h baseline seems to be
the best for minimizing residual temperature-polarization
correlations at intermediate scales. It is also better overall than
the $\tbase = 1$ min and 4 min cases for $\ell < 300$.

If one is only interested in large-scale features (low $\ell$) there
are no big differences between any of the baseline lengths from 15 s
to 1 h, but 10 s or less should be avoided. For the lowest
multipoles there is some randomness in the results, since we studied
only one noise realization, so one should not try to draw
conclusions from the small differences seen there for $\tbase = 15$
s to 1 h.  (The noise residuals, for the cases $\tbase = 1$ min and
$\tbase = 1.25$ s with noise prior, have been studied via Monte
Carlo in Keskitalo et al.~\cite{Randsvangen}.)

The signal baseline contributions appear a minor effect at all
scales. In this study the signal contained only the CMB. In reality,
the gradients in the signal are often dominated by foregrounds, and
therefore the signal baseline effect is larger. Foreground signals
are considered in Keih\"anen et al.~(\cite{Madam09}). Foregrounds
were also included in the map-making studies of Ashdown et al.
(\cite{Paris,Trieste}), and especially in the former there was a
detailed study on the signal baseline contribution and how it could
be minimized.

Map residuals influence the precision at which we are able to
determine the angular power spectrum of the CMB map. We can subtract
the expectation value of the $C_\ell$ of the residual from the map
spectrum, but individual realizations deviate from this expectation
value, leading to an error in the CMB $C_\ell$ estimate. A multipole
$\ell$ of a map has at the best $2\ell+1$ degrees of freedom.
Statistically isotropic signals (e.g. CMB or a white noise map of
uniform pixel variance) have these degrees of freedom. Deviations
from the statistical isotropy may lead to correlations in the
$m$-modes of a multipole which in turn decreases the degrees of
freedom and therefore increase the error in $C_\ell$. Fortunately
the strongest correlations of the map residuals occur nearly along
the ecliptic meridians.  Correlations along meridians do not lead to
correlations in the $m$-modes. Therefore we can expect that the
$m$-mode couplings due to map residuals are weak and the excess
spectrum error is small.  For now, we did not investigate these
errors any further, but decided to leave this for future studies.

\subsection{Low multipoles of I, Q, and U maps}

\begin{figure}[!tbp]
  \begin{center}
    \includegraphics*[width=0.5\textwidth]
      {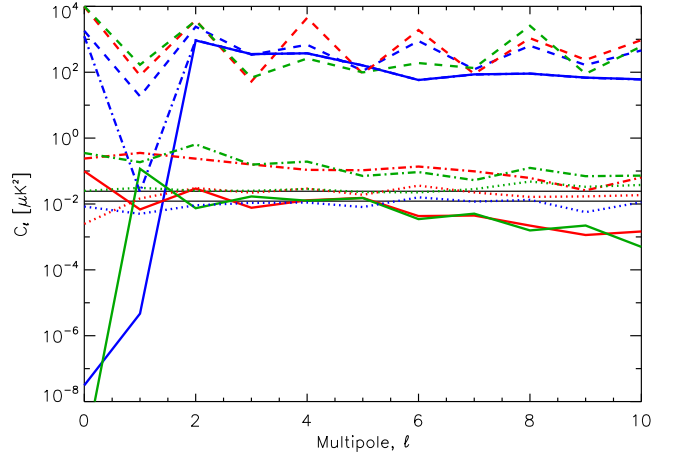}
    \caption{
The effect of destriping on the low multipoles of $I$ (\emph{blue}),
$Q$ (\emph{red}), and $U$ (\emph{green}) maps. The \emph{solid}
lines give the multipoles of the binned signal (CMB) map, where the
monopoles of $I$ and $U$, and the dipole of $I$ should ideally
vanish in our case. The \emph{dashed} lines are for the naive binned
map including the noise, and the \emph{dash-dotted} lines are for
the destriped ($\tbase = 1$ min) map.  Destriping is not able to
remove the noise monopole of the $I$ map (the monopole does change
but does not become small), but the noise monopoles of $Q$ and $U$
maps are removed, about equally well as the other low multipoles.
The \emph{dotted} lines are for the binned white noise map. The
horizontal \emph{black} lines give the expected white noise levels.
    }
    \label{fig:IQU_cl_lin}
  \end{center}
\end{figure}

\begin{figure}[!tbp]
  \begin{center}
    \vspace{-0.8cm}
    \includegraphics*[width=0.5\textwidth]
      {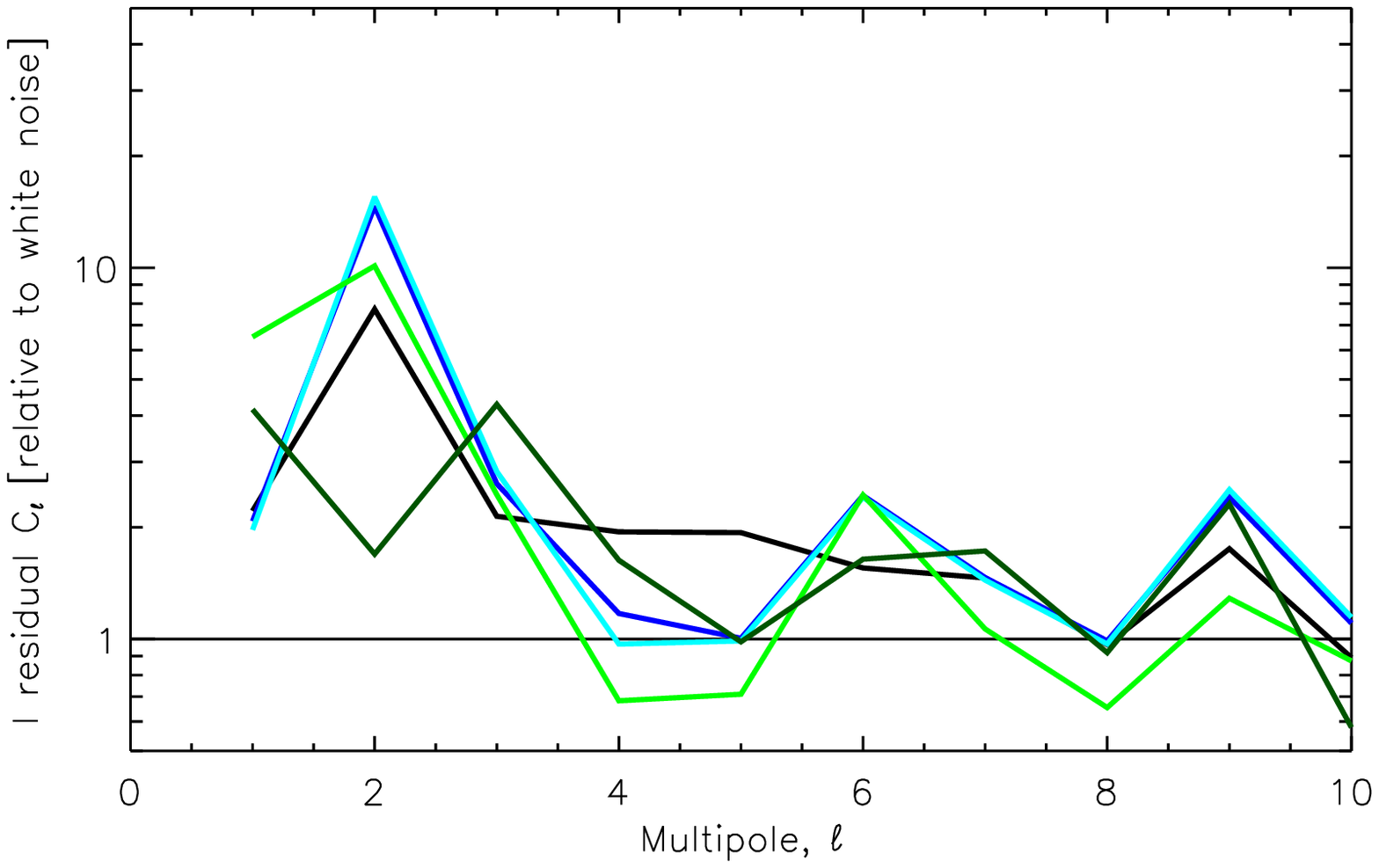}
    \vspace{-0.2cm}
    \includegraphics*[width=0.5\textwidth]
      {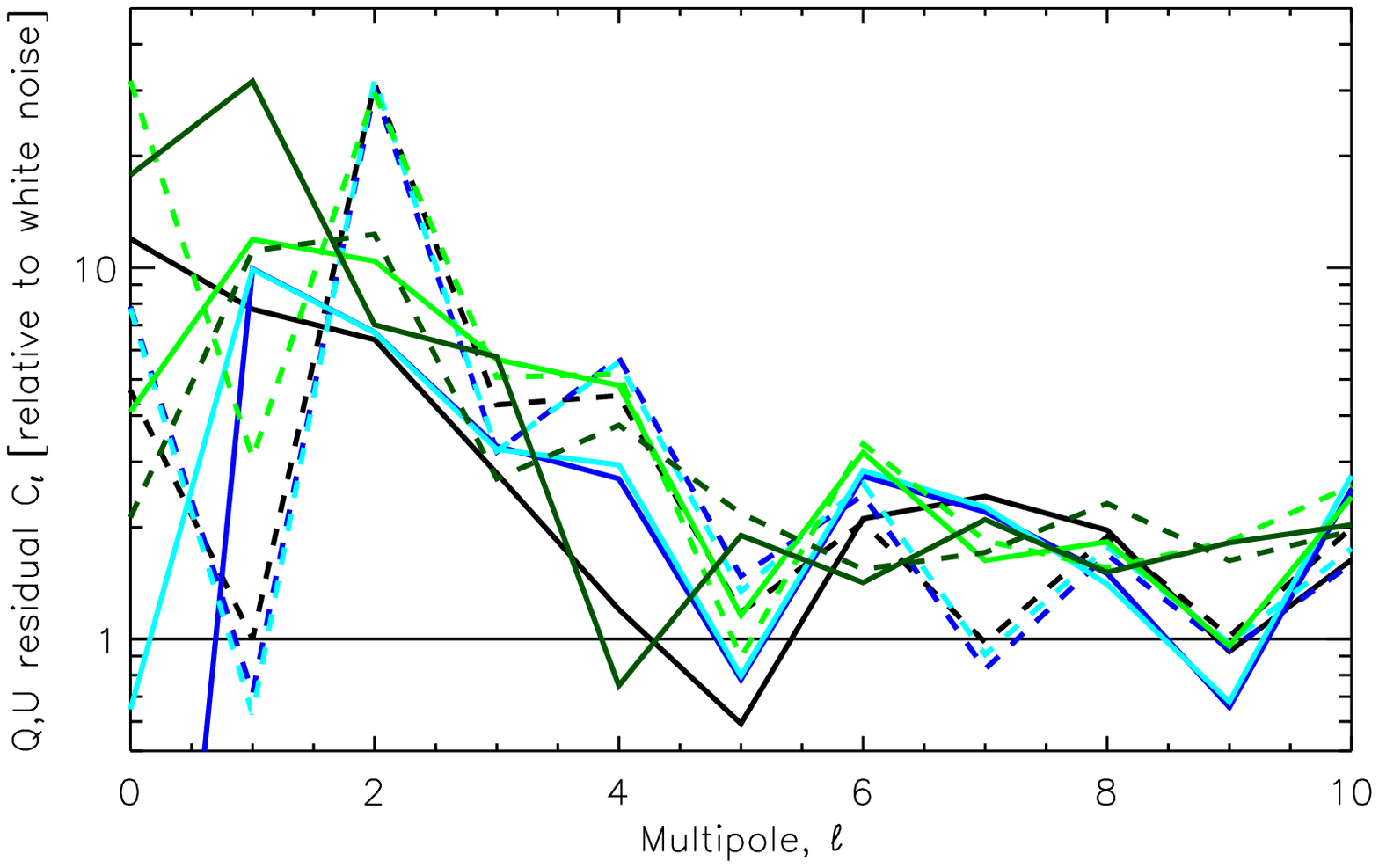}
    \caption{
The $I$ , $Q$, and $U$ low multipoles of the residual maps for
different baseline lengths: 1 h (\emph{black}, 4 min (\emph{blue}),
1 min (\emph{light blue}), 30 s (\emph{light green}), and 15 s
(\emph{dark green}). For this plot they are divided by the white
noise level. The \emph{top} panel shows $I$, the \emph{bottom} panel
$Q$ (\emph{solid}) and $U$ (\emph{dashed}).
    }
    \label{fig:IQU_cl_lin_tbase}
  \end{center}
\end{figure}

For cosmological purposes, one calculates the $C_\ell^{TT}$,
$C_\ell^{TE}$, $C_\ell^{EE}$, and $C_\ell^{BB}$ angular power
spectra of the output maps, which represent the fundamental
properties of the temperature and polarization field, and are
coordinate independent. However, for analyzing residual map
structure, it may be more intuitive to consider the $Q$ and $U$ maps
as two separate maps of a scalar quantity, and calculate their
ordinary (spin-0) angular power spectra.

The $Q$ and $U$ are given in terms of the $a^E_{\ell m}$ and
$a^B_{\ell m}$ as (Zaldarriaga \& Seljak \cite{ZS97})
 \beq
    Q + iU = - \sum\left(a^E_{\ell m} + ia^B_{\ell m}\right)
    {}_2{Y}_\ell^m \,,
 \eeq
where the
 \beq
    {}_2{Y}_\ell^m(\theta,\phi)  \equiv
    \sqrt{\frac{2\ell+1}{4\pi}}e^{im\phi}d^\ell_{m,-2}(\theta)
 \eeq
are the spin-2 spherical harmonics (Newman \& Penrose
\cite{Newman66}, Goldberg et al.~\cite{Goldberg67}) and the
$d^\ell_{m,m'}(\theta)$ are the Wigner $d$-functions (we follow
Varshalovich et al. \cite{VMK}), which are real.

Thus the monopoles $\bar{Q}$ and $\bar{U}$ of the $Q$ and $U$ maps
are given by
 \beq
  \bar{Q}+i\bar{U} = - \sum \left(a^E_{\ell m} + ia^B_{\ell m}\right)
  {}_2\bar{Y}_\ell^m  \,,
 \eeq
where
 \bea
  {}_2\bar{Y}_\ell^m & \equiv &
  \frac{1}{4\pi}\int{}_2\bar{Y}_\ell^m(\theta,\phi)d\Omega \nn\\
  & = & \delta_{m0}\sqrt{\frac{2\ell+1}{16\pi}}\int_0^\pi d_{0,-2}^\ell(\theta)\sin\theta
  d\theta \,,
 \eea
which are real. Thus the $Q$ and $U$ monopoles are given by
  \beq
      \bar{Q} = -\sum_\ell a^E_{\ell 0}\, {}_2\bar{Y}_\ell^0
      \qquad\mbox{and}\qquad
      \bar{U} = -\sum_\ell a^B_{\ell 0}\, {}_2\bar{Y}_\ell^0 \,,
  \eeq
Since in our case the input spectrum contained no $B$ mode, the
monopole vanishes in the input $U$ map.

Fig.~\ref{fig:IQU_cl_lin} illustrates the effect of destriping on
the $I$, $Q$, and $U$ maps.

The low multipoles of the residual maps are shown in
Fig.~\ref{fig:IQU_cl_lin_tbase}, divided by the white noise level.
We see that the residuals at lowest multipoles are larger than the
white noise level.  The baseline length does not make a large
difference for these low multipoles, except that too short baselines
(less than 15 s, not included in Fig.~\ref{fig:IQU_cl_lin_tbase})
should be avoided.

\section{Effect of noise knee frequency}
\label{sec:knee_frequency}

\begin{figure}[!tbp]
  \begin{center}
    \includegraphics*[width=0.5\textwidth]
      {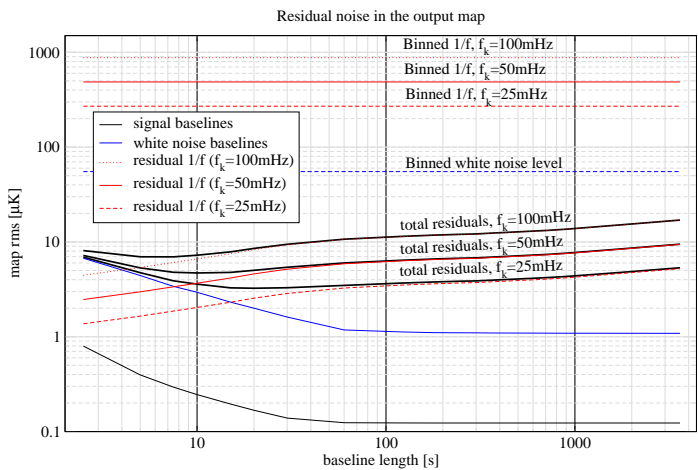}
    \caption[Effect of $\fk$]{
      The knee frequency directly relates to the amount of residual
      $1/f$ noise in the output map. Reducing the amount of correlated
      noise (and hence the knee frequency) causes the white noise
      baseline contribution to exceed residual $1/f$ contribution at
      longer baselines. Therefore lower $\fk$ means longer
      optimal baseline length.
    }
    \label{fig:fknee}
  \end{center}
\end{figure}

\begin{figure}[!tbp]
  \begin{center}
     \vspace{-0.7cm}
     \includegraphics*[width=0.5\textwidth]
      {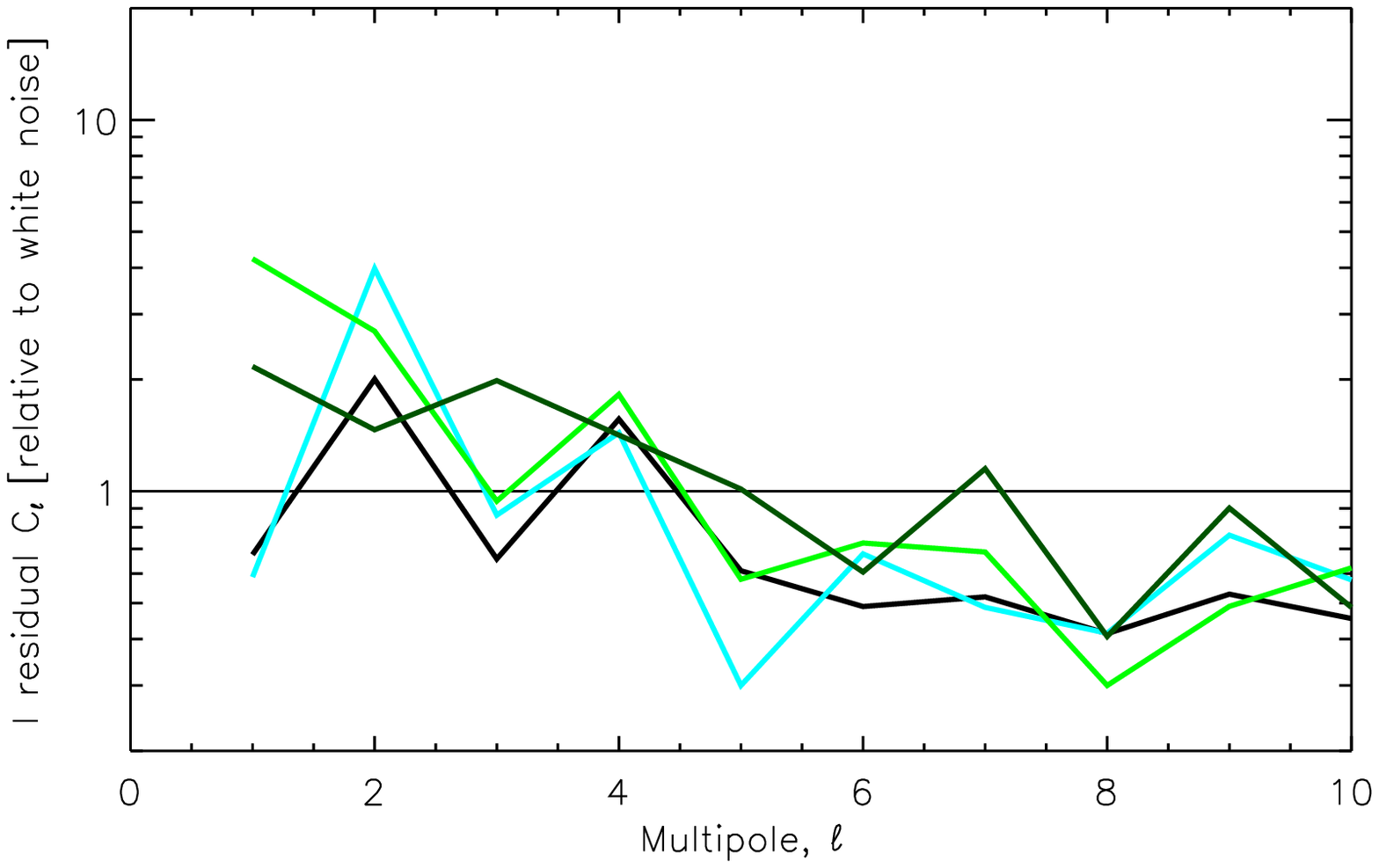}
    \vspace{-0.2cm}
    \includegraphics*[width=0.5\textwidth]
      {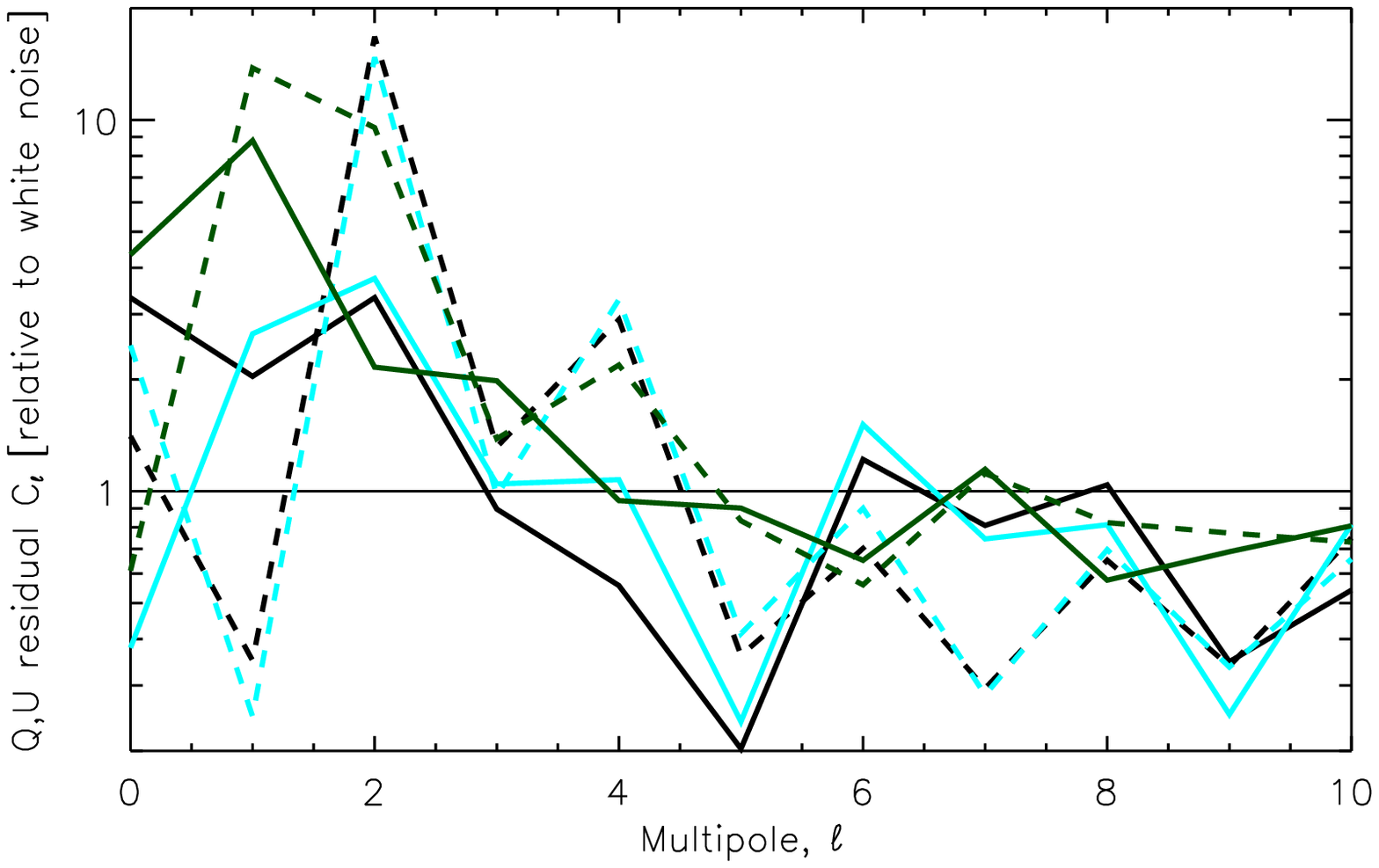}
    \caption{
Same as Fig.~\ref{fig:IQU_cl_lin_tbase}, but for a lower knee
frequency $\fk = 25$ mHz (and a smaller selection of baseline
lengths).
    }
    \label{fig:IQU_cl_lin_tbase_25mHz}
  \end{center}
\end{figure}

The importance of the correlated $1/f$ noise depends on its
amplitude and spectrum.  In this paper we do not consider the effect
of possible spectral features in the noise, and we have
parameterized the noise just by the $1/f$ slope $-\alpha$, knee
frequency $\fk$, and white noise level. Since we produced the
simulated $1/f$ noise separately from the white noise, we can change
$\fk$ simply by multiplying the $1/f$ part by
 \beq
    r_{1\rightarrow2} =
    \left(\frac{f_{k,2}}{f_{k,1}}\right)^{\alpha/2} \,.
 \eeq
This will change the residual $1/f$ contribution to the residual map
by the same factor, while the white noise baseline and signal
baseline contributions are unaffected (unless one changes to a
different $\tbase$). For angular power spectra the $1/f$
contributions are rescaled by the square of this factor. See
Fig.~\ref{fig:fknee}, where we have plotted the residual map rms as
a function of $\tbase$ for $\fk = 100$ mHz and $25$ mHz, besides the
case of $\fk = 50$ mHz, which we have so far considered. A higher
$\fk$ favors shorter baselines, since the stronger $1/f$ noise at
relatively high frequencies needs to be modeled better. A lower
$\fk$, on the other hand, favors longer baselines, since then the
white noise and signal baselines are relatively more important. For
$\fk \sim 10$ mHz the minimum rms would move to $\tbase \sim 1$ min.

In Fig.~\ref{fig:fknee} we show also the rms of the binned $1/f$
noise map. As we lower $\fk$ it moves down.  For very small $\fk$ it
would fall below the white noise baseline rms.  At this point simple
binning would produce a better result than destriping.  For our
simulated $1/f$ noise, and for long baselines, this would happen at
the extremely low $\fk = 21.6\,\mu$K.  It should be noted however,
that in our case the binned $1/f$ rms is heavily dominated by the
lowest frequencies, and in other cases (smaller slope $\alpha$,
larger $\fmin$) the relevant $1/f$ rms could be smaller for a given
$\fk$, so that simple binning could become superior already at a
higher $\fk$.

In Fig.~\ref{fig:IQU_cl_lin_tbase_25mHz} we show the low multipoles
of the residual $I$, $Q$, and $U$ maps recalculated for $\fk = 25$
mHz.

\section{Effect of noise prior}
\label{sec:noise_prior}

\begin{figure}[!tbp]
  \begin{center}
    \includegraphics*[width=0.5\textwidth]
             {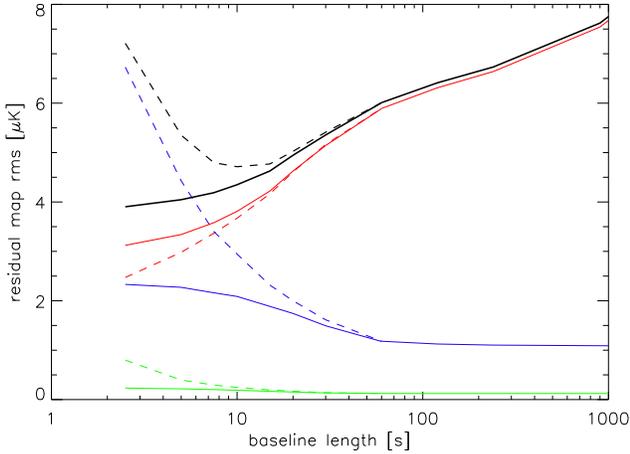}
    \caption{Same as the bottom panel of
Fig.~\ref{fig:component_residuals}, but the solid lines correspond
to using a noise prior.  The dashed lines are from
Fig.~\ref{fig:component_residuals}. The noise prior has practically
no effect for $\tbase \geq 1$ min.
 }
    \label{fig:map_rms_tbase_prior}
  \end{center}
\end{figure}

\begin{figure}[!tbp]
  \begin{center}
    \vspace{-0.5cm}
    \includegraphics*[width=0.5\textwidth]
      {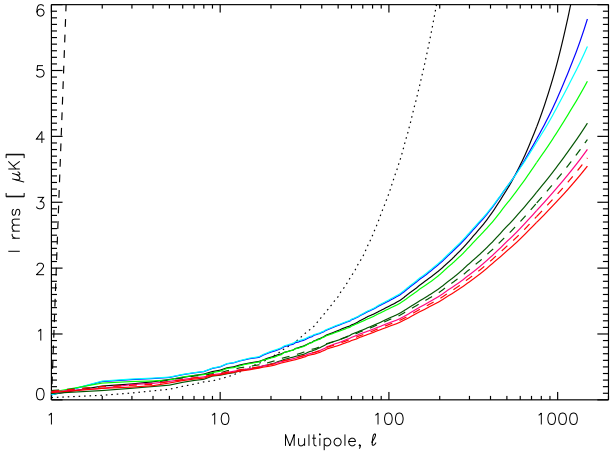}
    \vspace{-0.5cm}
    \caption{Same as top panel of Fig.~\ref{fig:cum_cl} but for
the case of using the noise prior. The colors and linestyles are the
same as in Fig.~\ref{fig:cum_cl}, except we have added the $\tbase =
2.5$ s case (\emph{solid red}), and the $\tbase = 5$ s case appears
now as the \emph{dashed red} line.
    }
    \label{fig:cum_cl_prior}
  \end{center}
\end{figure}

\begin{figure}[!tbp]
  \begin{center}
     \vspace{-0.8cm}
    \includegraphics*[width=0.5\textwidth]
      {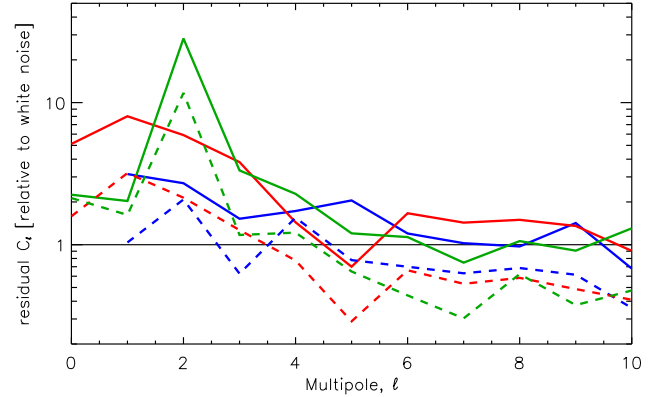}
    \caption{
Same as Fig.~\ref{fig:IQU_cl_lin_tbase}, but with the noise prior,
and using a short baseline $\tbase = 2.5$ s. The Stokes parameters
are now indicated by the colors: \emph{blue} for $I$, \emph{red} for
$Q$, and \emph{green} for $U$. The \emph{solid} lines are for $\fk =
50$ mHz, and the \emph{dashed} lines for $\fk = 25$ mHz.
    }
    \label{fig:IQU_cl_lin_f25_2fknee}
  \end{center}
\end{figure}

Although short baselines can potentially model correlated noise
better, they fail because of the large random amplitudes they pick
from white noise.  This can be remedied by applying prior
information on the noise spectrum to prevent too large differences
between amplitudes of nearby baselines. This is discussed in detail
in Keih\"{a}nen at al.~(\cite{Madam09}), but we give a short preview
of the results here. See Figs.~\ref{fig:map_rms_tbase_prior},
\ref{fig:cum_cl_prior}, and \ref{fig:IQU_cl_lin_f25_2fknee}.  The
noise prior has little effect for $\tbase \gg 1/\fk$, but for short
baselines the effect is dramatic. We note that now the results keep
improving as the baseline is shortened, at least until $\tbase =
2.5$ s, the shortest we tried. (Using very short baselines with the
noise prior makes the code more resource intensive.)  For the very
lowest multipoles, the results with the short baselines do not,
however, become much better than the ones obtained with longer
baselines (with or without noise filter). Compare
Fig.~\ref{fig:IQU_cl_lin_f25_2fknee} to
Figs.~\ref{fig:IQU_cl_lin_tbase} and
\ref{fig:IQU_cl_lin_tbase_25mHz}.

\section{Maps from shorter survey segments}
\label{sec:partial_surveys}

\begin{figure}[!tbp]
  \begin{center}
    \includegraphics*[width=0.5\textwidth]
      {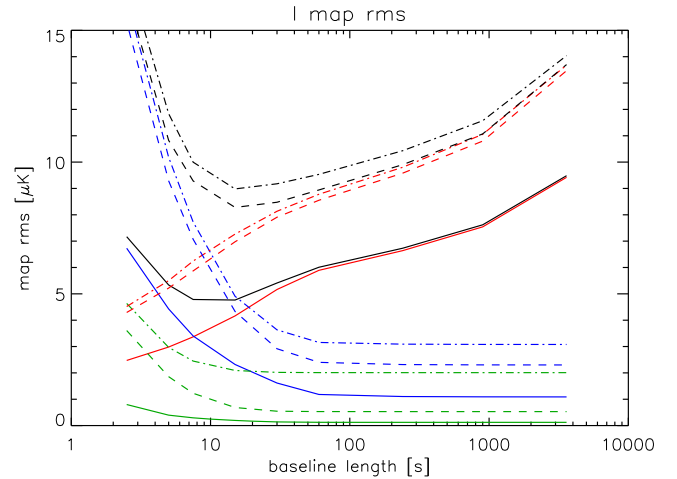}
    \caption{
Same as the bottom panel of Fig.~\ref{fig:component_residuals}, but
now we have included results from 7-month (\emph{dashed}) and
1-month (\emph{dash-dotted}) survey segments also.  The 1-month
results show averages from all 12 months of the survey.
    }
    \label{fig:rms_tbase_survey}
  \end{center}
\end{figure}

We now consider making maps from a shorter time segment of the data.
In a full year of observations, for a {\sc Planck}-like scanning
strategy, all parts of the sky are looked at two different seasons.
About 7 months is needed to observe the full sky. Maps from a
shorter segment cover just a part of the sky.

When maps are made from a shorter observation period, the number of
crossing points is reduced and their pattern is different.
Destriping 7 months of data differs qualitatively from the case of a
full year, since for a large part of the sky the second-season
observations are missing. We expect the loss of the corresponding
crossing points to result in a loss of output map quality due to
larger baseline errors. Destriping relies then more on the crossing
points which are near the ecliptic poles. When the observation
period is shortened further, only part of the sky is covered.  Since
the crossing point structure is not necessarily changed
qualitatively, we expect the map quality to worsen more slowly as a
function of survey duration. Because of the cycloidal scanning
strategy, the crossing point structure, however, changes with a
6-month period, so it will be different for different weeks or
months.

In Fig.~\ref{fig:rms_tbase_survey} we show the residual $I$ map rms
as a function of $\tbase$ for 1-month, 7-month, and 1-year surveys.
The results for the $Q$ and $U$ maps look qualitatively the same,
except that the signal baseline contribution is, of course, much
lower. We have excluded all pixels with {\em rcond} $\leq 0.4$ or
$\nhit \leq 400$ from the residual maps.

There are basically two kinds of effects contributing to these
results.  The more trivial effect is that of the lower hit count.
The average number of hits per pixel in the 7-month survey is 7/12
of that of the full-year survey, so we expect that alone to increase
the residual map rms by a factor of $\sqrt{12/7} = 1.31$.  This,
indeed, accounts for most of the change in the unmodeled $1/f$
contribution to the residual map rms. When the survey segment is
shortened below 7 months, the number of included pixels falls almost
in line with the number of samples, so the hit count per pixel stays
now roughly constant, and there is not much additional effect when
going from 7 months to 1 month.

The other effect is that of the change in the pattern of crossing
points on solving the baselines. This shows clearly in the white
noise baseline and signal baseline contributions, where baseline
errors were important already for the full 1-year survey. Here the
change from 7 months to 1 month brings also a significant change.  A
closer inspection of the different months reveals, however, that it
is only some of the months that are clearly worse than the 7-month
case.

\begin{figure}[!tbp]
\begin{center}
    \includegraphics*[trim=100 70 70 250, clip, width=0.24\textwidth]
    {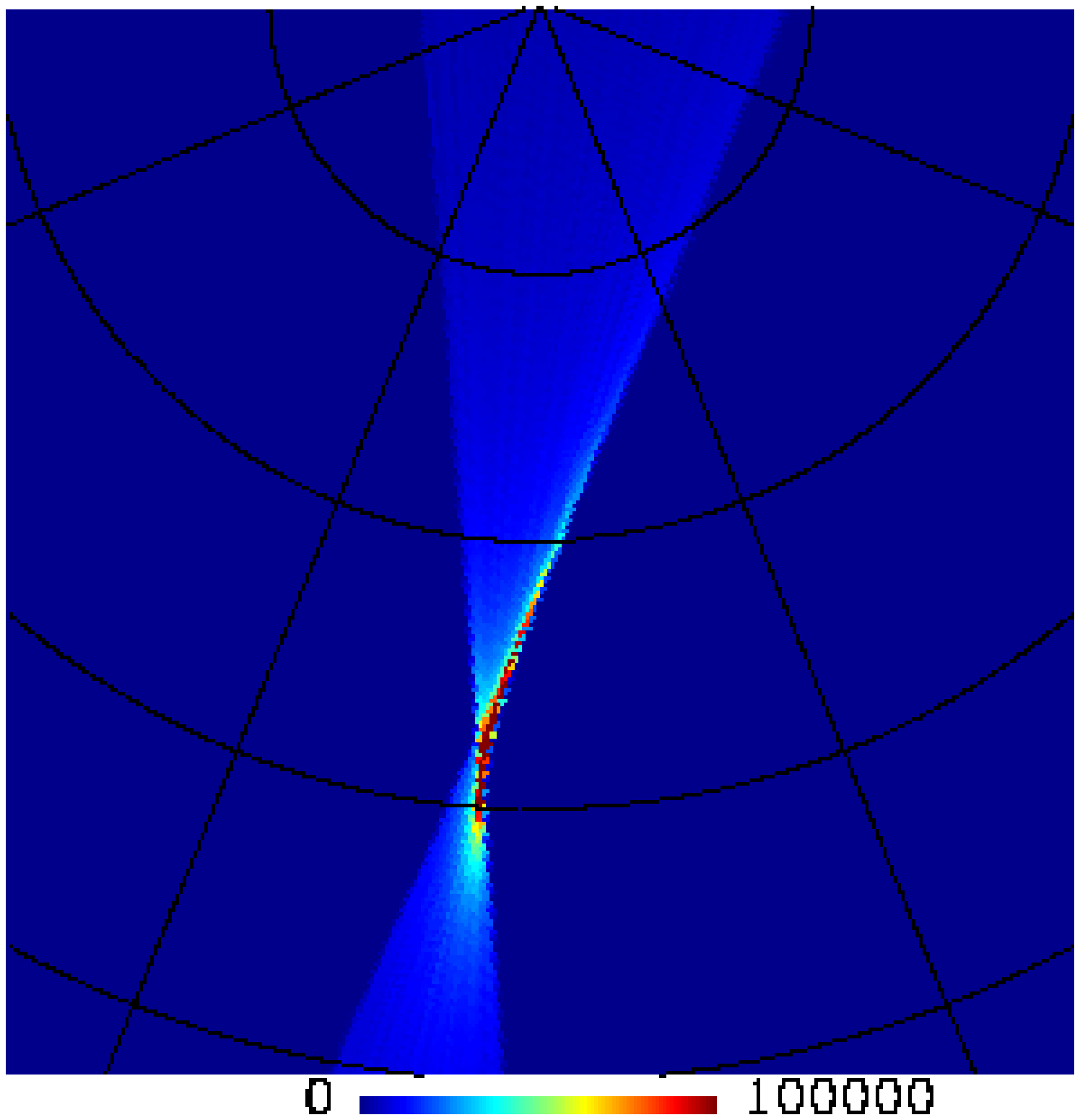}
    \includegraphics*[trim=100 70 70 250, clip, width=0.24\textwidth]
    {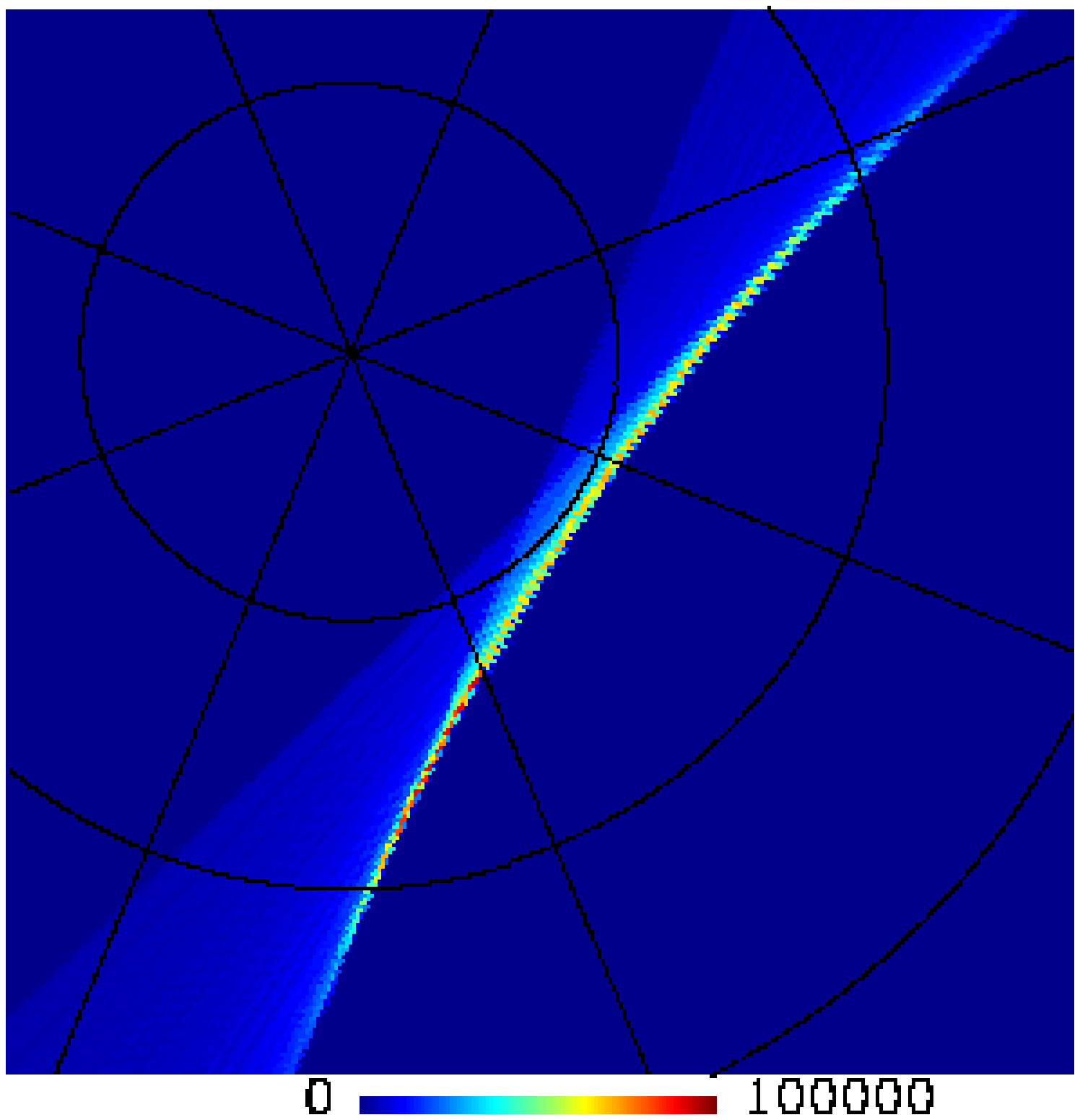}
    \vspace{-0.2cm}
    \caption{
Linear hit maps near the ecliptic North Pole for the one-month
surveys corresponding to the first (\emph{left}) and second
(\emph{right}) month of the full year survey.  Note how different is
the structure of crossing points between these two months.  The same
happens near the South Pole. In the first month all circles cross
each other at almost the same pixel, whereas in the second month the
crossings are spread over a wide arc.
    }
    \label{fig:hits_for_different_months}
\end{center}
\end{figure}

The first and the second month represent the two extreme cases. See
Fig.~\ref{fig:hits_for_different_months}. For the first month we are
close to the situation where all scanning circles cross each other
at the same pixel. The effect of this is the most striking for the
signal baseline component, for which it couples to the signal
gradients in those few pixels where the circles cross and the
pattern of hits within those pixels.

\begin{figure}[!tbp]
\begin{center}
   \vspace{-0.5cm}
    \includegraphics*[width=0.5\textwidth]
    {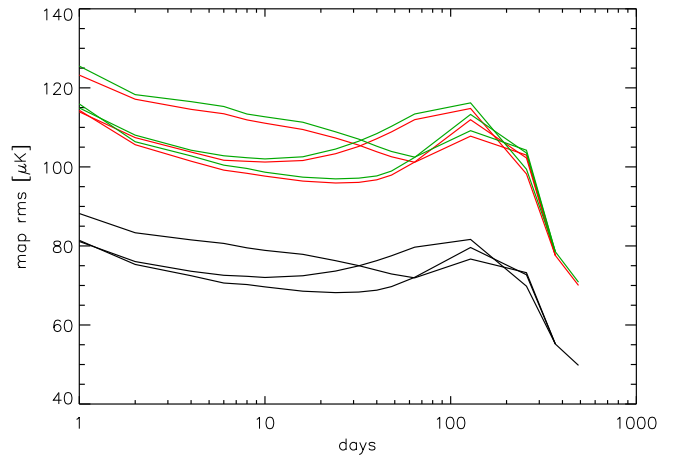}
    \vspace{-0.2cm}
    \caption{
The binned white noise map rms as a function of survey length for
$\tbase = 1$ min. Three different starting points were considered:
from the beginning of the simulated mission, starting 30 days later,
and starting 61 days later.  The \emph{black} line is for the Stokes
$I$ map, \emph{red} for $Q$, and \emph{green} for $U$.
    }
    \label{fig:survey_binned_white}
\end{center}
\end{figure}

To study the effect of the length of the survey segment, and also
its timewise location with respect to the cycloid, we fixed the
baseline length to $\tbase = 1$ min, and considered survey segments
of 1, 2, 4, 6, 8, 10, 16, 24, 32, 40, 48, 64, 128, 256, and 366
days. We also considered the effect of extending the mission to 488
days (16 months). Moreover, for each survey segment length (except
the 16 month one) we considered three different starting points for
the segment. To separate the effect of the change in hit count from
the effect of crossing point structure on solving baselines, we plot
first the rms of the binned white noise map, which has only the
first effect, in Fig.~\ref{fig:survey_binned_white}.

\begin{figure}[!tbp]
\begin{center}
   \vspace{-0.5cm}
    \includegraphics*[width=0.5\textwidth]
    {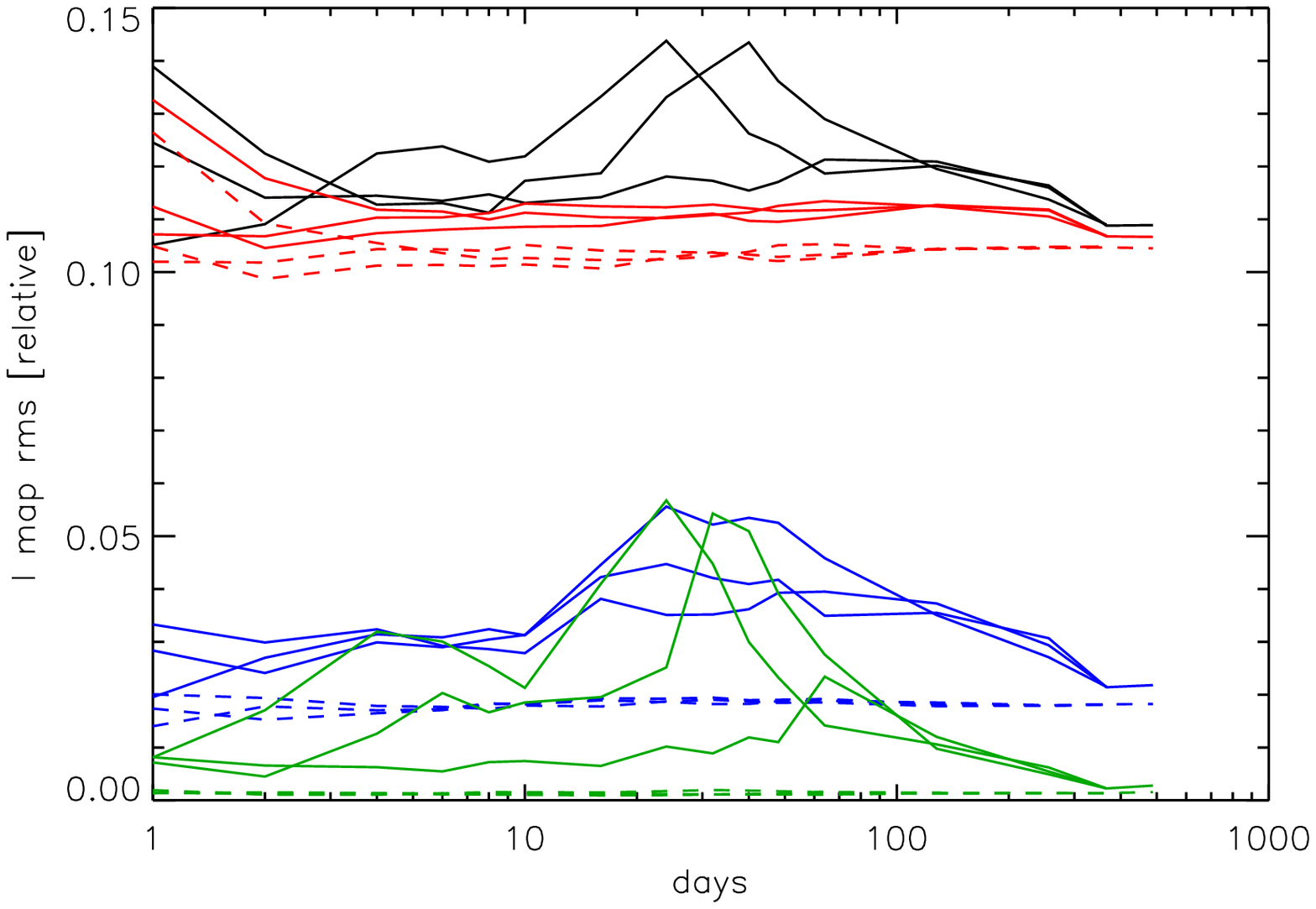}
    \includegraphics*[width=0.5\textwidth]
    {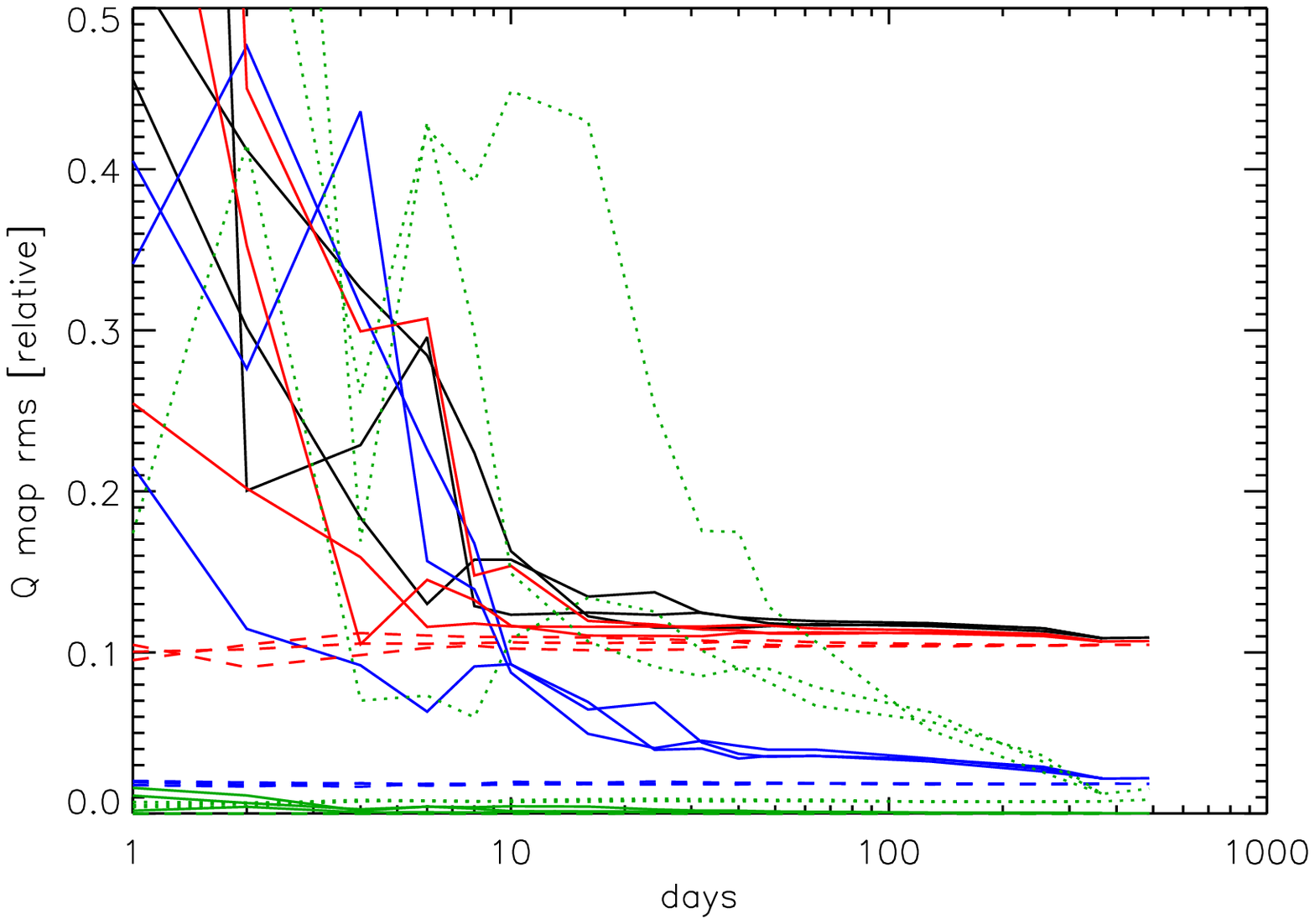}
    \vspace{-0.2cm}
    \caption{
The ratio of the residual map rms to the binned white noise map rms
(see Fig.~\ref{fig:survey_binned_white}) as a function of survey
length for $\tbase = 1$ min.  The \emph{black} line is for the full
residual (excluding binned white noise), \emph{blue} for the white
noise baseline map, \emph{red} for residual $1/f$, and \emph{green}
for signal baseline map.  We show also separately the reference
baseline contributions (\emph{dashed}). The \emph{top} panel is for
the Stokes parameter $I$, the \emph{bottom} panel for $Q$. Results
for $U$ are qualitatively the same as for $Q$. For $Q$, the signal
baseline component is plotted multiplied by a factor of 100 also
(\emph{dotted}).
    }
    \label{fig:survey_length_comp}
\end{center}
\end{figure}

\begin{figure}[!tbp]
\begin{center}
   \vspace{-0.5cm}
    \includegraphics*[width=0.5\textwidth]
    {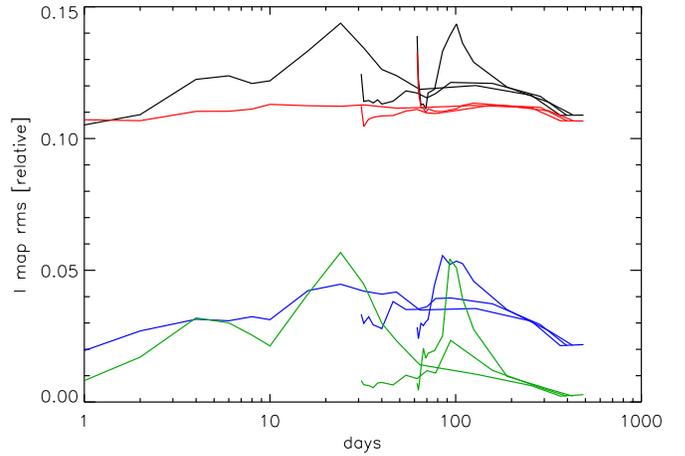}
    \vspace{-0.2cm}
    \caption{
The \emph{top} panel of Fig.~\ref{fig:survey_length_comp} replotted
so that the horizontal axis corresponds to the distance of the end
of the included data from the beginning of the simulated mission, so
only for the case of the first starting point does it correspond to
the length of the data used.
    }
    \label{fig:survey_length}
\end{center}
\end{figure}

To see the second effect, we consider the ratio of the residual map
rms to the rms of the binned white noise map from the same data.
This tells us how well we are doing compared to the white noise
level. See Fig.~\ref{fig:survey_length_comp}.  We see that for all
three starting points, the relative map quality consistently
improves after 45 days with further increase in the survey length up
to a full year. The relative quality (with respect to binned white
noise) of the 16-month map is the same as the 12-month map.

For shorter survey pieces the behavior as a function of survey
length depends on the starting point.  For $Q$ and $U$, short
segments, 10 days or less, are clearly much worse than longer ones.
For the $I$ map,  the relative quality in some cases worsens with
time up to about 40 days. This is related to the changing pattern of
crossing points in the cycloidal strategy. For some periods of time
the pattern is more ill-conditioned and adding data from such a
period to another short period makes things worse rather than
better.  From Fig.~\ref{fig:survey_length} we can conclude that one
such period is somewhere between days 1 and 24 and another somewhere
between days 70 and 101 of the simulated mission, since the residual
$I$ map rms is increasing during those periods. These coincide with
the times when the cycloidal scanning strategy is producing a
clustering of crossing points of nearby scanning rings at the
corners of caustics around ecliptic poles. The baselines can be
better solved from a more widely distributed set of crossing points
(Wright \cite{Wri96}). We can see from
Fig.~\ref{fig:survey_length_comp}, that the unmodeled $1/f$ and
reference baseline contributions stay relatively constant, so the
variation indeed comes from baseline error.

\section{Conclusions}
\label{sec:conclusion}

We have described our destriping map-making method (Polar/Madam) for
CMB surveys. The method has a parameter, the baseline length
$\tbase$, that affects the performance of the method. With long
baselines, the Madam code is faster and requires less computer
memory. The computer time and memory requirements of the code are
discussed in Ashdown et al.~(\cite{Trieste}), where it is compared
to other codes and methods.

Here we have done a detailed analysis of the residual errors in maps
produced with the method. In this paper we concentrated on
destriping without a noise prior. For short baselines the results
can be improved by utilizing prior information on the noise
spectrum.  This will be described in (Keih\"{a}nen et
al.~\cite{Madam09}).

We have divided the destriping residuals into six components. Three
of them, white noise reference baselines, unmodeled $1/f$ noise, and
pixelization noise reference baselines, are easy to estimate
analytically from the noise power spectrum and the signal angular
power spectrum.  In the map domain the baseline components appear as
a superposition of thin constant stripes of length
$\theta_\mathrm{base}$, whereas the unmodeled $1/f$ noise varies
along such stripes with a period comparable to
$\theta_\mathrm{base}$.

The three other components are related to how accurately baselines
can be solved from crossing points, and depend on the scanning
strategy. These baseline error components are minimized when there
are very many crossing points widely distributed. Especially when
making partial sky maps from short survey segments around a time
when ring crossings cluster in the same small region of the sky
these errors may blow up.  Since the baseline errors are correlated
over long time scales, they produce wide bands in the map domain and
are important for low multipoles.

The relevance of this analysis is that it guides us in what kind of
noise residuals to expect in the maps for given detector noise
spectra, and what baseline length to use in destriping map-making.

For long surveys, with a good distribution of crossing points, the
dominant residual error components are the white noise baselines and
unmodeled $1/f$ noise. Their combined effect can be minimized, when
the baseline length is chosen according to Eqs.~(\ref{opt_tbase}) or
(\ref{opt_tbase_1}), which put it close to $\tbase \approx
1/(2\fk)$, where $\fk$ is the knee frequency of the noise.  Because
of the other error components, one should choose a somewhat longer
baseline than this. (When a noise prior is used, shorter baselines
are better.)

For a {\sc Planck}-like scanning strategy, where the same circle of
the sky is observed many times, the difference between baseline
lengths from the spin period to the repointing period is mainly due
to nutation, and is small when the nutation is small compared to the
map pixel size.  If the knee frequency is comparable to, or smaller
than, the spin frequency, then the baseline length should be chosen
from this range, if no noise prior is used.  For a higher knee
frequency the residual errors are larger, and a shorter baseline is
better.

\begin{acknowledgements}
The work reported in this paper was done by the CTP Working Group of
the {\sc Planck} Consortia. {\sc Planck} is a mission of the
European Space Agency.  We thank K.~G\'{o}rski, C.R.~Lawrence, and
J.P.~Leahy for useful comments. This work was supported by the
Academy of Finland grants 205800, 213984, 214598, 121703, and
121962.  We acknowledge the support by the ASI contracts ``{\sc
Planck} LFI Activity of Phase E2'' and I/016/07/0 ``COFIS''. RK is
supported by the Jenny and Antti Wihuri Foundation. HKS thanks
Waldemar von Frenckells stiftelse, HKS and TP thank the Magnus
Ehrnrooth Foundation, and EK and TP thank the V\"{a}is\"{a}l\"{a}
Foundation for financial support. This work was supported by the
European Union through the Marie Curie Research and Training Network
``UniverseNet'' (MRTN-CT-2006-035863). We thank CSC (Finland) for
computational resources. We acknowledge use of the {\sc CAMB} code
for the computation of the theoretical CMB angular power spectrum.
This work has made use of the {\sc Planck} satellite simulation
package (level S), which is assembled by the Max Planck Institute
for Astrophysics {\sc Planck} Analysis Centre (MPAC). Some of the
results in this paper have been derived using the HEALPix package
(G\'orski et al. \cite{Gorski05}).

\end{acknowledgements}


\clearpage

\end{document}